\documentclass[%
reprint,%
 amssymb, amsmath,%
 aip,jcp,%
groupedaddress,%
]{revtex4-1}

\usepackage[colorlinks=true,linkcolor=blue]{hyperref}%
\usepackage{graphicx}
\usepackage{dcolumn}
\usepackage{bm}
\usepackage{amsbsy}
\usepackage{natbib}

\expandafter\ifx\csname package@font\endcsname\relax\else
 \expandafter\expandafter
 \expandafter\usepackage
 \expandafter\expandafter
 \expandafter{\csname package@font\endcsname}%
\fi
\newcommand{\lla}{\left\langle}
\newcommand{\rra}{\right\rangle}

\newcommand{\vphi}{\varphi}
\newcommand{\vtheta}{\vartheta}

\newcommand{\bs}[1]{\bm{#1}}

\DeclareRobustCommand{\ocite}[1]{\hspace{-1 ex} \nocite{#1}\citenum{#1}}

\let\Gamma\varGamma
\let\Delta\varDelta
\let\Theta\varTheta
\let\Lambda\varLambda
\let\Xi\varXi
\let\Pi\varPi
\let\Sigma\varSigma
\let\Upsilon\varUpsilon
\let\Phi\varPhi
\let\Psi\varPsi
\let\Omega\varOmega

\graphicspath{{./figures/}}

\begin{document}


\title{The physics of active polymers and filaments}

\author{Roland G. Winkler and Gerhard Gompper}
 \altaffiliation{Theoretical Physics of Living Matter, Institute of Biological Information Processing and Institute for Advanced Simulation, Forschungszentrum J\"ulich, D-52425 J\"ulich, Germany}
 \email{r.winkler@fz-juelich.de, g.gompper@fz-juelich.de}

\date{\today}

\begin{abstract}
Active matter agents consume internal energy or extract energy from the environment for locomotion and  force generation. Already  rather generic models, such as ensembles of active Brownian particles, exhibit phenomena, which are absent at equilibrium, in particular motility-induced phase separation and collective motion. Further  intriguing nonequilibrium effects emerge in assemblies of bound active agents as in linear polymers or filaments. The interplay of activity and conformational degrees of freedom gives rise to novel structural and dynamical features of individual polymers as well as in interacting ensembles. Such out-of-equilibrium polymers  are an integral part of living matter, ranging from  biological cells with filaments propelled by motor proteins in the cytoskeleton, and  RNA/DNA in the transcription process, to long swarming bacteria and worms such as  {\em Proteus mirabilis} and {\em Caenorhabditis elegans}, respectively.  
Even artificial active polymers have been synthesized. The emergent properties of active polymers or filaments depend on the coupling of the active process to their conformational degrees of freedom, aspects which are addressed in  this article. The theoretical models for tangentially and isotropically self-propelled or active-bath  driven polymers are presented, both in presence and absence of hydrodynamic interactions. The consequences for their conformational and dynamical properties are examined, emphasizing  the strong influence of the coupling between activity and hydrodynamic interactions.  Particular features of  emerging phenomena,  induced by steric and hydrodynamic interactions, are highlighted.  Various  important, yet theoretically unexplored, aspects are featured and future challenges are discussed.   
\end{abstract}

\maketitle

\section{Introduction} \label{sec:introduction}

Living matter is characterized by a multitude of complex dynamical processes maintaining its out-of-equilibrium nature.\cite{demi:10,fang:19} Molecular machines such as (motor) proteins and  ribosomes undergo conformational changes fueled by Adenosine Triphosphate (ATP), which drive and stir the cell interior. \cite{kapr:13,kapr:16} This triggers a hierarchy of dynamical processes, movements, and transport, resulting in a nonequilibrium state of the  cell---from the molecular to the whole-cell level---, \cite{fang:19}  with intriguing  collective phenomena emerging by migration and  locomotion also on scales much larger than individual cells.\cite{marc:13, elge:15,juli:18,haki:17,beer:19} The nature of living and active matter systems implies nonthermal fluctuations, broken detailed balance, and a violation of the dissipation-fluctuation relation, which renders their theoretical description particularly challenging.\cite{fang:19} 

Filaments and polymers are an integral part of biological systems, and their conformational and dynamical properties are substantially affected by the active processes to a yet unresolved extent.  

{\em Biological active polymers and filaments} --- Enzymatic conformational changes are considered to induce fluctuating hydrodynamic flows in the cytoplasm, which lead to an enhanced diffusion of dissolved colloidal and polymeric objects.\cite{casp:00,lau:03,bran:08,robe:10,fakh:14,guo:14,parr:14,gole:15,mikh:15,webe:12,kapr:16,gnes:18,wu:19}
In addition, kinesin motors walking along microtubule filaments  generate  forces that affect the dynamics of the cytoskeletal network, the transport properties of species in the cell, and the organization of the cell interior.\cite{lau:03,mack:08,lu:16,ravi:17} Even more, molecular motors give rise to nonequilibrium conformational fluctuations of actin filaments and microtubules. \cite{bran:08,webe:15}

 Within the nucleus, 
ATPases such as DNA or RNA polymerase (RNAP, DNAP) are involved in DNA transcription, where the information coded in the base-pair sequence of DNA is transcribed into DNA or RNA. This proceeds in several steps, where the ATPases locally unzip the two DNA strands, nucleotides are added to the synthesized molecules, and the ATPases move along the DNA.\cite{guth:99} Hence, every RNAP/DNAP translocation step is a complex process, which generates nonthermal fluctuations for both, RNAP/DNAP  as well as the transcribed DNA.\cite{guth:99,meji:15,beli:19}

Various of these active process are considered to  be involved in the spatial arrangement of the eukaryotic genome, controlling their dynamical properties, and being essential for the cell function.\cite{dipi:18} In particular, ATP-dependent processes affect the dynamics of chromosomal loci\cite{webe:12,jave:13} and chromatin.\cite{zido:13} Moreover, spatial segregation of active (euchromatin) and passive (heterochromatin) chromatin has been found. \cite{lieb:09,crem:15,solo:16} The detailed mechanism needs to be unraveled, possibly taking active processes into account.\cite{gana:14,smre:17}

Microswimmers are a particular class of active matter, where individual agents move autonomously by either converting internal chemical energy into directed motion, or utilizing energy from the environment.\cite{laug:09,rama:10,marc:13,elge:15,bech:16,gomp:20} Biological microswimmers, like  algae, sperm, and bacteria, are omnipresent. \cite{berg:04,elge:15,rama:10}
Numerous biological microswimmers are rather elongated and  polymer- or filament-like, undergoing shape changes during migration. A particular example are elongated swarming bacteria, propelled by flagella, such as {\em Proteus mirabilis}, {\em Vibrio parahaemolyticus},\cite{weib:14,auer:19} or {\em Serratia marcescens},\cite{li:19} which exhibit an intriguing collective behavior. Other microswimmers organize in chain-like structures, for example planktonic dinoflagellates \cite{sohn:11,sela:11} or {\em Bacillus subtilis} bacteria during biofilm formation.\cite{yama:19}  On a more macroscopic scale, nematodes such as {\em Caenorhabditis elegans} swim\cite{gagn:16} and collectively organize into dynamical networks.\cite{sugi:19} 

Similarly, biological polar filaments like actin and microtubules driven by molecular motors exhibit filament bundling and the emergence of active turbulence, which is characterized by high vorticity and presence of motile topological defects.\cite{nedl:97,howa:01,krus:04,baus:06,juel:07,hara:87,scha:10,rama:10,sanc:12,sumi:12,marc:13,pros:15,ravi:17,doos:18}

These diverse  aspects illustrate the relevance of activity for the function of cells and other biological polymer-like active objects. However,  a cell is a very complex system with a multitude of concurrent processes, hence, a classification or the separation of individual active processes is difficult. Here, a systematic study of emergent phenomena due to activity by synthetic model systems may be useful. More importantly, the understanding of active processes and their suitable implementation may be essential in the rational design of synthetic cells. \cite{blai:14,stan:18,gopf:18}

{\em Synthetic active polymers} --- Synthetic active or activated colloidal molecules\cite{loew:18} or polymers\cite{wink:17} are nowadays obtained in several ways.\cite{gomp:16,niu:18} Various concepts have been put forward for the design of synthetic active particles, which can serve as monomers. Typically, propulsion is based on phoretic effects, relying on local gradients of, e.g.,  electric fields (electrophoresis), concentration (diffusiophoresis),  and temperature (thermophoresis).\cite{hows:07,jian:10,vala:10,wurg:10,volp:11,thut:11,butt:13,hage:14,bech:16,maas:16}
Synthetic active  systems exhibit a wide spectrum of novel and fascinating phenomena, specifically activity-driven phase separation or large-scale collective motion. \cite{laug:09,rama:10,vics:12,roma:12,marc:13,elge:15,bech:16,marc:16.1,zoet:16}

Catalytic Janus particles have been shown to spontaneously self-assembly into  autonomously  swimming dimers  with a wide variety of morphologies.\cite{ebbe:10,gibb:17} Assembly of metal–dielectric Janus colloids (monomers) into  active chains  can be achieved by imbalanced interactions, where  the motility and the colloid interactions are simultaneously controlled by an AC electric field.\cite{yan:16,dile:16,zhan:16,zhan:16.1,nish:18,mart:19} Through the application of strong AC electric fields, linear chains of bound Janus particles can be achieved either by van-der-Waals forces or polymer linkers.\cite{vutu:17} Electrohydrodynamic convection rolls lead to self-assembled colloidal chains in a nematic liquid crystal matrix and directed movement.\cite{sasa:14} Linear self-assembly of dielectric colloidal particles is achieved by alternating magnetic fields, where the chain length can be controlled by the external field.\cite{mart:15,koko:17}
Moreover, chains of linked colloids, which are uniformly coated with catalytic nanoparticles, have been synthesizes.\cite{bisw:17} Hydrogen peroxide decomposition on the surfaces of the colloidal monomers generates phoretic flows, and activity induced hydrodynamic interactions between monomers results in an enhanced diffusive motion.\cite{bisw:17}

{\em Objectives} ---Filamentous and polymeric structures play a major role in biological  systems and are heavily involved in nonequilibrium processes. However, we are far from understanding  the interplay between out-of-equilibrium fluctuations,  corresponding polymer conformations,  and  emerging activity-driven self-organized structures.  Studies of molecular/polymeric active matter will reveal original physical phenomena, and promote the development of novel smart devices and materials.

In this perspective article, we  address the distinctive features emerging from the coupling of activity and conformational degrees of freedom of filamentous and polymeric structures.  
Assemblies of active colloidal particles have been denoted as {\em active colloidal molecules}.\cite{loew:18} Since we focus not only on synthetic colloidal systems, but also on polymers and filaments in biological systems, we will use the notion {\em active polymers}.\cite{wink:17}

Different concepts for the active dynamics of  polymers can be imagined or realized. For polymers composed of linearly connected monomers, the latter can be assumed to be themselves  active, i.e., self-propelling, or to be activated, i.e., externally driven by a nonthermal force. Self-propulsion typically emerges by the interaction of a microswimmer with the embedding  fluid. Here, hydrodynamics plays a major role and momentum is conserved. This is denoted as wet active matter. In contrast, dry active matter is characterized by the absence of local momentum conservation and, hence,  hydrodynamic interactions. Here, other effects, such as strong particle-particle interactions or contact with  a momentum-absorbing medium, as in bacteria gliding or granular beads vibrating on frictional surfaces, dominate over fluid-mediated interactions.\cite{shae:20} Moreover, externally-driven polymers are  subject to forces, which, when they are dissolved in a fluid, give rise to Stokeslet flows and  hydrodynamic interactions.   
Depending on the nature of the polymer environment, the emerging properties by the coupling of activity and the internal degrees of freedom can be rather different. \cite{eise:17,mart:19,mart:20} Moreover, the nature of the active force determines the driving of a  monomer. The forces on filaments driven by molecular motors are typically assumed to push them along the local filament tangent.\cite{elge:15,need:17,doos:18} Another realization are spatially independent, but time-correlated,  active forces on monomers. We will highlight the impact of the various aspects---self-propelled vs. actuated, dry vs. wet, tangentially vs. randomly driven---on the polymer conformations, dynamics, and collective effects.

The article is organized as follows. Section~\ref{sec:forces} describes realizations of dry and wet active forces on the level of monomers and bonds. The properties of dry polymers composed of active Brownian particles are discussed in Sec.~\ref{sec:dry_abpo}. Effects of hydrodynamic interactions are included in Sec.~\ref{sec:wet_abpo}. Section~\ref{sec:driven_filaments} is devoted to  properties of tangentially driven dry polymers, and Sec.~\ref{sec:self_propelled_colloid}  presents aspects of  hydrodynamically self-propelled polymers. Collective effects are discussed in Sec.~\ref{sec:collective}.
 Finally, Sec.~\ref{sec:summary} summarizes the major aspects and indicates  possible future research directions.

\section{Active forces} \label{sec:forces}

 In models of active polymers,  forces can directly be assigned to individual monomers, or to the bond connecting two monomers. This defines the minimal active units---either a monomer or a dumbbell of two connected monomers.  
 
\subsection{Active Brownian particles} \label{sec:abp}

A generic model of a self-propelled particle, especially suitable for dry active matter, is the well-know  active Brownian particle (ABP).\cite{hows:07,peru:10,roma:12,fily:12,bial:12,redn:13,elge:15,hage:15,marc:16.1,bech:16,loew:20} Its overdamped equations of motion  are given by
\begin{align} \label{eq:abp_trans}
\dot {\bm r}(t) = & \  \frac{1}{\gamma_T}  \bm F^a  + \frac{1}{\gamma_T} \left(\bm F(t) + \bm{\varGamma} (t) \right) , \\ \label{eq:abp_rot}
\dot {\bm e} (t) = & \ \bm \varTheta (t) \times \bm e(t)
\end{align}
for the translational and rotational motion of the position $\bm r$ and the orientation $\bm e$ in two or three dimensions (2D, 3D). The active force is $\bm F^a =\gamma_T v_0 \bm e$, with  $v_0$ the self-propulsion velocity and $\gamma_T$ the translational friction coefficient; $\bm F$ accounts for all non-active external forces, $\bm \varGamma$ is the translational thermal noise, and $\bm \varTheta$ the rotational noise. The latter are Gaussian and Markovian random processes with zero mean and the second moments
\begin{align}
\lla {\varGamma}_{\alpha}(t)  {\varGamma}_{\beta}(t') \rra = & \ 2 k_B  T \gamma_T \delta_{\alpha \beta} \delta(t-t') , \\
\lla \varTheta_{\alpha}(t)  \varTheta_{\beta}(t') \rra = & \ 2 D_R \delta_{\alpha \beta} \delta(t-t') ,
\end{align}
with $k_B$ the Boltzmann factor, $T$ the temperature, $D_R$ the rotational diffusion coefficient, and $\alpha, \beta \in \{x,y,z\}$. Equation~\eqref{eq:abp_rot} yields the correlation function
\begin{align} \label{eq:corr_e}
\lla \bm e (t) \cdot \bm e(0) \rra = e^{- (d-1)D_R t} 
\end{align}
in $d$ dimensions.
Hence, the particle [Eq.~\eqref{eq:abp_trans}] can be considered as exposed to thermal and colored noise with the correlation function \eqref{eq:corr_e}.

\begin{figure}[t]
	\begin{center}
		\includegraphics[width=\columnwidth]{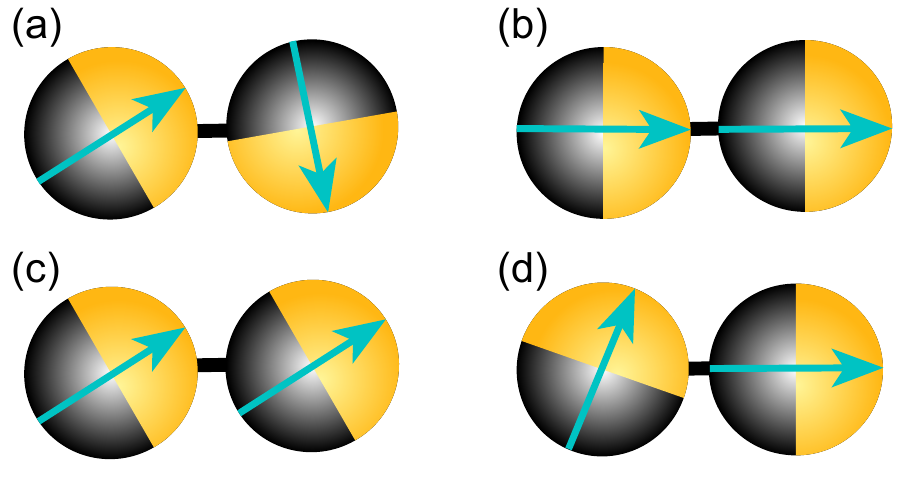}
	\end{center}\caption{Illustration of possible combinations of propulsion directions of active dumbbells. (a) The two ABPs rotate independently, \cite{wink:16,sieb:17} (b) the beads are propelled along the bond,\cite{suma:14} (c) correlated propulsion in a common direction oblique to the bond, where the propulsion direction may change in a diffusive manner,\cite{sieb:17} and (d) propulsion in non-parallel directions with fixed angles.\cite{nour:16} } \label{fig:sketch_dumbbell}
\end{figure}

\subsection{Active dumbbells and rods} \label{sec:dumbbell}

The combination of active monomers into linear assemblies provides a wide spectrum of possible combinations of active forces. This is illustrated  for two bound active particles forming a dumbbell. A bond, not restricting the orientational motion of two ABPs, leads to an active Brownian dumbbell. The dynamical properties and the phase behavior of such dumbbells have been studied.\cite{wink:16,sieb:17} Since  active dumbbells are a special case of active Brownian polymers, we refer to the general discussion is Sec.~\ref{sec:dry_abpo}  for  results. Specific features by  hydrodynamic interactions are addressed in Sec.~\ref{sec:dumbbell_self_prop}.  

Another extreme case is propulsion along the bond vector only. The equations of motion of the monomers for a such a dumbbell are ($i \in \{1,2\}$) \cite{suma:14}
\begin{align} \label{eq:dumb_tan}
\gamma_T \dot{\bm r}_i = \bm F^a(t) + \bm F_i(t) + \bm{\varGamma}_i (t),
\end{align}
with the active force $\bm F^a = F^a \bm u$ on every monomer  in the direction of the (unit) bond vector $\bm u = (\bm r_2 - \bm r_1)/|\bm r_2 - \bm r_1|$. Note that $\bm F_i$ contains the bond force. 
In the absence of any external force, thermal noise leads to a rotational diffusive motion of $\bm u$.\cite{cugl:15}  A generalization allows the active force to vary within a certain angle with respect to the bond vector.\cite{bian:18}
Oblique arrangement of the propulsion direction  leads to spiral trajectories in two dimensions, as has been demonstrated for Janus particles.\cite{nour:16}  Naturally many other options are  possible, and may, in modeling, be chosen according to experimental needs.

As indicated, the particular propulsion mechanism leads to a distinct dynamics and also effects on the emergent collective behavior can be expect, e.g., motility-induced phase separation (MIPS). In fact,
MIPS has been found for dumbbells propelled along their bond\cite{suma:14} as well as those propelled oblique with respect to the bond,\cite{sieb:17}  where spontaneously formed aggregates break chiral symmetry and rotate.\cite{suma:14} Such a rotation  is also obtained for systems of self-propelled particles\cite{yang:14.3} or rods\cite{kais:13} moving in two dimensions  on circles as well as for bend rigid self-propelled filaments.\cite{denk:16}

We focus on uniform systems of active particles of the same size, not addressing effects appearing in systems of dumbbells with different radii with correspondingly asymmetric flow fields.\cite{tao:08.1,vala:10,mich:14,wagn:17}  


\subsection{Self-propelled  force-free and torque-free  monomers in fluids} \label{sec:squirmer}

Synthetic and biological microswimmers are typically immersed in a fluid and hydrodynamic interactions (HI) are an integral part of their propulsion. Self-propelled particles move autonomously, with no external force or torque applied, and hence the total force/torque of the swimmer on the fluid and vice versa vanishes.\cite{elge:15,wink:18} The flow field generated by the microswimmer can be represented in terms of a multipole expansion.\cite{kim:91,pozr:92,spag:12,lask:15,math:16.2} The far field is dominated  by the force dipole (FD), source dipole (SD), force quadrupole (FQ), source quadrupole (SQ), and rotlet dipole (RD) contributions.\cite{spag:12,wink:18a}

In  theory and simulations, such an expansion can be exploited to calculate the flow field of  individual swimmers. Alternatively, the squirmer model can be applied, which was originally designed to model ciliated microswimmers. \cite{ligh:52,blak:71,ishi:06}  It is a rather generic model, which captures the essential swimming aspects and is nowadays applied to a broad class of microswimmers,\cite{ishi:06,llop:10,goet:10,moli:13,zoet:14,thee:16.1,yosh:17,thee:18} ranging from  diffusiophoretic particles to biological cells.\cite{pope:18}

A linear arrangement of such monomers leads to an intricate coupling of their flow fields and novel emergent conformational and dynamical features.\cite{jaya:12,lask:13,lask:15}

\subsubsection{Squirmers} \label{sec:squirmer_model}

A squirmer is  an axisymmetric rigid colloid with a prescribed surface fluid (slip) velocity. \cite{blak:71,ligh:52,ishi:06,pedl:16}
For a purely tangential fluid displacement,  the surface slip velocity of a sphere can be described as
\begin{align}
\bm v^{sq} = \sum_{n=1}^{\infty} \frac{2}{n(n+1)} B_n \sin \vartheta  P_{n}'(\cos \vartheta) \bm e_{\vartheta} 
\end{align} 
in terms of derivatives of the $n$-th order Legendre polynomial, $P_n(\cos \vartheta)$. Here,  $\vartheta$ is the angle between the body-fixed propulsion direction $\bm e$ and the considered point on the colloid surface with  tangent vector $\bm e_\vtheta$, and  $B_n$ is the amplitude  of the respective mode.  Typically  only two modes are considered, i.e., $B_n=0$ for $n \ge 3$.  \cite{ishi:06,llop:10,thee:16.1} Explicitly, the leading contributions yield the slip velocity\cite{ishi:06,llop:10,thee:16.1}
\begin{align} \label{eq:squirmer_velo}
\bm v^{sq}  =B _1 \sin \vtheta ( 1 +\beta  \cos \vtheta) \bm e_{\vtheta}. 
\end{align}
The parameter $B_1 = 2v_0/3$ determines the swimming velocity  and $\beta  = B_2/B_1$ characterizes the nature of the swimmer, namely a  pusher ($\beta < 0$), puller ($\beta  > 0$), and neutral squirmer ($\beta  = 0$), corresponding, e.g., to {\em E. coli}, {\em Chlamydomonas}, and {\em Volvox}, respectively.  The far field of a squirmer is well described by the flow fields of a force dipole (FD), a source dipole (SD), and a source quadrupole (SQ). \cite{thee:16.1,wink:18a}  Various extensions to spheroidal squirmers have been proposed, \cite{kell:77,ishi:13,thee:16.1} where some allow for an analytical calculation of the flow field.\cite{thee:16.1} 

The propulsion direction is not affected by the squirmer flow field. In case of several squirmers, but in absence of thermal fluctuations, interference of their flow fields leads to particular hydrodynamic collective effects and structure formation. \cite{alar:13,yosh:17,alar:17} In simulation approaches, which account for thermal fluctuations, often different structures are observed, and in dilute solution the propulsion orientational correlation function agrees with Eq.~\eqref{eq:corr_e} with the diffusion coefficient determined by the fluid viscosity.\cite{thee:16.1} 

\subsubsection{Non-axisymmetric swimmers} \label{sec:swimmer_expansion}

For a more general description and an extension to non-axisymmetric swimmers,  the swimmer  equations of motion can be expressed in terms of mobilities. \cite{sing:15,feld:94,pak:14,lask:15,sing:18.1}   Such an extension captures the many-body nature of the surface-force density in a suspension of many colloids,\cite{sing:18.1} whereas the original Stokes law applies to infinite dilution only. The center-of-mass translational velocity and the rotational frequency of a single force- and torque-free spherical particle in an unbounded fluid are
\begin{align}
\dot {\bm r} = \bm v^a(t)  , \hspace*{1cm} \dot{\bm \omega} = \bm \Omega^a(t) ,
\end{align}
where the active velocity and active torque are given by 
\begin{align}
\bm v^a = & \ - \frac{1}{4 \pi R^2} \oint \bm v^{s}(\bm r) d^2r ,  \\
 \bm \Omega^a = \ &  - \frac{3}{8 \pi R^2} \oint \bm r \times  \bm v^{s}(\bm r) d^2r ,
\end{align}
with the surface slip velocity $\bm v^s$, e.g., the squirmer velocity of Eq.~\eqref{eq:squirmer_velo}.\cite{sing:18.1,sing:15} More general expressions for ellipsoidal particles are presented in Ref.~\ocite{fair:89}. This is a rather obvious result, but it shows that a colloid  translates and rotates independently in response to the active surface velocity. 
The case of linearly connected active spheres is discussed in Sec.~\ref{sec:filament_self_prop}. 

\section{Dry Active Brownian Polymers (D-ABPO)} \label{sec:dry_abpo}

The  conformational and dynamical properties of dry (free-draining) active Brownian polymers, which will be denoted as D-ABPO, are typically studied analytically by  the well-known  Rouse model\cite{doi:86} of equilibrium polymer physics.\cite{vand:15,osma:17,sama:16,chak:19} However, the deficiencies of the model, in particular the extensibility of  bonds in the standard formulation, leads to inadequate  predictions of activity effects. In contrast, valuable quantitative predictions have been  obtained by computer simulations \cite{hard:14,sark:14,ghos:14,shin:15,eise:16,eise:16,mart:19} and more adequate analytical models, which will be introduced in the following.\cite{ghos:14,eise:16,eise:16,mous:19,mart:19}   

\subsection{Discrete model of active polymers}

Dry semiflexible active polymers  can be modeled as a linear chain of  $N_m$ 
linked active Brownian particles, with their dynamics described by the overdamped equations of motion \eqref{eq:abp_trans}  and \eqref{eq:abp_rot}. The force $\bm F_i$ on particle $i$ ($i=1,\ldots,N_m$)  includes bond, bending, and excluded-volume contributions.\cite{eise:16,eise:17,mart:19} The two independent parameters $v_0$ and $D_R$, characterizing activity, are combined in the dimensionless quantities \cite{eise:16,das:18.1,mart:19}
\begin{align} \label{eq:peclet}
 Pe = \frac{v_0}{l D_R} , \hspace*{5mm}   \Delta = \frac{D_T}{d_H^2 D_R}  ,
\end{align}
with $l$ the equilibrium bond length.
The P\'eclet number $Pe$ compares the time for the reorientation of an ABP monomer with that for its translation with velocity $v_0$ over the monomer radius, and $\Delta$ is the  ratio 
between the translational, $D_T=k_BT/3\pi \eta d_H$,  and rotational, $D_R$, diffusion coefficient of an individual monomer,
where $d_H$ denotes the monomer hydrodynamic diameter.\cite{eise:16,eise:17,mart:19} For a tangent hard-sphere-type polymer,  $d_H=l$ and $\Delta = 1/3$.   In order to  avoid artifacts in the polymer structures by activity, the force constant for the bond, $\kappa_{l}$, and that of the excluded-volume Lennard-Jones potential, $\varepsilon$, can be adjusted according to   $\kappa_{l} l^2 /k_BT = (10  +  2 Pe ) \times 10^3$ and  $\varepsilon /(k_BTPe) = 1$ as a function of $Pe$.   This  ensures a finite bond length within $3\%$  of the equilibrium value for a harmonic bond potential, and  an activity-independent overlap between monomers.\cite{eise:16,eise:17,mart:19}

\subsection{Continuum model of active polymers} \label{sec:abp_cont}

An analytical description of  D-ABPO properties is achieved by  a mean-field model for semiflexible polymers\cite{bawe:85,batt:87,lang:91,wink:94,ha:95,wink:03,harn:95,wink:10,wink:06.1}  augmented by the  active velocity $\bm v(s,t)$, which yields the Langevin equation 
\cite{eise:16,eise:17.1,mous:19}
\begin{align} \label{eq:equation_motion}
& \frac{\partial}{\partial t}{ \bm r}(s,t) = \bm v(s,t) \\ \nonumber
 & + \frac{1}{\gamma} \left( 2 \lambda k_B T \frac{\partial^2}{\partial s^2} { \bm r}(s,t)  - \epsilon k_B T
\frac{\partial^4} {\partial s^4} { \bm r}(s,t) + {\bm \Gamma} (s,t) \right) \:,
\end{align}
with suitable boundary conditions for the free ends of linear polymers  \cite{harn:95,wink:10,wink:06.1,eise:16} or periodic boundary conditions for ring polymers.\cite{mous:19}  Here, $\gamma$ is the translational friction coefficient per unit length and  $\epsilon =3/4p$  for  polymers in three dimensions,\cite{krat:49,arag:85,wink:07.1} where $p$ is related to the persistence length $l_p$ via $p=1/(2l_p)$. \cite{wink:92,wink:03} The terms with the second and fourth derivative in Eq.~(\ref{eq:equation_motion}) account for the entropic degrees of freedom and bending elasticity, respectively. The Lagrangian multiplier $\lambda$ is determined in a  mean-field manner by the global constraint of a finite contour length, $L$, \cite{eise:16,wink:92,wink:03,wink:06,wink:10}
\begin{align} \label{eq:constraint}
\int_{-L/2}^{L/2} \lla \left(\frac{\partial \bm r (s,t)}{\partial s}\right)^2 \rra d s = L ,
\end{align}

From Eq.~\eqref{eq:corr_e}, the correlation function of the velocity $\bm v(s,t)$ follows as
\begin{align} \label{eq:corr_colored}
\lla \bm v(s,t) \cdot \bm v(s',t')\rra = v_0^2 l e^{-\gamma_R(t-t')} \delta(s-s') \ ,
\end{align}
with $\gamma_R = 2 D_R$ the damping factor of the rotational motion. For the analytical solution, only the first and second moment of the distribution of the active velocity are needed.\cite{eise:16}

Note that the parameter $l$ in Eq.~\eqref{eq:corr_colored}  defines the number of active sites $L/l$ along the polymer. \cite{eise:16}

\subsection{Results}

\begin{figure}[t]
	\begin{center}
		\includegraphics[width=\columnwidth]{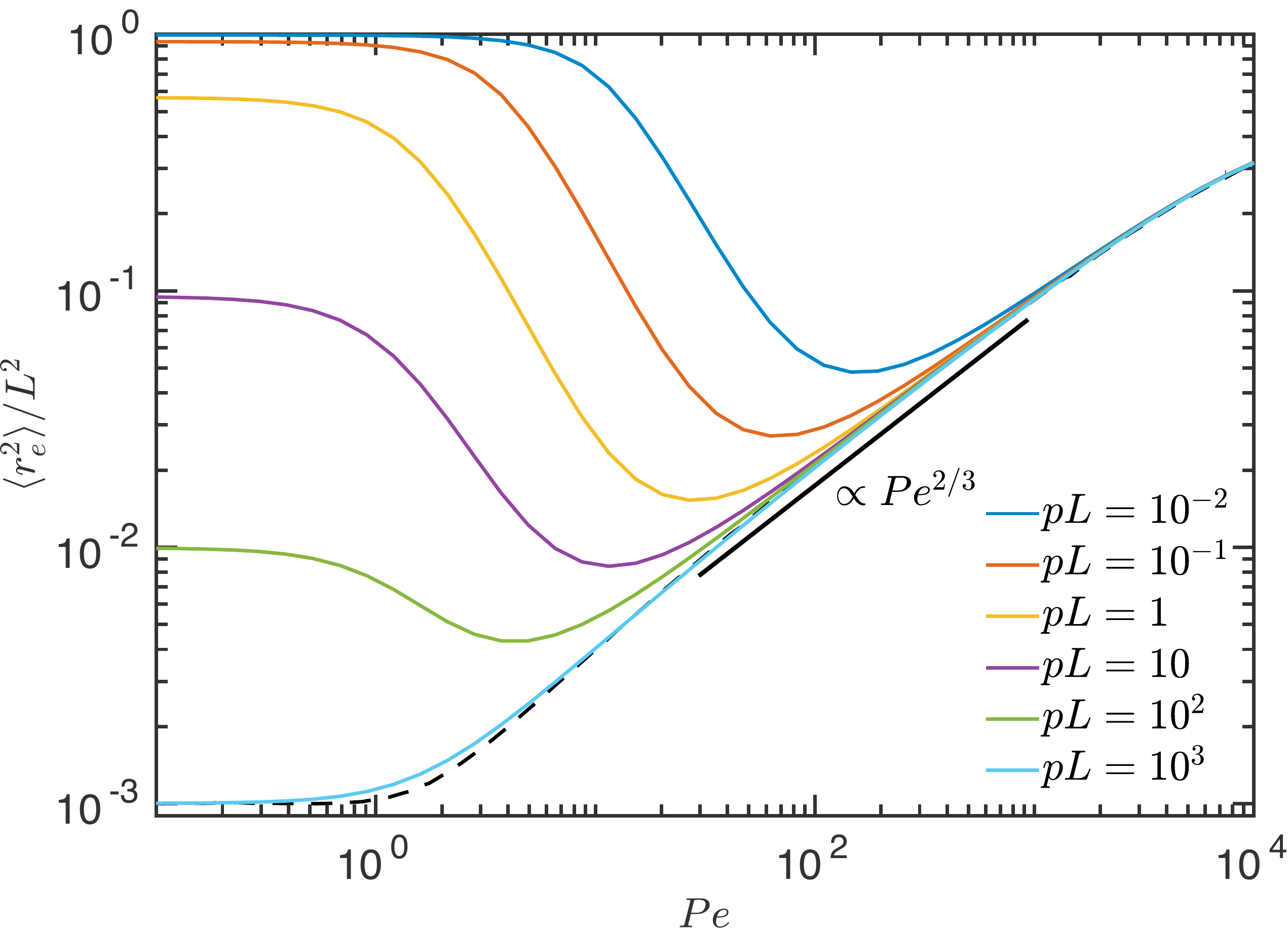}
	\end{center}\caption{Mean square end-to-end distances as a function of the Péclet number for semiflexible  D-ABPO with  $pL = L/2l_p = 10^3, \ 10^2, \ 10, \ 1, \ 10^{-1}, \ 10^{-2}$ (bottom to top at $Pe =10^{-1}$), and $\Delta=1/3$. \cite{eise:16}  The dashed line is the flexible polymer limit according to Eq.~\eqref{eq:end-to-end_flex}. ``T. Eisenstecken, G. Gompper, and R. G. Winkler, Polymers {\bf 8}, 304 (2016); licensed under a Creative Commons Attribution (CC BY) license.''
	} \label{fig:msqe_dry}
\end{figure}

\begin{figure}[t]
\begin{center}
\includegraphics[width=\columnwidth]{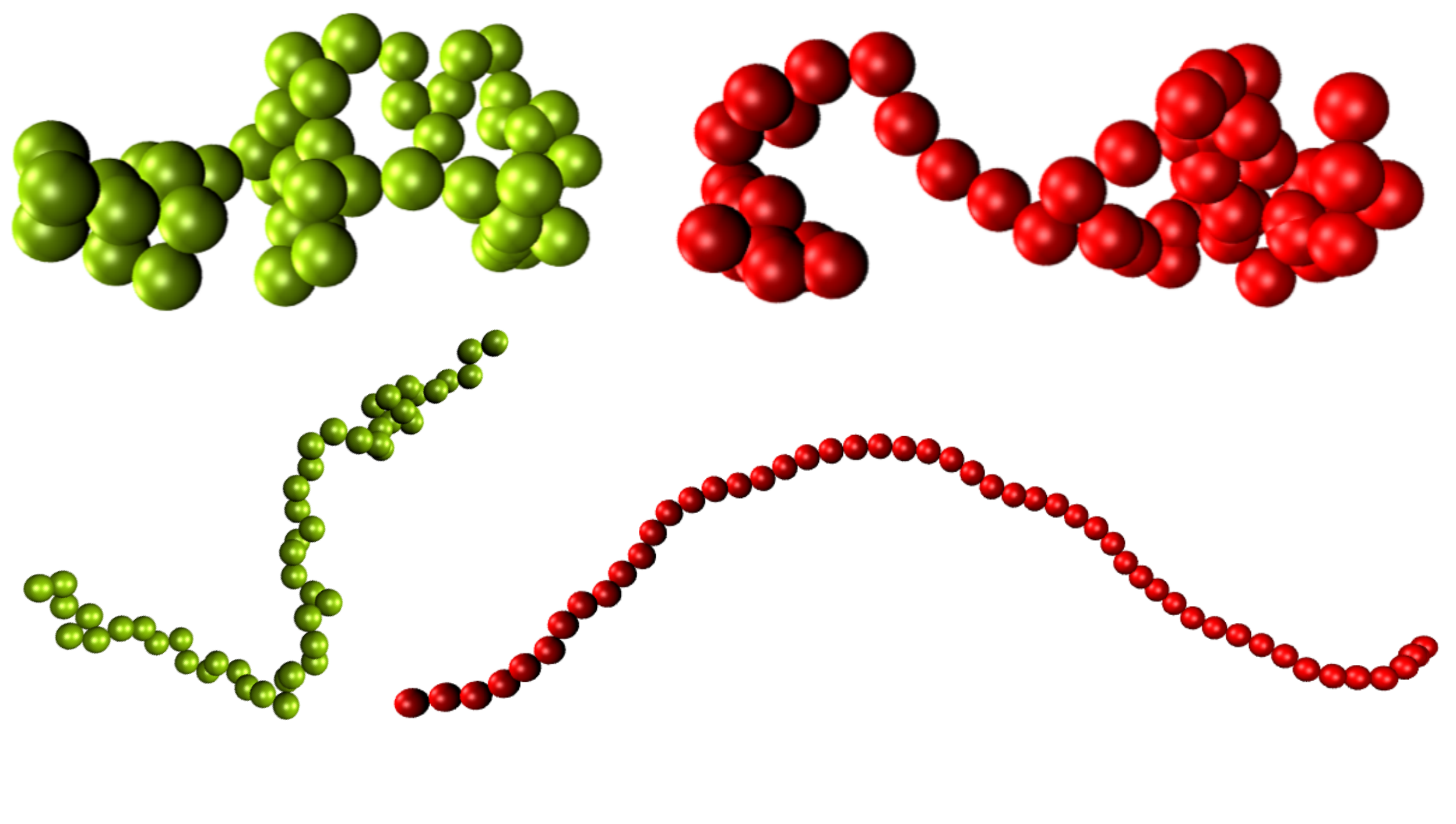}
\end{center}
\caption{Configurations of flexible phantom D-ABPO (red) and active polymers with self-propelled monomers in presence of HI (S-ABPO) (green) (cf. Sec.~\ref{sec:sabpo}) of length $N_m=50$ for the P\'eclet numbers $Pe=1$ (top) and $Pe =10^3$ (bottom).\cite{mart:19} ``Reproduced with permission from Soft Matter {\bf 12}, 8316 (2016). Copyright 2016 The Royal Society of Chemistry.''} \label{fig:snapshot_polymer}
\end{figure}

\subsubsection{Conformational properties}

The conformational properties of linear polymers are characterized by their mean square end-to-end distance\cite{eise:16}
\begin{align} \label{eq:end-to-end}
\lla \bm r_e^2\rra = 4 \sum_{n=1}^\infty \lla \bm \chi_{2n-1}^2 \rra \vphi_{2n-1}^2(L/2) ,
\end{align}
in terms of the eigenfunctions $\varphi_n$ of the differential operator on the rhs of Eq.~\eqref{eq:equation_motion} and the fluctuations of the  normal-mode amplitudes
\begin{align} \label{eq:mode_amplitude}
\lla \bs \chi_n^2 \rra = \frac{3 k_BT}{\gamma} \tau_n + \frac{v_0^2}{p(1+\gamma_R \tau_n)} \tau_n^2 \ .
\end{align}
The relaxation times $\tau_n$  as well as the eigenfunctions depend on the activity.
The closed expression
\begin{align} \label{eq:end-to-end_flex}
 \lla \bm r_e^2 \rra & \ =  \frac{L}{p \mu} \\ \nonumber  & \  + \frac{Pe^2 L}{6 p \mu \Delta} \left[
1 - \frac{\sqrt{1 + 6 \mu^2 \Delta }}{pL \sqrt{\mu}} \tanh \left( \frac{pL \sqrt{\mu}}{\sqrt{1 + 6  \mu^2 \Delta }} \right) \right]
\end{align}
is obtained in the limit of flexible polymers, $pL \gg 1$, with free ends, \cite{eise:16} with the relaxation times
\begin{align} \label{eq:tau_asym_flex}
\tau_n = \frac{\tau_R}{\mu n^2} \ ,
\end{align}
where $\tau_R = \gamma L^2/(3 \pi^2 k_B T p)$ is the  (passive) Rouse relaxation time. \cite{harn:95,doi:86}  The activity-depend factor $\mu$ accounts for the constraint \eqref{eq:constraint}. A detailed discussion of $\mu(Pe)$ for linear and ring polymers is presented in Refs.~\ocite{eise:17}, \ocite{eise:16}, and \ocite{mous:19}, respectively

The effect of activity on the conformational properties of active  polymers is illustrated in Fig.~\ref{fig:msqe_dry}. Figure~\ref{fig:snapshot_polymer} shows polymer  conformations for various P\'eclet numbers. For flexible polymers, activity causes polymer swelling with increasing $Pe$, which saturates at $L^2/2$ in the limit $pL \to \infty$ as a consequence of the finite contour length. The reason for the swelling is an increase of the persistence length $L_p$ of the active motion with increasing $Pe$, where  $L_p$ is defined as the distance displaced by activity with velocity $v_0$ in the time $1/2D_R$ of the decay of the correlation function \eqref{eq:corr_colored}, i.e.,  $L_p/l=Pe/2$. For $Pe \gg 1$, two groups of monomers moving in opposite directions will move far before  changing their direction substantially, which leads to stretching of the in-between part of the polymer. In contrast, semiflexible polymers shrink at weak $Pe$ and swell for large $Pe$ similarly to flexible polymers. Shrinkage is caused by enhanced fluctuations transverse to the polymer contour by activity. The bonds prevent fluctuations along the contour, which leads to an apparent softening of the semiflexible polymer.  At large P\'eclet numbers, tension in the polymer contour, which increases with activity, dominates over the energetic contribution of bending, so that the latter can be neglected and a semiflexible polymer appears flexible.
The comparison of the theoretical predictions of the polymer conformations with simulation results yields excellent agreement.\cite{eise:17.1}

In general, ring polymers exhibit similar features as linear polymers.\cite{mous:19}  Specifically,  active and thermal fluctuations attempt to shrink and crumple a ring-like structure.  However, this is opposed by a negative internal tension, whereas the tension in always positive for linear polymers.  In general, activity implies enhanced fluctuations of the normal-mode amplitudes.\cite{mous:19}   The fluctuation spectrum is dominated by activity already for moderate P\'eclet numbers ($Pe \gtrsim 10$)---the contribution with $v_0^2$ in Eq.~\eqref{eq:mode_amplitude}---and thermal fluctuations matter only for large mode numbers, i.e., at very small length scales. Notably, the major part of the spectrum is determined by tension, with a crossover from a $1/n^2$ to a $1/n^4$ power-law with increasing mode number (cf. Eq.~\eqref{eq:mode_amplitude} for $\tau_n \sim 1/n^2$). Hence, conclusions from the exponent on the underlying fluctuation mechanisms have to be drawn with care, and a $1/n^4$ dependence is not necessarily a sign of dominating bending modes. 

Qualitatively, the obtained  fluctuation spectrum for rings agrees with that of fluctuating membranes, where also an increase of the fluctuations at small wave vectors by activity compared to an equilibrium system has been observed experimentally,\cite{fari:09,turl:16,vutu:19,taka:20} in simulations,\cite{turl:16,vutu:19} and described theoretically.\cite{loub:12,taka:20} 

\begin{figure}[t]
	\begin{center}
		\includegraphics[width=\columnwidth]{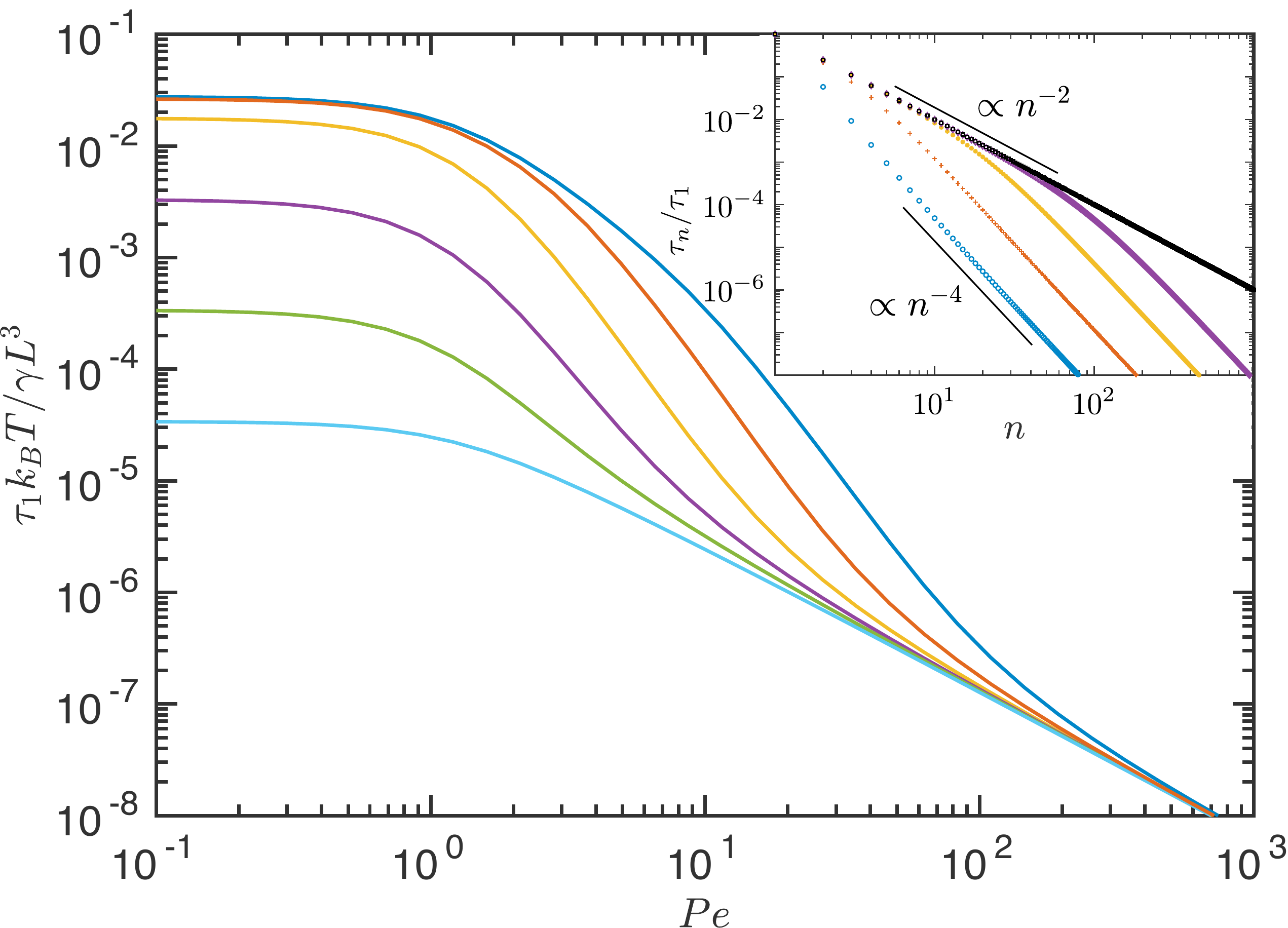}
	\end{center}
\caption{Longest relaxation times of semiflexible D-ABPO as a function of the P\'eclet number for  $pL = L/2lp = 10^3, 10^2, 10, 1, 10^{-1}$, and $10^{-2}$ (bottom to top). Inset: Mode-number dependence of the relaxation times of active polymers with $pL=10^{-2}$ for the  P\'eclet numbers $Pe = 10^1$, $3 \times 10^1$,  $10^2$, and $5 \times 10^2$ (bottom to top). The black squares (top) show the mode-number dependence of flexible polymers with $pL=10^3$.  The solid lines indicate the relations for flexible ($\sim n^{-2}$) and semiflexible ($\sim (2n-1)^{-4}$) polymers, respectively. $\tau_1$ is the longest relaxation time. \cite{eise:16} ``T. Eisenstecken, G. Gompper, and R. G. Winkler, Polymers {\bf 8}, 304 (2016); licensed under a Creative Commons Attribution (CC BY) license.''
} \label{fig:relax_dry}
\end{figure}

\subsubsection{Relaxation times}

The polymer relaxation times strongly depend on the activity, as displayed in Fig.~\ref{fig:relax_dry}. The longest time, $\tau_1$, decreases with increasing P\'eclet number for  $Pe \gtrsim 1$. According to Eq.~\eqref{eq:tau_asym_flex}, in the flexible limit, $\tau_1 \sim 1/\mu$ is solely determined by the stretching coefficient $\mu$, and its decrease  is  a consequence of the finite polymer contour length. With increasing stiffness, the initial drop is stronger, but the same asymptotic dependence is obtained for $Pe \to \infty$.

The inset of Fig.~\ref{fig:relax_dry} shows the dependence of the relaxation times $\tau_n$ of stiff polymers on the mode number. At low $Pe$, we find the well-know dependence $\tau_n/\tau_1 \sim (2n-1)^{-4}$ valid for semiflexible polymers. \cite{wink:07.1,harn:95,arag:85} With increasing $Pe$, the ratio $\tau_n/\tau_1$ increases, and for $Pe \gtrsim 50$ the small-mode-number relaxation times exhibit the dependence $\tau_n/\tau_1 \sim n^{-2}$ of flexible polymers.  At larger $n$, the relaxation times cross over to the semiflexible behavior again. However, the crossover point shifts to larger mode numbers with increasing activity. Hence, active polymers at large P\'eclet numbers appear flexible on large length and long time scales and only exhibit semiflexible behavior on small length scales. 

The theoretically predicted dependence of $\tau_1$ on $Pe$ has not been directly confirmed by simulations yet. However, the very good quantitative agreement between theoretical and simulations results of  polymer mean-square displacements \cite{mart:20}   supports the reliability of the theoretically obtained activity dependence.

\begin{figure}[t]
	\begin{center}
		\includegraphics[width=\columnwidth]{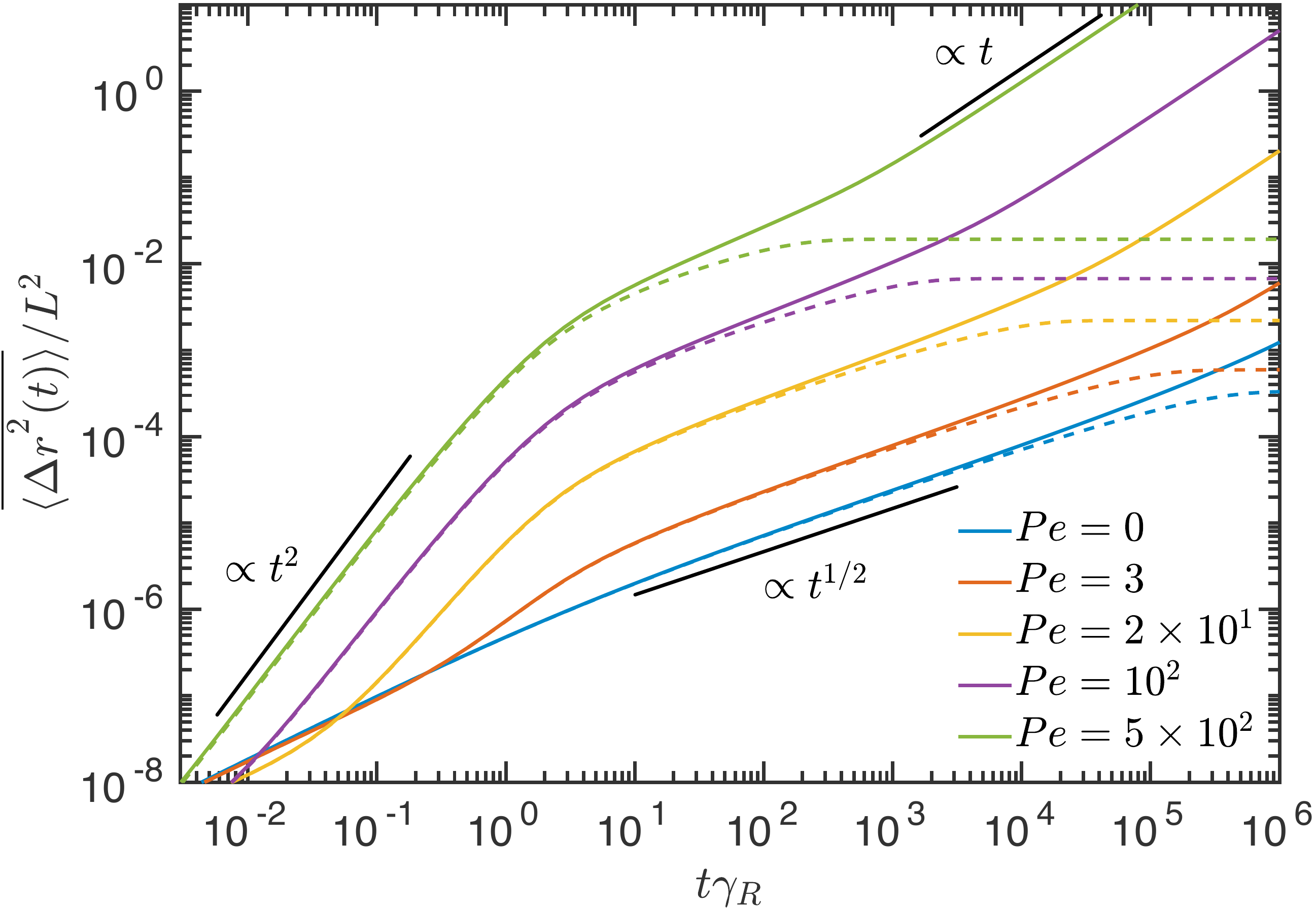}
	\end{center}
\caption{Mean-square displacements (MSD) of flexible D-ABPO with $pL=10^3$. The P\'eclet numbers are
$Pe=0$, $3$, $20$, $10^2$, and $5\times 10^2$ (bottom to top).
The time is scaled by the factor $\gamma_R=2D_R$ of the rotational diffusion. The dashed lines correspond to the MSD in the polymer center-of-mass reference frame. \cite{eise:17} ``Reproduced from T. Eisenstecken, G. Gompper, and R. G. Winkler, J. Chem. Phys. {\bf 146}, 154903 (2017), with the permission of AIP Publishing''
} \label{fig:msd_dry}
\end{figure}

\subsubsection{Mean-square displacement}

The contour-length averaged mean-square displacement (MSD)  of  active polymers is \cite{eise:17}
\begin{align} \label{eq:meansquaredisp} \nonumber
\overline{\lla \Delta  \bm r^2(t) \rra} = & \lla \Delta  \bm r_{cm}^2(t) \rra   + \frac{1}{L} \sum_{n=1}^{\infty}  \left[  \frac{6 k_BT \tau_n}{\gamma} \left(1 - e^{-t/\tau_n}\right)  \right. \\ & \left. + \frac{2 v_0^2 l \tau_n ^2}{1+ \gamma_R \tau_n } \left( 1 - \frac{e^{-\gamma_R t} - \gamma_R \tau_n e^{-t/\tau_n}  }{1 - \gamma_R \tau_n}  \right) \right] \ ,
\end{align}
with the center-of-mass mean-square displacement \cite{eise:16,shin:15,vand:15,chak:19}
\begin{align} \label{eq:msd_cm_dry}
\lla \Delta  \bm r_{cm}^2(t) \rra = \frac{6k_BT }{\gamma L} t + \frac{2 v_0^2 l }{\gamma_R^2 L}\left( \gamma_R t -1 + e^{-\gamma_R t} \right) \ .
\end{align}
Equation~\eqref{eq:msd_cm_dry} resembles the MSD  of a single active Brownian particle, with a ballistic time regime for short times  and a diffusive regime for long times (cf. Fig.~\ref{fig:msd_dry}) with the diffusion coefficient $D =k_BT[1+3 Pe^2/(2\Delta)]/(\gamma L)$. \cite{hows:07,wink:16,elge:15,marc:16.1}  The polymer nature is reflected in the total friction coefficient $\gamma L$,  in the Brownian motion (first term on rhs of Eq.~\eqref{eq:msd_cm_dry}),   and the number of active sites $L/l$,  in the active term. \cite{eise:16}  The center-of-mass motion of polymers is free of internal forces, hence, the  $L/l$  ABPs contribute independently to the MSD. In the limit $t \to \infty$, the center-of-mass MSD is proportional to $v_0^2l/L$. For an isotropic and homogeneous orientation of the propulsion directions, we expect no contribution of the active force to the MSD. However, by the Gaussian nature of the stochastic process, the orientation fluctuations are proportional to $\sqrt{l/L}$, which vanish in the limit $L \to \infty$, but lead to a finite contribution to the diffusion coefficient for $L< \infty$.

The site MSD (monomer MSD), cf. Fig.~\ref{fig:msd_dry}, is strongly affected by activity. It increases with increasing $Pe$,  and exhibits up to four different time regimes.\cite{eise:16} In the limit $t \to 0$, the  second term on the right hand side of Eq.~\eqref{eq:meansquaredisp} dominates and $\overline{\langle \Delta  \bm r^2(t) \rangle }\sim (t/\tau_1)^{1/2}$, exhibiting Rouse dynamics, however, with activity-dependent relaxation time $\tau_1$. For $t/\tau_1 \ll 1$,  $\gamma_R t \ll 1$, and $Pe \gg 1$, the active contribution dominates with a quadratic time dependence  $\overline{\langle \Delta  \bm r^2(t) \rangle} \sim t^2$.  On time scales $1/\gamma_R \ll t \ll \tau_1$, the internal polymer dynamics is most important, with the Rouse-like time dependence    $\overline{\langle \Delta  \bm r^2(t) \rangle} \sim \gamma_R Pe^2 (\tau_1 t)^{1/2}$ in the center-of-mass reference frame. Since $\tau_1$ decrease with increasing activity, this regime shortens with increasing $Pe$. At longer times, the center-of-mass MSD \eqref{eq:msd_cm_dry} dominates.\cite{eise:16,mart:19}

\section{Wet Active Brownian Polymers} \label{sec:wet_abpo}

Hydrodynamic interactions lead to a qualitative different polymer dynamics, as is well established for passive polymers,\cite{doi:86,harn:96} and  first studies on active polymers immersed in a fluid indicate even pronounced effects on their stationary-state conformational properties.\cite{mart:19,mart:20} The  influence  of hydrodynamics depends on the nature of the active force, i.e., self-propelled or bath-driven monomers.

Hydrodynamic interactions between monomers embedded in a fluid can be taken into account implicitly by the Oseen hydrodynamic tensor for point particles  and the Rotne-Prager-Yamakawa (RPY)  tensor for a sphere of diameter $l$.\cite{rotn:69,yama:70,jain:12.1,doi:86,dhon:96,mart:19} 

Alternatively, hydrodynamics can explicitly be taken into account by mesoscale hydrodynamics simulation techniques, such as the lattice Boltzmann method (LB),\cite{mcna:88,shan:93,succ:01,duen:09} the dissipative particle dynamics (DPD),\cite{hoog:92,espa:95} and the multiparticle collision dynamics (MPC) approach.\cite{male:99,kapr:08,gomp:09} Externally driven active polymers have been implemented in MPC.\cite{mart:19}

As mentioned above, the active, nonthermal process affecting the monomer dynamics can originate from internal sources or be imposed externally. In the first case, the monomer is force and torque free, whereas in the second case it is not. This leads to different equations of motion and consequently different behaviors.

\subsection{Discrete model of self-propelled active polymers (S-ABPO)} \label{sec:sabpo}

For active polymers with self-propelled monomers  (S-ABPO), the equations of motion are ($i=1,\ldots, N_m$) \cite{mart:19}
\begin{align} \label{eq:langevin_hi_sp}
  \dot{\bm{r} }_i(t) = & \ v_0 \bm e_i(t)+ \sum^{N_m}_{j=1} \mathbf{H}_{ij} \left[ \bm{ F}_j(t) + \bm {\varGamma}_j(t) \right] ,
\end{align}
where $\bm F_i$ comprises all inter- and intramolecular forces as for D-ABPO. The second moment of the Gaussian and Markovian random force is now given by
\begin{align}
\lla \bm \varGamma_i (t) \bm \varGamma_j^T (t')\rra & = 2 k_BT \mathbf{H}_{ij}^{-1} \delta(t-t') \ ,
\end{align}
where  $\mathbf{H}_{ij}^{-1}$ is the inverse of the hydrodynamic tensor 
\begin{align} \label{eq:htensor}
\mathbf{H}_{ij}(\bm r_{ij}) = \frac{\delta_{ij}}{3 \pi \eta l} \mathbf{I} + (1-\delta_{ij}) {\bf G}(\bm r_{ij}) ,
\end{align}
and ${\bf G}(\bm r)$ the Oseen or RPY tensor. 
The rotational motion of the monomers, Eq.~\eqref{eq:abp_rot}, is not affected by hydrodynamics. No Stokeslet due to self-propulsion is taken into account, only Stokeslets arising from bond, bending, and excluded-volume interactions between monomers, as well as thermal forces are considered in this description.  Higher-order multipole contributions of the active monomers are neglect, especially the force dipole. Since point particles are considered, source multipoles are  absent. All these multipoles decay faster than a Stokeslet. Hence,   the long-range character of HI in polymers of a broad class of active monomers are captured. As far as near-field hydrodynamic effects are concerned, this model is closest to polymers composed of neutral squirmers, \cite{llop:10,goet:10,thee:16,wink:16.1} where particular effects by higher multipole interactions between monomers are not resolved.  \cite{jaya:12,lask:15}
Section~\ref{sec:dumbbell_self_prop} discusses orientational correlations of the propulsion directions of squirmer monomers of a dumbbell by higher-order multipoles. 

The equations of motion \eqref{eq:langevin_hi_sp} are solved by the Ermak–McCammon algorithm.\cite{alle:87,erma:78}

\begin{figure}[t]
	\begin{center}
		\includegraphics[width=\columnwidth]{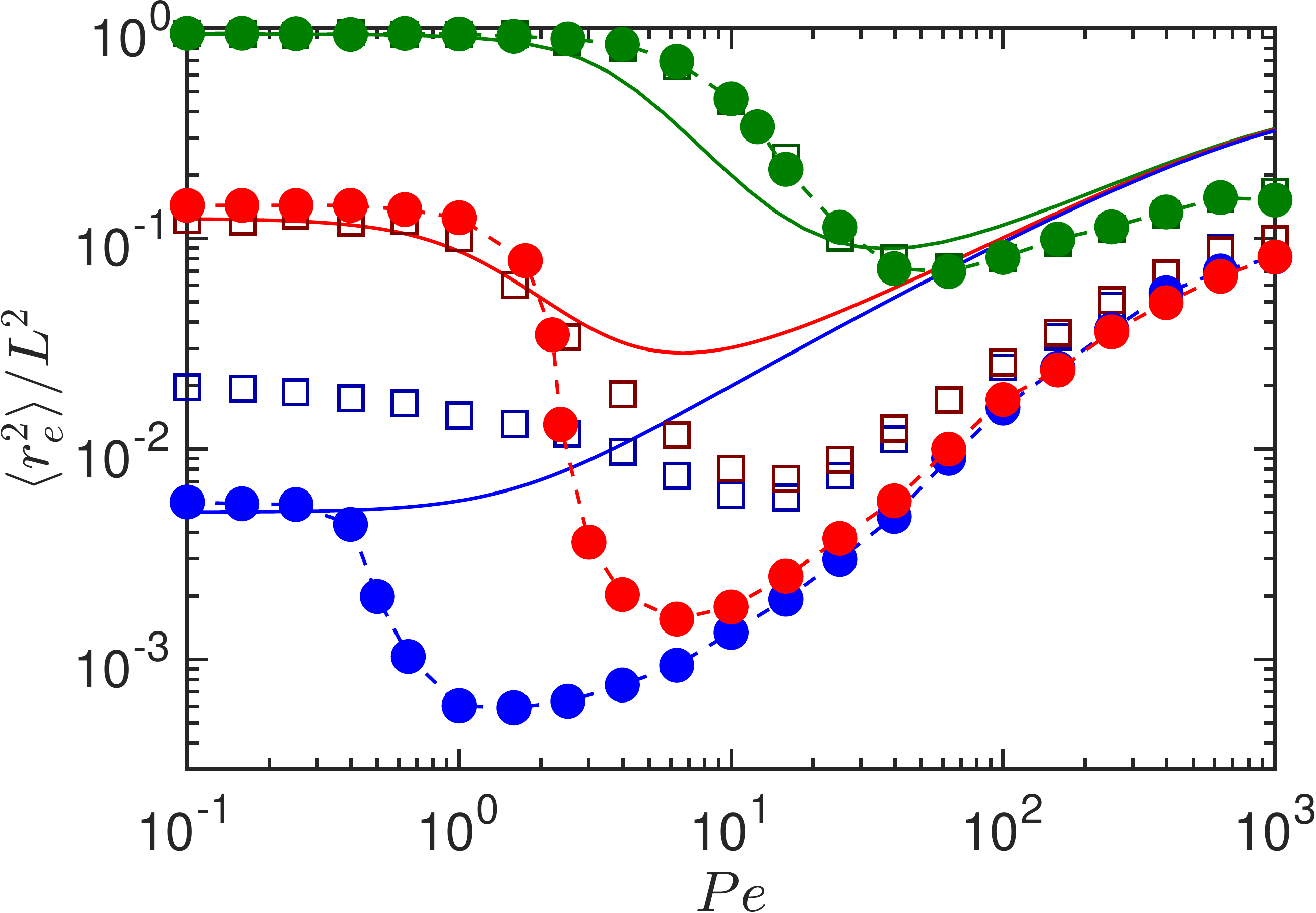}
	\end{center}
\caption{
Mean square end-to-end distance  as a function of  P\'eclet number for semiflexible S-ABPO  with  $N_m=200$ ($L=199l$)  monomers   for
$pL = 2 \times 10^2$ (blue),  $10^1$ (red), and $10^{-1}$ (green). The bullets correspond to phantom and open squares to  self-avoiding polymers. The dashed lines are guides for the eye.  \cite{mart:19} The solid lines are analytical results for semiflexible free-draining polymers (D-ABPO). ``Reproduced with permission from Soft Matter {\bf 12}, 8316 (2016). Copyright 2016 The Royal Society of Chemistry.'' } \label{fig:msqe_self}
\end{figure}

\begin{figure}[t]
	\begin{center}
		\includegraphics[width=\columnwidth]{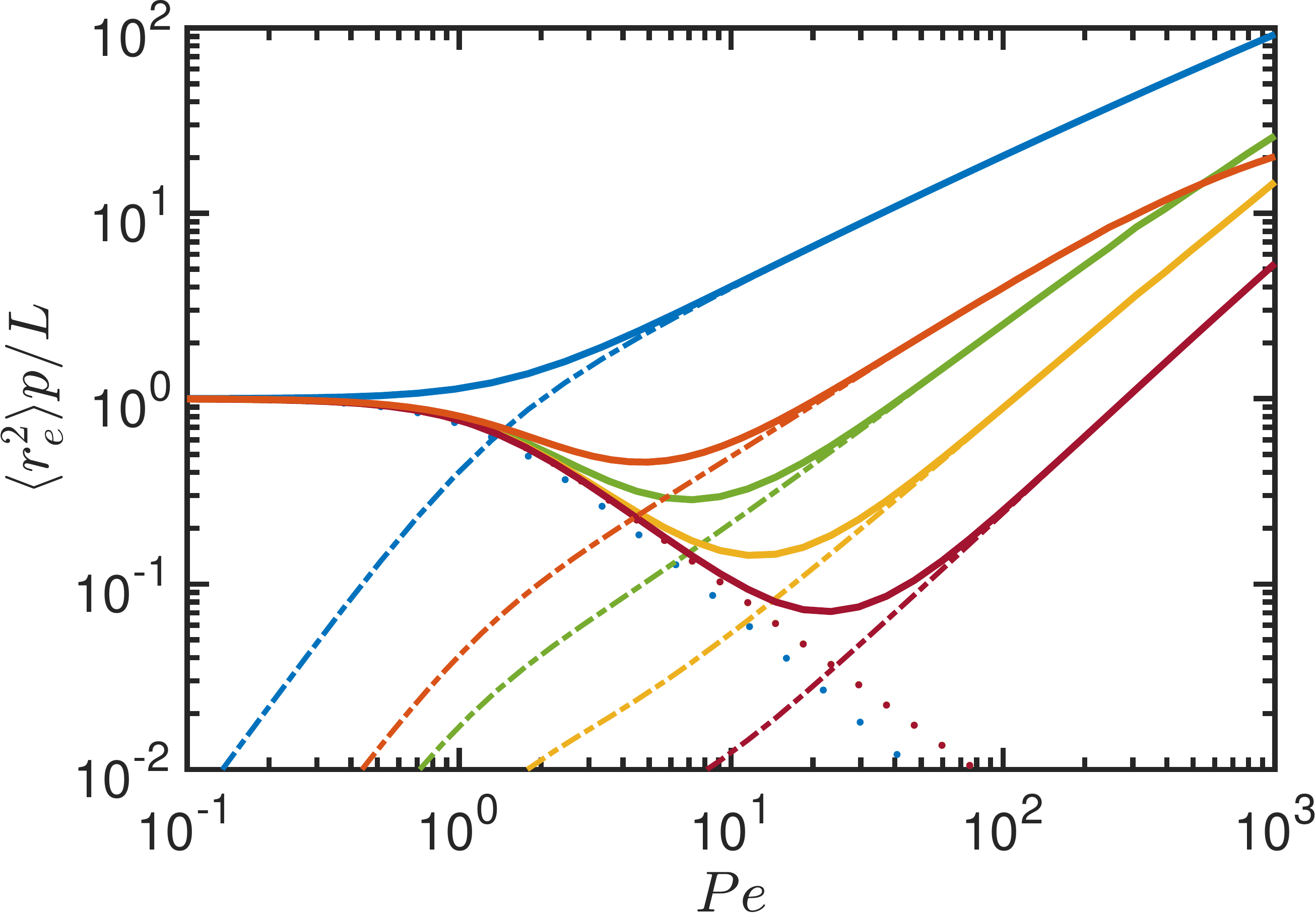}
	\end{center}
\caption{Polymer mean square end-to-end distance $\langle \bm r_e^2 \rangle$, Eq. ~\eqref{eq:end-to-end_self}, as a function of the P\'eclet number $Pe$ for flexible S-ABPOs of length $pL = 2\times 10^2$ (orange), $10^3$ (green), $10^4$ (yellow), and $10^5$ (magenta). The blue lines correspond to a free-draining flexible polymers (D-ABPO) with $pL=50$.  The dotted curves represent the contribution with the relaxation times $\tau_n$ and the dashed-dotted curves that with  $v_0^2$ of Eq.~\eqref{eq:end-to-end_self}, respectively. \cite{mart:19} ``Reproduced with permission from Soft Matter {\bf 12}, 8316 (2016). Copyright 2016 The Royal Society of Chemistry.''} \label{fig:msqe_anal_self}
\end{figure}

\subsubsection{Conformational properties}

Figure~\ref{fig:msqe_self} depicts the confirmations of the active polymers as a function of $Pe$, which are strongly influence by HI. Compared to free-draining polymers (D-ABPO), hydrodynamics leads to a substantial shrinkage of the polymers in the range $1 < Pe \lesssim 10^2$ and a reduced swelling for larger $Pe$. This qualitative difference is illustrated by the snapshots of Fig.~\ref{fig:snapshot_polymer}. The shrinkage depends on polymer length and is substantially stronger for longer polymers.\cite{mart:19} Also semiflexible polymers in presence of HI shrink stronger than those in its absence, but the effect vanishes gradually as $pL \to 0$. This is a consequence of the reduced influence of hydrodynamic interactions for rather stiff polymers. \cite{wink:07.1} Yet, the asymptotic  value for $Pe \to \infty$ is smaller than that  for D-ABPO, because the conformational properties  are determined by polymer entropy rather than stiffness in this limit, with a substantial hydrodynamic effect.

Self-avoidance reduces the extent of shrinkage, specifically for flexible polymers, but the excluded-volume effects vanish with decreasing $pL$, and for $pL<1$ there is hardly any difference between  phantom and  self-avoiding polymers. Moreover, the swelling behavior with and without excluded-volume interactions is rather similar in the limit $Pe \gg 1$. Interestingly, phantom and self-avoiding polymers show a different  but universal dependence on $Pe$ as they start to swell. Here, active forces exceed both, excluded-volume interactions and bending forces.

Analytical theory predicts a stronger increase of $\langle \bm r_e^2 \rangle$ with increasing $Pe$ in the swelling regime compared to D-ABPO (cf. Fig.~\ref{fig:msqe_anal_self}).  The dependence on $Pe$ changes from $\langle \bm r_e^2 \rangle \sim Pe$ for $pL\approx 10^2 - 10^3$ to  $\langle \bm r_e^2 \rangle \sim Pe^{3/2}$ for $pL \approx 10^5$.  Hence, hydrodynamic interactions lead to a qualitative different $Pe$ dependence. 

The polymer collapse is a consequence of the time-scale separation of the thermal and the active contribution to the mean square end-to-end distance.\cite{mart:19}  Hydrodynamic interactions enhance the polymer dynamics and shorten the relaxation times, as is well-know for the Zimm model of flexible polymers \cite{zimm:56,yama:70,doi:86} and also for semiflexible polymers.\cite{harn:96,kroy:97,wink:99} Within the preaveraging approximation, \cite{doi:86} the relaxation times are given by
\begin{align} \label{eq:relax_time_hi}
\tilde{\tau}_n =  \frac{\tau_n}{1 + 3 \pi \eta G_{nn}} ,
\end{align}
with $\tau_n$ the relaxation times in absence of hydrodynamic interactions, and $G_{nn}$ the matrix elements of the Oseen tensor in terms of eigenfunctions of the semiflexible polymers. \cite{harn:95,eise:17,mart:19} Since $ G_{nn} \geqslant 0$, $\tilde \tau_n \leqslant \tau_n$.  Analytical theory yields the mean square end-to-end distance
\begin{align} \label{eq:end-to-end_self}
\lla \bm r_e^2\rra = \frac{8}{L} \sum_{n, \, \text{odd}} \left( \frac{k_BT \tau_n}{\pi \eta} + \frac{v_0^2l \tilde \tau_n^2}{1+\gamma_R \tilde \tau_n} \right) ,
\end{align}
where the first term in the brackets arises from thermal fluctuations and the  second from  activity. However, also the relaxation time $\tau_n$ and the elements $G_{nn}$ depend on activity.
Figure~\ref{fig:msqe_anal_self} displays the different contributions to $\langle \bm r_e^2 \rangle$ for various polymer lengths. The initial shrinkage of $\langle \bm r_e^2 \rangle$ with increasing $Pe$ is caused by the decreasing relaxation times $\tau_n \sim 1/Pe$ with increasing activity.\cite{mart:19} The $v_0^2$-dependent term causes a swelling of the polymers. For  D-ABPO, the competing effects lead to an overall swelling, since swelling exceeds shrinkage. For  S-ABPO, swelling is weaker due to fluid-induced collective motion compared to the random motion of  D-ABPO, and  $\langle \bm r_e^2 \rangle $ assumes a minimum. This is a consequence of  $\tilde \tau_n \leqslant \tau_n$.

Here, we like to mention that a certain amount of shrinkage has also been observed for self-avoiding D-ABPO over a certain range of P\'eclet numbers, which is attributed to  specific monomer packing.\cite{eise:17.1,anan:20}  

\subsubsection{Dynamical properties}

The mean-square displacement of flexible and semiflexible polymers exhibits the Zimm behavior $t^{2/3}$  or the dependence $t^{3/4}$ ($t/\tilde \tau_1 \ll 1$), respectively, for $Pe <1$ in presence of hydrodynamic interactions.\cite{doi:86,petr:06,kroy:97} As for the D-ABPO, activity yields a ballistic time regime for $t/\tilde \tau_1 \ll 1$ and  $\gamma_R t \ll 1$, and a regime dominated by internal polymer dynamics at times $1/\gamma_R \ll t \ll \tilde \tau_1$ for long polymers, $pL \gg 1$, and $Pe \gg1 $.  Interestingly, in the latter regime analytical calculations yield the power-law behavior $\langle \bm r_e^2 \rangle \sim t^{2/5}$, i.e., an exponent even smaller than that of D-ABPO.\cite{mart:19}  For long times, $t/\tilde \tau_1 \gg 1$, the diffusive regime \eqref{eq:msd_cm_dry} of D-ABPO is assumed, with the same active diffusion coefficient.\cite{mart:19}

\subsection{Discrete model of externally actuated polymers (E-ABPO)} \label{sec:abepo_model}

In case of externally actuated monomers (E-ABPO), the polymer equations of motion are given by \cite{mart:20}
\begin{align} \label{eq:langevin_hi_ex}
  \dot{\bm{r} }_i(t) =   \sum^{N_m}_{j=1} \mathbf{H}_{ij} \left[ \gamma_T v_0 \bm e_j(t)+ \bm{ F}_j(t) + \bm {\varGamma}_j(t) \right] .
\end{align}
Here, also the external active force generates a Stokeslet for each monomer. 

A possible realization of {E-ABPO} is a passive polymer embedded in an active bath. Experimentally, an externally driven polymer can in principle be realized by forcing a chain of colloidal particles by optical tweezers.\cite{grie:03} Optical forces are very well suited to manipulate objects as small as $5 nm$ and up to hundreds of micrometers.\cite{grie:03}  Combined with computer-generated holograms, many particles can be manipulated with a single laser beam at the same time.

\subsubsection{Conformational properties}

The  mean square end-to-end distance of flexible {E-ABPO}  increases monotonically with increasing $Pe$, \cite{mart:20} in contrast to S-ABPO.\cite{mart:19} As for D-ABPO and S-ABPO, semiflexible polymers shrink initially with increasing $Pe$ and swell again for large P\'eclet numbers in a universal manner. The comparison with the  mean square end-to-end distance curves for D-ABPO  reveals a stronger impact of activity on the conformations of E-ABPO. In particular, flexible polymers swell  and semiflexible polymers shrink  already for smaller P\'eclet numbers. However, the same asymptotic limit is assumed for $Pe \to \infty$, which follows from Eq.~\eqref{eq:end-to-end_ex} for $\gamma_R \tilde \tau_n \ll 1$. Hence, S-ABPO exhibit the more compact structures compared to both, D-ABPO and E-ABPO. In the universal, stiffness- and excluded-volume-independent high-$Pe$ regime, the  mean square end-to-end distance exhibits the power-law  increase $\langle \bm r_e^2 \rangle \sim Pe^{1/2}$  with increasing $Pe$. This increase is weaker than that of D-ABPO, with the dependence $\langle \bm r_e^2 \rangle \sim Pe^{2/3}$ (cf. Fig.~\ref{fig:msd_dry}), which emphasizes the strong effect of the internal  dynamics on the active polymer conformational properties.\cite{mart:20}

As for the shrinkage of the S-ABPO, enhanced swelling is related to a particular dependence of the mean square end-to-end distance on the relaxation times, which is given by
\begin{align} \label{eq:end-to-end_ex}
\lla \bm r_e^2\rra = \frac{8}{L} \sum_{n, \, \text{odd}} \left( \frac{k_BT \tau_n}{\pi \eta} + \frac{v_0^2l \tau_n^2}{1+\gamma_R \tilde \tau_n} \right) .
\end{align}
Since
\begin{align}
\frac{\tau_n^2}{1+\gamma_R \tilde \tau_n} \ge  \frac{\tau_n^2}{1+\gamma_R  \tau_n} \ge \frac{\tilde \tau_n^2}{1+\gamma_R \tilde \tau_n} ,
\end{align}
the active contribution in  Eq.~\eqref{eq:end-to-end_ex} is larger than that of  D-ABPO and S-ABPO, which implies a stronger overall swelling.\cite{mart:20}

\begin{figure}[t]
	\begin{center}
		\includegraphics[width=\columnwidth]{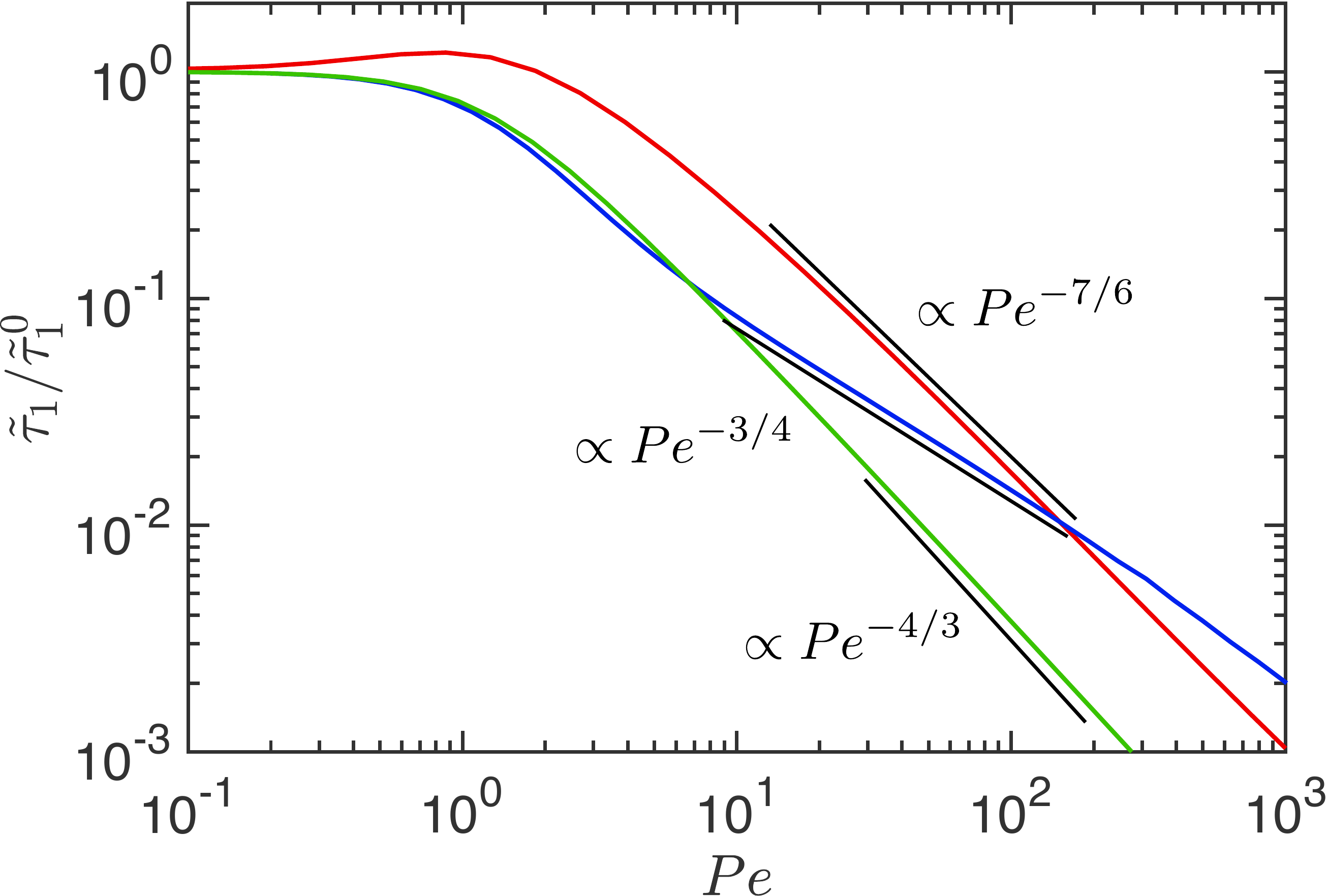}
	\end{center}
\caption{Longest polymer relaxation time $\tilde{\tau}_1$ normalized by the corresponding passive value  $\tilde{\tau}_1^0$ as a function of the P\'eclet number $Pe$ for flexible  D-ABPO (green), S-ABPO (blue), and E-ABPO (red) with $pL =10^3$.
} \label{fig:relax_comp}
\end{figure}

\subsubsection{Dynamical properties}

The distinct  coupling between activity and hydrodynamics changes the activity-dependence of the relaxation times  substantially, as illustrated in Fig.~\ref{fig:relax_comp}. Over a wide range of P\'eclet numbers, $\tilde \tau_1$ of  E-ABPO is larger than the relaxation time of D-ABPO and S-ABPO, which is reflected in the strong increase of the mean square end-to-end distance. The (longest) relaxation time decreases with in increasing $Pe$, hence, $\gamma_R \tilde \tau_n \ll 1$ in the limit $Pe \to \infty$. Moreover, $\tilde \tau_1$ exhibits the same dependence on $Pe$ as $\tau_1$ in absence of HI in that limit, which yields the same asymptotic value for  $\langle \bm r_e^2 \rangle$.\cite{mart:20} We like to emphasize that $\tilde \tau_1$ of semiflexible polymers assumes the same $Pe$ dependence and scaling for $Pe \gg 1$ as flexible polymers (cf. Fig.~\ref{fig:relax_dry}). 

The MSD exhibits the same sequence of time regimes as for D-ABPO and S-ABPO (cf. Fig.~\ref{fig:msd_dry}). Remarkably, long polymers exhibit a  subdiffusive behavior for times $1/\gamma_R \ll t \ll \tilde \tau_1$  dominated by their internal   dynamics, where analytical theory predicts the power-law time dependence $t^{5/7}$, which is close to $t^{3/4}$.

\section{Tangentially Driven Filaments} \label{sec:driven_filaments}

Cytoskeleton polymers  such as  microtubules are rather stiff and are typically denoted as filaments. They exhibit active motion fueled by molecular motors {\em in vivo}\cite{gang:12} and {\em in vitro}.\cite{hara:87,nedl:97,juel:07,scha:10,sumi:12,sanc:12,enno:16,need:17,wu:17,belm:17,hube:18,doos:18,guil:18}
By the nature of their propulsion mechanism, such filaments are typically modeled as tangentially driven rodlike\cite{peru:06,gine:10,wens:12,wens:12.1,abke:13,ravi:17,abau:18,baer:20} or semiflexible polymers\cite{live:01,jaya:12,jian:14,chel:14,isel:15,denk:16,bisw:17,duma:18,prat:18,bian:18,anan:18,pete:20} in molecular approaches.

\subsection{Continuum model} \label{sec:tang_anal}

The theoretical description of Sec.~\ref{sec:abp_cont}  can be adapted to the case of tangential driving semiflexible polymers, which yields the equation of motion 
\begin{align} \label{eq:equ_motion_polar}
& \gamma \frac{\partial}{\partial t}{ \bm r}(s,t) = f^a \frac{\partial}{\partial s} \bm r(s,t) \\ \nonumber
 & +  2 \lambda k_B T \frac{\partial^2}{\partial s^2} { \bm r}(s,t)  - \epsilon k_B T
\frac{\partial^4} {\partial s^4} { \bm r}(s,t) + {\bm \Gamma} (s,t)  ,
\end{align}
with the active force $f^a$  per unit length. This linear partial differential equation can be solved by an eigenfunction expansion in terms of the eigenfunctions $\psi_n(s)$ of the equation ${\cal O} \psi_n(s) = - \xi_n \psi_n(s)$ with the operator
\begin{align} \label{eq:eigen_polar}
{\cal O} = f^a
\frac{\partial}{\partial s} + 2 \lambda k_B T \frac{\partial^2}{\partial s^2}   - \epsilon k_B T
\frac{\partial^4} {\partial s^4}
\end{align}
and the  eigenvalues $\xi_n$. However, the operator ${\cal O}$ is non-Hermitian, hence, an expansion into a biorthogonal basis set has to be applied, with the adjoint eigenvalue equation \cite{risk:89}
${\cal O}^{\dag} \psi_n^{\dag}(s) = - \xi_n \psi_n^{\dag}(s) $
and
\begin{align} \label{eq:eigen_polar_adj}
{\cal O}^{\dag}  =  - f^a\frac{\partial}{\partial s} + 2 \lambda k_B T \frac{\partial^2}{\partial s^2}   - \epsilon k_B T
\frac{\partial^4} {\partial s^4}  .
\end{align}
In case of a flexible polymers ($\epsilon =0$), the eigenfunctions are\cite{phil:20}
\begin{align}
\psi_n(s)=\sqrt{\frac{2}{L}} \frac{1}{\sqrt{k_n^2+g^2}} e^{-gs}\left[ k_n \cos (k_ns)+g\sin(k_ns) \right] ,
\end{align}
with $k_n=n\pi/L$, $g = f^a/(4 \lambda k_BT)$, and the eigenvalues $\xi_n = 2 \lambda k_BT(k_n^2 + g^2)$ for the boundary condition $\partial \bm r / \partial s = 0$ at $s=0,L$, which implies $\partial \psi_n / \partial s = 0$ at $s=0,L$. The adjunct eigenfunctions are $\psi_n^\dag(s) = e^{2gs} \psi_n(s)$, and obey the boundary condition $d \psi_n^{\dag}(s)/ds - 2 g \psi_n^{\dag}(s) = 0$ at $s=0,L$. The eigenfunction for the eigenvalue $\xi_0=0$ is $\psi_0=(2g/(e^{2gL}-1))^{1/2}$.\cite{phil:20}

\subsection{Discrete model} \label{sec:polar_disc}

The dynamics of discrete polar active filaments is described by Eq.~\eqref{eq:dumb_tan}, with bond, bending, and excluded-volume forces.  However, for the active force on particle $i$, various nonequivalent representations have been applied. In most cases, the active force is assumed to be tangential to the chain at a monomer,\cite{bian:18}   
\begin{align} \label{eq:force_prop_tang_1}
\bm F_i^a = f^a  \frac{\bm r_{i+1}- \bm r_{i-1}}{|\bm r_{i+1}- \bm r_{i-1}|} = f^a \bm t_i ,
\end{align}
with the unit tangent vector $\bm t_i$. 
Other simulation studies utilized  push-pull type forces \cite{isel:15}
\begin{align} \label{eq:force_prop_tang_2}
\bm F_i^a = \frac{f^a}{l} [(\bm r_{i+1}- \bm r_{i})  + (\bm r_{i}- \bm r_{i-1})] = \frac{f^a}{l} (\bm r_{i+1} - \bm r_{i-1}) ,
\end{align}
 or\cite{jian:14,prat:18,anan:18}
\begin{align} \label{eq:force_prop_tang_3}
 \bm F_i^a = f^a \left[ \frac{\bm r_{i+1}- \bm r_{i}}{|\bm r_{i+1}- \bm r_{i}|}   + \frac{\bm r_{i}- \bm r_{i-1}}{|\bm r_{i}- \bm r_{i-1}|} \right] .
\end{align}
For strong bond potentials and small bond-length variations, the latter two variants are essentially identical, but differ from Eq.~\eqref{eq:force_prop_tang_1}. All these forces yield the same continuum limit (Eq.~\ref{eq:equ_motion_polar}).  In simulations, we can expect results to be independent of the adopted discretization for  not too strong activities.  However, deviations will emerge above a certain activity and, in particular, for a small number of monomers. This is reflected by the differences in the conformational properties of the continuous filaments (Eq.~\eqref{eq:equ_motion_polar}) \cite{pete:20} and of discrete models (cf. Sec.~\ref{sec:tangential_results}).\cite{bian:18,anan:18} 
 
 \subsection{Results} \label{sec:tangential_results}
 
 The theoretical analysis of Sec.~\ref{sec:tang_anal} suggest that the conformational properties of flexible polar polymers are unaffected by activity. In particular, the mean square end-to-end distance is equal to the value of the passive counterpart.\cite{pete:20,phil:20} This is in contrast to simulation results for a discrete model, which indicate a shrinkage of flexible  polymers at large active forces $f^a$, \cite{bian:18,anan:18}  for both,  the  active forces  \eqref{eq:force_prop_tang_1} as well as \eqref{eq:force_prop_tang_3}, although shrinkage is more pronounced for the force \eqref{eq:force_prop_tang_1}. The difference is a consequence of the particular discrete representation of the continuous polymers (cf. Sec.~\ref{sec:polar_disc}).
 
 So far, very little further analytical results for individual polar filaments have been derived, specifically for semiflexible polymers,  which might be related to the non-Hermitian character of the equations of motion. 
 
Confinement in two dimensions enhances the influence of  excluded-volume interactions on the emerging structures. Here,  semiflexible filaments exhibit a transition to a spiral phase with increasing activity at $Pe \sim l_p/L$.\cite{isel:15,isel:16,prat:18,duma:18} Tethered filaments start beating in the presence of a polar force, or assume spiral shapes when the fixed end is able to rotate freely.\cite{chel:14,anan:19}  

The filament dynamics depends on the composition of  its ends. As an example, a load at the leading end leads to  distinct  locomotion patterns such as beating and rotation, depending on stiffness and propulsion strength.\cite{isel:16}    

Simulations predict a substantial influence of hydrodynamic interactions  on the rotational dynamics of single  and multiple filaments as well  on their collective dynamics and structure formation  in 3D. \cite{jian:14,jian:14.1} According to the definition of Sec.~\ref{sec:abepo_model}, the latter polymers are externally actuated by a tangential force of the type of Eq.~\eqref{eq:force_prop_tang_3}.  

\section{Polymers with self-propelled  colloidal  monomers} \label{sec:self_propelled_colloid}

Polymers can be composed of  monomers with their self-propulsion modeled by the concepts described in Sec.~\ref{sec:squirmer}, i.e., as squirmers or a more general expansion of the monomer flow field. Although various studies have already been performed based on mobilities, \cite{lask:15,sing:18.1} only  squirmer dumbbells  have been considered so far. In contrast to the Brownian monomers of Secs.~\ref{sec:dry_abpo} and \ref{sec:wet_abpo}, the hydrodynamically self-propelled monomers include  higher order multipoles such as force dipole, source dipole, etc.,\cite{thee:16.1,wink:18a} which, by monomer-monomer interactions, give rise to additional features.

\begin{figure*}[t]
	\begin{center}
		\includegraphics[width=\textwidth]{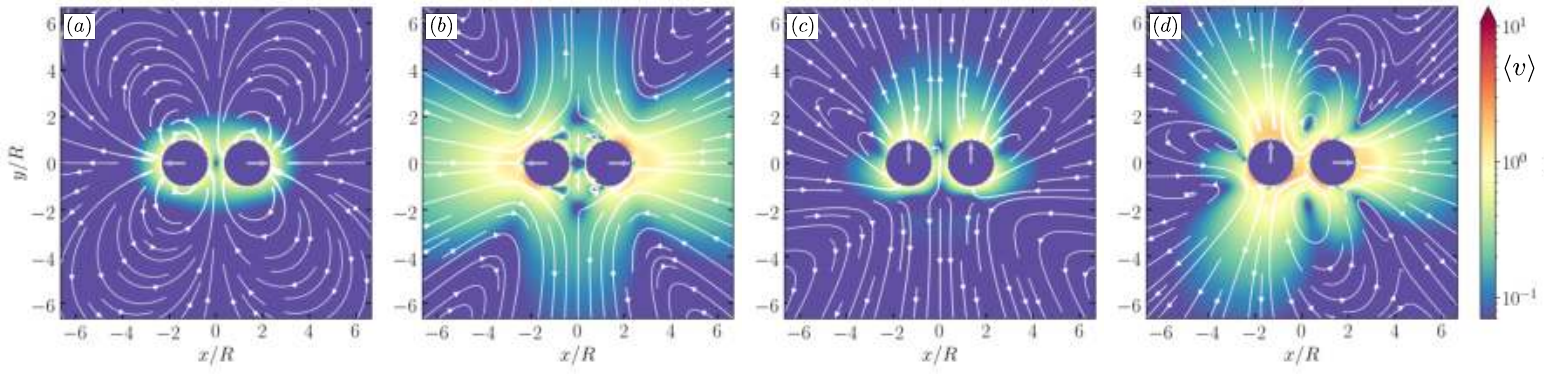}
	\end{center}\caption{Illustration of the flow field of a dumbbell composed of two spherical squirmers for their preferred propulsion directions. For (a) neutral squirmers ($\beta =0$) and (b) pullers ($\beta = 5$), the propulsion direction is preferentially antiparallel. In the case of (c) a weak pusher ($\beta = -1$),  the squirmer propulsion direction is mostly parallel and orthogonal to the bond.  Stronger pushers ($\beta =-5$) exhibit preferentially an arrangement as in (c) as well as an orthogonal arrangement as in (d). Note that the flow fields are  superpositions of the flow fields of the individual squirmers. \cite{clop:20}} \label{fig:squirm_dumb_flow}
\end{figure*}

\subsection{Linear squirmer assemblies} \label{sec:dumbbell_self_prop}

The swimming behavior of athermal squirmer dumbbells has been investigated by the boundary integral method,\cite{ishi:19} where, in this context, swimming refers to ballistic motion. The calculations show a strong effect of the interfering swimmer flow fields on their orientation and locomotion. In particular, a flow-induced torque prevents stable forward swimming of individually freely rotating spherical squirmer in a dumbbell. By introducing a (short) bond, which  restricts an independent rotation, circular trajectories for pushers and stable side-by-side swimming for pullers can be obtained. \cite{ishi:19}   

Athermal squirmer lack rotational diffusion, thus,  absence of locomotion of freely rotating spheres in a dumbbell is a consequence  of the missing thermal fluctuations of the fluid. Such fluctuations can be taken into account  by mesoscale hydrodynamic simulations, e.g., the multiparticle collision dynamics (MPC) method,\cite{male:99,kapr:08,gomp:09} and locomotion is obtained.   
At infinite dilution, rotation is an activity-independent degree of freedom of a squirmer (cf. Sec.~\ref{sec:squirmer_model}), which, however, strongly affects its long-time diffusion behavior, in the same way as for an ABP (see Eq.~\eqref{eq:msd_cm_dry} for comparison)   

Simulations of squirmer dumbbells yield preferred orientations of the propulsion directions, $\bm e_i$, with respect to each other and the bond connecting their centers, as illustrated in the snapshots of Fig.~\ref{fig:squirm_dumb_flow}.\cite{clop:20,nava:10} This preference can be quantified   by the average relative orientation of the propulsion direction, i.e., $\langle \bm e_1 \cdot \bm e_2\rangle$. As expected for the active stresses, in dumbbells of neutral squirmers, alignment is very weak, and only for high $Pe$ a slightly negative value  is obtained. Puller exhibit very strong orientational correlations, with a preferential antiparallel alignment (cf. Fig.~\ref{fig:squirm_dumb_flow} (a), (b)).  Strong correlations are also present for pushers, however, with a parallel alignment of the $\bm e_i$ orthogonal  with respect to the bond between the squirmers (Fig. ~\ref{fig:squirm_dumb_flow}(c)), and a near orthogonal arrangement as show in Fig.~\ref{fig:squirm_dumb_flow}(d).

The preference in orientation is reflected in the dumbbell dynamics. The locomotion of the pusher dumbbell is closest to the dynamics of an ABP, specifically for $\beta \approx -1$. Significant deviations to an ABP dumbbell appear for smaller and large $\beta$. In particular, the preferred antiparallel alignment  for neutral squirmers and pullers leads to a pronounced slow-down of the dumbbell  dynamics with a rather abrupt change at $\beta \gtrsim 0$.  Notably, the long-time diffusion coefficient of puller dumbbells at large active stresses, $\beta=5$, is reduced by orders of magnitude compared to that of  ABP dumbbells. This is inline with the theoretical studies, \cite{ishi:19} however, thermal fluctuations are responsible for a finite diffusion coefficient at long times.

Hydrodynamic interactions between the monomer flow fields strongly affect the conformational properties of squirmer polymers. Simulations of  dodecamers yield strongest swelling for neutral squirmers, and essentially no swelling for pullers ($\beta=5$), with a mean square end-to-end distance close to the passive value.

\subsection{Active filaments of tangentially propelled monomers} \label{sec:filament_self_prop}

\subsubsection{Colloidal monomers}

The translational and rotational  motion of chains of  linearly connected  monomers can be described by a set of Langevin equations.\cite{lask:15} The equations are based on the integral solution  of Stokes flow,\cite{pozr:92} which is evaluated  analytically by an expansion of the boundary fields in tensorial spherical harmonics. The omission of  hydrodynamic many-body effects by the assumption of  pair-wise additivity  yields a computationally tractable approximation in terms of  the Oseen tensor and its derivatives.\cite{kirk:48} For tangentially driven filaments, the orientation of a monomer is no longer a dynamical degree of freedom, and the equations of motion reduce to translation only. 

\begin{figure}[t]
	\begin{center}
		\includegraphics[width=\columnwidth]{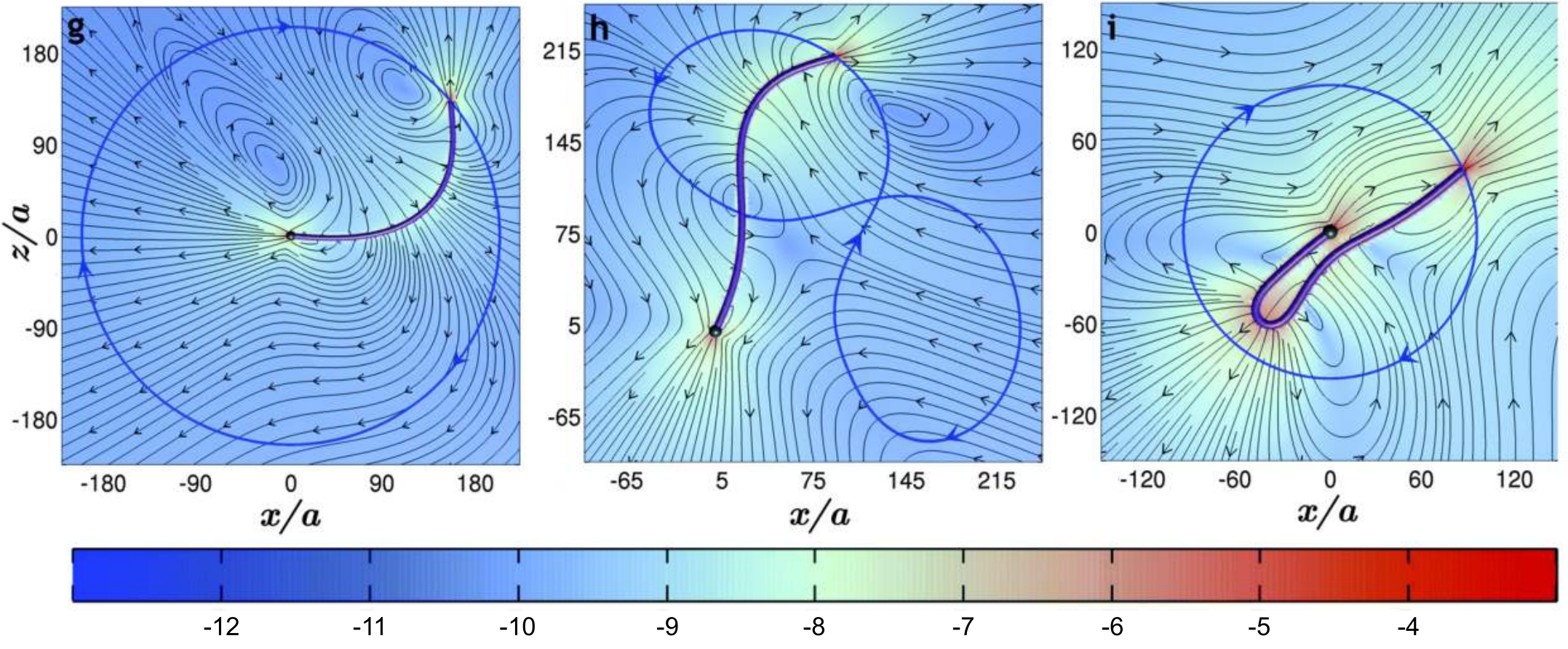}
	\end{center}\caption{Nonequilibrium stationary states of a single-end tethered filament. Increasing activity (from left to right)  leads  to particular stationary-state conformations and rotation (g, i), or filament beating (h). The background color indicates the logarithm of the magnitude of the fluid velocity.  Reproduced from Ref.~\ocite{lask:15}  with permission from the Royal Society of Chemistry. ``Reproduced with permission from Soft Matter {\bf 11}, 9073 (2015). Copyright 2015 The Royal Society of Chemistry.''
	} \label{fig:filament_hyd}
\end{figure}

In contrast to the Brownian semiflexible polymers and filaments of Secs.~\ref{sec:dry_abpo}, \ref{sec:wet_abpo}, and \ref{sec:driven_filaments}, where activity originates  from an active velocity or force,  the  presence of higher order hydrodynamic  multipoles yields active motion even in  absence of a propulsion velocity. This is due to the generation of flows perpendicular to the tangent vector for a curved contour (cf. Fig.~\ref{fig:filament_hyd})  This gives rise to particular stationary-state filament conformations and dynamical states, such as rotation of free filaments or beating and rotation of tethered filaments (cf. Fig.~\ref{fig:filament_hyd}). \cite{jaya:12,lask:13,lask:15} Thermal noise will modify this behavior, to an extent which remains to be investigated.

\subsubsection{Pointlike monomers} \label{sec:point_monomes}

A somewhat simpler approach for hydrodynamically self-propelled, pointlike "monomers" has been suggested,\cite{saint:18}
where Eq.~\eqref{eq:abp_trans} is employed  for the monomer dynamics, however, with constraint rather than harmonic bond forces, and the force-dipole-type active force for a bond/link $\bm u_{i+1}$, \cite{elge:15}
 \begin{align}
 \bm F^a_i = - F_0 \bm u_{i+1} ,  \hspace*{5mm} \bm F^a_{i+1} = F_0 \bm u_{i+1} ,
 \end{align}
where $F_0$  is magnitude of the force, and its sign implies extensile ($F_0>0$) or contractile ($F_0<0$) dipoles.  The active forces are assumed to arise from stresses exerted by molecular motors temporarily  attached to a bond. The time interval of the  attached  and detached states are taken from an  exponential distribution.\cite{saint:18} The active velocity in the attached state is then 
\begin{align}
\bm v^a_i = \sum_j  {\bf G} (\bm r_i - \bm r_j) \bm F_j^a(t) .
\end{align}
Considering a single, long, and densely packed polymer in a spherical cavity, simulations show that hydrodynamic interactions between extensile dipoles can lead to large-scale coherent motion (cf. Sec.~\ref{sec:collective}).

\begin{figure*}[t]
	\begin{center}
		\includegraphics[width=\textwidth]{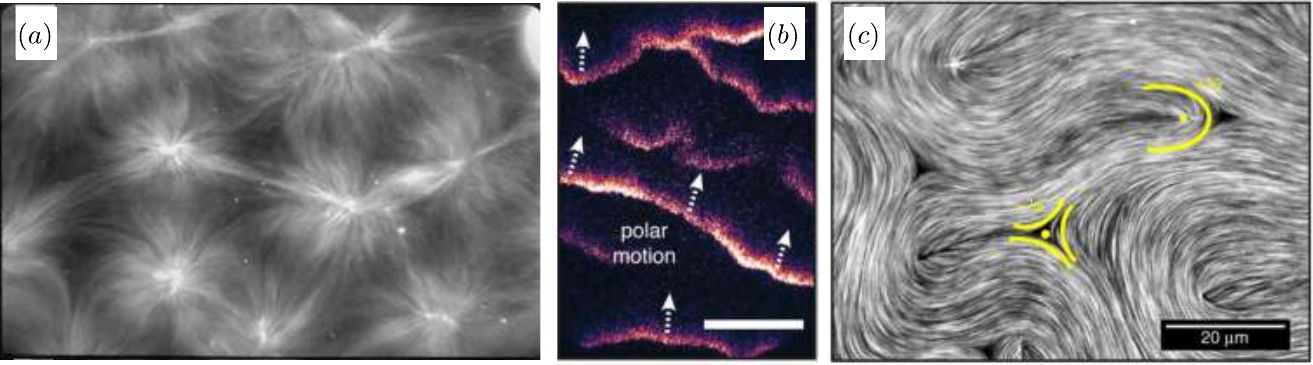}
	\end{center}
	\caption{Large-scale patterns formed by self-organization of biological filaments and molecular motors. (a) Lattice of asters and vortices formed by microtubules.\cite{nedl:97} (b) Motility assay with fronts of polar actin clusters  moving in the same direction as the filaments.\cite{hube:18} (c)  Network of nematically ordered microtubules exhibiting  turbulent motion with a  $+1/2$ (top right) and a $-1/2$ defect (bottom left).\cite{doos:18}  (a) ``Reproduced with permission from Nature {\bf 389}, 305 (1997). Copyright 1997  Macmillan Publishers Ltd.''  (b) ``Reproduced with permission from Science {\bf 361}, 255 (2018). Copyright 2018  AAAS.'' (c) ``A. Doostmohammadi, J. Ign{\'e}s-Mullol, J. M. Yeomans, and F. Sag{\'u}es, Nat. Commun. {\bf 9}, 3246 (2018); licensed under a Creative Commons Attribution (CC BY) license.''} \label{fig:aster}
\end{figure*}

\section{Collective behavior} \label{sec:collective}

Already ABPs and active dumbbells exhibit an intriguing collective behavior, especially motility-induced  phase separation.\cite{fily:12,redn:13,bial:12,marc:16.1,wyso:14,sten:14,wyso:16,cate:15,bech:16,suma:14,sieb:17,digr:18,thee:18} Steric interactions by the extended shape of filaments and polymers, and their conformational degrees of freedom lead to further novel phenomena and behaviors.    

Assays of microtubules mixed with molecular motors are paramount examples for collective effects emerging in out-of-equilibrium systems including polymers/filaments.\cite{sumi:12,need:17,doos:18} Simplified systems comprising microtubules and  kinesin motor constructs  are able to self-organize into stable structures.\cite{nedl:97}  The motors dynamically crosslink the microtubules and their directed motion along the polar filaments leads to formation of vortex-like structures and asters  with the microtubules arranged radially outward (Fig.~\ref{fig:aster}(a)).  \cite{nedl:97}  Along the same line, highly concentrated actin filaments propelled by molecular motors in a motility assay display emergent collective motion.\cite{scha:10,hube:18}   Above a critical density, the filaments self-organize into coherently moving structures with persistent density modulations, such as clusters, swirls, and interconnected bands (Fig.~\ref{fig:aster}(b)).\cite{scha:10} Furthermore, addition of  a depletion agent to a concentrated  microtubule-kinesin mixture leads to microtubule assemblance into bundles, hundreds of microns long.\cite{sanc:12,need:17,doos:18}  Kinesin clusters in bundles of microtubules of different polarity induce filament sliding and trigger their extension.\cite{doos:18}
At high enough concentration, the microtubules form a percolating active network characterized by internally driven chaotic flows, hydrodynamic instabilities, enhanced transport, and fluid mixing.\cite{sanc:12}   Activity destroys long-range nematic ordering and leads to active turbulence, with short-range nematic order and dynamically creation and annihilation of topological  defects (Fig.~\ref{fig:aster}(c)).\cite{sanc:12,doos:18}

\begin{figure*}[t]
	\begin{center}
		\includegraphics[width=\textwidth]{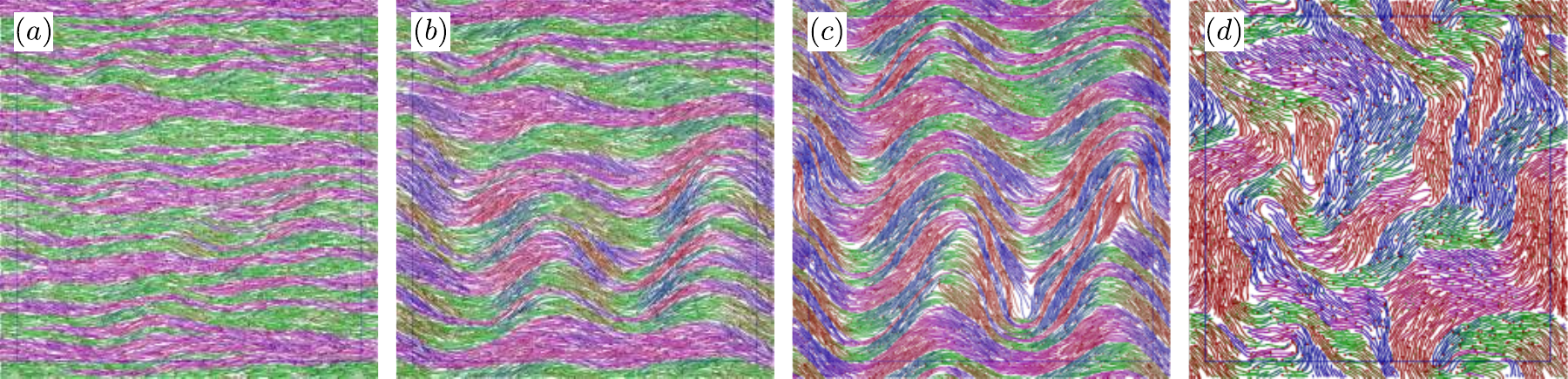}
	\end{center}
	\caption{Euler-like buckling instability of polar bands by semiflexible, propelled filaments. The snapshots depict a time sequence from  (a) the development of polarity-sorted bands, (b), (c) progressive bending and breaking of bands, and finally (d) the formation of the long-time disordered structure.\cite{vlie:19} ``G. Vliegenthart, A. Ravichandran, M. Ripoll, T. Auth, and G. Gompper, Sci. Adv. in press (2020); licensed under a Creative Commons Attribution (CC BY) license.''} \label{fig:nematic}
\end{figure*}

The instability of the nematic phase of long polar filaments is usually attributed to the pusher-like hydrodynamics of systems with extensile force dipoles. A sinusoidal perturbation of the nematic order, with a wave vector along the filament direction, is amplified by the dipole flow field. \cite{sain:08}
However, a similar type of undulation instability is  observed in simulations
of semiflexible polar filaments, which are temporarily connected by motor proteins sliding them against each other---without any hydrodynamic interactions (cf.  Fig.~\ref{fig:nematic}). This system  is extensile, because motors mainly exert forces on antiparallel filaments. For an
initially nematic phase with random filament orientation, first a polarity-sorting into
bands occurs, followed by buckling of the bands, and a transition to an isotropic phase
with polar domains (cf.  Fig.~\ref{fig:nematic}).\cite{vlie:19} The isotropic phase is characterized by
nematic-like $+1/2$ and $-1/2$ defects, as depicted in Fig.~\ref{fig:aster}(c).
\cite{marc:13} Here, it is interesting to note that (i) active nematics are 
actually quite difficult to model on the filament level, because filaments move actively,
but cannot have a preferred direction, and (ii) the filaments are actually {\em polar},
so that the system has both polar and nematic characteristics at the same time. This 
system has therefore been termed a {\em polar active nematic} in 
Ref.~\ocite{vlie:19}.

\begin{figure}[t]
	\begin{center}
		\includegraphics[width=0.7\columnwidth]{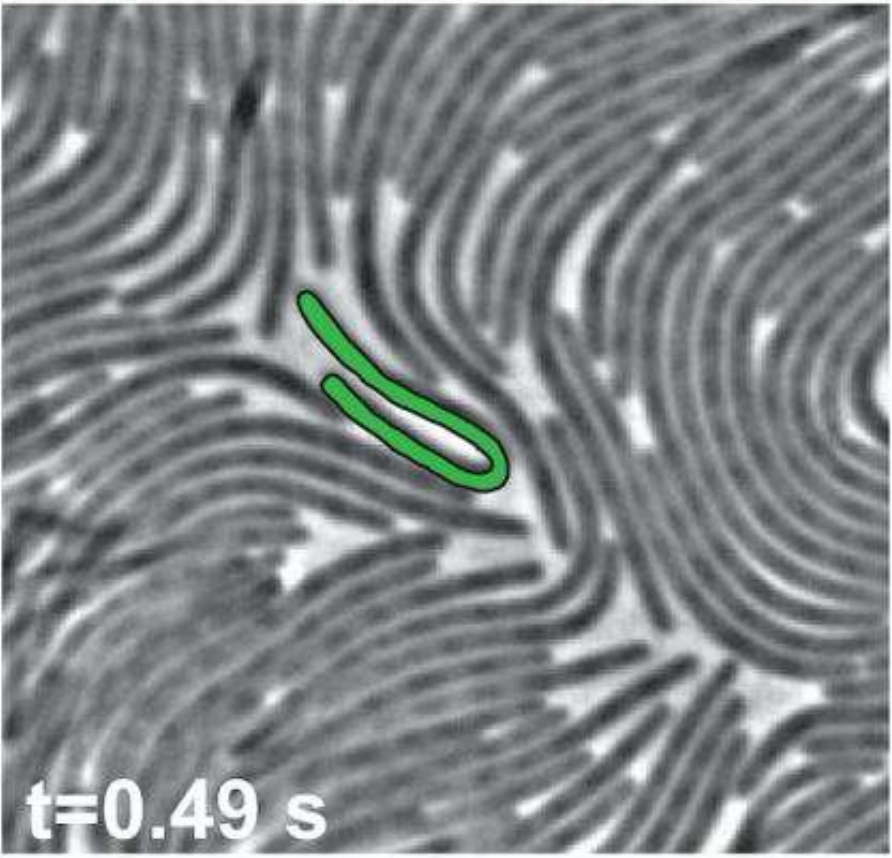}
	\end{center}
	\caption{Image  of {\em P. mirabilis}  swarmer cells in a colony actively moving across a surface.\cite{auer:19}  The  green,  generally straight cell illustrates the ability of  swarmer cells to bend substantially. ``G. K. Auer, P. M. Oliver, M. Rajendram, T.-Y. Lin, Q. Yao, G. J. Jensen, and D. B. Weibel, mBio {\bf  10}, e00210 (2019); licensed under a Creative Commons Attribution (CC BY) license.''} \label{fig:swarmer}
\end{figure}

\begin{figure}[t]
	\begin{center}
		\includegraphics[width=0.9\columnwidth]{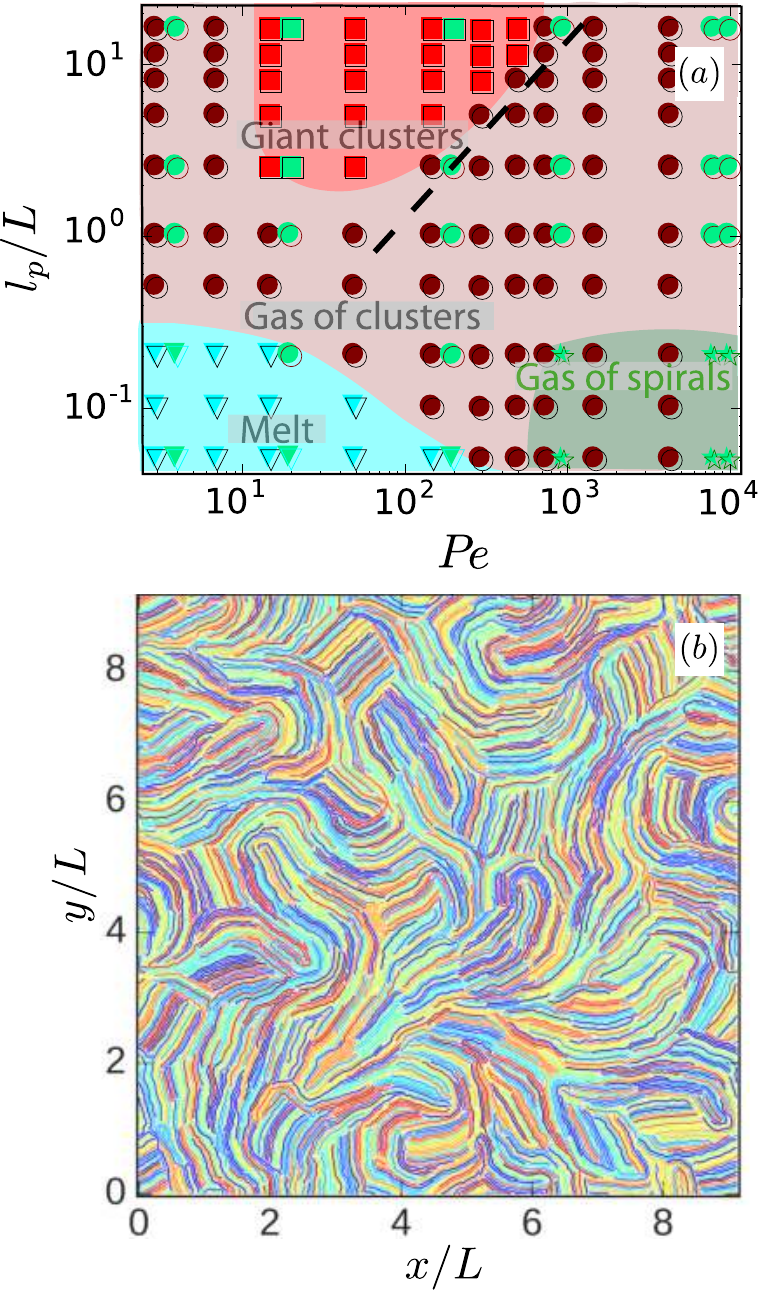}
	\end{center}
	\caption{(a) Phase diagram of tangentially driven semiflexible filaments as a function of $l_p/L =1/(2pL)$ and $Pe$, where the P\'eclet numbers is defined as $Pe=f^aL^2/(k_BT)$.  Cyan triangles indicate the melt phase, magenta circles the gas of clusters, and the red squares the giant clusters phase. Green filled symbols depict the simulations with high aspect ratio filaments, $L/l= 100$, while the other points correspond to $L/l= 25$.
	(b) Snapshot of a  high density turbulent  phase at $Pe = 90$. The filament colors are chosen randomly.\cite{duma:18} 
	``Reproduced with permission from Soft Matter {\bf 14}, 4483 (2018). Copyright 2018 The Royal Society of Chemistry.''} \label{fig:turbulence}
\end{figure}

\begin{figure*}[t]
	\begin{center}
		\includegraphics[width=\textwidth]{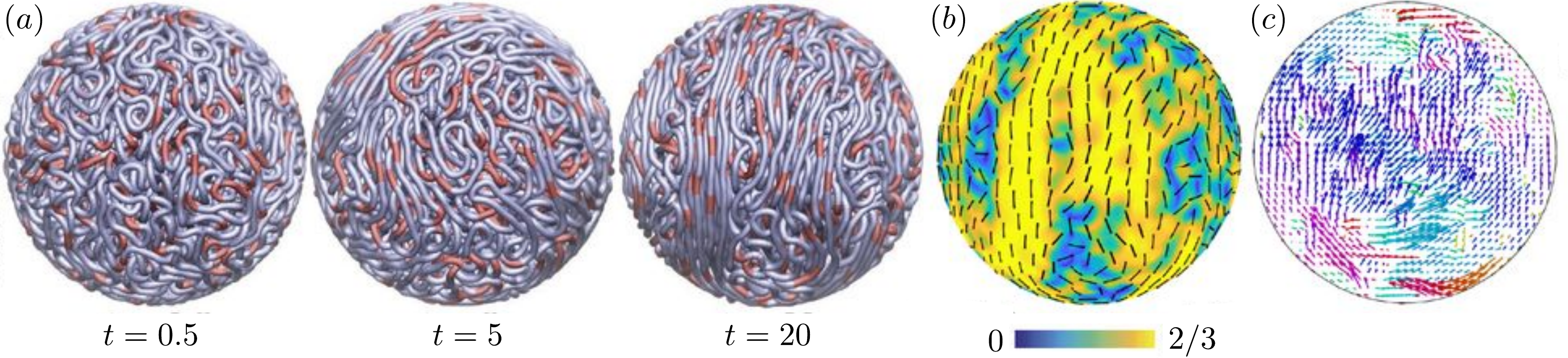}
	\end{center}
	\caption{Simulations of active chains with  extensile dipoles. (a) Filament configurations at different times; the red segments
 indicate instantaneous dipole locations. (b) The nematic structure of the filaments in a shell adjacent to the surface of the confining sphere. The color code  indicates  the  nematic order tensor.  (c) Chromatin displacement map calculated in a plane across the spherical domain over a time interval $\Delta t = 0.2$.   ``Reproduced with permission from  {\bf 115}, 1144263 (2018). Copyright 2018 National Academy of Sciences.''  } \label{fig:chromation}
\end{figure*}

At surfaces, various flagellated bacteria alter their morphologies as they become more elongated by suppression of cell division. These so called swarmer cells migrate collectively over surfaces and
are able to form stable aggregates, which can be highly motile (swarming).\cite{cope:09,kear:10}  Some cells, such as {\em Proteus mirabilis} or {\em Vibrio parahaemolyticus} become rather elongated  ($>10 - 100$ $\mu m$) and polymer-like.\cite{auer:19} Since they are propelled by flagella, which are shorter than the body length and distributed all over the body, locomotion of individual cell is mainly tangential.   Figure~\ref{fig:swarmer} shows that such cells can undergo large conformational changes from nearly rodlike to strongly bend structures while moving collectively in a film with other cells. As actin and microtubule filaments in dense suspensions,  the swarmer cells  exhibit local nematic order with defects and  large-scale collective motion.

The collective effects of tangentially driven filaments have been studied by simulations in 2D, and new phases depending on density, 
stiffness, activity, and aspect ratio have been identified.\cite{duma:18,prat:18}  At moderate densities and activities, 
stiff filaments organize into mobile clusters, structures which are reminiscent to those formed by self-propelled 
rods.\cite{peru:06,gine:10,wens:12,abke:13,abau:18,baer:20} High activities and low bending rigidity yield spiral formation of individual filaments, which then 
translate as compact disc-like objects.\cite{isel:15} With increasing density, spiral collisions yield again motile clusters.\cite{duma:18} 
At moderate densities, a reentrant behavior is observed with increasing activity, from small clusters to giant clusters and back to small
clusters.\cite{duma:18} Here, the disintegration of the giant cluster occurs when $Pe \sim \l_p/L$, i.e. when active forces
become comparable to bending forces. A similar threshold of $Pe \sim \l_p/L$ is seen for the transition from open (equilibrium-polymer-like)
conformations to spirals in the dilute case, due to the same force competition.
At higher densities,  a rich collective behavior is 
observed, such as a melt-like structures with topological defects.\cite{prat:18} Even higher densities lead to jammed 
states of semiflexible filaments at small P\'eclet numbers, followed by laning and active turbulence for higher activities.\cite{duma:18} 
It is important to emphasize that this active turbulence occurs in a system without momentum conservation, and therefore
at zero Reynolds number. Nevertheless, the turbulent state is characterized by a power-law decay of the kinetic-energy spectrum as a function
of wave vector, reminiscent of Kolmogorov turbulence.
The turbulent phase exhibits topological defects and a dynamics similar to the behavior seen theoretically in  active nematics and 
experimentally in solutions of microtubules in presence of molecular motors.\cite{sanc:12,doos:18,guil:18}

A flexible polymer composed of hydrodynamically self-propelled, pointlike, and extensile force dipoles (cf.  Sec.~\ref{sec:point_monomes}),  confined in a sphere, as a model of chromatin organized in the cell nucleus, shows large-scale coherent motion (cf. Fig.~\ref{fig:chromation}).  The extensile force, mimicking a local molecular motor,  reorganizes the polymer by stretching  it into  long and mutually aligned segments. This process is driven by the long-range  flows generated along the chain by motor activity, which tends to both straighten the polymer locally and to align nearby regions.\cite{saint:18}  As a result, large patches with high nematic order  appear. Studies of passive and contractile polymers show that this transition to a highly coherent state is linked with the extensile dipole, which emphasizes the interplay between connectivity of dipoles in a chain   and  hydrodynamic interactions mediated by the embedding fluid.\cite{saint:18}

\section{Conclusions and Outlook} \label{sec:summary}

We have demonstrated that the coupling of polymer degrees of freedom and active driving leads to many novel phenomena with respect to polymer conformational and  dynamical properties. Depending on the nature of the active process---random vs. tangential driving---polymers swell or collapse, and their  dynamics is typically enhanced. Moreover, hydrodynamic coupling of the flow field by the active monomeric units leads to long-range correlations and coherent motion. In ensembles of active polymers, steric and hydrodynamic interactions imply  novel long-range collective turbulent motion with large-scale patterns. The various aspects emphasize the uniqueness of systems comprised of active polymers and filaments, and  renders active soft matter a promising class of new materials.\cite{need:17,cate:11} 

We only begin to understand  and to unravel the properties of polymeric active-matter systems. The current perspective article describes theoretical and modeling approaches and major results of  mainly individual polymers and filaments. These approaches  establish the basis for further studies in various directions.

{\em Phase behavior}---As mentioned, active particles and filaments  exhibit novel collective phenomena and new phases such as MIPS. Typically two dimensional (2D) systems are considered, much less attention has been paid to three dimensional (3D) realizations. However, we can expect the appearance of novel structures and phases in 3D. First studies of active nematics  consisting of microtubules and kinesin molecular motor yield a  chaotic dynamics of the entire system.\cite{ducl:20} The analysis of spatial gradients of the director field show regions with large elastic distortions, which  mainly form  curvilinear structures as either isolated loops or belong to a complex network of system-spanning lines. These distortions are topological disclination lines characteristic of 3D nematics.\cite{ducl:20}   

Far less is know about the phase behavior of active Brownian polymers, both in 2D and 3D. Comparison of results for ABPs and active dumbbells in 2D indicate a shift of the critical activity for the onset of the MIPS to higher values.\cite{sieb:17} We can expect a further shift of the critical point for longer polymers, and a possible suppression of  the phase transition beyond a particular polymer length. 

{\em Confinement}---Confinement will alter the properties of the active polymers substantially. Already spherical ABPs accumulate at walls. We expect an even stronger accumulation for active polymers. As an example, simulations of D-ABPO confined in a slit yield  deviations in the scaling behavior of the wall force established for passive polymers.\cite{das:19.2}    

{\em Active-passive mixtures}---New phenomena appear in mixtures of active and passive components. Already systems of ABPs mixed with passive colloids exhibit phase separation\cite{sten:15,wyso:16,roge:20} and a collective interface dynamics.\cite{wyso:16}  Passive semiflexible polymers embedded in an active bath of ABPs  exhibit novel transient states in 2D,\cite{hard:14,niko:16,xia:19,xia:19.1} where an activity-induced bending of the polymers implies an asymmetric exposure to active particles, with  ABPs accumulating in regions of highest curvature, as has been observed for ABPs in confinement.\cite{fily:14} This leads to particular polymer conformations such as hairpins,  structures which  are only temporarily stable and dissolve and rebuild in the course of time.

The presence of both active and passive components is a hallmark of living systems. So far, the impact of activity on the properties of the passive components is unclear. Simulations  of two-component mixtures of polymers at  different temperatures, which is an alternative approach to establish a nonequilibrium state,\cite{gros:15} yield phase separation of the two components.\cite{smre:17} Here, the two temperatures  account for example for the activity of 
hetero- and euchromatin, which could play a role in chromation separation in the cell nucleus.\cite{lieb:09,crem:15,solo:16,gana:14,smre:17} Although temperature control is an acceptable way of implementing activity differences, other out-of-equilibrium processes should be considered as well. Cells exhibit coherent structures---so-called membraneless organelles or condensates---encompassing and concentrating specific molecules such as proteins and RNA in the cytoplasm.\cite{bran:11,fala:17,babi:20} {\em In-vivo} experiments suggest that active processes, which occur constantly within such organelles, play a role in their formation.\cite{bran:11,fala:17} An understanding of  the interplay between equilibrium thermodynamic driving forces and nonequilibrium activity in  organelle  formation  is fundamental in the strive to elucidate their functional properties, and their contribution to cell physiology and diseases.

{\em Active-passive copolymers}---Passive copolymer systems self-assemble  into ordered and tuneable structures in a wide range of morphologies, including spheres, cylinders, bicontinuous phases, lamellae, vesicles, and many other complex or hierarchical assemblies, driven by the  preferred attractive and repulsive interactions between the different constituents of the copolymers.\cite{bate:90,mai:12} Copolymers composed of monomers of  different  activity will show an even richer phase and dynamical behavior, and will provide additional means to control structure formation and transport. Since biological macromolecules are rather intrinsically disorder than homogeneous, studies of copolymers will improve our understanding of structure formation in biological cells, for active polymers as well as for passive copolymers in an active bath.  Specifically for polymeric assemblies of hydrodynamically self-propelled monomers, a  strong hydrodynamical monomer-monomer coupling can be expected, e.g., by monomers with different active stress or diameter, leading to a rich dynamics and particular self-organized structures.\cite{wagn:17} Assembling several active/passive monomers into an unit may even be a route to design intelligent active particles, which sense and respond autonomously to other units and show swarm-intelligent behavior as schools of fish or swarms of birds.\cite{marc:13,cava:14,popk:16,gomp:20}          

{\em Active turbulence}---As mentioned several times, active nematics and bacteria exhibit chaotic, turbulent behavior. Up to now, similarities or differences to  Kolmogorov-type turbulence\cite{kolm:91} of large Reynolds number fluids have not been resolved satisfactorily.  In particular, the impact of polymer degrees of freedom needs to be investigated.

{\em Hydrodynamic interactions}---Various of the discussed examples highlight the relevance  of hydrodynamic interactions for the conformational, dynamical, and collective effects of active polymers. Here, a broad range of studies are  required  to unravel hydrodynamics effects on the phase behavior of active polymers, their properties under confinement and under flow.   

{\em Rheology}---Experiments show that the viscosity of a dilute suspension of swimming bacteria is lower than that of the fluid medium.\cite{lope:15,mart:20.1} This is a consequence of the interplay between bacteria (pushers) alignment with the flow and stress generation by the microswimmers, which enhances the applied shear stress and leads to an apparent viscosity reduction that increases with increasing volume fraction of cells.\cite{mart:20.1} Even  ``superfluidity'' is obtained in bacterial suspensions at the onset of nonlinear flow and bacteria collective motion.\cite{lope:15,mart:20.1} Certainly, various of the observed aspects are of hydrodynamic origin and depend crucially on the flow field of the microswimmer. Analytical calculations of dilute dumbbell and D-ABPO systems already reveal an influence of activity on the zero-shear viscosity and  on the shear thinning behavior.\cite{wink:16,mart:18.1}  On a more macroscopic scale, shear experiments on long, slender, and entangled living worms ({\em Tubifex tubifex}) show  that shear thinning is reduced by activity and the concentration dependence of the low-shear viscosity exhibits a different scaling from that of regular polymers.\cite{bebl:20} This highlights the wide range of rheological phenomena in active matter. Clearly, systematic studies of various kinds of active polymers are needed to shed light onto their rheological behavior. Ultimately, such polymers might be useful as rheological modifiers.    

In summary, there is a wide range of interesting aspects of active polymers, ranging from  individual polymers with their different driving mechanisms,  to their collective properties as a function of  concentration, to rheological aspects. Simulations and analytical theories, which are based on the described models and approaches, will certainly play a decisive role in the elucidation of the many novel and unexpected nonequilibrium features of these systems.    
     
\section*{Data Availability}
The data that support the findings of this study are available from the corresponding author upon reasonable request.


\begin{acknowledgments}
We thank C. Abaurrea-Velasco, T. Auth, J. Clopes, O. Duman, T. Eisenstecken, J. Elgeti, D. A. Fedosov,  A. Ghavami, R. Isele-Holder, A. Martin Gomez, C. Philipps, A. Ravichandran,  M. Ripoll,  and G. A. Vliegenthart, for enjoyable collaborations and stimulating discussions. We gratefully acknowledge partial support from the DFG within the priority program SPP 1726 on “Microswimmers--from Single Particle Motion to Collective Behaviour”. A computing-time grant on the supercomputer JURECA at J\"ulich Supercomputing Centre (JSC) is thankfully acknowledged. 
\end{acknowledgments}
 
\section*{bibliography}
%

\begin{thebibliography}{263}%
\makeatletter
\providecommand \@ifxundefined [1]{%
 \@ifx{#1\undefined}
}%
\providecommand \@ifnum [1]{%
 \ifnum #1\expandafter \@firstoftwo
 \else \expandafter \@secondoftwo
 \fi
}%
\providecommand \@ifx [1]{%
 \ifx #1\expandafter \@firstoftwo
 \else \expandafter \@secondoftwo
 \fi
}%
\providecommand \natexlab [1]{#1}%
\providecommand \enquote  [1]{``#1''}%
\providecommand \bibnamefont  [1]{#1}%
\providecommand \bibfnamefont [1]{#1}%
\providecommand \citenamefont [1]{#1}%
\providecommand \href@noop [0]{\@secondoftwo}%
\providecommand \href [0]{\begingroup \@sanitize@url \@href}%
\providecommand \@href[1]{\@@startlink{#1}\@@href}%
\providecommand \@@href[1]{\endgroup#1\@@endlink}%
\providecommand \@sanitize@url [0]{\catcode `\\12\catcode `\$12\catcode
  `\&12\catcode `\#12\catcode `\^12\catcode `\_12\catcode `\%12\relax}%
\providecommand \@@startlink[1]{}%
\providecommand \@@endlink[0]{}%
\providecommand \url  [0]{\begingroup\@sanitize@url \@url }%
\providecommand \@url [1]{\endgroup\@href {#1}{\urlprefix }}%
\providecommand \urlprefix  [0]{URL }%
\providecommand \Eprint [0]{\href }%
\providecommand \doibase [0]{http://dx.doi.org/}%
\providecommand \selectlanguage [0]{\@gobble}%
\providecommand \bibinfo  [0]{\@secondoftwo}%
\providecommand \bibfield  [0]{\@secondoftwo}%
\providecommand \translation [1]{[#1]}%
\providecommand \BibitemOpen [0]{}%
\providecommand \bibitemStop [0]{}%
\providecommand \bibitemNoStop [0]{.\EOS\space}%
\providecommand \EOS [0]{\spacefactor3000\relax}%
\providecommand \BibitemShut  [1]{\csname bibitem#1\endcsname}%
\let\auto@bib@innerbib\@empty
\bibitem [{\citenamefont {Demirel}(2010)}]{demi:10}%
  \BibitemOpen
  \bibfield  {author} {\bibinfo {author} {\bibfnamefont {Y.}~\bibnamefont
  {Demirel}},\ }\bibfield  {title} {\enquote {\bibinfo {title} {Nonequilibrium
  thermodynamics modeling of coupled biochemical cycles in living cells},}\
  }\href {\doibase https://doi.org/10.1016/j.jnnfm.2010.02.006} {\bibfield
  {journal} {\bibinfo  {journal} {J. Non-Newtonian Fluid Mech.}\ }\textbf
  {\bibinfo {volume} {165}},\ \bibinfo {pages} {953} (\bibinfo {year}
  {2010})}\BibitemShut {NoStop}%
\bibitem [{\citenamefont {Fang}\ \emph {et~al.}(2019)\citenamefont {Fang},
  \citenamefont {Kruse}, \citenamefont {Lu},\ and\ \citenamefont
  {Wang}}]{fang:19}%
  \BibitemOpen
  \bibfield  {author} {\bibinfo {author} {\bibfnamefont {X.}~\bibnamefont
  {Fang}}, \bibinfo {author} {\bibfnamefont {K.}~\bibnamefont {Kruse}},
  \bibinfo {author} {\bibfnamefont {T.}~\bibnamefont {Lu}}, \ and\ \bibinfo
  {author} {\bibfnamefont {J.}~\bibnamefont {Wang}},\ }\bibfield  {title}
  {\enquote {\bibinfo {title} {Nonequilibrium physics in biology},}\ }\href
  {\doibase 10.1103/RevModPhys.91.045004} {\bibfield  {journal} {\bibinfo
  {journal} {Rev. Mod. Phys.}\ }\textbf {\bibinfo {volume} {91}},\ \bibinfo
  {pages} {045004} (\bibinfo {year} {2019})}\BibitemShut {NoStop}%
\bibitem [{\citenamefont {Kapral}(2013)}]{kapr:13}%
  \BibitemOpen
  \bibfield  {author} {\bibinfo {author} {\bibfnamefont {R.}~\bibnamefont
  {Kapral}},\ }\bibfield  {title} {\enquote {\bibinfo {title} {Perspective:
  Nanomotors without moving parts that propel themselves in solution},}\ }\href
  {https://doi.org/10.1063/1.4773981} {\bibfield  {journal} {\bibinfo
  {journal} {J. Chem. Phys.}\ }\textbf {\bibinfo {volume} {138}},\ \bibinfo
  {pages} {020901} (\bibinfo {year} {2013})}\BibitemShut {NoStop}%
\bibitem [{\citenamefont {Kapral}\ and\ \citenamefont
  {Mikhailov}(2016)}]{kapr:16}%
  \BibitemOpen
  \bibfield  {author} {\bibinfo {author} {\bibfnamefont {R.}~\bibnamefont
  {Kapral}}\ and\ \bibinfo {author} {\bibfnamefont {A.~S.}\ \bibnamefont
  {Mikhailov}},\ }\bibfield  {title} {\enquote {\bibinfo {title} {Stirring a
  fluid at low {R}eynolds numbers: Hydrodynamic collective effects of active
  proteins in biological cells},}\ }\href {\doibase
  10.1016/j.physd.2015.10.024} {\bibfield  {journal} {\bibinfo  {journal}
  {Physica D}\ }\textbf {\bibinfo {volume} {318-319}},\ \bibinfo {pages} {100}
  (\bibinfo {year} {2016})}\BibitemShut {NoStop}%
\bibitem [{\citenamefont {Marchetti}\ \emph {et~al.}(2013)\citenamefont
  {Marchetti}, \citenamefont {Joanny}, \citenamefont {Ramaswamy}, \citenamefont
  {Liverpool}, \citenamefont {Prost}, \citenamefont {Rao},\ and\ \citenamefont
  {Simha}}]{marc:13}%
  \BibitemOpen
  \bibfield  {author} {\bibinfo {author} {\bibfnamefont {M.~C.}\ \bibnamefont
  {Marchetti}}, \bibinfo {author} {\bibfnamefont {J.~F.}\ \bibnamefont
  {Joanny}}, \bibinfo {author} {\bibfnamefont {S.}~\bibnamefont {Ramaswamy}},
  \bibinfo {author} {\bibfnamefont {T.~B.}\ \bibnamefont {Liverpool}}, \bibinfo
  {author} {\bibfnamefont {J.}~\bibnamefont {Prost}}, \bibinfo {author}
  {\bibfnamefont {M.}~\bibnamefont {Rao}}, \ and\ \bibinfo {author}
  {\bibfnamefont {R.~A.}\ \bibnamefont {Simha}},\ }\bibfield  {title} {\enquote
  {\bibinfo {title} {Hydrodynamics of soft active matter},}\ }\href {\doibase
  10.1103/RevModPhys.85.1143} {\bibfield  {journal} {\bibinfo  {journal} {Rev.
  Mod. Phys.}\ }\textbf {\bibinfo {volume} {85}},\ \bibinfo {pages} {1143}
  (\bibinfo {year} {2013})}\BibitemShut {NoStop}%
\bibitem [{\citenamefont {Elgeti}, \citenamefont {Winkler},\ and\ \citenamefont
  {Gompper}(2015)}]{elge:15}%
  \BibitemOpen
  \bibfield  {author} {\bibinfo {author} {\bibfnamefont {J.}~\bibnamefont
  {Elgeti}}, \bibinfo {author} {\bibfnamefont {R.~G.}\ \bibnamefont {Winkler}},
  \ and\ \bibinfo {author} {\bibfnamefont {G.}~\bibnamefont {Gompper}},\
  }\bibfield  {title} {\enquote {\bibinfo {title} {Physics of
  microswimmers---single particle motion and collective behavior: a review},}\
  }\href {\doibase 10.1088/0034-4885/78/5/056601} {\bibfield  {journal}
  {\bibinfo  {journal} {Rep. Prog. Phys.}\ }\textbf {\bibinfo {volume} {78}},\
  \bibinfo {pages} {056601} (\bibinfo {year} {2015})}\BibitemShut {NoStop}%
\bibitem [{\citenamefont {J{\"u}licher}, \citenamefont {Grill},\ and\
  \citenamefont {Salbreux}(2018)}]{juli:18}%
  \BibitemOpen
  \bibfield  {author} {\bibinfo {author} {\bibfnamefont {F.}~\bibnamefont
  {J{\"u}licher}}, \bibinfo {author} {\bibfnamefont {S.~W.}\ \bibnamefont
  {Grill}}, \ and\ \bibinfo {author} {\bibfnamefont {G.}~\bibnamefont
  {Salbreux}},\ }\bibfield  {title} {\enquote {\bibinfo {title} {Hydrodynamic
  theory of active matter},}\ }\href {\doibase 10.1088/1361-6633/aab6bb}
  {\bibfield  {journal} {\bibinfo  {journal} {Rep. Prog. Phys.}\ }\textbf
  {\bibinfo {volume} {81}},\ \bibinfo {pages} {076601} (\bibinfo {year}
  {2018})}\BibitemShut {NoStop}%
\bibitem [{\citenamefont {Hakim}\ and\ \citenamefont
  {Silberzan}(2017)}]{haki:17}%
  \BibitemOpen
  \bibfield  {author} {\bibinfo {author} {\bibfnamefont {V.}~\bibnamefont
  {Hakim}}\ and\ \bibinfo {author} {\bibfnamefont {P.}~\bibnamefont
  {Silberzan}},\ }\bibfield  {title} {\enquote {\bibinfo {title} {Collective
  cell migration: a physics perspective},}\ }\href {\doibase
  10.1088/1361-6633/aa65ef} {\bibfield  {journal} {\bibinfo  {journal} {Rep.
  Prog. Phys.}\ }\textbf {\bibinfo {volume} {80}},\ \bibinfo {pages} {076601}
  (\bibinfo {year} {2017})}\BibitemShut {NoStop}%
\bibitem [{\citenamefont {Be'er}\ and\ \citenamefont {Ariel}(2019)}]{beer:19}%
  \BibitemOpen
  \bibfield  {author} {\bibinfo {author} {\bibfnamefont {A.}~\bibnamefont
  {Be'er}}\ and\ \bibinfo {author} {\bibfnamefont {G.}~\bibnamefont {Ariel}},\
  }\bibfield  {title} {\enquote {\bibinfo {title} {A statistical physics view
  of swarming bacteria},}\ }\href {\doibase 10.1186/s40462-019-0153-9}
  {\bibfield  {journal} {\bibinfo  {journal} {Mov. Ecol.}\ }\textbf {\bibinfo
  {volume} {7}},\ \bibinfo {pages} {9} (\bibinfo {year} {2019})}\BibitemShut
  {NoStop}%
\bibitem [{\citenamefont {Caspi}, \citenamefont {Granek},\ and\ \citenamefont
  {Elbaum}(2000)}]{casp:00}%
  \BibitemOpen
  \bibfield  {author} {\bibinfo {author} {\bibfnamefont {A.}~\bibnamefont
  {Caspi}}, \bibinfo {author} {\bibfnamefont {R.}~\bibnamefont {Granek}}, \
  and\ \bibinfo {author} {\bibfnamefont {M.}~\bibnamefont {Elbaum}},\
  }\bibfield  {title} {\enquote {\bibinfo {title} {Enhanced diffusion in active
  intracellular transport},}\ }\href {\doibase 10.1103/PhysRevLett.85.5655}
  {\bibfield  {journal} {\bibinfo  {journal} {Phys. Rev. Lett.}\ }\textbf
  {\bibinfo {volume} {85}},\ \bibinfo {pages} {5655} (\bibinfo {year}
  {2000})}\BibitemShut {NoStop}%
\bibitem [{\citenamefont {Lau}\ \emph {et~al.}(2003)\citenamefont {Lau},
  \citenamefont {Hoffman}, \citenamefont {Davies}, \citenamefont {Crocker},\
  and\ \citenamefont {Lubensky}}]{lau:03}%
  \BibitemOpen
  \bibfield  {author} {\bibinfo {author} {\bibfnamefont {A.~W.~C.}\
  \bibnamefont {Lau}}, \bibinfo {author} {\bibfnamefont {B.~D.}\ \bibnamefont
  {Hoffman}}, \bibinfo {author} {\bibfnamefont {A.}~\bibnamefont {Davies}},
  \bibinfo {author} {\bibfnamefont {J.~C.}\ \bibnamefont {Crocker}}, \ and\
  \bibinfo {author} {\bibfnamefont {T.~C.}\ \bibnamefont {Lubensky}},\
  }\bibfield  {title} {\enquote {\bibinfo {title} {Microrheology, stress
  fluctuations, and active behavior of living cells},}\ }\href {\doibase
  10.1103/PhysRevLett.91.198101} {\bibfield  {journal} {\bibinfo  {journal}
  {Phys. Rev. Lett.}\ }\textbf {\bibinfo {volume} {91}},\ \bibinfo {pages}
  {198101} (\bibinfo {year} {2003})}\BibitemShut {NoStop}%
\bibitem [{\citenamefont {Brangwynne}\ \emph {et~al.}(2008)\citenamefont
  {Brangwynne}, \citenamefont {Koenderink}, \citenamefont {MacKintosh},\ and\
  \citenamefont {Weitz}}]{bran:08}%
  \BibitemOpen
  \bibfield  {author} {\bibinfo {author} {\bibfnamefont {C.~P.}\ \bibnamefont
  {Brangwynne}}, \bibinfo {author} {\bibfnamefont {G.~H.}\ \bibnamefont
  {Koenderink}}, \bibinfo {author} {\bibfnamefont {F.~C.}\ \bibnamefont
  {MacKintosh}}, \ and\ \bibinfo {author} {\bibfnamefont {D.~A.}\ \bibnamefont
  {Weitz}},\ }\bibfield  {title} {\enquote {\bibinfo {title} {Cytoplasmic
  diffusion: molecular motors mix it up},}\ }\href {\doibase
  10.1083/jcb.200806149} {\bibfield  {journal} {\bibinfo  {journal} {J. Cell.
  Biol.}\ }\textbf {\bibinfo {volume} {183}},\ \bibinfo {pages} {583} (\bibinfo
  {year} {2008})}\BibitemShut {NoStop}%
\bibitem [{\citenamefont {Robert}\ \emph {et~al.}(2010)\citenamefont {Robert},
  \citenamefont {Nguyen}, \citenamefont {Gallet},\ and\ \citenamefont
  {Wilhelm}}]{robe:10}%
  \BibitemOpen
  \bibfield  {author} {\bibinfo {author} {\bibfnamefont {D.}~\bibnamefont
  {Robert}}, \bibinfo {author} {\bibfnamefont {T.-H.}\ \bibnamefont {Nguyen}},
  \bibinfo {author} {\bibfnamefont {F.}~\bibnamefont {Gallet}}, \ and\ \bibinfo
  {author} {\bibfnamefont {C.}~\bibnamefont {Wilhelm}},\ }\bibfield  {title}
  {\enquote {\bibinfo {title} {In vivo determination of fluctuating forces
  during endosome trafficking using a combination of active and passive
  microrheology},}\ }\href {\doibase 10.1371/journal.pone.0010046} {\bibfield
  {journal} {\bibinfo  {journal} {PLOS ONE}\ }\textbf {\bibinfo {volume} {5}},\
  \bibinfo {pages} {e10046} (\bibinfo {year} {2010})}\BibitemShut {NoStop}%
\bibitem [{\citenamefont {Fakhri}\ \emph {et~al.}(2014)\citenamefont {Fakhri},
  \citenamefont {Wessel}, \citenamefont {Willms}, \citenamefont {Pasquali},
  \citenamefont {Klopfenstein}, \citenamefont {MacKintosh},\ and\ \citenamefont
  {Schmidt}}]{fakh:14}%
  \BibitemOpen
  \bibfield  {author} {\bibinfo {author} {\bibfnamefont {N.}~\bibnamefont
  {Fakhri}}, \bibinfo {author} {\bibfnamefont {A.~D.}\ \bibnamefont {Wessel}},
  \bibinfo {author} {\bibfnamefont {C.}~\bibnamefont {Willms}}, \bibinfo
  {author} {\bibfnamefont {M.}~\bibnamefont {Pasquali}}, \bibinfo {author}
  {\bibfnamefont {D.~R.}\ \bibnamefont {Klopfenstein}}, \bibinfo {author}
  {\bibfnamefont {F.~C.}\ \bibnamefont {MacKintosh}}, \ and\ \bibinfo {author}
  {\bibfnamefont {C.~F.}\ \bibnamefont {Schmidt}},\ }\bibfield  {title}
  {\enquote {\bibinfo {title} {High-resolution mapping of intracellular
  fluctuations using carbon nanotubes},}\ }\href {\doibase
  10.1126/science.1250170} {\bibfield  {journal} {\bibinfo  {journal}
  {Science}\ }\textbf {\bibinfo {volume} {344}},\ \bibinfo {pages} {1031}
  (\bibinfo {year} {2014})}\BibitemShut {NoStop}%
\bibitem [{\citenamefont {Guo}\ \emph {et~al.}(2014)\citenamefont {Guo},
  \citenamefont {Ehrlicher}, \citenamefont {Jensen}, \citenamefont {Renz},
  \citenamefont {Moore}, \citenamefont {Goldman}, \citenamefont
  {Lippincott-Schwartz}, \citenamefont {Mackintosh},\ and\ \citenamefont
  {Weitz}}]{guo:14}%
  \BibitemOpen
  \bibfield  {author} {\bibinfo {author} {\bibfnamefont {M.}~\bibnamefont
  {Guo}}, \bibinfo {author} {\bibfnamefont {A.~J.}\ \bibnamefont {Ehrlicher}},
  \bibinfo {author} {\bibfnamefont {M.~H.}\ \bibnamefont {Jensen}}, \bibinfo
  {author} {\bibfnamefont {M.}~\bibnamefont {Renz}}, \bibinfo {author}
  {\bibfnamefont {J.~R.}\ \bibnamefont {Moore}}, \bibinfo {author}
  {\bibfnamefont {R.~D.}\ \bibnamefont {Goldman}}, \bibinfo {author}
  {\bibfnamefont {J.}~\bibnamefont {Lippincott-Schwartz}}, \bibinfo {author}
  {\bibfnamefont {F.~C.}\ \bibnamefont {Mackintosh}}, \ and\ \bibinfo {author}
  {\bibfnamefont {D.~A.}\ \bibnamefont {Weitz}},\ }\bibfield  {title} {\enquote
  {\bibinfo {title} {Probing the stochastic, motor-driven properties of the
  cytoplasm using force spectrum microscopy},}\ }\href {\doibase
  https://doi.org/10.1016/j.cell.2014.06.051} {\bibfield  {journal} {\bibinfo
  {journal} {Cell}\ }\textbf {\bibinfo {volume} {158}},\ \bibinfo {pages} {822}
  (\bibinfo {year} {2014})}\BibitemShut {NoStop}%
\bibitem [{\citenamefont {Parry}\ \emph {et~al.}(2014)\citenamefont {Parry},
  \citenamefont {Surovtsev}, \citenamefont {Cabeen}, \citenamefont {O'Hern},
  \citenamefont {Dufresne},\ and\ \citenamefont {Jacobs-Wagner}}]{parr:14}%
  \BibitemOpen
  \bibfield  {author} {\bibinfo {author} {\bibfnamefont {B.~R.}\ \bibnamefont
  {Parry}}, \bibinfo {author} {\bibfnamefont {I.~V.}\ \bibnamefont
  {Surovtsev}}, \bibinfo {author} {\bibfnamefont {M.~T.}\ \bibnamefont
  {Cabeen}}, \bibinfo {author} {\bibfnamefont {C.~S.}\ \bibnamefont {O'Hern}},
  \bibinfo {author} {\bibfnamefont {E.~R.}\ \bibnamefont {Dufresne}}, \ and\
  \bibinfo {author} {\bibfnamefont {C.}~\bibnamefont {Jacobs-Wagner}},\
  }\bibfield  {title} {\enquote {\bibinfo {title} {The bacterial cytoplasm has
  glass-like properties and is fluidized by metabolic activity},}\ }\href
  {\doibase https://doi.org/10.1016/j.cell.2013.11.028} {\bibfield  {journal}
  {\bibinfo  {journal} {Cell}\ }\textbf {\bibinfo {volume} {156}},\ \bibinfo
  {pages} {183} (\bibinfo {year} {2014})}\BibitemShut {NoStop}%
\bibitem [{\citenamefont {Golestanian}(2015)}]{gole:15}%
  \BibitemOpen
  \bibfield  {author} {\bibinfo {author} {\bibfnamefont {R.}~\bibnamefont
  {Golestanian}},\ }\bibfield  {title} {\enquote {\bibinfo {title} {Enhanced
  diffusion of enzymes that catalyze exothermic reactions},}\ }\href {\doibase
  10.1103/PhysRevLett.115.108102} {\bibfield  {journal} {\bibinfo  {journal}
  {Phys. Rev. Lett.}\ }\textbf {\bibinfo {volume} {115}},\ \bibinfo {pages}
  {108102} (\bibinfo {year} {2015})}\BibitemShut {NoStop}%
\bibitem [{\citenamefont {Mikhailov}\ and\ \citenamefont
  {Kapral}(2015)}]{mikh:15}%
  \BibitemOpen
  \bibfield  {author} {\bibinfo {author} {\bibfnamefont {A.~S.}\ \bibnamefont
  {Mikhailov}}\ and\ \bibinfo {author} {\bibfnamefont {R.}~\bibnamefont
  {Kapral}},\ }\bibfield  {title} {\enquote {\bibinfo {title} {Hydrodynamic
  collective effects of active protein machines in solution and lipid
  bilayers},}\ }\href {\doibase 10.1073/pnas.1506825112} {\bibfield  {journal}
  {\bibinfo  {journal} {Proc. Natl. Acad. Sci. USA}\ }\textbf {\bibinfo
  {volume} {112}},\ \bibinfo {pages} {E3639} (\bibinfo {year}
  {2015})}\BibitemShut {NoStop}%
\bibitem [{\citenamefont {Weber}, \citenamefont {Spakowitz},\ and\
  \citenamefont {Theriot}(2012)}]{webe:12}%
  \BibitemOpen
  \bibfield  {author} {\bibinfo {author} {\bibfnamefont {S.~C.}\ \bibnamefont
  {Weber}}, \bibinfo {author} {\bibfnamefont {A.~J.}\ \bibnamefont
  {Spakowitz}}, \ and\ \bibinfo {author} {\bibfnamefont {J.~A.}\ \bibnamefont
  {Theriot}},\ }\bibfield  {title} {\enquote {\bibinfo {title} {Nonthermal
  {ATP}-dependent fluctuations contribute to the in vivo motion of chromosomal
  loci},}\ }\href {\doibase 10.1073/pnas.1119505109} {\bibfield  {journal}
  {\bibinfo  {journal} {Proc. Natl. Acad. Sci. USA}\ }\textbf {\bibinfo
  {volume} {109}},\ \bibinfo {pages} {7338} (\bibinfo {year}
  {2012})}\BibitemShut {NoStop}%
\bibitem [{\citenamefont {Gnesotto}\ \emph {et~al.}(2018)\citenamefont
  {Gnesotto}, \citenamefont {Mura}, \citenamefont {Gladrow},\ and\
  \citenamefont {Broedersz}}]{gnes:18}%
  \BibitemOpen
  \bibfield  {author} {\bibinfo {author} {\bibfnamefont {F.~S.}\ \bibnamefont
  {Gnesotto}}, \bibinfo {author} {\bibfnamefont {F.}~\bibnamefont {Mura}},
  \bibinfo {author} {\bibfnamefont {J.}~\bibnamefont {Gladrow}}, \ and\
  \bibinfo {author} {\bibfnamefont {C.~P.}\ \bibnamefont {Broedersz}},\
  }\bibfield  {title} {\enquote {\bibinfo {title} {Broken detailed balance and
  non-equilibrium dynamics in living systems: a review},}\ }\href {\doibase
  10.1088/1361-6633/aab3ed} {\bibfield  {journal} {\bibinfo  {journal} {Rep.
  Prog. Phys.}\ }\textbf {\bibinfo {volume} {81}},\ \bibinfo {pages} {066601}
  (\bibinfo {year} {2018})}\BibitemShut {NoStop}%
\bibitem [{\citenamefont {Wu}\ \emph {et~al.}(2019)\citenamefont {Wu},
  \citenamefont {Japaridze}, \citenamefont {Zheng}, \citenamefont {Wiktor},
  \citenamefont {Kerssemakers},\ and\ \citenamefont {Dekker}}]{wu:19}%
  \BibitemOpen
  \bibfield  {author} {\bibinfo {author} {\bibfnamefont {F.}~\bibnamefont
  {Wu}}, \bibinfo {author} {\bibfnamefont {A.}~\bibnamefont {Japaridze}},
  \bibinfo {author} {\bibfnamefont {X.}~\bibnamefont {Zheng}}, \bibinfo
  {author} {\bibfnamefont {J.}~\bibnamefont {Wiktor}}, \bibinfo {author}
  {\bibfnamefont {J.~W.~J.}\ \bibnamefont {Kerssemakers}}, \ and\ \bibinfo
  {author} {\bibfnamefont {C.}~\bibnamefont {Dekker}},\ }\bibfield  {title}
  {\enquote {\bibinfo {title} {Direct imaging of the circular chromosome in a
  live bacterium},}\ }\href {\doibase 10.1038/s41467-019-10221-0} {\bibfield
  {journal} {\bibinfo  {journal} {Nat. Commun.}\ }\textbf {\bibinfo {volume}
  {10}},\ \bibinfo {pages} {2194} (\bibinfo {year} {2019})}\BibitemShut
  {NoStop}%
\bibitem [{\citenamefont {MacKintosh}\ and\ \citenamefont
  {Levine}(2008)}]{mack:08}%
  \BibitemOpen
  \bibfield  {author} {\bibinfo {author} {\bibfnamefont {F.~C.}\ \bibnamefont
  {MacKintosh}}\ and\ \bibinfo {author} {\bibfnamefont {A.~J.}\ \bibnamefont
  {Levine}},\ }\bibfield  {title} {\enquote {\bibinfo {title} {Nonequilibrium
  mechanics and dynamics of motor-activated gels},}\ }\href {\doibase
  10.1103/PhysRevLett.100.018104} {\bibfield  {journal} {\bibinfo  {journal}
  {Phys. Rev. Lett.}\ }\textbf {\bibinfo {volume} {100}},\ \bibinfo {pages}
  {018104} (\bibinfo {year} {2008})}\BibitemShut {NoStop}%
\bibitem [{\citenamefont {Lu}\ \emph {et~al.}(2016)\citenamefont {Lu},
  \citenamefont {Winding}, \citenamefont {Lakonishok}, \citenamefont
  {Wildonger},\ and\ \citenamefont {Gelfand}}]{lu:16}%
  \BibitemOpen
  \bibfield  {author} {\bibinfo {author} {\bibfnamefont {W.}~\bibnamefont
  {Lu}}, \bibinfo {author} {\bibfnamefont {M.}~\bibnamefont {Winding}},
  \bibinfo {author} {\bibfnamefont {M.}~\bibnamefont {Lakonishok}}, \bibinfo
  {author} {\bibfnamefont {J.}~\bibnamefont {Wildonger}}, \ and\ \bibinfo
  {author} {\bibfnamefont {V.~I.}\ \bibnamefont {Gelfand}},\ }\bibfield
  {title} {\enquote {\bibinfo {title} {Microtubule--microtubule sliding by
  kinesin-1 is essential for normal cytoplasmic streaming in
  \emph{{D}rosophila} oocytes},}\ }\href {\doibase 10.1073/pnas.1522424113}
  {\bibfield  {journal} {\bibinfo  {journal} {Proc. Natl. Acad. Sci. USA}\
  }\textbf {\bibinfo {volume} {113}},\ \bibinfo {pages} {E4995} (\bibinfo
  {year} {2016})}\BibitemShut {NoStop}%
\bibitem [{\citenamefont {Ravichandran}\ \emph {et~al.}(2017)\citenamefont
  {Ravichandran}, \citenamefont {Vliegenthart}, \citenamefont {Saggiorato},
  \citenamefont {Auth},\ and\ \citenamefont {Gompper}}]{ravi:17}%
  \BibitemOpen
  \bibfield  {author} {\bibinfo {author} {\bibfnamefont {A.}~\bibnamefont
  {Ravichandran}}, \bibinfo {author} {\bibfnamefont {G.~A.}\ \bibnamefont
  {Vliegenthart}}, \bibinfo {author} {\bibfnamefont {G.}~\bibnamefont
  {Saggiorato}}, \bibinfo {author} {\bibfnamefont {T.}~\bibnamefont {Auth}}, \
  and\ \bibinfo {author} {\bibfnamefont {G.}~\bibnamefont {Gompper}},\
  }\bibfield  {title} {\enquote {\bibinfo {title} {Enhanced dynamics of
  confined cytoskeletal filaments driven by asymmetric motors},}\ }\href
  {\doibase https://doi.org/10.1016/j.bpj.2017.07.016} {\bibfield  {journal}
  {\bibinfo  {journal} {Biophys. J.}\ }\textbf {\bibinfo {volume} {113}},\
  \bibinfo {pages} {1121} (\bibinfo {year} {2017})}\BibitemShut {NoStop}%
\bibitem [{\citenamefont {Weber}\ \emph {et~al.}(2015)\citenamefont {Weber},
  \citenamefont {Suzuki}, \citenamefont {Schaller}, \citenamefont {Aranson},
  \citenamefont {Bausch},\ and\ \citenamefont {Frey}}]{webe:15}%
  \BibitemOpen
  \bibfield  {author} {\bibinfo {author} {\bibfnamefont {C.~A.}\ \bibnamefont
  {Weber}}, \bibinfo {author} {\bibfnamefont {R.}~\bibnamefont {Suzuki}},
  \bibinfo {author} {\bibfnamefont {V.}~\bibnamefont {Schaller}}, \bibinfo
  {author} {\bibfnamefont {I.~S.}\ \bibnamefont {Aranson}}, \bibinfo {author}
  {\bibfnamefont {A.~R.}\ \bibnamefont {Bausch}}, \ and\ \bibinfo {author}
  {\bibfnamefont {E.}~\bibnamefont {Frey}},\ }\bibfield  {title} {\enquote
  {\bibinfo {title} {Random bursts determine dynamics of active filaments},}\
  }\href {\doibase 10.1073/pnas.1421322112} {\bibfield  {journal} {\bibinfo
  {journal} {Proc. Natl. Acad. Sci. USA}\ }\textbf {\bibinfo {volume} {112}},\
  \bibinfo {pages} {10703} (\bibinfo {year} {2015})}\BibitemShut {NoStop}%
\bibitem [{\citenamefont {Guthold}\ \emph {et~al.}(1999)\citenamefont
  {Guthold}, \citenamefont {Zhu}, \citenamefont {Rivetti}, \citenamefont
  {Yang}, \citenamefont {Thomson}, \citenamefont {Kasas}, \citenamefont
  {Hansma}, \citenamefont {Smith}, \citenamefont {Hansma},\ and\ \citenamefont
  {Bustamante}}]{guth:99}%
  \BibitemOpen
  \bibfield  {author} {\bibinfo {author} {\bibfnamefont {M.}~\bibnamefont
  {Guthold}}, \bibinfo {author} {\bibfnamefont {X.}~\bibnamefont {Zhu}},
  \bibinfo {author} {\bibfnamefont {C.}~\bibnamefont {Rivetti}}, \bibinfo
  {author} {\bibfnamefont {G.}~\bibnamefont {Yang}}, \bibinfo {author}
  {\bibfnamefont {N.~H.}\ \bibnamefont {Thomson}}, \bibinfo {author}
  {\bibfnamefont {S.}~\bibnamefont {Kasas}}, \bibinfo {author} {\bibfnamefont
  {H.~G.}\ \bibnamefont {Hansma}}, \bibinfo {author} {\bibfnamefont
  {B.}~\bibnamefont {Smith}}, \bibinfo {author} {\bibfnamefont {P.~K.}\
  \bibnamefont {Hansma}}, \ and\ \bibinfo {author} {\bibfnamefont
  {C.}~\bibnamefont {Bustamante}},\ }\bibfield  {title} {\enquote {\bibinfo
  {title} {Direct observation of one-dimensional diffusion and transcription by
  \emph{{E}scherichia coli} {RNA} polymerase},}\ }\href {\doibase
  https://doi.org/10.1016/S0006-3495(99)77067-0} {\bibfield  {journal}
  {\bibinfo  {journal} {Biophys. J.}\ }\textbf {\bibinfo {volume} {77}},\
  \bibinfo {pages} {2284} (\bibinfo {year} {1999})}\BibitemShut {NoStop}%
\bibitem [{\citenamefont {Mejia}, \citenamefont {Nudler},\ and\ \citenamefont
  {Bustamante}(2015)}]{meji:15}%
  \BibitemOpen
  \bibfield  {author} {\bibinfo {author} {\bibfnamefont {Y.~X.}\ \bibnamefont
  {Mejia}}, \bibinfo {author} {\bibfnamefont {E.}~\bibnamefont {Nudler}}, \
  and\ \bibinfo {author} {\bibfnamefont {C.}~\bibnamefont {Bustamante}},\
  }\bibfield  {title} {\enquote {\bibinfo {title} {Trigger loop folding
  determines transcription rate of \emph{{E}scherichia coli's} {RNA}
  polymerase},}\ }\href {\doibase 10.1073/pnas.1421067112} {\bibfield
  {journal} {\bibinfo  {journal} {Proc. Natl. Acad. Sci. USA}\ }\textbf
  {\bibinfo {volume} {112}},\ \bibinfo {pages} {743} (\bibinfo {year}
  {2015})}\BibitemShut {NoStop}%
\bibitem [{\citenamefont {Belitsky}\ and\ \citenamefont
  {Sch{\"u}tz}(2019)}]{beli:19}%
  \BibitemOpen
  \bibfield  {author} {\bibinfo {author} {\bibfnamefont {V.}~\bibnamefont
  {Belitsky}}\ and\ \bibinfo {author} {\bibfnamefont {G.~M.}\ \bibnamefont
  {Sch{\"u}tz}},\ }\bibfield  {title} {\enquote {\bibinfo {title} {Stationary
  {RNA} polymerase fluctuations during transcription elongation},}\ }\href
  {\doibase 10.1103/PhysRevE.99.012405} {\bibfield  {journal} {\bibinfo
  {journal} {Phys. Rev. E}\ }\textbf {\bibinfo {volume} {99}},\ \bibinfo
  {pages} {012405} (\bibinfo {year} {2019})}\BibitemShut {NoStop}%
\bibitem [{\citenamefont {Di~Pierro}\ \emph {et~al.}(2018)\citenamefont
  {Di~Pierro}, \citenamefont {Potoyan}, \citenamefont {Wolynes},\ and\
  \citenamefont {Onuchic}}]{dipi:18}%
  \BibitemOpen
  \bibfield  {author} {\bibinfo {author} {\bibfnamefont {M.}~\bibnamefont
  {Di~Pierro}}, \bibinfo {author} {\bibfnamefont {D.~A.}\ \bibnamefont
  {Potoyan}}, \bibinfo {author} {\bibfnamefont {P.~G.}\ \bibnamefont
  {Wolynes}}, \ and\ \bibinfo {author} {\bibfnamefont {J.}~\bibnamefont
  {Onuchic}},\ }\bibfield  {title} {\enquote {\bibinfo {title} {Anomalous
  diffusion, spatial coherence, and viscoelasticity from the energy landscape
  of human chromosomes},}\ }\href {\doibase 10.1073/pnas.1806297115} {\bibfield
   {journal} {\bibinfo  {journal} {Proc. Natl. Acad. Sci. USA}\ }\textbf
  {\bibinfo {volume} {115}},\ \bibinfo {pages} {7753} (\bibinfo {year}
  {2018})}\BibitemShut {NoStop}%
\bibitem [{\citenamefont {Javer}\ \emph {et~al.}(2013)\citenamefont {Javer},
  \citenamefont {Long}, \citenamefont {Nugent}, \citenamefont {Grisi},
  \citenamefont {Siriwatwetchakul}, \citenamefont {Dorfman}, \citenamefont
  {Cicuta},\ and\ \citenamefont {Cosentino~Lagomarsino}}]{jave:13}%
  \BibitemOpen
  \bibfield  {author} {\bibinfo {author} {\bibfnamefont {A.}~\bibnamefont
  {Javer}}, \bibinfo {author} {\bibfnamefont {Z.}~\bibnamefont {Long}},
  \bibinfo {author} {\bibfnamefont {E.}~\bibnamefont {Nugent}}, \bibinfo
  {author} {\bibfnamefont {M.}~\bibnamefont {Grisi}}, \bibinfo {author}
  {\bibfnamefont {K.}~\bibnamefont {Siriwatwetchakul}}, \bibinfo {author}
  {\bibfnamefont {K.~D.}\ \bibnamefont {Dorfman}}, \bibinfo {author}
  {\bibfnamefont {P.}~\bibnamefont {Cicuta}}, \ and\ \bibinfo {author}
  {\bibfnamefont {M.}~\bibnamefont {Cosentino~Lagomarsino}},\ }\bibfield
  {title} {\enquote {\bibinfo {title} {Short-time movement of {{E.}} {coli}
  chromosomal loci depends on coordinate and subcellular localization},}\
  }\href {\doibase 10.1038/ncomms3003} {\bibfield  {journal} {\bibinfo
  {journal} {Nat. Commun.}\ }\textbf {\bibinfo {volume} {4}},\ \bibinfo {pages}
  {3003} (\bibinfo {year} {2013})}\BibitemShut {NoStop}%
\bibitem [{\citenamefont {Zidovska}, \citenamefont {Weitz},\ and\ \citenamefont
  {Mitchison}(2013)}]{zido:13}%
  \BibitemOpen
  \bibfield  {author} {\bibinfo {author} {\bibfnamefont {A.}~\bibnamefont
  {Zidovska}}, \bibinfo {author} {\bibfnamefont {D.~A.}\ \bibnamefont {Weitz}},
  \ and\ \bibinfo {author} {\bibfnamefont {T.~J.}\ \bibnamefont {Mitchison}},\
  }\bibfield  {title} {\enquote {\bibinfo {title} {Micron-scale coherence in
  interphase chromatin dynamics},}\ }\href
  {http://www.pnas.org/content/110/39/15555.abstract} {\bibfield  {journal}
  {\bibinfo  {journal} {Proc. Natl. Acad. Sci. USA}\ }\textbf {\bibinfo
  {volume} {110}},\ \bibinfo {pages} {15555} (\bibinfo {year}
  {2013})}\BibitemShut {NoStop}%
\bibitem [{\citenamefont {Lieberman-Aiden}\ \emph {et~al.}(2009)\citenamefont
  {Lieberman-Aiden}, \citenamefont {van Berkum}, \citenamefont {Williams},
  \citenamefont {Imakaev}, \citenamefont {Ragoczy}, \citenamefont {Telling},
  \citenamefont {Amit}, \citenamefont {Lajoie}, \citenamefont {Sabo},
  \citenamefont {Dorschner}, \citenamefont {Sandstrom}, \citenamefont
  {Bernstein}, \citenamefont {Bender}, \citenamefont {Groudine}, \citenamefont
  {Gnirke}, \citenamefont {Stamatoyannopoulos}, \citenamefont {Mirny},
  \citenamefont {Lander},\ and\ \citenamefont {Dekker}}]{lieb:09}%
  \BibitemOpen
  \bibfield  {author} {\bibinfo {author} {\bibfnamefont {E.}~\bibnamefont
  {Lieberman-Aiden}}, \bibinfo {author} {\bibfnamefont {N.~L.}\ \bibnamefont
  {van Berkum}}, \bibinfo {author} {\bibfnamefont {L.}~\bibnamefont
  {Williams}}, \bibinfo {author} {\bibfnamefont {M.}~\bibnamefont {Imakaev}},
  \bibinfo {author} {\bibfnamefont {T.}~\bibnamefont {Ragoczy}}, \bibinfo
  {author} {\bibfnamefont {A.}~\bibnamefont {Telling}}, \bibinfo {author}
  {\bibfnamefont {I.}~\bibnamefont {Amit}}, \bibinfo {author} {\bibfnamefont
  {B.~R.}\ \bibnamefont {Lajoie}}, \bibinfo {author} {\bibfnamefont {P.~J.}\
  \bibnamefont {Sabo}}, \bibinfo {author} {\bibfnamefont {M.~O.}\ \bibnamefont
  {Dorschner}}, \bibinfo {author} {\bibfnamefont {R.}~\bibnamefont
  {Sandstrom}}, \bibinfo {author} {\bibfnamefont {B.}~\bibnamefont
  {Bernstein}}, \bibinfo {author} {\bibfnamefont {M.~A.}\ \bibnamefont
  {Bender}}, \bibinfo {author} {\bibfnamefont {M.}~\bibnamefont {Groudine}},
  \bibinfo {author} {\bibfnamefont {A.}~\bibnamefont {Gnirke}}, \bibinfo
  {author} {\bibfnamefont {J.}~\bibnamefont {Stamatoyannopoulos}}, \bibinfo
  {author} {\bibfnamefont {L.~A.}\ \bibnamefont {Mirny}}, \bibinfo {author}
  {\bibfnamefont {E.~S.}\ \bibnamefont {Lander}}, \ and\ \bibinfo {author}
  {\bibfnamefont {J.}~\bibnamefont {Dekker}},\ }\bibfield  {title} {\enquote
  {\bibinfo {title} {Comprehensive mapping of long-range interactions reveals
  folding principles of the human genome},}\ }\href {\doibase
  10.1126/science.1181369} {\bibfield  {journal} {\bibinfo  {journal}
  {Science}\ }\textbf {\bibinfo {volume} {326}},\ \bibinfo {pages} {289}
  (\bibinfo {year} {2009})}\BibitemShut {NoStop}%
\bibitem [{\citenamefont {Cremer}\ \emph {et~al.}(2015)\citenamefont {Cremer},
  \citenamefont {Cremer}, \citenamefont {H{\"u}bner}, \citenamefont
  {Strickfaden}, \citenamefont {Smeets}, \citenamefont {Popken}, \citenamefont
  {Sterr}, \citenamefont {Markaki}, \citenamefont {Rippe},\ and\ \citenamefont
  {Cremer}}]{crem:15}%
  \BibitemOpen
  \bibfield  {author} {\bibinfo {author} {\bibfnamefont {T.}~\bibnamefont
  {Cremer}}, \bibinfo {author} {\bibfnamefont {M.}~\bibnamefont {Cremer}},
  \bibinfo {author} {\bibfnamefont {B.}~\bibnamefont {H{\"u}bner}}, \bibinfo
  {author} {\bibfnamefont {H.}~\bibnamefont {Strickfaden}}, \bibinfo {author}
  {\bibfnamefont {D.}~\bibnamefont {Smeets}}, \bibinfo {author} {\bibfnamefont
  {J.}~\bibnamefont {Popken}}, \bibinfo {author} {\bibfnamefont
  {M.}~\bibnamefont {Sterr}}, \bibinfo {author} {\bibfnamefont
  {Y.}~\bibnamefont {Markaki}}, \bibinfo {author} {\bibfnamefont
  {K.}~\bibnamefont {Rippe}}, \ and\ \bibinfo {author} {\bibfnamefont
  {C.}~\bibnamefont {Cremer}},\ }\bibfield  {title} {\enquote {\bibinfo {title}
  {The 4d nucleome: Evidence for a dynamic nuclear landscape based on
  co-aligned active and inactive nuclear compartments},}\ }\href {\doibase
  10.1016/j.febslet.2015.05.037} {\bibfield  {journal} {\bibinfo  {journal}
  {FEBS Lett.}\ }\textbf {\bibinfo {volume} {589}},\ \bibinfo {pages} {2931}
  (\bibinfo {year} {2015})}\BibitemShut {NoStop}%
\bibitem [{\citenamefont {Solovei}, \citenamefont {Thanisch},\ and\
  \citenamefont {Feodorova}(2016)}]{solo:16}%
  \BibitemOpen
  \bibfield  {author} {\bibinfo {author} {\bibfnamefont {I.}~\bibnamefont
  {Solovei}}, \bibinfo {author} {\bibfnamefont {K.}~\bibnamefont {Thanisch}}, \
  and\ \bibinfo {author} {\bibfnamefont {Y.}~\bibnamefont {Feodorova}},\
  }\bibfield  {title} {\enquote {\bibinfo {title} {How to rule the nucleus:
  divide et impera},}\ }\bibfield  {booktitle} {\emph {\bibinfo {booktitle}
  {Cell nucleus}},\ }\href {\doibase https://doi.org/10.1016/j.ceb.2016.02.014}
  {\bibfield  {journal} {\bibinfo  {journal} {Current Opinion in Cell Biology}\
  }\textbf {\bibinfo {volume} {40}},\ \bibinfo {pages} {47} (\bibinfo {year}
  {2016})}\BibitemShut {NoStop}%
\bibitem [{\citenamefont {Ganai}, \citenamefont {Sengupta},\ and\ \citenamefont
  {Menon}(2014)}]{gana:14}%
  \BibitemOpen
  \bibfield  {author} {\bibinfo {author} {\bibfnamefont {N.}~\bibnamefont
  {Ganai}}, \bibinfo {author} {\bibfnamefont {S.}~\bibnamefont {Sengupta}}, \
  and\ \bibinfo {author} {\bibfnamefont {G.~I.}\ \bibnamefont {Menon}},\
  }\bibfield  {title} {\enquote {\bibinfo {title} {Chromosome positioning from
  activity-based segregation},}\ }\href {\doibase 10.1093/nar/gkt1417}
  {\bibfield  {journal} {\bibinfo  {journal} {Nucleic Acids Res.}\ }\textbf
  {\bibinfo {volume} {42}},\ \bibinfo {pages} {4145} (\bibinfo {year}
  {2014})}\BibitemShut {NoStop}%
\bibitem [{\citenamefont {Smrek}\ and\ \citenamefont {Kremer}(2017)}]{smre:17}%
  \BibitemOpen
  \bibfield  {author} {\bibinfo {author} {\bibfnamefont {J.}~\bibnamefont
  {Smrek}}\ and\ \bibinfo {author} {\bibfnamefont {K.}~\bibnamefont {Kremer}},\
  }\bibfield  {title} {\enquote {\bibinfo {title} {Small activity differences
  drive phase separation in active-passive polymer mixtures},}\ }\href
  {\doibase 10.1103/PhysRevLett.118.098002} {\bibfield  {journal} {\bibinfo
  {journal} {Phys. Rev. Lett.}\ }\textbf {\bibinfo {volume} {118}},\ \bibinfo
  {pages} {098002} (\bibinfo {year} {2017})}\BibitemShut {NoStop}%
\bibitem [{\citenamefont {Lauga}\ and\ \citenamefont {Powers}(2009)}]{laug:09}%
  \BibitemOpen
  \bibfield  {author} {\bibinfo {author} {\bibfnamefont {E.}~\bibnamefont
  {Lauga}}\ and\ \bibinfo {author} {\bibfnamefont {T.~R.}\ \bibnamefont
  {Powers}},\ }\bibfield  {title} {\enquote {\bibinfo {title} {{T}he
  hydrodynamics of swimming microorganisms},}\ }\href {\doibase
  10.1088/0034-4885/72/9/096601} {\bibfield  {journal} {\bibinfo  {journal}
  {Rep. Prog. Phys.}\ }\textbf {\bibinfo {volume} {72}},\ \bibinfo {pages}
  {096601} (\bibinfo {year} {2009})}\BibitemShut {NoStop}%
\bibitem [{\citenamefont {Ramaswamy}(2010)}]{rama:10}%
  \BibitemOpen
  \bibfield  {author} {\bibinfo {author} {\bibfnamefont {S.}~\bibnamefont
  {Ramaswamy}},\ }\bibfield  {title} {\enquote {\bibinfo {title} {The mechanics
  and statistics of active matter},}\ }\href {\doibase
  10.1146/annurev-conmatphys-070909-104101} {\bibfield  {journal} {\bibinfo
  {journal} {Annu. Rev. Cond. Mat. Phys.}\ }\textbf {\bibinfo {volume} {1}},\
  \bibinfo {pages} {323} (\bibinfo {year} {2010})}\BibitemShut {NoStop}%
\bibitem [{\citenamefont {Bechinger}\ \emph {et~al.}(2016)\citenamefont
  {Bechinger}, \citenamefont {Di~Leonardo}, \citenamefont {L{\"o}wen},
  \citenamefont {Reichhardt}, \citenamefont {Volpe},\ and\ \citenamefont
  {Volpe}}]{bech:16}%
  \BibitemOpen
  \bibfield  {author} {\bibinfo {author} {\bibfnamefont {C.}~\bibnamefont
  {Bechinger}}, \bibinfo {author} {\bibfnamefont {R.}~\bibnamefont
  {Di~Leonardo}}, \bibinfo {author} {\bibfnamefont {H.}~\bibnamefont
  {L{\"o}wen}}, \bibinfo {author} {\bibfnamefont {C.}~\bibnamefont
  {Reichhardt}}, \bibinfo {author} {\bibfnamefont {G.}~\bibnamefont {Volpe}}, \
  and\ \bibinfo {author} {\bibfnamefont {G.}~\bibnamefont {Volpe}},\ }\bibfield
   {title} {\enquote {\bibinfo {title} {Active particles in complex and crowded
  environments},}\ }\href {\doibase 10.1103/RevModPhys.88.045006} {\bibfield
  {journal} {\bibinfo  {journal} {Rev. Mod. Phys.}\ }\textbf {\bibinfo {volume}
  {88}},\ \bibinfo {pages} {045006} (\bibinfo {year} {2016})}\BibitemShut
  {NoStop}%
\bibitem [{\citenamefont {Gompper}\ \emph {et~al.}(2020)\citenamefont
  {Gompper}, \citenamefont {Winkler}, \citenamefont {Speck}, \citenamefont
  {Solon}, \citenamefont {Nardini}, \citenamefont {Peruani}, \citenamefont
  {L{\"o}wen}, \citenamefont {Golestanian}, \citenamefont {Kaupp},
  \citenamefont {Alvarez}, \citenamefont {Ki{\o}rboe}, \citenamefont {Lauga},
  \citenamefont {Poon}, \citenamefont {DeSimone}, \citenamefont
  {Mui{\~n}os-Landin}, \citenamefont {Fischer}, \citenamefont {S{\"o}ker},
  \citenamefont {Cichos}, \citenamefont {Kapral}, \citenamefont {Gaspard},
  \citenamefont {Ripoll}, \citenamefont {Sagues}, \citenamefont
  {Doostmohammadi}, \citenamefont {Yeomans}, \citenamefont {Aranson},
  \citenamefont {Bechinger}, \citenamefont {Stark}, \citenamefont {Hemelrijk},
  \citenamefont {Nedelec}, \citenamefont {Sarkar}, \citenamefont {Aryaksama},
  \citenamefont {Lacroix}, \citenamefont {Duclos}, \citenamefont {Yashunsky},
  \citenamefont {Silberzan}, \citenamefont {Arroyo},\ and\ \citenamefont
  {Kale}}]{gomp:20}%
  \BibitemOpen
  \bibfield  {author} {\bibinfo {author} {\bibfnamefont {G.}~\bibnamefont
  {Gompper}}, \bibinfo {author} {\bibfnamefont {R.~G.}\ \bibnamefont
  {Winkler}}, \bibinfo {author} {\bibfnamefont {T.}~\bibnamefont {Speck}},
  \bibinfo {author} {\bibfnamefont {A.}~\bibnamefont {Solon}}, \bibinfo
  {author} {\bibfnamefont {C.}~\bibnamefont {Nardini}}, \bibinfo {author}
  {\bibfnamefont {F.}~\bibnamefont {Peruani}}, \bibinfo {author} {\bibfnamefont
  {H.}~\bibnamefont {L{\"o}wen}}, \bibinfo {author} {\bibfnamefont
  {R.}~\bibnamefont {Golestanian}}, \bibinfo {author} {\bibfnamefont {U.~B.}\
  \bibnamefont {Kaupp}}, \bibinfo {author} {\bibfnamefont {L.}~\bibnamefont
  {Alvarez}}, \bibinfo {author} {\bibfnamefont {T.}~\bibnamefont {Ki{\o}rboe}},
  \bibinfo {author} {\bibfnamefont {E.}~\bibnamefont {Lauga}}, \bibinfo
  {author} {\bibfnamefont {W.~C.~K.}\ \bibnamefont {Poon}}, \bibinfo {author}
  {\bibfnamefont {A.}~\bibnamefont {DeSimone}}, \bibinfo {author}
  {\bibfnamefont {S.}~\bibnamefont {Mui{\~n}os-Landin}}, \bibinfo {author}
  {\bibfnamefont {A.}~\bibnamefont {Fischer}}, \bibinfo {author} {\bibfnamefont
  {N.~A.}\ \bibnamefont {S{\"o}ker}}, \bibinfo {author} {\bibfnamefont
  {F.}~\bibnamefont {Cichos}}, \bibinfo {author} {\bibfnamefont
  {R.}~\bibnamefont {Kapral}}, \bibinfo {author} {\bibfnamefont
  {P.}~\bibnamefont {Gaspard}}, \bibinfo {author} {\bibfnamefont
  {M.}~\bibnamefont {Ripoll}}, \bibinfo {author} {\bibfnamefont
  {F.}~\bibnamefont {Sagues}}, \bibinfo {author} {\bibfnamefont
  {A.}~\bibnamefont {Doostmohammadi}}, \bibinfo {author} {\bibfnamefont
  {J.~M.}\ \bibnamefont {Yeomans}}, \bibinfo {author} {\bibfnamefont {I.~S.}\
  \bibnamefont {Aranson}}, \bibinfo {author} {\bibfnamefont {C.}~\bibnamefont
  {Bechinger}}, \bibinfo {author} {\bibfnamefont {H.}~\bibnamefont {Stark}},
  \bibinfo {author} {\bibfnamefont {C.~K.}\ \bibnamefont {Hemelrijk}}, \bibinfo
  {author} {\bibfnamefont {F.~J.}\ \bibnamefont {Nedelec}}, \bibinfo {author}
  {\bibfnamefont {T.}~\bibnamefont {Sarkar}}, \bibinfo {author} {\bibfnamefont
  {T.}~\bibnamefont {Aryaksama}}, \bibinfo {author} {\bibfnamefont
  {M.}~\bibnamefont {Lacroix}}, \bibinfo {author} {\bibfnamefont
  {G.}~\bibnamefont {Duclos}}, \bibinfo {author} {\bibfnamefont
  {V.}~\bibnamefont {Yashunsky}}, \bibinfo {author} {\bibfnamefont
  {P.}~\bibnamefont {Silberzan}}, \bibinfo {author} {\bibfnamefont
  {M.}~\bibnamefont {Arroyo}}, \ and\ \bibinfo {author} {\bibfnamefont
  {S.}~\bibnamefont {Kale}},\ }\bibfield  {title} {\enquote {\bibinfo {title}
  {The 2020 motile active matter roadmap},}\ }\href {\doibase
  10.1088/1361-648x/ab6348} {\bibfield  {journal} {\bibinfo  {journal} {J.
  Phys: Condens. Matter}\ }\textbf {\bibinfo {volume} {32}},\ \bibinfo {pages}
  {193001} (\bibinfo {year} {2020})}\BibitemShut {NoStop}%
\bibitem [{\citenamefont {Berg}(2004)}]{berg:04}%
  \BibitemOpen
  \bibfield  {author} {\bibinfo {author} {\bibfnamefont {H.~C.}\ \bibnamefont
  {Berg}},\ }\href {\doibase 10.1007/b97370} {\emph {\bibinfo {title} {{E}.
  {C}oli in {M}otion}}},\ Biological and Medical Physics Series\ (\bibinfo
  {publisher} {Springer},\ \bibinfo {address} {New York},\ \bibinfo {year}
  {2004})\BibitemShut {NoStop}%
\bibitem [{\citenamefont {WeibelLab}(2012)}]{weib:14}%
  \BibitemOpen
  \bibfield  {author} {\bibinfo {author} {\bibnamefont {WeibelLab}},\
  }\href@noop {} {\enquote {\bibinfo {title} {Swarm timelapse},}\ }\bibinfo
  {howpublished} {https://www.youtube.com/watch?v=0YYAtPCIMc0} (\bibinfo {year}
  {August 3, 2012})\BibitemShut {NoStop}%
\bibitem [{\citenamefont {Auer}\ \emph {et~al.}(2019)\citenamefont {Auer},
  \citenamefont {Oliver}, \citenamefont {Rajendram}, \citenamefont {Lin},
  \citenamefont {Yao}, \citenamefont {Jensen},\ and\ \citenamefont
  {Weibel}}]{auer:19}%
  \BibitemOpen
  \bibfield  {author} {\bibinfo {author} {\bibfnamefont {G.~K.}\ \bibnamefont
  {Auer}}, \bibinfo {author} {\bibfnamefont {P.~M.}\ \bibnamefont {Oliver}},
  \bibinfo {author} {\bibfnamefont {M.}~\bibnamefont {Rajendram}}, \bibinfo
  {author} {\bibfnamefont {T.-Y.}\ \bibnamefont {Lin}}, \bibinfo {author}
  {\bibfnamefont {Q.}~\bibnamefont {Yao}}, \bibinfo {author} {\bibfnamefont
  {G.~J.}\ \bibnamefont {Jensen}}, \ and\ \bibinfo {author} {\bibfnamefont
  {D.~B.}\ \bibnamefont {Weibel}},\ }\bibfield  {title} {\enquote {\bibinfo
  {title} {Bacterial swarming reduces \emph{{P}roteus mirabilis} and
  \emph{{V}ibrio parahaemolyticus} cell stiffness and increases $\beta$-lactam
  susceptibility},}\ }\href {\doibase 10.1128/mBio.00210-19} {\bibfield
  {journal} {\bibinfo  {journal} {mBio}\ }\textbf {\bibinfo {volume} {10}},\
  \bibinfo {pages} {e00210} (\bibinfo {year} {2019})}\BibitemShut {NoStop}%
\bibitem [{\citenamefont {Li}\ \emph {et~al.}(2019)\citenamefont {Li},
  \citenamefont {Shi}, \citenamefont {Huang}, \citenamefont {Chen},
  \citenamefont {Xiao}, \citenamefont {Liu}, \citenamefont {Chat{\'e}},\ and\
  \citenamefont {Zhang}}]{li:19}%
  \BibitemOpen
  \bibfield  {author} {\bibinfo {author} {\bibfnamefont {H.}~\bibnamefont
  {Li}}, \bibinfo {author} {\bibfnamefont {X.-q.}\ \bibnamefont {Shi}},
  \bibinfo {author} {\bibfnamefont {M.}~\bibnamefont {Huang}}, \bibinfo
  {author} {\bibfnamefont {X.}~\bibnamefont {Chen}}, \bibinfo {author}
  {\bibfnamefont {M.}~\bibnamefont {Xiao}}, \bibinfo {author} {\bibfnamefont
  {C.}~\bibnamefont {Liu}}, \bibinfo {author} {\bibfnamefont {H.}~\bibnamefont
  {Chat{\'e}}}, \ and\ \bibinfo {author} {\bibfnamefont {H.~P.}\ \bibnamefont
  {Zhang}},\ }\bibfield  {title} {\enquote {\bibinfo {title} {Data-driven
  quantitative modeling of bacterial active nematics},}\ }\href {\doibase
  10.1073/pnas.1812570116} {\bibfield  {journal} {\bibinfo  {journal} {Proc.
  Natl. Acad. Sci. USA}\ }\textbf {\bibinfo {volume} {116}},\ \bibinfo {pages}
  {777} (\bibinfo {year} {2019})}\BibitemShut {NoStop}%
\bibitem [{\citenamefont {Sohn}\ \emph {et~al.}(2011)\citenamefont {Sohn},
  \citenamefont {Seo}, \citenamefont {Choi}, \citenamefont {Lee}, \citenamefont
  {Kang},\ and\ \citenamefont {Kang}}]{sohn:11}%
  \BibitemOpen
  \bibfield  {author} {\bibinfo {author} {\bibfnamefont {M.~H.}\ \bibnamefont
  {Sohn}}, \bibinfo {author} {\bibfnamefont {K.~W.}\ \bibnamefont {Seo}},
  \bibinfo {author} {\bibfnamefont {Y.~S.}\ \bibnamefont {Choi}}, \bibinfo
  {author} {\bibfnamefont {S.~J.}\ \bibnamefont {Lee}}, \bibinfo {author}
  {\bibfnamefont {Y.~S.}\ \bibnamefont {Kang}}, \ and\ \bibinfo {author}
  {\bibfnamefont {Y.~S.}\ \bibnamefont {Kang}},\ }\bibfield  {title} {\enquote
  {\bibinfo {title} {{Determination of the swimming trajectory and speed of
  chain-forming dinoflagellate {\em Cochlodinium} {\em polykrikoides} with
  digital holographic particle tracking velocimetry}},}\ }\href {\doibase
  10.1007/s00227-010-1581-7} {\bibfield  {journal} {\bibinfo  {journal} {Mar.
  Biol. Biology}\ }\textbf {\bibinfo {volume} {158}},\ \bibinfo {pages} {561}
  (\bibinfo {year} {2011})}\BibitemShut {NoStop}%
\bibitem [{\citenamefont {Selander}\ \emph {et~al.}(2011)\citenamefont
  {Selander}, \citenamefont {Jakobsen}, \citenamefont {Lombard},\ and\
  \citenamefont {Ki{\o}rboe}}]{sela:11}%
  \BibitemOpen
  \bibfield  {author} {\bibinfo {author} {\bibfnamefont {E.}~\bibnamefont
  {Selander}}, \bibinfo {author} {\bibfnamefont {H.~H.}\ \bibnamefont
  {Jakobsen}}, \bibinfo {author} {\bibfnamefont {F.}~\bibnamefont {Lombard}}, \
  and\ \bibinfo {author} {\bibfnamefont {T.}~\bibnamefont {Ki{\o}rboe}},\
  }\bibfield  {title} {\enquote {\bibinfo {title} {Grazer cues induce stealth
  behavior in marine dinoflagellates},}\ }\href {\doibase
  10.1073/pnas.1011870108} {\bibfield  {journal} {\bibinfo  {journal} {Proc.
  Natl. Acad. Sci. USA}\ }\textbf {\bibinfo {volume} {108}},\ \bibinfo {pages}
  {4030} (\bibinfo {year} {2011})}\BibitemShut {NoStop}%
\bibitem [{\citenamefont {Yaman}\ \emph {et~al.}(2019)\citenamefont {Yaman},
  \citenamefont {Demir}, \citenamefont {Vetter},\ and\ \citenamefont
  {Kocabas}}]{yama:19}%
  \BibitemOpen
  \bibfield  {author} {\bibinfo {author} {\bibfnamefont {Y.~I.}\ \bibnamefont
  {Yaman}}, \bibinfo {author} {\bibfnamefont {E.}~\bibnamefont {Demir}},
  \bibinfo {author} {\bibfnamefont {R.}~\bibnamefont {Vetter}}, \ and\ \bibinfo
  {author} {\bibfnamefont {A.}~\bibnamefont {Kocabas}},\ }\bibfield  {title}
  {\enquote {\bibinfo {title} {Emergence of active nematics in chaining
  bacterial biofilms},}\ }\href {\doibase 10.1038/s41467-019-10311-z}
  {\bibfield  {journal} {\bibinfo  {journal} {Nat. Commun.}\ }\textbf {\bibinfo
  {volume} {10}},\ \bibinfo {pages} {2285} (\bibinfo {year}
  {2019})}\BibitemShut {NoStop}%
\bibitem [{\citenamefont {Gagnon}\ and\ \citenamefont
  {Arratia}(2016)}]{gagn:16}%
  \BibitemOpen
  \bibfield  {author} {\bibinfo {author} {\bibfnamefont {D.~A.}\ \bibnamefont
  {Gagnon}}\ and\ \bibinfo {author} {\bibfnamefont {P.~E.}\ \bibnamefont
  {Arratia}},\ }\bibfield  {title} {\enquote {\bibinfo {title} {The cost of
  swimming in generalized newtonian fluids: experiments with \emph{{C}.
  elegans}},}\ }\href {\doibase DOI: 10.1017/jfm.2016.420} {\bibfield
  {journal} {\bibinfo  {journal} {J. Fluid Mech.}\ }\textbf {\bibinfo {volume}
  {800}},\ \bibinfo {pages} {753} (\bibinfo {year} {2016})}\BibitemShut
  {NoStop}%
\bibitem [{\citenamefont {Sugi}\ \emph {et~al.}(2019)\citenamefont {Sugi},
  \citenamefont {Ito}, \citenamefont {Nishimura},\ and\ \citenamefont
  {Nagai}}]{sugi:19}%
  \BibitemOpen
  \bibfield  {author} {\bibinfo {author} {\bibfnamefont {T.}~\bibnamefont
  {Sugi}}, \bibinfo {author} {\bibfnamefont {H.}~\bibnamefont {Ito}}, \bibinfo
  {author} {\bibfnamefont {M.}~\bibnamefont {Nishimura}}, \ and\ \bibinfo
  {author} {\bibfnamefont {K.~H.}\ \bibnamefont {Nagai}},\ }\bibfield  {title}
  {\enquote {\bibinfo {title} {\emph{{C}. elegans} collectively forms dynamical
  networks},}\ }\href {\doibase 10.1038/s41467-019-08537-y} {\bibfield
  {journal} {\bibinfo  {journal} {Nat. Commun.}\ }\textbf {\bibinfo {volume}
  {10}},\ \bibinfo {pages} {683} (\bibinfo {year} {2019})}\BibitemShut
  {NoStop}%
\bibitem [{\citenamefont {N{\'e}d{\'e}lec}\ \emph {et~al.}(1997)\citenamefont
  {N{\'e}d{\'e}lec}, \citenamefont {Surrey}, \citenamefont {Maggs},\ and\
  \citenamefont {Leibler}}]{nedl:97}%
  \BibitemOpen
  \bibfield  {author} {\bibinfo {author} {\bibfnamefont {F.~J.}\ \bibnamefont
  {N{\'e}d{\'e}lec}}, \bibinfo {author} {\bibfnamefont {T.}~\bibnamefont
  {Surrey}}, \bibinfo {author} {\bibfnamefont {A.~C.}\ \bibnamefont {Maggs}}, \
  and\ \bibinfo {author} {\bibfnamefont {S.}~\bibnamefont {Leibler}},\
  }\bibfield  {title} {\enquote {\bibinfo {title} {Self-organization of
  microtubules and motors},}\ }\href {\doibase 10.1038/38532} {\bibfield
  {journal} {\bibinfo  {journal} {Nature}\ }\textbf {\bibinfo {volume} {389}},\
  \bibinfo {pages} {305} (\bibinfo {year} {1997})}\BibitemShut {NoStop}%
\bibitem [{\citenamefont {Howard}(2001)}]{howa:01}%
  \BibitemOpen
  \bibfield  {author} {\bibinfo {author} {\bibfnamefont {J.}~\bibnamefont
  {Howard}},\ }\href@noop {} {\emph {\bibinfo {title} {Mechanics of motor
  proteins and the cytoskeleton}}}\ (\bibinfo  {publisher} {Sinauer Associates
  Sunderland, MA},\ \bibinfo {year} {2001})\BibitemShut {NoStop}%
\bibitem [{\citenamefont {Kruse}\ \emph {et~al.}(2004)\citenamefont {Kruse},
  \citenamefont {Joanny}, \citenamefont {J{\"u}licher}, \citenamefont {Prost},\
  and\ \citenamefont {Sekimoto}}]{krus:04}%
  \BibitemOpen
  \bibfield  {author} {\bibinfo {author} {\bibfnamefont {K.}~\bibnamefont
  {Kruse}}, \bibinfo {author} {\bibfnamefont {J.~F.}\ \bibnamefont {Joanny}},
  \bibinfo {author} {\bibfnamefont {F.}~\bibnamefont {J{\"u}licher}}, \bibinfo
  {author} {\bibfnamefont {J.}~\bibnamefont {Prost}}, \ and\ \bibinfo {author}
  {\bibfnamefont {K.}~\bibnamefont {Sekimoto}},\ }\bibfield  {title} {\enquote
  {\bibinfo {title} {Asters, vortices, and rotating spirals in active gels of
  polar filaments},}\ }\href {\doibase 10.1103/PhysRevLett.92.078101}
  {\bibfield  {journal} {\bibinfo  {journal} {Phys. Rev. Lett.}\ }\textbf
  {\bibinfo {volume} {92}},\ \bibinfo {pages} {078101} (\bibinfo {year}
  {2004})}\BibitemShut {NoStop}%
\bibitem [{\citenamefont {Bausch}\ and\ \citenamefont {Kroy}(2006)}]{baus:06}%
  \BibitemOpen
  \bibfield  {author} {\bibinfo {author} {\bibfnamefont {A.~R.}\ \bibnamefont
  {Bausch}}\ and\ \bibinfo {author} {\bibfnamefont {K.}~\bibnamefont {Kroy}},\
  }\bibfield  {title} {\enquote {\bibinfo {title} {A bottom-up approach to cell
  mechanics},}\ }\href {\doibase 10.1038/nphys260} {\bibfield  {journal}
  {\bibinfo  {journal} {Nat. Phys.}\ }\textbf {\bibinfo {volume} {2}},\
  \bibinfo {pages} {231} (\bibinfo {year} {2006})}\BibitemShut {NoStop}%
\bibitem [{\citenamefont {J{\"u}licher}\ \emph {et~al.}(2007)\citenamefont
  {J{\"u}licher}, \citenamefont {Kruse}, \citenamefont {Prost},\ and\
  \citenamefont {Joanny}}]{juel:07}%
  \BibitemOpen
  \bibfield  {author} {\bibinfo {author} {\bibfnamefont {F.}~\bibnamefont
  {J{\"u}licher}}, \bibinfo {author} {\bibfnamefont {K.}~\bibnamefont {Kruse}},
  \bibinfo {author} {\bibfnamefont {J.}~\bibnamefont {Prost}}, \ and\ \bibinfo
  {author} {\bibfnamefont {J.-F.}\ \bibnamefont {Joanny}},\ }\bibfield  {title}
  {\enquote {\bibinfo {title} {Active behavior of the cytoskeleton},}\ }\href
  {\doibase 10.1016/j.physrep.2007.02.018} {\bibfield  {journal} {\bibinfo
  {journal} {Phys. Rep.}\ }\textbf {\bibinfo {volume} {449}},\ \bibinfo {pages}
  {3} (\bibinfo {year} {2007})}\BibitemShut {NoStop}%
\bibitem [{\citenamefont {Harada}\ \emph {et~al.}(1987)\citenamefont {Harada},
  \citenamefont {Noguchi}, \citenamefont {Kishino},\ and\ \citenamefont
  {Yanagida}}]{hara:87}%
  \BibitemOpen
  \bibfield  {author} {\bibinfo {author} {\bibfnamefont {Y.}~\bibnamefont
  {Harada}}, \bibinfo {author} {\bibfnamefont {A.}~\bibnamefont {Noguchi}},
  \bibinfo {author} {\bibfnamefont {A.}~\bibnamefont {Kishino}}, \ and\
  \bibinfo {author} {\bibfnamefont {T.}~\bibnamefont {Yanagida}},\ }\bibfield
  {title} {\enquote {\bibinfo {title} {Sliding movement of single actin
  filaments on one-headed myosin filaments},}\ }\href {\doibase
  10.1038/326805a0} {\bibfield  {journal} {\bibinfo  {journal} {Nature}\
  }\textbf {\bibinfo {volume} {326}},\ \bibinfo {pages} {805} (\bibinfo {year}
  {1987})}\BibitemShut {NoStop}%
\bibitem [{\citenamefont {Schaller}\ \emph {et~al.}(2010)\citenamefont
  {Schaller}, \citenamefont {Weber}, \citenamefont {Semmrich}, \citenamefont
  {Frey},\ and\ \citenamefont {Bausch}}]{scha:10}%
  \BibitemOpen
  \bibfield  {author} {\bibinfo {author} {\bibfnamefont {V.}~\bibnamefont
  {Schaller}}, \bibinfo {author} {\bibfnamefont {C.}~\bibnamefont {Weber}},
  \bibinfo {author} {\bibfnamefont {C.}~\bibnamefont {Semmrich}}, \bibinfo
  {author} {\bibfnamefont {E.}~\bibnamefont {Frey}}, \ and\ \bibinfo {author}
  {\bibfnamefont {A.~R.}\ \bibnamefont {Bausch}},\ }\bibfield  {title}
  {\enquote {\bibinfo {title} {Polar patterns of driven filaments},}\ }\href
  {\doibase 10.1038/nature09312} {\bibfield  {journal} {\bibinfo  {journal}
  {Nature}\ }\textbf {\bibinfo {volume} {467}},\ \bibinfo {pages} {73}
  (\bibinfo {year} {2010})}\BibitemShut {NoStop}%
\bibitem [{\citenamefont {Sanchez}\ \emph {et~al.}(2012)\citenamefont
  {Sanchez}, \citenamefont {Chen}, \citenamefont {DeCamp}, \citenamefont
  {Heymann},\ and\ \citenamefont {Dogic}}]{sanc:12}%
  \BibitemOpen
  \bibfield  {author} {\bibinfo {author} {\bibfnamefont {T.}~\bibnamefont
  {Sanchez}}, \bibinfo {author} {\bibfnamefont {D.~T.~N.}\ \bibnamefont
  {Chen}}, \bibinfo {author} {\bibfnamefont {S.~J.}\ \bibnamefont {DeCamp}},
  \bibinfo {author} {\bibfnamefont {M.}~\bibnamefont {Heymann}}, \ and\
  \bibinfo {author} {\bibfnamefont {Z.}~\bibnamefont {Dogic}},\ }\bibfield
  {title} {\enquote {\bibinfo {title} {Spontaneous motion in hierarchically
  assembled active matter},}\ }\href {\doibase 10.1038/nature11591} {\bibfield
  {journal} {\bibinfo  {journal} {Nature}\ }\textbf {\bibinfo {volume} {491}},\
  \bibinfo {pages} {431} (\bibinfo {year} {2012})}\BibitemShut {NoStop}%
\bibitem [{\citenamefont {Sumino}\ \emph {et~al.}(2012)\citenamefont {Sumino},
  \citenamefont {Nagai}, \citenamefont {Shitaka}, \citenamefont {Tanaka},
  \citenamefont {Yoshikawa}, \citenamefont {Chate},\ and\ \citenamefont
  {Oiwa}}]{sumi:12}%
  \BibitemOpen
  \bibfield  {author} {\bibinfo {author} {\bibfnamefont {Y.}~\bibnamefont
  {Sumino}}, \bibinfo {author} {\bibfnamefont {K.~H.}\ \bibnamefont {Nagai}},
  \bibinfo {author} {\bibfnamefont {Y.}~\bibnamefont {Shitaka}}, \bibinfo
  {author} {\bibfnamefont {D.}~\bibnamefont {Tanaka}}, \bibinfo {author}
  {\bibfnamefont {K.}~\bibnamefont {Yoshikawa}}, \bibinfo {author}
  {\bibfnamefont {H.}~\bibnamefont {Chate}}, \ and\ \bibinfo {author}
  {\bibfnamefont {K.}~\bibnamefont {Oiwa}},\ }\bibfield  {title} {\enquote
  {\bibinfo {title} {Large-scale vortex lattice emerging from collectively
  moving microtubules},}\ }\href {\doibase 10.1038/nature10874} {\bibfield
  {journal} {\bibinfo  {journal} {Nature}\ }\textbf {\bibinfo {volume} {483}},\
  \bibinfo {pages} {448} (\bibinfo {year} {2012})}\BibitemShut {NoStop}%
\bibitem [{\citenamefont {Prost}, \citenamefont {J{\"u}licher},\ and\
  \citenamefont {Joanny}(2015)}]{pros:15}%
  \BibitemOpen
  \bibfield  {author} {\bibinfo {author} {\bibfnamefont {J.}~\bibnamefont
  {Prost}}, \bibinfo {author} {\bibfnamefont {F.}~\bibnamefont {J{\"u}licher}},
  \ and\ \bibinfo {author} {\bibfnamefont {J.-F.}\ \bibnamefont {Joanny}},\
  }\bibfield  {title} {\enquote {\bibinfo {title} {Active gel physics},}\
  }\href {\doibase 10.1038/nphys3224} {\bibfield  {journal} {\bibinfo
  {journal} {Nat. Phys.}\ }\textbf {\bibinfo {volume} {11}},\ \bibinfo {pages}
  {111} (\bibinfo {year} {2015})}\BibitemShut {NoStop}%
\bibitem [{\citenamefont {Doostmohammadi}\ \emph {et~al.}(2018)\citenamefont
  {Doostmohammadi}, \citenamefont {Ign{\'e}s-Mullol}, \citenamefont {Yeomans},\
  and\ \citenamefont {Sagu{\'e}s}}]{doos:18}%
  \BibitemOpen
  \bibfield  {author} {\bibinfo {author} {\bibfnamefont {A.}~\bibnamefont
  {Doostmohammadi}}, \bibinfo {author} {\bibfnamefont {J.}~\bibnamefont
  {Ign{\'e}s-Mullol}}, \bibinfo {author} {\bibfnamefont {J.~M.}\ \bibnamefont
  {Yeomans}}, \ and\ \bibinfo {author} {\bibfnamefont {F.}~\bibnamefont
  {Sagu{\'e}s}},\ }\bibfield  {title} {\enquote {\bibinfo {title} {Active
  nematics},}\ }\href {\doibase 10.1038/s41467-018-05666-8} {\bibfield
  {journal} {\bibinfo  {journal} {Nat. Commun.}\ }\textbf {\bibinfo {volume}
  {9}},\ \bibinfo {pages} {3246} (\bibinfo {year} {2018})}\BibitemShut
  {NoStop}%
\bibitem [{\citenamefont {Blain}\ and\ \citenamefont
  {Szostak}(2014)}]{blai:14}%
  \BibitemOpen
  \bibfield  {author} {\bibinfo {author} {\bibfnamefont {J.~C.}\ \bibnamefont
  {Blain}}\ and\ \bibinfo {author} {\bibfnamefont {J.~W.}\ \bibnamefont
  {Szostak}},\ }\bibfield  {title} {\enquote {\bibinfo {title} {Progress toward
  synthetic cells},}\ }\href {\doibase 10.1146/annurev-biochem-080411-124036}
  {\bibfield  {journal} {\bibinfo  {journal} {Ann. Rev. Biochem.}\ }\textbf
  {\bibinfo {volume} {83}},\ \bibinfo {pages} {615} (\bibinfo {year}
  {2014})}\BibitemShut {NoStop}%
\bibitem [{\citenamefont {Stano}(2018)}]{stan:18}%
  \BibitemOpen
  \bibfield  {author} {\bibinfo {author} {\bibfnamefont {P.}~\bibnamefont
  {Stano}},\ }\bibfield  {title} {\enquote {\bibinfo {title} {Is research on
  ``synthetic cells'' moving to the next level?}}\ }\href {\doibase
  10.3390/life9010003} {\bibfield  {journal} {\bibinfo  {journal} {Life}\
  }\textbf {\bibinfo {volume} {9}},\ \bibinfo {pages} {3} (\bibinfo {year}
  {2018})}\BibitemShut {NoStop}%
\bibitem [{\citenamefont {G\"opfrich}, \citenamefont {Platzman},\ and\
  \citenamefont {Spatz}(2018)}]{gopf:18}%
  \BibitemOpen
  \bibfield  {author} {\bibinfo {author} {\bibfnamefont {K.}~\bibnamefont
  {G\"opfrich}}, \bibinfo {author} {\bibfnamefont {I.}~\bibnamefont
  {Platzman}}, \ and\ \bibinfo {author} {\bibfnamefont {J.~P.}\ \bibnamefont
  {Spatz}},\ }\bibfield  {title} {\enquote {\bibinfo {title} {Mastering
  complexity: Towards bottom-up construction of multifunctional eukaryotic
  synthetic cells},}\ }\href {\doibase
  https://doi.org/10.1016/j.tibtech.2018.03.008} {\bibfield  {journal}
  {\bibinfo  {journal} {Trends in Biotechnology}\ }\textbf {\bibinfo {volume}
  {36}},\ \bibinfo {pages} {938} (\bibinfo {year} {2018})}\BibitemShut
  {NoStop}%
\bibitem [{\citenamefont {L{\"o}wen}(2018)}]{loew:18}%
  \BibitemOpen
  \bibfield  {author} {\bibinfo {author} {\bibfnamefont {H.}~\bibnamefont
  {L{\"o}wen}},\ }\bibfield  {title} {\enquote {\bibinfo {title} {Active
  colloidal molecules},}\ }\href {\doibase 10.1209/0295-5075/121/58001}
  {\bibfield  {journal} {\bibinfo  {journal} {EPL}\ }\textbf {\bibinfo {volume}
  {121}},\ \bibinfo {pages} {58001} (\bibinfo {year} {2018})}\BibitemShut
  {NoStop}%
\bibitem [{\citenamefont {Winkler}, \citenamefont {Elgeti},\ and\ \citenamefont
  {Gompper}(2017)}]{wink:17}%
  \BibitemOpen
  \bibfield  {author} {\bibinfo {author} {\bibfnamefont {R.~G.}\ \bibnamefont
  {Winkler}}, \bibinfo {author} {\bibfnamefont {J.}~\bibnamefont {Elgeti}}, \
  and\ \bibinfo {author} {\bibfnamefont {G.}~\bibnamefont {Gompper}},\
  }\bibfield  {title} {\enquote {\bibinfo {title} {Active polymers---emergent
  conformational and dynamical properties: A brief review},}\ }\href {\doibase
  10.7566/JPSJ.86.101014} {\bibfield  {journal} {\bibinfo  {journal} {J. Phys.
  Soc. Jpn.}\ }\textbf {\bibinfo {volume} {86}},\ \bibinfo {pages} {101014}
  (\bibinfo {year} {2017})}\BibitemShut {NoStop}%
\bibitem [{\citenamefont {Gompper}\ \emph {et~al.}(2016)\citenamefont
  {Gompper}, \citenamefont {Bechinger}, \citenamefont {Herminghaus},
  \citenamefont {Isele-Holder}, \citenamefont {Kaupp}, \citenamefont
  {L{\"o}wen}, \citenamefont {Stark},\ and\ \citenamefont {Winkler}}]{gomp:16}%
  \BibitemOpen
  \bibfield  {author} {\bibinfo {author} {\bibfnamefont {G.}~\bibnamefont
  {Gompper}}, \bibinfo {author} {\bibfnamefont {C.}~\bibnamefont {Bechinger}},
  \bibinfo {author} {\bibfnamefont {S.}~\bibnamefont {Herminghaus}}, \bibinfo
  {author} {\bibfnamefont {R.}~\bibnamefont {Isele-Holder}}, \bibinfo {author}
  {\bibfnamefont {U.~B.}\ \bibnamefont {Kaupp}}, \bibinfo {author}
  {\bibfnamefont {H.}~\bibnamefont {L{\"o}wen}}, \bibinfo {author}
  {\bibfnamefont {H.}~\bibnamefont {Stark}}, \ and\ \bibinfo {author}
  {\bibfnamefont {R.~G.}\ \bibnamefont {Winkler}},\ }\bibfield  {title}
  {\enquote {\bibinfo {title} {Microswimmers--from single particle motion to
  collective behavior},}\ }\href {\doibase 10.1140/epjst/e2016-60095-3}
  {\bibfield  {journal} {\bibinfo  {journal} {Eur. Phys. J. Spec. Top.}\
  }\textbf {\bibinfo {volume} {225}},\ \bibinfo {pages} {2061} (\bibinfo {year}
  {2016})}\BibitemShut {NoStop}%
\bibitem [{\citenamefont {Niu}\ and\ \citenamefont {Palberg}(2018)}]{niu:18}%
  \BibitemOpen
  \bibfield  {author} {\bibinfo {author} {\bibfnamefont {R.}~\bibnamefont
  {Niu}}\ and\ \bibinfo {author} {\bibfnamefont {T.}~\bibnamefont {Palberg}},\
  }\bibfield  {title} {\enquote {\bibinfo {title} {Modular approach to
  microswimming},}\ }\href {\doibase 10.1039/C8SM00995C} {\bibfield  {journal}
  {\bibinfo  {journal} {Soft Matter}\ }\textbf {\bibinfo {volume} {14}},\
  \bibinfo {pages} {7554} (\bibinfo {year} {2018})}\BibitemShut {NoStop}%
\bibitem [{\citenamefont {Howse}\ \emph {et~al.}(2007)\citenamefont {Howse},
  \citenamefont {Jones}, \citenamefont {Ryan}, \citenamefont {Gough},
  \citenamefont {Vafabakhsh},\ and\ \citenamefont {Golestanian}}]{hows:07}%
  \BibitemOpen
  \bibfield  {author} {\bibinfo {author} {\bibfnamefont {J.~R.}\ \bibnamefont
  {Howse}}, \bibinfo {author} {\bibfnamefont {R.~A.~L.}\ \bibnamefont {Jones}},
  \bibinfo {author} {\bibfnamefont {A.~J.}\ \bibnamefont {Ryan}}, \bibinfo
  {author} {\bibfnamefont {T.}~\bibnamefont {Gough}}, \bibinfo {author}
  {\bibfnamefont {R.}~\bibnamefont {Vafabakhsh}}, \ and\ \bibinfo {author}
  {\bibfnamefont {R.}~\bibnamefont {Golestanian}},\ }\bibfield  {title}
  {\enquote {\bibinfo {title} {Self-motile colloidal particles: From directed
  propulsion to random walk},}\ }\href {\doibase 10.1103/PhysRevLett.99.048102}
  {\bibfield  {journal} {\bibinfo  {journal} {Phys. Rev. Lett.}\ }\textbf
  {\bibinfo {volume} {99}},\ \bibinfo {pages} {048102} (\bibinfo {year}
  {2007})}\BibitemShut {NoStop}%
\bibitem [{\citenamefont {Jiang}, \citenamefont {Yoshinaga},\ and\
  \citenamefont {Sano}(2010)}]{jian:10}%
  \BibitemOpen
  \bibfield  {author} {\bibinfo {author} {\bibfnamefont {H.-R.}\ \bibnamefont
  {Jiang}}, \bibinfo {author} {\bibfnamefont {N.}~\bibnamefont {Yoshinaga}}, \
  and\ \bibinfo {author} {\bibfnamefont {M.}~\bibnamefont {Sano}},\ }\bibfield
  {title} {\enquote {\bibinfo {title} {Active motion of a janus particle by
  self-thermophoresis in a defocused laser beam},}\ }\href {\doibase
  10.1103/PhysRevLett.105.268302} {\bibfield  {journal} {\bibinfo  {journal}
  {Physical Review Letters}\ }\textbf {\bibinfo {volume} {105}},\ \bibinfo
  {pages} {268302} (\bibinfo {year} {2010})}\BibitemShut {NoStop}%
\bibitem [{\citenamefont {Valadares}\ \emph {et~al.}(2010)\citenamefont
  {Valadares}, \citenamefont {Tao}, \citenamefont {Zacharia}, \citenamefont
  {Kitaev}, \citenamefont {Galembeck}, \citenamefont {Kapral},\ and\
  \citenamefont {Ozin}}]{vala:10}%
  \BibitemOpen
  \bibfield  {author} {\bibinfo {author} {\bibfnamefont {L.~F.}\ \bibnamefont
  {Valadares}}, \bibinfo {author} {\bibfnamefont {Y.-G.}\ \bibnamefont {Tao}},
  \bibinfo {author} {\bibfnamefont {N.~S.}\ \bibnamefont {Zacharia}}, \bibinfo
  {author} {\bibfnamefont {V.}~\bibnamefont {Kitaev}}, \bibinfo {author}
  {\bibfnamefont {F.}~\bibnamefont {Galembeck}}, \bibinfo {author}
  {\bibfnamefont {R.}~\bibnamefont {Kapral}}, \ and\ \bibinfo {author}
  {\bibfnamefont {G.~A.}\ \bibnamefont {Ozin}},\ }\bibfield  {title} {\enquote
  {\bibinfo {title} {Catalytic nanomotors: Self-propelled sphere dimers},}\
  }\href {\doibase 10.1002/smll.200901976} {\bibfield  {journal} {\bibinfo
  {journal} {Small}\ }\textbf {\bibinfo {volume} {6}},\ \bibinfo {pages} {565}
  (\bibinfo {year} {2010})}\BibitemShut {NoStop}%
\bibitem [{\citenamefont {W{\"u}rger}(2010)}]{wurg:10}%
  \BibitemOpen
  \bibfield  {author} {\bibinfo {author} {\bibfnamefont {A.}~\bibnamefont
  {W{\"u}rger}},\ }\bibfield  {title} {\enquote {\bibinfo {title} {Thermal
  non-equilibrium transport in colloids},}\ }\href {\doibase
  10.1088/0034-4885/73/12/126601} {\bibfield  {journal} {\bibinfo  {journal}
  {Rep. Prog. Phys.}\ }\textbf {\bibinfo {volume} {73}},\ \bibinfo {pages}
  {126601} (\bibinfo {year} {2010})}\BibitemShut {NoStop}%
\bibitem [{\citenamefont {Volpe}\ \emph {et~al.}(2011)\citenamefont {Volpe},
  \citenamefont {Buttinoni}, \citenamefont {Vogt}, \citenamefont
  {K{\"u}mmerer},\ and\ \citenamefont {Bechinger}}]{volp:11}%
  \BibitemOpen
  \bibfield  {author} {\bibinfo {author} {\bibfnamefont {G.}~\bibnamefont
  {Volpe}}, \bibinfo {author} {\bibfnamefont {I.}~\bibnamefont {Buttinoni}},
  \bibinfo {author} {\bibfnamefont {D.}~\bibnamefont {Vogt}}, \bibinfo {author}
  {\bibfnamefont {H.~J.}\ \bibnamefont {K{\"u}mmerer}}, \ and\ \bibinfo
  {author} {\bibfnamefont {C.}~\bibnamefont {Bechinger}},\ }\bibfield  {title}
  {\enquote {\bibinfo {title} {Microswimmers in patterned environments},}\
  }\href {\doibase 10.1039/C1SM05960B} {\bibfield  {journal} {\bibinfo
  {journal} {Soft Matter}\ }\textbf {\bibinfo {volume} {7}},\ \bibinfo {pages}
  {8810} (\bibinfo {year} {2011})}\BibitemShut {NoStop}%
\bibitem [{\citenamefont {Thutupalli}, \citenamefont {Seemann},\ and\
  \citenamefont {Herminghaus}(2011)}]{thut:11}%
  \BibitemOpen
  \bibfield  {author} {\bibinfo {author} {\bibfnamefont {S.}~\bibnamefont
  {Thutupalli}}, \bibinfo {author} {\bibfnamefont {R.}~\bibnamefont {Seemann}},
  \ and\ \bibinfo {author} {\bibfnamefont {S.}~\bibnamefont {Herminghaus}},\
  }\bibfield  {title} {\enquote {\bibinfo {title} {Swarming behavior of simple
  model squirmers},}\ }\href {\doibase 10.1088/1367-2630/13/7/073021}
  {\bibfield  {journal} {\bibinfo  {journal} {New J. Phys.}\ }\textbf {\bibinfo
  {volume} {13}},\ \bibinfo {pages} {073021} (\bibinfo {year}
  {2011})}\BibitemShut {NoStop}%
\bibitem [{\citenamefont {Buttinoni}\ \emph {et~al.}(2013)\citenamefont
  {Buttinoni}, \citenamefont {Bialk{\'e}}, \citenamefont {K{\"u}mmel},
  \citenamefont {L{\"o}wen}, \citenamefont {Bechinger},\ and\ \citenamefont
  {Speck}}]{butt:13}%
  \BibitemOpen
  \bibfield  {author} {\bibinfo {author} {\bibfnamefont {I.}~\bibnamefont
  {Buttinoni}}, \bibinfo {author} {\bibfnamefont {J.}~\bibnamefont
  {Bialk{\'e}}}, \bibinfo {author} {\bibfnamefont {F.}~\bibnamefont
  {K{\"u}mmel}}, \bibinfo {author} {\bibfnamefont {H.}~\bibnamefont
  {L{\"o}wen}}, \bibinfo {author} {\bibfnamefont {C.}~\bibnamefont
  {Bechinger}}, \ and\ \bibinfo {author} {\bibfnamefont {T.}~\bibnamefont
  {Speck}},\ }\bibfield  {title} {\enquote {\bibinfo {title} {Dynamical
  clustering and phase separation in suspensions of self-propelled colloidal
  particles},}\ }\href {\doibase 10.1103/PhysRevLett.110.238301} {\bibfield
  {journal} {\bibinfo  {journal} {Phys. Rev. Lett.}\ }\textbf {\bibinfo
  {volume} {110}},\ \bibinfo {pages} {238301} (\bibinfo {year}
  {2013})}\BibitemShut {NoStop}%
\bibitem [{\citenamefont {ten Hagen}\ \emph {et~al.}(2014)\citenamefont {ten
  Hagen}, \citenamefont {K{\"u}mmel}, \citenamefont {Wittkowski}, \citenamefont
  {Takagi}, \citenamefont {L{\"o}wen},\ and\ \citenamefont
  {Bechinger}}]{hage:14}%
  \BibitemOpen
  \bibfield  {author} {\bibinfo {author} {\bibfnamefont {B.}~\bibnamefont {ten
  Hagen}}, \bibinfo {author} {\bibfnamefont {F.}~\bibnamefont {K{\"u}mmel}},
  \bibinfo {author} {\bibfnamefont {R.}~\bibnamefont {Wittkowski}}, \bibinfo
  {author} {\bibfnamefont {D.}~\bibnamefont {Takagi}}, \bibinfo {author}
  {\bibfnamefont {H.}~\bibnamefont {L{\"o}wen}}, \ and\ \bibinfo {author}
  {\bibfnamefont {C.}~\bibnamefont {Bechinger}},\ }\bibfield  {title} {\enquote
  {\bibinfo {title} {Gravitaxis of asymmetric self-propelled colloidal
  particles},}\ }\href {\doibase 10.1038/ncomms5829} {\bibfield  {journal}
  {\bibinfo  {journal} {Nat. Commun.}\ }\textbf {\bibinfo {volume} {5}},\
  \bibinfo {pages} {4829} (\bibinfo {year} {2014})}\BibitemShut {NoStop}%
\bibitem [{\citenamefont {Maass}\ \emph {et~al.}(2016)\citenamefont {Maass},
  \citenamefont {Kr{\"u}ger}, \citenamefont {Herminghaus},\ and\ \citenamefont
  {Bahr}}]{maas:16}%
  \BibitemOpen
  \bibfield  {author} {\bibinfo {author} {\bibfnamefont {C.~C.}\ \bibnamefont
  {Maass}}, \bibinfo {author} {\bibfnamefont {C.}~\bibnamefont {Kr{\"u}ger}},
  \bibinfo {author} {\bibfnamefont {S.}~\bibnamefont {Herminghaus}}, \ and\
  \bibinfo {author} {\bibfnamefont {C.}~\bibnamefont {Bahr}},\ }\bibfield
  {title} {\enquote {\bibinfo {title} {Swimming droplets},}\ }\href {\doibase
  10.1146/annurev-conmatphys-031115-011517} {\bibfield  {journal} {\bibinfo
  {journal} {Annu. Rev. Cond. Mat. Phys.}\ }\textbf {\bibinfo {volume} {7}},\
  \bibinfo {pages} {171} (\bibinfo {year} {2016})}\BibitemShut {NoStop}%
\bibitem [{\citenamefont {Vicsek}\ and\ \citenamefont
  {Zafeiris}(2012)}]{vics:12}%
  \BibitemOpen
  \bibfield  {author} {\bibinfo {author} {\bibfnamefont {T.}~\bibnamefont
  {Vicsek}}\ and\ \bibinfo {author} {\bibfnamefont {A.}~\bibnamefont
  {Zafeiris}},\ }\bibfield  {title} {\enquote {\bibinfo {title} {Collective
  motion},}\ }\href {\doibase 10.1016/j.physrep.2012.03.00} {\bibfield
  {journal} {\bibinfo  {journal} {Phys. Rep.}\ }\textbf {\bibinfo {volume}
  {517}},\ \bibinfo {pages} {71} (\bibinfo {year} {2012})}\BibitemShut
  {NoStop}%
\bibitem [{\citenamefont {Romanczuk}\ \emph {et~al.}(2012)\citenamefont
  {Romanczuk}, \citenamefont {B{\"a}r}, \citenamefont {Ebeling}, \citenamefont
  {Lindner},\ and\ \citenamefont {Schimansky-Geier}}]{roma:12}%
  \BibitemOpen
  \bibfield  {author} {\bibinfo {author} {\bibfnamefont {P.}~\bibnamefont
  {Romanczuk}}, \bibinfo {author} {\bibfnamefont {M.}~\bibnamefont {B{\"a}r}},
  \bibinfo {author} {\bibfnamefont {W.}~\bibnamefont {Ebeling}}, \bibinfo
  {author} {\bibfnamefont {B.}~\bibnamefont {Lindner}}, \ and\ \bibinfo
  {author} {\bibfnamefont {L.}~\bibnamefont {Schimansky-Geier}},\ }\bibfield
  {title} {\enquote {\bibinfo {title} {Active {B}rownian particles},}\ }\href
  {\doibase 10.1140/epjst/e2012-01529-y} {\bibfield  {journal} {\bibinfo
  {journal} {Eur. Phys. J. Spec. Top.}\ }\textbf {\bibinfo {volume} {202}},\
  \bibinfo {pages} {1} (\bibinfo {year} {2012})}\BibitemShut {NoStop}%
\bibitem [{\citenamefont {Marchetti}\ \emph {et~al.}(2016)\citenamefont
  {Marchetti}, \citenamefont {Fily}, \citenamefont {Henkes}, \citenamefont
  {Patch},\ and\ \citenamefont {Yllanes}}]{marc:16.1}%
  \BibitemOpen
  \bibfield  {author} {\bibinfo {author} {\bibfnamefont {M.~C.}\ \bibnamefont
  {Marchetti}}, \bibinfo {author} {\bibfnamefont {Y.}~\bibnamefont {Fily}},
  \bibinfo {author} {\bibfnamefont {S.}~\bibnamefont {Henkes}}, \bibinfo
  {author} {\bibfnamefont {A.}~\bibnamefont {Patch}}, \ and\ \bibinfo {author}
  {\bibfnamefont {D.}~\bibnamefont {Yllanes}},\ }\bibfield  {title} {\enquote
  {\bibinfo {title} {Minimal model of active colloids highlights the role of
  mechanical interactions in controlling the emergent behavior of active
  matter},}\ }\href {\doibase 10.1016/j.cocis.2016.01.003} {\bibfield
  {journal} {\bibinfo  {journal} {Curr. Opin. Colloid Interface Sci.}\ }\textbf
  {\bibinfo {volume} {21}},\ \bibinfo {pages} {34} (\bibinfo {year}
  {2016})}\BibitemShut {NoStop}%
\bibitem [{\citenamefont {Z{\"o}ttl}\ and\ \citenamefont
  {Stark}(2016)}]{zoet:16}%
  \BibitemOpen
  \bibfield  {author} {\bibinfo {author} {\bibfnamefont {A.}~\bibnamefont
  {Z{\"o}ttl}}\ and\ \bibinfo {author} {\bibfnamefont {H.}~\bibnamefont
  {Stark}},\ }\bibfield  {title} {\enquote {\bibinfo {title} {Emergent behavior
  in active colloids},}\ }\href {\doibase 10.1088/0953-8984/28/25/253001}
  {\bibfield  {journal} {\bibinfo  {journal} {J. Phys.: Condens. Matter}\
  }\textbf {\bibinfo {volume} {28}},\ \bibinfo {pages} {253001} (\bibinfo
  {year} {2016})}\BibitemShut {NoStop}%
\bibitem [{\citenamefont {Ebbens}\ \emph {et~al.}(2010)\citenamefont {Ebbens},
  \citenamefont {Jones}, \citenamefont {Ryan}, \citenamefont {Golestanian},\
  and\ \citenamefont {Howse}}]{ebbe:10}%
  \BibitemOpen
  \bibfield  {author} {\bibinfo {author} {\bibfnamefont {S.}~\bibnamefont
  {Ebbens}}, \bibinfo {author} {\bibfnamefont {R.~A.~L.}\ \bibnamefont
  {Jones}}, \bibinfo {author} {\bibfnamefont {A.~J.}\ \bibnamefont {Ryan}},
  \bibinfo {author} {\bibfnamefont {R.}~\bibnamefont {Golestanian}}, \ and\
  \bibinfo {author} {\bibfnamefont {J.~R.}\ \bibnamefont {Howse}},\ }\bibfield
  {title} {\enquote {\bibinfo {title} {Self-assembled autonomous runners and
  tumblers},}\ }\href {\doibase 10.1103/PhysRevE.82.015304} {\bibfield
  {journal} {\bibinfo  {journal} {Phys. Rev. E}\ }\textbf {\bibinfo {volume}
  {82}},\ \bibinfo {pages} {015304} (\bibinfo {year} {2010})}\BibitemShut
  {NoStop}%
\bibitem [{\citenamefont {Gibbs}\ \emph {et~al.}(2017)\citenamefont {Gibbs},
  \citenamefont {Nourhani}, \citenamefont {Johnson},\ and\ \citenamefont
  {Lammert}}]{gibb:17}%
  \BibitemOpen
  \bibfield  {author} {\bibinfo {author} {\bibfnamefont {J.~G.}\ \bibnamefont
  {Gibbs}}, \bibinfo {author} {\bibfnamefont {A.}~\bibnamefont {Nourhani}},
  \bibinfo {author} {\bibfnamefont {J.~N.}\ \bibnamefont {Johnson}}, \ and\
  \bibinfo {author} {\bibfnamefont {P.~E.}\ \bibnamefont {Lammert}},\
  }\bibfield  {title} {\enquote {\bibinfo {title} {Spiral diffusion of
  self-assembled dimers of {J}anus spheres},}\ }\href {\doibase DOI:
  10.1557/adv.2017.383} {\bibfield  {journal} {\bibinfo  {journal} {MRS Adv.}\
  }\textbf {\bibinfo {volume} {2}},\ \bibinfo {pages} {3471} (\bibinfo {year}
  {2017})}\BibitemShut {NoStop}%
\bibitem [{\citenamefont {Yan}\ \emph {et~al.}(2016)\citenamefont {Yan},
  \citenamefont {Han}, \citenamefont {Zhang}, \citenamefont {Xu}, \citenamefont
  {Luijten},\ and\ \citenamefont {Granick}}]{yan:16}%
  \BibitemOpen
  \bibfield  {author} {\bibinfo {author} {\bibfnamefont {J.}~\bibnamefont
  {Yan}}, \bibinfo {author} {\bibfnamefont {M.}~\bibnamefont {Han}}, \bibinfo
  {author} {\bibfnamefont {J.}~\bibnamefont {Zhang}}, \bibinfo {author}
  {\bibfnamefont {C.}~\bibnamefont {Xu}}, \bibinfo {author} {\bibfnamefont
  {E.}~\bibnamefont {Luijten}}, \ and\ \bibinfo {author} {\bibfnamefont
  {S.}~\bibnamefont {Granick}},\ }\bibfield  {title} {\enquote {\bibinfo
  {title} {Reconfiguring active particles by electrostatic imbalance},}\ }\href
  {\doibase 10.1038/nmat4696} {\bibfield  {journal} {\bibinfo  {journal} {Nat.
  Mat.}\ }\textbf {\bibinfo {volume} {15}},\ \bibinfo {pages} {1095} (\bibinfo
  {year} {2016})}\BibitemShut {NoStop}%
\bibitem [{\citenamefont {Di~Leonardo}(2016)}]{dile:16}%
  \BibitemOpen
  \bibfield  {author} {\bibinfo {author} {\bibfnamefont {R.}~\bibnamefont
  {Di~Leonardo}},\ }\bibfield  {title} {\enquote {\bibinfo {title} {Active
  colloids: Controlled collective motions},}\ }\href {\doibase
  10.1038/nmat4761} {\bibfield  {journal} {\bibinfo  {journal} {Nat. Mat.}\
  }\textbf {\bibinfo {volume} {15}},\ \bibinfo {pages} {1057} (\bibinfo {year}
  {2016})}\BibitemShut {NoStop}%
\bibitem [{\citenamefont {Zhang}, \citenamefont {Yan},\ and\ \citenamefont
  {Granick}(2016)}]{zhan:16}%
  \BibitemOpen
  \bibfield  {author} {\bibinfo {author} {\bibfnamefont {J.}~\bibnamefont
  {Zhang}}, \bibinfo {author} {\bibfnamefont {J.}~\bibnamefont {Yan}}, \ and\
  \bibinfo {author} {\bibfnamefont {S.}~\bibnamefont {Granick}},\ }\bibfield
  {title} {\enquote {\bibinfo {title} {Directed self-assembly pathways of
  active colloidal clusters},}\ }\href {\doibase 10.1002/anie.201509978}
  {\bibfield  {journal} {\bibinfo  {journal} {Angew. Chem. Int. Ed.}\ }\textbf
  {\bibinfo {volume} {55}},\ \bibinfo {pages} {5166} (\bibinfo {year}
  {2016})}\BibitemShut {NoStop}%
\bibitem [{\citenamefont {Zhang}\ and\ \citenamefont
  {Granick}(2016)}]{zhan:16.1}%
  \BibitemOpen
  \bibfield  {author} {\bibinfo {author} {\bibfnamefont {J.}~\bibnamefont
  {Zhang}}\ and\ \bibinfo {author} {\bibfnamefont {S.}~\bibnamefont
  {Granick}},\ }\bibfield  {title} {\enquote {\bibinfo {title} {Natural
  selection in the colloid world: active chiral spirals},}\ }\href {\doibase
  10.1039/C6FD00077K} {\bibfield  {journal} {\bibinfo  {journal} {Faraday
  Discuss.}\ }\textbf {\bibinfo {volume} {191}},\ \bibinfo {pages} {35}
  (\bibinfo {year} {2016})}\BibitemShut {NoStop}%
\bibitem [{\citenamefont {Nishiguchi}\ \emph {et~al.}(2018)\citenamefont
  {Nishiguchi}, \citenamefont {Iwasawa}, \citenamefont {Jiang},\ and\
  \citenamefont {Sano}}]{nish:18}%
  \BibitemOpen
  \bibfield  {author} {\bibinfo {author} {\bibfnamefont {D.}~\bibnamefont
  {Nishiguchi}}, \bibinfo {author} {\bibfnamefont {J.}~\bibnamefont {Iwasawa}},
  \bibinfo {author} {\bibfnamefont {H.-R.}\ \bibnamefont {Jiang}}, \ and\
  \bibinfo {author} {\bibfnamefont {M.}~\bibnamefont {Sano}},\ }\bibfield
  {title} {\enquote {\bibinfo {title} {Flagellar dynamics of chains of active
  {J}anus particles fueled by an {AC} electric field},}\ }\href {\doibase
  10.1088/1367-2630/aa9b48} {\bibfield  {journal} {\bibinfo  {journal} {New. J.
  Phys.}\ }\textbf {\bibinfo {volume} {20}},\ \bibinfo {pages} {015002}
  (\bibinfo {year} {2018})}\BibitemShut {NoStop}%
\bibitem [{\citenamefont {Mart{{\'\i}}n-G{{\'o}}mez}\ \emph
  {et~al.}(2019)\citenamefont {Mart{{\'\i}}n-G{{\'o}}mez}, \citenamefont
  {Eisenstecken}, \citenamefont {Gompper},\ and\ \citenamefont
  {Winkler}}]{mart:19}%
  \BibitemOpen
  \bibfield  {author} {\bibinfo {author} {\bibfnamefont {A.}~\bibnamefont
  {Mart{{\'\i}}n-G{{\'o}}mez}}, \bibinfo {author} {\bibfnamefont
  {T.}~\bibnamefont {Eisenstecken}}, \bibinfo {author} {\bibfnamefont
  {G.}~\bibnamefont {Gompper}}, \ and\ \bibinfo {author} {\bibfnamefont
  {R.~G.}\ \bibnamefont {Winkler}},\ }\bibfield  {title} {\enquote {\bibinfo
  {title} {Active {B}rownian filaments with hydrodynamic interactions:
  conformations and dynamics},}\ }\href {\doibase 10.1039/C9SM00391F}
  {\bibfield  {journal} {\bibinfo  {journal} {Soft Matter}\ }\textbf {\bibinfo
  {volume} {15}},\ \bibinfo {pages} {3957} (\bibinfo {year}
  {2019})}\BibitemShut {NoStop}%
\bibitem [{\citenamefont {Vutukuri}\ \emph {et~al.}(2017)\citenamefont
  {Vutukuri}, \citenamefont {Bet}, \citenamefont {van Roij}, \citenamefont
  {Dijkstra},\ and\ \citenamefont {Huck}}]{vutu:17}%
  \BibitemOpen
  \bibfield  {author} {\bibinfo {author} {\bibfnamefont {H.~R.}\ \bibnamefont
  {Vutukuri}}, \bibinfo {author} {\bibfnamefont {B.}~\bibnamefont {Bet}},
  \bibinfo {author} {\bibfnamefont {R.}~\bibnamefont {van Roij}}, \bibinfo
  {author} {\bibfnamefont {M.}~\bibnamefont {Dijkstra}}, \ and\ \bibinfo
  {author} {\bibfnamefont {W.~T.~S.}\ \bibnamefont {Huck}},\ }\bibfield
  {title} {\enquote {\bibinfo {title} {Rational design and dynamics of
  self-propelled colloidal bead chains: from rotators to flagella},}\ }\href
  {\doibase 10.1038/s41598-017-16731-5} {\bibfield  {journal} {\bibinfo
  {journal} {Sci. Rep.}\ }\textbf {\bibinfo {volume} {7}},\ \bibinfo {pages}
  {16758} (\bibinfo {year} {2017})}\BibitemShut {NoStop}%
\bibitem [{\citenamefont {Sasaki}\ \emph {et~al.}(2014)\citenamefont {Sasaki},
  \citenamefont {Takikawa}, \citenamefont {Jampani}, \citenamefont {Hoshikawa},
  \citenamefont {Seto}, \citenamefont {Bahr}, \citenamefont {Herminghaus},
  \citenamefont {Hidaka},\ and\ \citenamefont {Orihara}}]{sasa:14}%
  \BibitemOpen
  \bibfield  {author} {\bibinfo {author} {\bibfnamefont {Y.}~\bibnamefont
  {Sasaki}}, \bibinfo {author} {\bibfnamefont {Y.}~\bibnamefont {Takikawa}},
  \bibinfo {author} {\bibfnamefont {V.~S.~R.}\ \bibnamefont {Jampani}},
  \bibinfo {author} {\bibfnamefont {H.}~\bibnamefont {Hoshikawa}}, \bibinfo
  {author} {\bibfnamefont {T.}~\bibnamefont {Seto}}, \bibinfo {author}
  {\bibfnamefont {C.}~\bibnamefont {Bahr}}, \bibinfo {author} {\bibfnamefont
  {S.}~\bibnamefont {Herminghaus}}, \bibinfo {author} {\bibfnamefont
  {Y.}~\bibnamefont {Hidaka}}, \ and\ \bibinfo {author} {\bibfnamefont
  {H.}~\bibnamefont {Orihara}},\ }\bibfield  {title} {\enquote {\bibinfo
  {title} {Colloidal caterpillars for cargo transportation},}\ }\href {\doibase
  10.1039/C4SM01354A} {\bibfield  {journal} {\bibinfo  {journal} {Soft Matter}\
  }\textbf {\bibinfo {volume} {10}},\ \bibinfo {pages} {8813} (\bibinfo {year}
  {2014})}\BibitemShut {NoStop}%
\bibitem [{\citenamefont {Martinez-Pedrero}\ \emph {et~al.}(2015)\citenamefont
  {Martinez-Pedrero}, \citenamefont {Ortiz-Ambriz}, \citenamefont
  {Pagonabarraga},\ and\ \citenamefont {Tierno}}]{mart:15}%
  \BibitemOpen
  \bibfield  {author} {\bibinfo {author} {\bibfnamefont {F.}~\bibnamefont
  {Martinez-Pedrero}}, \bibinfo {author} {\bibfnamefont {A.}~\bibnamefont
  {Ortiz-Ambriz}}, \bibinfo {author} {\bibfnamefont {I.}~\bibnamefont
  {Pagonabarraga}}, \ and\ \bibinfo {author} {\bibfnamefont {P.}~\bibnamefont
  {Tierno}},\ }\bibfield  {title} {\enquote {\bibinfo {title} {Colloidal
  microworms propelling via a cooperative hydrodynamic conveyor belt},}\ }\href
  {\doibase 10.1103/PhysRevLett.115.138301} {\bibfield  {journal} {\bibinfo
  {journal} {Phys. Rev. Lett.}\ }\textbf {\bibinfo {volume} {115}},\ \bibinfo
  {pages} {138301} (\bibinfo {year} {2015})}\BibitemShut {NoStop}%
\bibitem [{\citenamefont {Kokot}\ \emph {et~al.}(2017)\citenamefont {Kokot},
  \citenamefont {Das}, \citenamefont {Winkler}, \citenamefont {Gompper},
  \citenamefont {Aranson},\ and\ \citenamefont {Snezhko}}]{koko:17}%
  \BibitemOpen
  \bibfield  {author} {\bibinfo {author} {\bibfnamefont {G.}~\bibnamefont
  {Kokot}}, \bibinfo {author} {\bibfnamefont {S.}~\bibnamefont {Das}}, \bibinfo
  {author} {\bibfnamefont {R.~G.}\ \bibnamefont {Winkler}}, \bibinfo {author}
  {\bibfnamefont {G.}~\bibnamefont {Gompper}}, \bibinfo {author} {\bibfnamefont
  {I.~S.}\ \bibnamefont {Aranson}}, \ and\ \bibinfo {author} {\bibfnamefont
  {A.}~\bibnamefont {Snezhko}},\ }\bibfield  {title} {\enquote {\bibinfo
  {title} {Active turbulence in a gas of self-assembled spinners},}\ }\href
  {\doibase 10.1073/pnas.1710188114} {\bibfield  {journal} {\bibinfo  {journal}
  {Proc. Natl. Acad. Sci. USA}\ }\textbf {\bibinfo {volume} {114}},\ \bibinfo
  {pages} {12870} (\bibinfo {year} {2017})}\BibitemShut {NoStop}%
\bibitem [{\citenamefont {Biswas}\ \emph {et~al.}(2017)\citenamefont {Biswas},
  \citenamefont {Manna}, \citenamefont {Laskar}, \citenamefont {Kumar},
  \citenamefont {Adhikari},\ and\ \citenamefont {Kumaraswamy}}]{bisw:17}%
  \BibitemOpen
  \bibfield  {author} {\bibinfo {author} {\bibfnamefont {B.}~\bibnamefont
  {Biswas}}, \bibinfo {author} {\bibfnamefont {R.~K.}\ \bibnamefont {Manna}},
  \bibinfo {author} {\bibfnamefont {A.}~\bibnamefont {Laskar}}, \bibinfo
  {author} {\bibfnamefont {P.~B.~S.}\ \bibnamefont {Kumar}}, \bibinfo {author}
  {\bibfnamefont {R.}~\bibnamefont {Adhikari}}, \ and\ \bibinfo {author}
  {\bibfnamefont {G.}~\bibnamefont {Kumaraswamy}},\ }\bibfield  {title}
  {\enquote {\bibinfo {title} {Linking catalyst-coated isotropic colloids into
  ``active''flexible chains enhances their diffusivity},}\ }\href {\doibase
  10.1021/acsnano.7b04265} {\bibfield  {journal} {\bibinfo  {journal} {ACS
  Nano}\ }\textbf {\bibinfo {volume} {11}},\ \bibinfo {pages} {10025} (\bibinfo
  {year} {2017})}\BibitemShut {NoStop}%
\bibitem [{\citenamefont {Shaebani}\ \emph {et~al.}(2020)\citenamefont
  {Shaebani}, \citenamefont {Wysocki}, \citenamefont {Winkler}, \citenamefont
  {Gompper},\ and\ \citenamefont {Rieger}}]{shae:20}%
  \BibitemOpen
  \bibfield  {author} {\bibinfo {author} {\bibfnamefont {M.~R.}\ \bibnamefont
  {Shaebani}}, \bibinfo {author} {\bibfnamefont {A.}~\bibnamefont {Wysocki}},
  \bibinfo {author} {\bibfnamefont {R.~G.}\ \bibnamefont {Winkler}}, \bibinfo
  {author} {\bibfnamefont {G.}~\bibnamefont {Gompper}}, \ and\ \bibinfo
  {author} {\bibfnamefont {H.}~\bibnamefont {Rieger}},\ }\bibfield  {title}
  {\enquote {\bibinfo {title} {Computational models for active matter},}\
  }\href {\doibase 10.1038/s42254-020-0152-1} {\bibfield  {journal} {\bibinfo
  {journal} {Nat. Rev. Phys.}\ }\textbf {\bibinfo {volume} {2}},\ \bibinfo
  {pages} {181} (\bibinfo {year} {2020})}\BibitemShut {NoStop}%
\bibitem [{\citenamefont {Eisenstecken}, \citenamefont {Gompper},\ and\
  \citenamefont {Winkler}(2017)}]{eise:17}%
  \BibitemOpen
  \bibfield  {author} {\bibinfo {author} {\bibfnamefont {T.}~\bibnamefont
  {Eisenstecken}}, \bibinfo {author} {\bibfnamefont {G.}~\bibnamefont
  {Gompper}}, \ and\ \bibinfo {author} {\bibfnamefont {R.~G.}\ \bibnamefont
  {Winkler}},\ }\bibfield  {title} {\enquote {\bibinfo {title} {Internal
  dynamics of semiflexible polymers with active noise},}\ }\href {\doibase
  10.1063/1.4981012} {\bibfield  {journal} {\bibinfo  {journal} {J. Chem.
  Phys.}\ }\textbf {\bibinfo {volume} {146}},\ \bibinfo {pages} {154903}
  (\bibinfo {year} {2017})}\BibitemShut {NoStop}%
\bibitem [{\citenamefont {Martin-Gomez}\ \emph {et~al.}(2020)\citenamefont
  {Martin-Gomez}, \citenamefont {Eisenstecken}, \citenamefont {Gompper},\ and\
  \citenamefont {Winkler}}]{mart:20}%
  \BibitemOpen
  \bibfield  {author} {\bibinfo {author} {\bibfnamefont {A.}~\bibnamefont
  {Martin-Gomez}}, \bibinfo {author} {\bibfnamefont {T.}~\bibnamefont
  {Eisenstecken}}, \bibinfo {author} {\bibfnamefont {G.}~\bibnamefont
  {Gompper}}, \ and\ \bibinfo {author} {\bibfnamefont {R.~G.}\ \bibnamefont
  {Winkler}},\ }\bibfield  {title} {\enquote {\bibinfo {title} {Hydrodynamics
  of polymers in an active bath},}\ }\href {\doibase
  10.1103/PhysRevE.101.052612} {\bibfield  {journal} {\bibinfo  {journal}
  {Phys. Rev. E}\ }\textbf {\bibinfo {volume} {101}},\ \bibinfo {pages}
  {052612} (\bibinfo {year} {2020})}\BibitemShut {NoStop}%
\bibitem [{\citenamefont {Needleman}\ and\ \citenamefont
  {Dogic}(2017)}]{need:17}%
  \BibitemOpen
  \bibfield  {author} {\bibinfo {author} {\bibfnamefont {D.}~\bibnamefont
  {Needleman}}\ and\ \bibinfo {author} {\bibfnamefont {Z.}~\bibnamefont
  {Dogic}},\ }\bibfield  {title} {\enquote {\bibinfo {title} {Active matter at
  the interface between materials science and cell biology},}\ }\href {\doibase
  10.1038/natrevmats.2017.48} {\bibfield  {journal} {\bibinfo  {journal} {Nat.
  Rev. Mater.}\ }\textbf {\bibinfo {volume} {2}},\ \bibinfo {pages} {201748}
  (\bibinfo {year} {2017})}\BibitemShut {NoStop}%
\bibitem [{\citenamefont {Peruani}, \citenamefont {Schimansky-Geier},\ and\
  \citenamefont {B{\"a}r}(2010)}]{peru:10}%
  \BibitemOpen
  \bibfield  {author} {\bibinfo {author} {\bibfnamefont {F.}~\bibnamefont
  {Peruani}}, \bibinfo {author} {\bibfnamefont {L.}~\bibnamefont
  {Schimansky-Geier}}, \ and\ \bibinfo {author} {\bibfnamefont
  {M.}~\bibnamefont {B{\"a}r}},\ }\bibfield  {title} {\enquote {\bibinfo
  {title} {Cluster dynamics and cluster size distributions in systems of
  self-propelled particles},}\ }\href {\doibase 10.1140/epjst/e2010-01349-1}
  {\bibfield  {journal} {\bibinfo  {journal} {Eur. Phys. J. Spec. Top.}\
  }\textbf {\bibinfo {volume} {191}},\ \bibinfo {pages} {173} (\bibinfo {year}
  {2010})}\BibitemShut {NoStop}%
\bibitem [{\citenamefont {Fily}\ and\ \citenamefont
  {Marchetti}(2012)}]{fily:12}%
  \BibitemOpen
  \bibfield  {author} {\bibinfo {author} {\bibfnamefont {Y.}~\bibnamefont
  {Fily}}\ and\ \bibinfo {author} {\bibfnamefont {M.~C.}\ \bibnamefont
  {Marchetti}},\ }\bibfield  {title} {\enquote {\bibinfo {title} {Athermal
  phase separation of self-propelled particles with no alignment},}\ }\href
  {\doibase 10.1103/PhysRevLett.108.235702} {\bibfield  {journal} {\bibinfo
  {journal} {Phys. Rev. Lett.}\ }\textbf {\bibinfo {volume} {108}},\ \bibinfo
  {pages} {235702} (\bibinfo {year} {2012})}\BibitemShut {NoStop}%
\bibitem [{\citenamefont {Bialk{\'e}}, \citenamefont {Speck},\ and\
  \citenamefont {L{\"o}wen}(2012)}]{bial:12}%
  \BibitemOpen
  \bibfield  {author} {\bibinfo {author} {\bibfnamefont {J.}~\bibnamefont
  {Bialk{\'e}}}, \bibinfo {author} {\bibfnamefont {T.}~\bibnamefont {Speck}}, \
  and\ \bibinfo {author} {\bibfnamefont {H.}~\bibnamefont {L{\"o}wen}},\
  }\bibfield  {title} {\enquote {\bibinfo {title} {Crystallization in a dense
  suspension of self-propelled particles},}\ }\href {\doibase
  10.1103/PhysRevLett.108.168301} {\bibfield  {journal} {\bibinfo  {journal}
  {Phys. Rev. Lett.}\ }\textbf {\bibinfo {volume} {108}},\ \bibinfo {pages}
  {168301} (\bibinfo {year} {2012})}\BibitemShut {NoStop}%
\bibitem [{\citenamefont {Redner}, \citenamefont {Hagan},\ and\ \citenamefont
  {Baskaran}(2013)}]{redn:13}%
  \BibitemOpen
  \bibfield  {author} {\bibinfo {author} {\bibfnamefont {G.~S.}\ \bibnamefont
  {Redner}}, \bibinfo {author} {\bibfnamefont {M.~F.}\ \bibnamefont {Hagan}}, \
  and\ \bibinfo {author} {\bibfnamefont {A.}~\bibnamefont {Baskaran}},\
  }\bibfield  {title} {\enquote {\bibinfo {title} {Structure and dynamics of a
  phase-separating active colloidal fluid},}\ }\href {\doibase
  10.1103/PhysRevLett.110.055701} {\bibfield  {journal} {\bibinfo  {journal}
  {Phys. Rev. Lett.}\ }\textbf {\bibinfo {volume} {110}},\ \bibinfo {pages}
  {055701} (\bibinfo {year} {2013})}\BibitemShut {NoStop}%
\bibitem [{\citenamefont {ten Hagen}\ \emph {et~al.}(2015)\citenamefont {ten
  Hagen}, \citenamefont {Wittkowski}, \citenamefont {Takagi}, \citenamefont
  {K{\"u}mmel}, \citenamefont {Bechinger},\ and\ \citenamefont
  {L{\"o}wen}}]{hage:15}%
  \BibitemOpen
  \bibfield  {author} {\bibinfo {author} {\bibfnamefont {B.}~\bibnamefont {ten
  Hagen}}, \bibinfo {author} {\bibfnamefont {R.}~\bibnamefont {Wittkowski}},
  \bibinfo {author} {\bibfnamefont {D.}~\bibnamefont {Takagi}}, \bibinfo
  {author} {\bibfnamefont {F.}~\bibnamefont {K{\"u}mmel}}, \bibinfo {author}
  {\bibfnamefont {C.}~\bibnamefont {Bechinger}}, \ and\ \bibinfo {author}
  {\bibfnamefont {H.}~\bibnamefont {L{\"o}wen}},\ }\bibfield  {title} {\enquote
  {\bibinfo {title} {Can the self-propulsion of anisotropic microswimmers be
  described by using forces and torques?}}\ }\href {\doibase
  10.1088/0953-8984/27/19/194110} {\bibfield  {journal} {\bibinfo  {journal}
  {J. Phys.: Condens. Matter}\ }\textbf {\bibinfo {volume} {27}},\ \bibinfo
  {pages} {194110} (\bibinfo {year} {2015})}\BibitemShut {NoStop}%
\bibitem [{\citenamefont {L{\"o}wen}(2020)}]{loew:20}%
  \BibitemOpen
  \bibfield  {author} {\bibinfo {author} {\bibfnamefont {H.}~\bibnamefont
  {L{\"o}wen}},\ }\bibfield  {title} {\enquote {\bibinfo {title} {Inertial
  effects of self-propelled particles: From active {B}rownian to active
  {L}angevin motion},}\ }\href {\doibase 10.1063/1.5134455} {\bibfield
  {journal} {\bibinfo  {journal} {J. Chem. Phys.}\ }\textbf {\bibinfo {volume}
  {152}},\ \bibinfo {pages} {040901} (\bibinfo {year} {2020})}\BibitemShut
  {NoStop}%
\bibitem [{\citenamefont {Winkler}(2016{\natexlab{a}})}]{wink:16}%
  \BibitemOpen
  \bibfield  {author} {\bibinfo {author} {\bibfnamefont {R.~G.}\ \bibnamefont
  {Winkler}},\ }\bibfield  {title} {\enquote {\bibinfo {title} {Dynamics of
  flexible active {B}rownian dumbbells in the absence and the presence of shear
  flow},}\ }\href {\doibase 10.1039/C5SM02965A} {\bibfield  {journal} {\bibinfo
   {journal} {Soft Matter}\ }\textbf {\bibinfo {volume} {12}},\ \bibinfo
  {pages} {3737} (\bibinfo {year} {2016}{\natexlab{a}})}\BibitemShut {NoStop}%
\bibitem [{\citenamefont {Siebert}\ \emph {et~al.}(2017)\citenamefont
  {Siebert}, \citenamefont {Letz}, \citenamefont {Speck},\ and\ \citenamefont
  {Virnau}}]{sieb:17}%
  \BibitemOpen
  \bibfield  {author} {\bibinfo {author} {\bibfnamefont {J.~T.}\ \bibnamefont
  {Siebert}}, \bibinfo {author} {\bibfnamefont {J.}~\bibnamefont {Letz}},
  \bibinfo {author} {\bibfnamefont {T.}~\bibnamefont {Speck}}, \ and\ \bibinfo
  {author} {\bibfnamefont {P.}~\bibnamefont {Virnau}},\ }\bibfield  {title}
  {\enquote {\bibinfo {title} {Phase behavior of active {B}rownian disks,
  spheres, and dimers},}\ }\href {\doibase 10.1039/C6SM02622B} {\bibfield
  {journal} {\bibinfo  {journal} {Soft Matter}\ }\textbf {\bibinfo {volume}
  {13}},\ \bibinfo {pages} {1020} (\bibinfo {year} {2017})}\BibitemShut
  {NoStop}%
\bibitem [{\citenamefont {Suma}\ \emph {et~al.}(2014)\citenamefont {Suma},
  \citenamefont {Gonnella}, \citenamefont {Marenduzzo},\ and\ \citenamefont
  {Orlandini}}]{suma:14}%
  \BibitemOpen
  \bibfield  {author} {\bibinfo {author} {\bibfnamefont {A.}~\bibnamefont
  {Suma}}, \bibinfo {author} {\bibfnamefont {G.}~\bibnamefont {Gonnella}},
  \bibinfo {author} {\bibfnamefont {D.}~\bibnamefont {Marenduzzo}}, \ and\
  \bibinfo {author} {\bibfnamefont {E.}~\bibnamefont {Orlandini}},\ }\bibfield
  {title} {\enquote {\bibinfo {title} {Motility-induced phase separation in an
  active dumbbell fluid},}\ }\href {\doibase 10.1209/0295-5075/108/56004}
  {\bibfield  {journal} {\bibinfo  {journal} {EPL}\ }\textbf {\bibinfo {volume}
  {108}},\ \bibinfo {pages} {56004} (\bibinfo {year} {2014})}\BibitemShut
  {NoStop}%
\bibitem [{\citenamefont {Nourhani}\ \emph {et~al.}(2016)\citenamefont
  {Nourhani}, \citenamefont {Ebbens}, \citenamefont {Gibbs},\ and\
  \citenamefont {Lammert}}]{nour:16}%
  \BibitemOpen
  \bibfield  {author} {\bibinfo {author} {\bibfnamefont {A.}~\bibnamefont
  {Nourhani}}, \bibinfo {author} {\bibfnamefont {S.~J.}\ \bibnamefont
  {Ebbens}}, \bibinfo {author} {\bibfnamefont {J.~G.}\ \bibnamefont {Gibbs}}, \
  and\ \bibinfo {author} {\bibfnamefont {P.~E.}\ \bibnamefont {Lammert}},\
  }\bibfield  {title} {\enquote {\bibinfo {title} {Spiral diffusion of rotating
  self-propellers with stochastic perturbation},}\ }\href {\doibase
  10.1103/PhysRevE.94.030601} {\bibfield  {journal} {\bibinfo  {journal} {Phys.
  Rev. E}\ }\textbf {\bibinfo {volume} {94}},\ \bibinfo {pages} {030601}
  (\bibinfo {year} {2016})}\BibitemShut {NoStop}%
\bibitem [{\citenamefont {Cugliandolo}, \citenamefont {Gonnella},\ and\
  \citenamefont {Suma}(2015)}]{cugl:15}%
  \BibitemOpen
  \bibfield  {author} {\bibinfo {author} {\bibfnamefont {L.~F.}\ \bibnamefont
  {Cugliandolo}}, \bibinfo {author} {\bibfnamefont {G.}~\bibnamefont
  {Gonnella}}, \ and\ \bibinfo {author} {\bibfnamefont {A.}~\bibnamefont
  {Suma}},\ }\bibfield  {title} {\enquote {\bibinfo {title} {Rotational and
  translational diffusion in an interacting active dumbbell system},}\ }\href
  {\doibase 10.1103/PhysRevE.91.062124} {\bibfield  {journal} {\bibinfo
  {journal} {Phys. Rev. E}\ }\textbf {\bibinfo {volume} {91}},\ \bibinfo
  {pages} {062124} (\bibinfo {year} {2015})}\BibitemShut {NoStop}%
\bibitem [{\citenamefont {Bianco}, \citenamefont {Locatelli},\ and\
  \citenamefont {Malgaretti}(2018)}]{bian:18}%
  \BibitemOpen
  \bibfield  {author} {\bibinfo {author} {\bibfnamefont {V.}~\bibnamefont
  {Bianco}}, \bibinfo {author} {\bibfnamefont {E.}~\bibnamefont {Locatelli}}, \
  and\ \bibinfo {author} {\bibfnamefont {P.}~\bibnamefont {Malgaretti}},\
  }\bibfield  {title} {\enquote {\bibinfo {title} {Globulelike conformation and
  enhanced diffusion of active polymers},}\ }\href {\doibase
  10.1103/PhysRevLett.121.217802} {\bibfield  {journal} {\bibinfo  {journal}
  {Phys. Rev. Lett.}\ }\textbf {\bibinfo {volume} {121}},\ \bibinfo {pages}
  {217802} (\bibinfo {year} {2018})}\BibitemShut {NoStop}%
\bibitem [{\citenamefont {Yang}, \citenamefont {Qiu},\ and\ \citenamefont
  {Gompper}(2014)}]{yang:14.3}%
  \BibitemOpen
  \bibfield  {author} {\bibinfo {author} {\bibfnamefont {Y.}~\bibnamefont
  {Yang}}, \bibinfo {author} {\bibfnamefont {F.}~\bibnamefont {Qiu}}, \ and\
  \bibinfo {author} {\bibfnamefont {G.}~\bibnamefont {Gompper}},\ }\bibfield
  {title} {\enquote {\bibinfo {title} {Self-organized vortices of circling
  self-propelled particles and curved active flagella},}\ }\href {\doibase
  10.1103/PhysRevE.89.012720} {\bibfield  {journal} {\bibinfo  {journal} {Phys.
  Rev. E}\ }\textbf {\bibinfo {volume} {89}},\ \bibinfo {pages} {012720}
  (\bibinfo {year} {2014})}\BibitemShut {NoStop}%
\bibitem [{\citenamefont {Kaiser}\ and\ \citenamefont
  {L\"owen}(2013)}]{kais:13}%
  \BibitemOpen
  \bibfield  {author} {\bibinfo {author} {\bibfnamefont {A.}~\bibnamefont
  {Kaiser}}\ and\ \bibinfo {author} {\bibfnamefont {H.}~\bibnamefont
  {L\"owen}},\ }\bibfield  {title} {\enquote {\bibinfo {title} {Vortex arrays
  as emergent collective phenomena for circle swimmers},}\ }\href {\doibase
  10.1103/PhysRevE.87.032712} {\bibfield  {journal} {\bibinfo  {journal} {Phys.
  Rev. E}\ }\textbf {\bibinfo {volume} {87}},\ \bibinfo {pages} {032712}
  (\bibinfo {year} {2013})}\BibitemShut {NoStop}%
\bibitem [{\citenamefont {Denk}\ \emph {et~al.}(2016)\citenamefont {Denk},
  \citenamefont {Huber}, \citenamefont {Reithmann},\ and\ \citenamefont
  {Frey}}]{denk:16}%
  \BibitemOpen
  \bibfield  {author} {\bibinfo {author} {\bibfnamefont {J.}~\bibnamefont
  {Denk}}, \bibinfo {author} {\bibfnamefont {L.}~\bibnamefont {Huber}},
  \bibinfo {author} {\bibfnamefont {E.}~\bibnamefont {Reithmann}}, \ and\
  \bibinfo {author} {\bibfnamefont {E.}~\bibnamefont {Frey}},\ }\bibfield
  {title} {\enquote {\bibinfo {title} {Active curved polymers form vortex
  patterns on membranes},}\ }\href {\doibase 10.1103/PhysRevLett.116.178301}
  {\bibfield  {journal} {\bibinfo  {journal} {Phys. Rev. Lett.}\ }\textbf
  {\bibinfo {volume} {116}},\ \bibinfo {pages} {178301} (\bibinfo {year}
  {2016})}\BibitemShut {NoStop}%
\bibitem [{\citenamefont {Tao}\ and\ \citenamefont {Kapral}(2008)}]{tao:08.1}%
  \BibitemOpen
  \bibfield  {author} {\bibinfo {author} {\bibfnamefont {Y.-G.}\ \bibnamefont
  {Tao}}\ and\ \bibinfo {author} {\bibfnamefont {R.}~\bibnamefont {Kapral}},\
  }\bibfield  {title} {\enquote {\bibinfo {title} {Design of chemically
  propelled nanodimer motors},}\ }\href {\doibase 10.1063/1.2908078} {\bibfield
   {journal} {\bibinfo  {journal} {J. Chem. Phys.}\ }\textbf {\bibinfo {volume}
  {128}},\ \bibinfo {pages} {164518} (\bibinfo {year} {2008})}\BibitemShut
  {NoStop}%
\bibitem [{\citenamefont {Michelin}\ and\ \citenamefont
  {Lauga}(2014)}]{mich:14}%
  \BibitemOpen
  \bibfield  {author} {\bibinfo {author} {\bibfnamefont {S.}~\bibnamefont
  {Michelin}}\ and\ \bibinfo {author} {\bibfnamefont {E.}~\bibnamefont
  {Lauga}},\ }\bibfield  {title} {\enquote {\bibinfo {title} {Phoretic
  self-propulsion at finite {P{\'e}clet} numbers},}\ }\href {\doibase
  10.1017/jfm.2014.158} {\bibfield  {journal} {\bibinfo  {journal} {J. Fluid
  Mech.}\ }\textbf {\bibinfo {volume} {747}},\ \bibinfo {pages} {572} (\bibinfo
  {year} {2014})}\BibitemShut {NoStop}%
\bibitem [{\citenamefont {Wagner}\ and\ \citenamefont
  {Ripoll}(2017)}]{wagn:17}%
  \BibitemOpen
  \bibfield  {author} {\bibinfo {author} {\bibfnamefont {M.}~\bibnamefont
  {Wagner}}\ and\ \bibinfo {author} {\bibfnamefont {M.}~\bibnamefont
  {Ripoll}},\ }\bibfield  {title} {\enquote {\bibinfo {title} {Hydrodynamic
  front-like swarming of phoretically active dimeric colloids},}\ }\href
  {\doibase 10.1209/0295-5075/119/66007} {\bibfield  {journal} {\bibinfo
  {journal} {EPL}\ }\textbf {\bibinfo {volume} {119}},\ \bibinfo {pages}
  {66007} (\bibinfo {year} {2017})}\BibitemShut {NoStop}%
\bibitem [{\citenamefont {Winkler}\ and\ \citenamefont
  {Gompper}(2018{\natexlab{a}})}]{wink:18}%
  \BibitemOpen
  \bibfield  {author} {\bibinfo {author} {\bibfnamefont {R.~G.}\ \bibnamefont
  {Winkler}}\ and\ \bibinfo {author} {\bibfnamefont {G.}~\bibnamefont
  {Gompper}},\ }\bibfield  {title} {\enquote {\bibinfo {title} {Hydrodynamics
  in motile active matter},}\ }\bibfield  {booktitle} {\emph {\bibinfo
  {booktitle} {Handbook of Materials Modeling: Methods: Theory and Modeling}},\
  }\href {\doibase 10.1007/978-3-319-42913-7{\_}35-1} {\bibfield  {journal}
  {\bibinfo  {journal} {Handbook of Materials Modeling: Methods: Theory and
  Modeling. Springer}\ }\bibinfo {series} {Handbook of Materials Modeling},\
  \bibinfo {pages} {1} (\bibinfo {year} {2018}{\natexlab{a}})}\BibitemShut
  {NoStop}%
\bibitem [{\citenamefont {Kim}\ and\ \citenamefont {Karrila}(1991)}]{kim:91}%
  \BibitemOpen
  \bibfield  {author} {\bibinfo {author} {\bibfnamefont {S.}~\bibnamefont
  {Kim}}\ and\ \bibinfo {author} {\bibfnamefont {S.~J.}\ \bibnamefont
  {Karrila}},\ }\href@noop {} {\emph {\bibinfo {title} {Microhydrodynamics:
  principles and selected applications}}}\ (\bibinfo  {publisher}
  {Butterworth-Heinemann, Boston},\ \bibinfo {year} {1991})\BibitemShut
  {NoStop}%
\bibitem [{\citenamefont {Pozrikidis}(1992)}]{pozr:92}%
  \BibitemOpen
  \bibfield  {author} {\bibinfo {author} {\bibfnamefont {C.}~\bibnamefont
  {Pozrikidis}},\ }\href {\doibase 10.1002/zamm.19940740208} {\emph {\bibinfo
  {title} {Boundary integral and singularity methods for linearized viscous
  flow}}}\ (\bibinfo  {publisher} {Cambridge University Press},\ \bibinfo
  {address} {Cambridge},\ \bibinfo {year} {1992})\BibitemShut {NoStop}%
\bibitem [{\citenamefont {Spagnolie}\ and\ \citenamefont
  {Lauga}(2012)}]{spag:12}%
  \BibitemOpen
  \bibfield  {author} {\bibinfo {author} {\bibfnamefont {S.~E.}\ \bibnamefont
  {Spagnolie}}\ and\ \bibinfo {author} {\bibfnamefont {E.}~\bibnamefont
  {Lauga}},\ }\bibfield  {title} {\enquote {\bibinfo {title} {Hydrodynamics of
  self-propulsion near a boundary: predictions and accuracy of far-field
  approximations},}\ }\href {\doibase 10.1017/jfm.2012.101} {\bibfield
  {journal} {\bibinfo  {journal} {J. Fluid Mech.}\ }\textbf {\bibinfo {volume}
  {700}},\ \bibinfo {pages} {105} (\bibinfo {year} {2012})}\BibitemShut
  {NoStop}%
\bibitem [{\citenamefont {Laskar}\ and\ \citenamefont
  {Adhikari}(2015)}]{lask:15}%
  \BibitemOpen
  \bibfield  {author} {\bibinfo {author} {\bibfnamefont {A.}~\bibnamefont
  {Laskar}}\ and\ \bibinfo {author} {\bibfnamefont {R.}~\bibnamefont
  {Adhikari}},\ }\bibfield  {title} {\enquote {\bibinfo {title} {Brownian
  microhydrodynamics of active filaments},}\ }\href {\doibase
  10.1039/C5SM02021B} {\bibfield  {journal} {\bibinfo  {journal} {Soft Matter}\
  }\textbf {\bibinfo {volume} {11}},\ \bibinfo {pages} {9073} (\bibinfo {year}
  {2015})}\BibitemShut {NoStop}%
\bibitem [{\citenamefont {Mathijssen}\ \emph {et~al.}(2016)\citenamefont
  {Mathijssen}, \citenamefont {Doostmohammadi}, \citenamefont {Yeomans},\ and\
  \citenamefont {Shendruk}}]{math:16.2}%
  \BibitemOpen
  \bibfield  {author} {\bibinfo {author} {\bibfnamefont {A.~J. T.~M.}\
  \bibnamefont {Mathijssen}}, \bibinfo {author} {\bibfnamefont
  {A.}~\bibnamefont {Doostmohammadi}}, \bibinfo {author} {\bibfnamefont
  {J.~M.}\ \bibnamefont {Yeomans}}, \ and\ \bibinfo {author} {\bibfnamefont
  {T.~N.}\ \bibnamefont {Shendruk}},\ }\bibfield  {title} {\enquote {\bibinfo
  {title} {Hotspots of boundary accumulation: dynamics and statistics of
  micro-swimmers in flowing films},}\ }\href {\doibase 10.1098/rsif.2015.0936}
  {\bibfield  {journal} {\bibinfo  {journal} {J. R. Soc. Interface}\ }\textbf
  {\bibinfo {volume} {13}},\ \bibinfo {pages} {20150936} (\bibinfo {year}
  {2016})}\BibitemShut {NoStop}%
\bibitem [{\citenamefont {Winkler}\ and\ \citenamefont
  {Gompper}(2018{\natexlab{b}})}]{wink:18a}%
  \BibitemOpen
  \bibfield  {author} {\bibinfo {author} {\bibfnamefont {R.~G.}\ \bibnamefont
  {Winkler}}\ and\ \bibinfo {author} {\bibfnamefont {G.}~\bibnamefont
  {Gompper}},\ }\bibfield  {title} {\enquote {\bibinfo {title} {Hydrodynamics
  in motile active matter},}\ }in\ \href {\doibase
  10.1007/978-3-319-42913-7{\_}35-1} {\emph {\bibinfo {booktitle} {Handbook of
  Materials Modeling: Methods: Theory and Modeling}}},\ \bibinfo {editor}
  {edited by\ \bibinfo {editor} {\bibfnamefont {W.}~\bibnamefont {Andreoni}}\
  and\ \bibinfo {editor} {\bibfnamefont {S.}~\bibnamefont {Yip}}}\ (\bibinfo
  {publisher} {Springer International Publishing},\ \bibinfo {address} {Cham},\
  \bibinfo {year} {2018})\ pp.\ \bibinfo {pages} {1--20}\BibitemShut {NoStop}%
\bibitem [{\citenamefont {Lighthill}(1952)}]{ligh:52}%
  \BibitemOpen
  \bibfield  {author} {\bibinfo {author} {\bibfnamefont {M.~J.}\ \bibnamefont
  {Lighthill}},\ }\bibfield  {title} {\enquote {\bibinfo {title} {On the
  squirming motion of nearly spherical deformable bodies through liquids at
  very small {R}eynolds numbers},}\ }\href {\doibase 10.1002/cpa.3160050201}
  {\bibfield  {journal} {\bibinfo  {journal} {Comm. Pure Appl. Math.}\ }\textbf
  {\bibinfo {volume} {5}},\ \bibinfo {pages} {109} (\bibinfo {year}
  {1952})}\BibitemShut {NoStop}%
\bibitem [{\citenamefont {Blake}(1971)}]{blak:71}%
  \BibitemOpen
  \bibfield  {author} {\bibinfo {author} {\bibfnamefont {J.~R.}\ \bibnamefont
  {Blake}},\ }\bibfield  {title} {\enquote {\bibinfo {title} {A spherical
  envelope approach to ciliary propulsion},}\ }\href {\doibase
  10.1017/S002211207100048X} {\bibfield  {journal} {\bibinfo  {journal} {J.
  Fluid Mech.}\ }\textbf {\bibinfo {volume} {46}},\ \bibinfo {pages} {199}
  (\bibinfo {year} {1971})}\BibitemShut {NoStop}%
\bibitem [{\citenamefont {Ishikawa}, \citenamefont {Simmonds},\ and\
  \citenamefont {Pedley}(2006)}]{ishi:06}%
  \BibitemOpen
  \bibfield  {author} {\bibinfo {author} {\bibfnamefont {T.}~\bibnamefont
  {Ishikawa}}, \bibinfo {author} {\bibfnamefont {M.~P.}\ \bibnamefont
  {Simmonds}}, \ and\ \bibinfo {author} {\bibfnamefont {T.~J.}\ \bibnamefont
  {Pedley}},\ }\bibfield  {title} {\enquote {\bibinfo {title} {Hydrodynamic
  interaction of two swimming model micro-organisms},}\ }\href {\doibase
  10.1017/S0022112006002631} {\bibfield  {journal} {\bibinfo  {journal} {J.
  Fluid Mech.}\ }\textbf {\bibinfo {volume} {568}},\ \bibinfo {pages} {119}
  (\bibinfo {year} {2006})}\BibitemShut {NoStop}%
\bibitem [{\citenamefont {Llopis}\ and\ \citenamefont
  {Pagonabarraga}(2010)}]{llop:10}%
  \BibitemOpen
  \bibfield  {author} {\bibinfo {author} {\bibfnamefont {I.}~\bibnamefont
  {Llopis}}\ and\ \bibinfo {author} {\bibfnamefont {I.}~\bibnamefont
  {Pagonabarraga}},\ }\bibfield  {title} {\enquote {\bibinfo {title}
  {Hydrodynamic interactions in squirmer motion: Swimming with a neighbour and
  close to a wall},}\ }\href {\doibase 10.1016/j.jnnfm.2010.01.023} {\bibfield
  {journal} {\bibinfo  {journal} {J. Non-Newtonian Fluid Mech.}\ }\textbf
  {\bibinfo {volume} {165}},\ \bibinfo {pages} {946} (\bibinfo {year}
  {2010})}\BibitemShut {NoStop}%
\bibitem [{\citenamefont {G{\"o}tze}\ and\ \citenamefont
  {Gompper}(2010)}]{goet:10}%
  \BibitemOpen
  \bibfield  {author} {\bibinfo {author} {\bibfnamefont {I.~O.}\ \bibnamefont
  {G{\"o}tze}}\ and\ \bibinfo {author} {\bibfnamefont {G.}~\bibnamefont
  {Gompper}},\ }\bibfield  {title} {\enquote {\bibinfo {title} {Mesoscale
  simulations of hydrodynamic squirmer interactions},}\ }\href {\doibase
  10.1103/PhysRevE.82.041921} {\bibfield  {journal} {\bibinfo  {journal} {Phys.
  Rev. E}\ }\textbf {\bibinfo {volume} {82}},\ \bibinfo {pages} {041921}
  (\bibinfo {year} {2010})}\BibitemShut {NoStop}%
\bibitem [{\citenamefont {Molina}, \citenamefont {Nakayama},\ and\
  \citenamefont {Yamamoto}(2013)}]{moli:13}%
  \BibitemOpen
  \bibfield  {author} {\bibinfo {author} {\bibfnamefont {J.~J.}\ \bibnamefont
  {Molina}}, \bibinfo {author} {\bibfnamefont {Y.}~\bibnamefont {Nakayama}}, \
  and\ \bibinfo {author} {\bibfnamefont {R.}~\bibnamefont {Yamamoto}},\
  }\bibfield  {title} {\enquote {\bibinfo {title} {Hydrodynamic interactions of
  self-propelled swimmers},}\ }\href {\doibase 10.1039/C3SM00140G} {\bibfield
  {journal} {\bibinfo  {journal} {Soft Matter}\ }\textbf {\bibinfo {volume}
  {9}},\ \bibinfo {pages} {4923} (\bibinfo {year} {2013})}\BibitemShut
  {NoStop}%
\bibitem [{\citenamefont {Z{\"o}ttl}\ and\ \citenamefont
  {Stark}(2014)}]{zoet:14}%
  \BibitemOpen
  \bibfield  {author} {\bibinfo {author} {\bibfnamefont {A.}~\bibnamefont
  {Z{\"o}ttl}}\ and\ \bibinfo {author} {\bibfnamefont {H.}~\bibnamefont
  {Stark}},\ }\bibfield  {title} {\enquote {\bibinfo {title} {Hydrodynamics
  determines collective motion and phase behavior of active colloids in
  quasi-two-dimensional confinement},}\ }\href {\doibase
  10.1103/PhysRevLett.112.118101} {\bibfield  {journal} {\bibinfo  {journal}
  {Phys. Rev. Lett.}\ }\textbf {\bibinfo {volume} {112}},\ \bibinfo {pages}
  {118101} (\bibinfo {year} {2014})}\BibitemShut {NoStop}%
\bibitem [{\citenamefont {Theers}\ \emph
  {et~al.}(2016{\natexlab{a}})\citenamefont {Theers}, \citenamefont {Westphal},
  \citenamefont {Gompper},\ and\ \citenamefont {Winkler}}]{thee:16.1}%
  \BibitemOpen
  \bibfield  {author} {\bibinfo {author} {\bibfnamefont {M.}~\bibnamefont
  {Theers}}, \bibinfo {author} {\bibfnamefont {E.}~\bibnamefont {Westphal}},
  \bibinfo {author} {\bibfnamefont {G.}~\bibnamefont {Gompper}}, \ and\
  \bibinfo {author} {\bibfnamefont {R.~G.}\ \bibnamefont {Winkler}},\
  }\bibfield  {title} {\enquote {\bibinfo {title} {Modeling a spheroidal
  microswimmer and cooperative swimming in a narrow slit},}\ }\href {\doibase
  10.1039/C6SM01424K} {\bibfield  {journal} {\bibinfo  {journal} {Soft Matter}\
  }\textbf {\bibinfo {volume} {12}},\ \bibinfo {pages} {7372} (\bibinfo {year}
  {2016}{\natexlab{a}})}\BibitemShut {NoStop}%
\bibitem [{\citenamefont {Yoshinaga}\ and\ \citenamefont
  {Liverpool}(2017)}]{yosh:17}%
  \BibitemOpen
  \bibfield  {author} {\bibinfo {author} {\bibfnamefont {N.}~\bibnamefont
  {Yoshinaga}}\ and\ \bibinfo {author} {\bibfnamefont {T.~B.}\ \bibnamefont
  {Liverpool}},\ }\bibfield  {title} {\enquote {\bibinfo {title} {Hydrodynamic
  interactions in dense active suspensions: From polar order to dynamical
  clusters},}\ }\href {\doibase 10.1103/PhysRevE.96.020603} {\bibfield
  {journal} {\bibinfo  {journal} {Phys. Rev. E}\ }\textbf {\bibinfo {volume}
  {96}},\ \bibinfo {pages} {020603} (\bibinfo {year} {2017})}\BibitemShut
  {NoStop}%
\bibitem [{\citenamefont {Theers}\ \emph {et~al.}(2018)\citenamefont {Theers},
  \citenamefont {Westphal}, \citenamefont {Qi}, \citenamefont {Winkler},\ and\
  \citenamefont {Gompper}}]{thee:18}%
  \BibitemOpen
  \bibfield  {author} {\bibinfo {author} {\bibfnamefont {M.}~\bibnamefont
  {Theers}}, \bibinfo {author} {\bibfnamefont {E.}~\bibnamefont {Westphal}},
  \bibinfo {author} {\bibfnamefont {K.}~\bibnamefont {Qi}}, \bibinfo {author}
  {\bibfnamefont {R.~G.}\ \bibnamefont {Winkler}}, \ and\ \bibinfo {author}
  {\bibfnamefont {G.}~\bibnamefont {Gompper}},\ }\bibfield  {title} {\enquote
  {\bibinfo {title} {Clustering of microswimmers: interplay of shape and
  hydrodynamics},}\ }\href {\doibase 10.1039/C8SM01390J} {\bibfield  {journal}
  {\bibinfo  {journal} {Soft Matter}\ }\textbf {\bibinfo {volume} {14}},\
  \bibinfo {pages} {8590} (\bibinfo {year} {2018})}\BibitemShut {NoStop}%
\bibitem [{\citenamefont {Popescu}\ \emph {et~al.}(2018)\citenamefont
  {Popescu}, \citenamefont {Uspal}, \citenamefont {Eskandari}, \citenamefont
  {Tasinkevych},\ and\ \citenamefont {Dietrich}}]{pope:18}%
  \BibitemOpen
  \bibfield  {author} {\bibinfo {author} {\bibfnamefont {M.~N.}\ \bibnamefont
  {Popescu}}, \bibinfo {author} {\bibfnamefont {W.~E.}\ \bibnamefont {Uspal}},
  \bibinfo {author} {\bibfnamefont {Z.}~\bibnamefont {Eskandari}}, \bibinfo
  {author} {\bibfnamefont {M.}~\bibnamefont {Tasinkevych}}, \ and\ \bibinfo
  {author} {\bibfnamefont {S.}~\bibnamefont {Dietrich}},\ }\bibfield  {title}
  {\enquote {\bibinfo {title} {Effective squirmer models for self-phoretic
  chemically active spherical colloids},}\ }\href {\doibase
  10.1140/epje/i2018-11753-1} {\bibfield  {journal} {\bibinfo  {journal} {Eur.
  Phys. J. E}\ }\textbf {\bibinfo {volume} {41}},\ \bibinfo {pages} {145}
  (\bibinfo {year} {2018})}\BibitemShut {NoStop}%
\bibitem [{\citenamefont {Jayaraman}\ \emph {et~al.}(2012)\citenamefont
  {Jayaraman}, \citenamefont {Ramachandran}, \citenamefont {Ghose},
  \citenamefont {Laskar}, \citenamefont {Bhamla}, \citenamefont {Kumar},\ and\
  \citenamefont {Adhikari}}]{jaya:12}%
  \BibitemOpen
  \bibfield  {author} {\bibinfo {author} {\bibfnamefont {G.}~\bibnamefont
  {Jayaraman}}, \bibinfo {author} {\bibfnamefont {S.}~\bibnamefont
  {Ramachandran}}, \bibinfo {author} {\bibfnamefont {S.}~\bibnamefont {Ghose}},
  \bibinfo {author} {\bibfnamefont {A.}~\bibnamefont {Laskar}}, \bibinfo
  {author} {\bibfnamefont {M.~S.}\ \bibnamefont {Bhamla}}, \bibinfo {author}
  {\bibfnamefont {P.~B.~S.}\ \bibnamefont {Kumar}}, \ and\ \bibinfo {author}
  {\bibfnamefont {R.}~\bibnamefont {Adhikari}},\ }\bibfield  {title} {\enquote
  {\bibinfo {title} {Autonomous motility of active filaments due to spontaneous
  flow-symmetry breaking},}\ }\href {\doibase 10.1103/PhysRevLett.109.158302}
  {\bibfield  {journal} {\bibinfo  {journal} {Phys. Rev. Lett.}\ }\textbf
  {\bibinfo {volume} {109}},\ \bibinfo {pages} {158302} (\bibinfo {year}
  {2012})}\BibitemShut {NoStop}%
\bibitem [{\citenamefont {Laskar}\ \emph {et~al.}(2013)\citenamefont {Laskar},
  \citenamefont {Singh}, \citenamefont {Ghose}, \citenamefont {Jayaraman},
  \citenamefont {Kumar},\ and\ \citenamefont {Adhikari}}]{lask:13}%
  \BibitemOpen
  \bibfield  {author} {\bibinfo {author} {\bibfnamefont {A.}~\bibnamefont
  {Laskar}}, \bibinfo {author} {\bibfnamefont {R.}~\bibnamefont {Singh}},
  \bibinfo {author} {\bibfnamefont {S.}~\bibnamefont {Ghose}}, \bibinfo
  {author} {\bibfnamefont {G.}~\bibnamefont {Jayaraman}}, \bibinfo {author}
  {\bibfnamefont {P.~B.~S.}\ \bibnamefont {Kumar}}, \ and\ \bibinfo {author}
  {\bibfnamefont {R.}~\bibnamefont {Adhikari}},\ }\bibfield  {title} {\enquote
  {\bibinfo {title} {Hydrodynamic instabilities provide a generic route to
  spontaneous biomimetic oscillations in chemomechanically active filaments},}\
  }\href {\doibase 10.1038/srep01964} {\bibfield  {journal} {\bibinfo
  {journal} {Sci. Rep.}\ }\textbf {\bibinfo {volume} {3}},\ \bibinfo {pages}
  {1964} (\bibinfo {year} {2013})}\BibitemShut {NoStop}%
\bibitem [{\citenamefont {Pedley}(2016)}]{pedl:16}%
  \BibitemOpen
  \bibfield  {author} {\bibinfo {author} {\bibfnamefont {T.~J.}\ \bibnamefont
  {Pedley}},\ }\bibfield  {title} {\enquote {\bibinfo {title} {Spherical
  squirmers: models for swimming micro-organisms},}\ }\href {\doibase doi:
  10.1093/imamat/hxw030} {\bibfield  {journal} {\bibinfo  {journal} {IMA J.
  Appl. Math.}\ }\textbf {\bibinfo {volume} {81}},\ \bibinfo {pages} {488}
  (\bibinfo {year} {2016})}\BibitemShut {NoStop}%
\bibitem [{\citenamefont {Keller}\ and\ \citenamefont {Wu}(1977)}]{kell:77}%
  \BibitemOpen
  \bibfield  {author} {\bibinfo {author} {\bibfnamefont {S.~R.}\ \bibnamefont
  {Keller}}\ and\ \bibinfo {author} {\bibfnamefont {T.~Y.}\ \bibnamefont
  {Wu}},\ }\bibfield  {title} {\enquote {\bibinfo {title} {A porous
  prolate-spheroidal model for ciliated micro-organisms},}\ }\href {\doibase
  10.1017/S0022112077001669} {\bibfield  {journal} {\bibinfo  {journal} {J.
  Fluid Mech.}\ }\textbf {\bibinfo {volume} {80}},\ \bibinfo {pages} {259}
  (\bibinfo {year} {1977})}\BibitemShut {NoStop}%
\bibitem [{\citenamefont {Ishimoto}\ and\ \citenamefont
  {Gaffney}(2013)}]{ishi:13}%
  \BibitemOpen
  \bibfield  {author} {\bibinfo {author} {\bibfnamefont {K.}~\bibnamefont
  {Ishimoto}}\ and\ \bibinfo {author} {\bibfnamefont {E.~A.}\ \bibnamefont
  {Gaffney}},\ }\bibfield  {title} {\enquote {\bibinfo {title} {Squirmer
  dynamics near a boundary},}\ }\href {\doibase 10.1103/PhysRevE.88.062702}
  {\bibfield  {journal} {\bibinfo  {journal} {Phys. Rev. E}\ }\textbf {\bibinfo
  {volume} {88}},\ \bibinfo {pages} {062702} (\bibinfo {year}
  {2013})}\BibitemShut {NoStop}%
\bibitem [{\citenamefont {Alarc{\'o}n}\ and\ \citenamefont
  {Pagonabarraga}(2013)}]{alar:13}%
  \BibitemOpen
  \bibfield  {author} {\bibinfo {author} {\bibfnamefont {F.}~\bibnamefont
  {Alarc{\'o}n}}\ and\ \bibinfo {author} {\bibfnamefont {I.}~\bibnamefont
  {Pagonabarraga}},\ }\bibfield  {title} {\enquote {\bibinfo {title}
  {Spontaneous aggregation and global polar ordering in squirmer
  suspensions},}\ }\href {\doibase 10.1016/j.molliq.2012.12.009} {\bibfield
  {journal} {\bibinfo  {journal} {J. Mol. Liq.}\ }\textbf {\bibinfo {volume}
  {185}},\ \bibinfo {pages} {56} (\bibinfo {year} {2013})}\BibitemShut
  {NoStop}%
\bibitem [{\citenamefont {Alarc{\'o}n}, \citenamefont {Valeriani},\ and\
  \citenamefont {Pagonabarraga}(2017)}]{alar:17}%
  \BibitemOpen
  \bibfield  {author} {\bibinfo {author} {\bibfnamefont {F.}~\bibnamefont
  {Alarc{\'o}n}}, \bibinfo {author} {\bibfnamefont {C.}~\bibnamefont
  {Valeriani}}, \ and\ \bibinfo {author} {\bibfnamefont {I.}~\bibnamefont
  {Pagonabarraga}},\ }\bibfield  {title} {\enquote {\bibinfo {title}
  {Morphology of clusters of attractive dry and wet self-propelled spherical
  particle suspensions},}\ }\href {\doibase 10.1039/C6SM01752E} {\bibfield
  {journal} {\bibinfo  {journal} {Soft Matter}\ }\textbf {\bibinfo {volume}
  {13}},\ \bibinfo {pages} {814} (\bibinfo {year} {2017})}\BibitemShut
  {NoStop}%
\bibitem [{\citenamefont {Singh}, \citenamefont {Ghose},\ and\ \citenamefont
  {Adhikari}(2015)}]{sing:15}%
  \BibitemOpen
  \bibfield  {author} {\bibinfo {author} {\bibfnamefont {R.}~\bibnamefont
  {Singh}}, \bibinfo {author} {\bibfnamefont {S.}~\bibnamefont {Ghose}}, \ and\
  \bibinfo {author} {\bibfnamefont {R.}~\bibnamefont {Adhikari}},\ }\bibfield
  {title} {\enquote {\bibinfo {title} {Many-body microhydrodynamics of
  colloidal particles with active boundary layers},}\ }\href {\doibase
  10.1088/1742-5468/2015/06/P06017} {\bibfield  {journal} {\bibinfo  {journal}
  {J. Stat. Mech. Theor. Exp.}\ }\textbf {\bibinfo {volume} {2015}},\ \bibinfo
  {pages} {P06017} (\bibinfo {year} {2015})}\BibitemShut {NoStop}%
\bibitem [{\citenamefont {Felderhof}\ and\ \citenamefont
  {Jones}(1994)}]{feld:94}%
  \BibitemOpen
  \bibfield  {author} {\bibinfo {author} {\bibfnamefont {B.~U.}\ \bibnamefont
  {Felderhof}}\ and\ \bibinfo {author} {\bibfnamefont {R.~B.}\ \bibnamefont
  {Jones}},\ }\bibfield  {title} {\enquote {\bibinfo {title} {Small-amplitude
  swimming of a sphere},}\ }\href {\doibase
  https://doi.org/10.1016/0378-4371(94)90170-8} {\bibfield  {journal} {\bibinfo
   {journal} {Physica A}\ }\textbf {\bibinfo {volume} {202}},\ \bibinfo {pages}
  {119} (\bibinfo {year} {1994})}\BibitemShut {NoStop}%
\bibitem [{\citenamefont {Pak}\ and\ \citenamefont {Lauga}(2014)}]{pak:14}%
  \BibitemOpen
  \bibfield  {author} {\bibinfo {author} {\bibfnamefont {O.~S.}\ \bibnamefont
  {Pak}}\ and\ \bibinfo {author} {\bibfnamefont {E.}~\bibnamefont {Lauga}},\
  }\bibfield  {title} {\enquote {\bibinfo {title} {Generalized squirming motion
  of a sphere},}\ }\href {\doibase 10.1007/s10665-014-9690-9} {\bibfield
  {journal} {\bibinfo  {journal} {J. Eng. Math.}\ }\textbf {\bibinfo {volume}
  {88}},\ \bibinfo {pages} {1} (\bibinfo {year} {2014})}\BibitemShut {NoStop}%
\bibitem [{\citenamefont {Singh}\ and\ \citenamefont
  {Adhikari}(2018)}]{sing:18.1}%
  \BibitemOpen
  \bibfield  {author} {\bibinfo {author} {\bibfnamefont {R.}~\bibnamefont
  {Singh}}\ and\ \bibinfo {author} {\bibfnamefont {R.}~\bibnamefont
  {Adhikari}},\ }\bibfield  {title} {\enquote {\bibinfo {title} {Generalized
  stokes laws for active colloids and their applications},}\ }\href {\doibase
  10.1088/2399-6528/aaab0d} {\bibfield  {journal} {\bibinfo  {journal} {J.
  Phys. Commun.}\ }\textbf {\bibinfo {volume} {2}},\ \bibinfo {pages} {025025}
  (\bibinfo {year} {2018})}\BibitemShut {NoStop}%
\bibitem [{\citenamefont {Fair}\ and\ \citenamefont
  {Anderson}(1989)}]{fair:89}%
  \BibitemOpen
  \bibfield  {author} {\bibinfo {author} {\bibfnamefont {M.~C.}\ \bibnamefont
  {Fair}}\ and\ \bibinfo {author} {\bibfnamefont {J.~L.}\ \bibnamefont
  {Anderson}},\ }\bibfield  {title} {\enquote {\bibinfo {title}
  {Electrophoresis of nonuniformly charged ellipsoidal particles},}\ }\href
  {\doibase https://doi.org/10.1016/0021-9797(89)90045-3} {\bibfield  {journal}
  {\bibinfo  {journal} {J. Colloid Interface Sci.}\ }\textbf {\bibinfo {volume}
  {127}},\ \bibinfo {pages} {388} (\bibinfo {year} {1989})}\BibitemShut
  {NoStop}%
\bibitem [{\citenamefont {Doi}\ and\ \citenamefont {Edwards}(1986)}]{doi:86}%
  \BibitemOpen
  \bibfield  {author} {\bibinfo {author} {\bibfnamefont {M.}~\bibnamefont
  {Doi}}\ and\ \bibinfo {author} {\bibfnamefont {S.~F.}\ \bibnamefont
  {Edwards}},\ }\href@noop {} {\emph {\bibinfo {title} {The Theory of Polymer
  Dynamics}}}\ (\bibinfo  {publisher} {Clarendon Press},\ \bibinfo {address}
  {Oxford},\ \bibinfo {year} {1986})\BibitemShut {NoStop}%
\bibitem [{\citenamefont {Vandebroek}\ and\ \citenamefont
  {Vanderzande}(2015)}]{vand:15}%
  \BibitemOpen
  \bibfield  {author} {\bibinfo {author} {\bibfnamefont {H.}~\bibnamefont
  {Vandebroek}}\ and\ \bibinfo {author} {\bibfnamefont {C.}~\bibnamefont
  {Vanderzande}},\ }\bibfield  {title} {\enquote {\bibinfo {title} {Dynamics of
  a polymer in an active and viscoelastic bath},}\ }\href {\doibase
  10.1103/PhysRevE.92.060601} {\bibfield  {journal} {\bibinfo  {journal} {Phys.
  Rev. E}\ }\textbf {\bibinfo {volume} {92}},\ \bibinfo {pages} {060601}
  (\bibinfo {year} {2015})}\BibitemShut {NoStop}%
\bibitem [{\citenamefont {Osmanovic}\ and\ \citenamefont
  {Rabin}(2017)}]{osma:17}%
  \BibitemOpen
  \bibfield  {author} {\bibinfo {author} {\bibfnamefont {D.}~\bibnamefont
  {Osmanovic}}\ and\ \bibinfo {author} {\bibfnamefont {Y.}~\bibnamefont
  {Rabin}},\ }\bibfield  {title} {\enquote {\bibinfo {title} {Dynamics of
  active {R}ouse chains},}\ }\href {\doibase 10.1039/C6SM02722A} {\bibfield
  {journal} {\bibinfo  {journal} {Soft Matter}\ }\textbf {\bibinfo {volume}
  {13}},\ \bibinfo {pages} {963} (\bibinfo {year} {2017})}\BibitemShut
  {NoStop}%
\bibitem [{\citenamefont {Samanta}\ and\ \citenamefont
  {Chakrabarti}(2016)}]{sama:16}%
  \BibitemOpen
  \bibfield  {author} {\bibinfo {author} {\bibfnamefont {N.}~\bibnamefont
  {Samanta}}\ and\ \bibinfo {author} {\bibfnamefont {R.}~\bibnamefont
  {Chakrabarti}},\ }\bibfield  {title} {\enquote {\bibinfo {title} {Chain
  reconfiguration in active noise},}\ }\href {\doibase
  10.1088/1751-8113/49/19/195601} {\bibfield  {journal} {\bibinfo  {journal}
  {J. Phys. A: Math. Theor.}\ }\textbf {\bibinfo {volume} {49}},\ \bibinfo
  {pages} {195601} (\bibinfo {year} {2016})}\BibitemShut {NoStop}%
\bibitem [{\citenamefont {Chaki}\ and\ \citenamefont
  {Chakrabarti}(2019)}]{chak:19}%
  \BibitemOpen
  \bibfield  {author} {\bibinfo {author} {\bibfnamefont {S.}~\bibnamefont
  {Chaki}}\ and\ \bibinfo {author} {\bibfnamefont {R.}~\bibnamefont
  {Chakrabarti}},\ }\bibfield  {title} {\enquote {\bibinfo {title} {Enhanced
  diffusion, swelling, and slow reconfiguration of a single chain in
  non-{Gaussian} active bath},}\ }\href {\doibase 10.1063/1.5086152} {\bibfield
   {journal} {\bibinfo  {journal} {J. Chem. Phys.}\ }\textbf {\bibinfo {volume}
  {150}},\ \bibinfo {pages} {094902} (\bibinfo {year} {2019})}\BibitemShut
  {NoStop}%
\bibitem [{\citenamefont {Harder}, \citenamefont {Valeriani},\ and\
  \citenamefont {Cacciuto}(2014)}]{hard:14}%
  \BibitemOpen
  \bibfield  {author} {\bibinfo {author} {\bibfnamefont {J.}~\bibnamefont
  {Harder}}, \bibinfo {author} {\bibfnamefont {C.}~\bibnamefont {Valeriani}}, \
  and\ \bibinfo {author} {\bibfnamefont {A.}~\bibnamefont {Cacciuto}},\
  }\bibfield  {title} {\enquote {\bibinfo {title} {Activity-induced collapse
  and reexpansion of rigid polymers},}\ }\href {\doibase
  10.1103/PhysRevE.90.062312} {\bibfield  {journal} {\bibinfo  {journal} {Phys.
  Rev. E}\ }\textbf {\bibinfo {volume} {90}},\ \bibinfo {pages} {062312}
  (\bibinfo {year} {2014})}\BibitemShut {NoStop}%
\bibitem [{\citenamefont {Sarkar}\ \emph {et~al.}(2014)\citenamefont {Sarkar},
  \citenamefont {Thakur}, \citenamefont {Tao},\ and\ \citenamefont
  {Kapral}}]{sark:14}%
  \BibitemOpen
  \bibfield  {author} {\bibinfo {author} {\bibfnamefont {D.}~\bibnamefont
  {Sarkar}}, \bibinfo {author} {\bibfnamefont {S.}~\bibnamefont {Thakur}},
  \bibinfo {author} {\bibfnamefont {Y.-G.}\ \bibnamefont {Tao}}, \ and\
  \bibinfo {author} {\bibfnamefont {R.}~\bibnamefont {Kapral}},\ }\bibfield
  {title} {\enquote {\bibinfo {title} {Ring closure dynamics for a chemically
  active polymer},}\ }\href {\doibase 10.1039/C4SM01941E} {\bibfield  {journal}
  {\bibinfo  {journal} {Soft Matter}\ }\textbf {\bibinfo {volume} {10}},\
  \bibinfo {pages} {9577} (\bibinfo {year} {2014})}\BibitemShut {NoStop}%
\bibitem [{\citenamefont {Ghosh}\ and\ \citenamefont {Gov}(2014)}]{ghos:14}%
  \BibitemOpen
  \bibfield  {author} {\bibinfo {author} {\bibfnamefont {A.}~\bibnamefont
  {Ghosh}}\ and\ \bibinfo {author} {\bibfnamefont {N.~S.}\ \bibnamefont
  {Gov}},\ }\bibfield  {title} {\enquote {\bibinfo {title} {Dynamics of active
  semiflexible polymers},}\ }\href {\doibase 10.1016/j.bpj.2014.07.034}
  {\bibfield  {journal} {\bibinfo  {journal} {Biophys. J.}\ }\textbf {\bibinfo
  {volume} {107}},\ \bibinfo {pages} {1065} (\bibinfo {year}
  {2014})}\BibitemShut {NoStop}%
\bibitem [{\citenamefont {Shin}\ \emph {et~al.}(2015)\citenamefont {Shin},
  \citenamefont {Cherstvy}, \citenamefont {Kim},\ and\ \citenamefont
  {Metzler}}]{shin:15}%
  \BibitemOpen
  \bibfield  {author} {\bibinfo {author} {\bibfnamefont {J.}~\bibnamefont
  {Shin}}, \bibinfo {author} {\bibfnamefont {A.~G.}\ \bibnamefont {Cherstvy}},
  \bibinfo {author} {\bibfnamefont {W.~K.}\ \bibnamefont {Kim}}, \ and\
  \bibinfo {author} {\bibfnamefont {R.}~\bibnamefont {Metzler}},\ }\bibfield
  {title} {\enquote {\bibinfo {title} {Facilitation of polymer looping and
  giant polymer diffusivity in crowded solutions of active particles},}\ }\href
  {\doibase 10.1088/1367-2630/17/11/113008} {\bibfield  {journal} {\bibinfo
  {journal} {New J. Phys.}\ }\textbf {\bibinfo {volume} {17}},\ \bibinfo
  {pages} {113008} (\bibinfo {year} {2015})}\BibitemShut {NoStop}%
\bibitem [{\citenamefont {Eisenstecken}, \citenamefont {Gompper},\ and\
  \citenamefont {Winkler}(2016)}]{eise:16}%
  \BibitemOpen
  \bibfield  {author} {\bibinfo {author} {\bibfnamefont {T.}~\bibnamefont
  {Eisenstecken}}, \bibinfo {author} {\bibfnamefont {G.}~\bibnamefont
  {Gompper}}, \ and\ \bibinfo {author} {\bibfnamefont {R.~G.}\ \bibnamefont
  {Winkler}},\ }\bibfield  {title} {\enquote {\bibinfo {title} {Conformational
  properties of active semiflexible polymers},}\ }\href {\doibase
  10.3390/polym8080304} {\bibfield  {journal} {\bibinfo  {journal} {Polymers}\
  }\textbf {\bibinfo {volume} {8}},\ \bibinfo {pages} {304} (\bibinfo {year}
  {2016})}\BibitemShut {NoStop}%
\bibitem [{\citenamefont {Mousavi}, \citenamefont {Gompper},\ and\
  \citenamefont {Winkler}(2019)}]{mous:19}%
  \BibitemOpen
  \bibfield  {author} {\bibinfo {author} {\bibfnamefont {S.~M.}\ \bibnamefont
  {Mousavi}}, \bibinfo {author} {\bibfnamefont {G.}~\bibnamefont {Gompper}}, \
  and\ \bibinfo {author} {\bibfnamefont {R.~G.}\ \bibnamefont {Winkler}},\
  }\bibfield  {title} {\enquote {\bibinfo {title} {Active {B}rownian ring
  polymers},}\ }\href {\doibase 10.1063/1.5082723} {\bibfield  {journal}
  {\bibinfo  {journal} {J. Chem. Phys.}\ }\textbf {\bibinfo {volume} {150}},\
  \bibinfo {pages} {064913} (\bibinfo {year} {2019})}\BibitemShut {NoStop}%
\bibitem [{\citenamefont {Das}, \citenamefont {Gompper},\ and\ \citenamefont
  {Winkler}(2018)}]{das:18.1}%
  \BibitemOpen
  \bibfield  {author} {\bibinfo {author} {\bibfnamefont {S.}~\bibnamefont
  {Das}}, \bibinfo {author} {\bibfnamefont {G.}~\bibnamefont {Gompper}}, \ and\
  \bibinfo {author} {\bibfnamefont {R.~G.}\ \bibnamefont {Winkler}},\
  }\bibfield  {title} {\enquote {\bibinfo {title} {Confined active {B}rownian
  particles: theoretical description of propulsion-induced accumulation},}\
  }\href {\doibase 10.1088/1367-2630/aa9d4b} {\bibfield  {journal} {\bibinfo
  {journal} {New J. Phys.}\ }\textbf {\bibinfo {volume} {20}},\ \bibinfo
  {pages} {015001} (\bibinfo {year} {2018})}\BibitemShut {NoStop}%
\bibitem [{\citenamefont {Bawendi}\ and\ \citenamefont
  {Freed}(1985)}]{bawe:85}%
  \BibitemOpen
  \bibfield  {author} {\bibinfo {author} {\bibfnamefont {M.~G.}\ \bibnamefont
  {Bawendi}}\ and\ \bibinfo {author} {\bibfnamefont {K.~F.}\ \bibnamefont
  {Freed}},\ }\bibfield  {title} {\enquote {\bibinfo {title} {A {W}iener
  integral model for stiff polymer chains},}\ }\href {\doibase
  10.1063/1.1675038} {\bibfield  {journal} {\bibinfo  {journal} {J. Chem.
  Phys.}\ }\textbf {\bibinfo {volume} {83}},\ \bibinfo {pages} {2491} (\bibinfo
  {year} {1985})}\BibitemShut {NoStop}%
\bibitem [{\citenamefont {Battacharjee}\ and\ \citenamefont
  {Muthukumar}(1987)}]{batt:87}%
  \BibitemOpen
  \bibfield  {author} {\bibinfo {author} {\bibfnamefont {S.~M.}\ \bibnamefont
  {Battacharjee}}\ and\ \bibinfo {author} {\bibfnamefont {M.}~\bibnamefont
  {Muthukumar}},\ }\bibfield  {title} {\enquote {\bibinfo {title} {Statistical
  mechanics of solutions of semiflexible chains: {A} path integral
  formulation},}\ }\href {\doibase 10.1063/1.452579} {\bibfield  {journal}
  {\bibinfo  {journal} {J. Chem. Phys.}\ }\textbf {\bibinfo {volume} {86}},\
  \bibinfo {pages} {411} (\bibinfo {year} {1987})}\BibitemShut {NoStop}%
\bibitem [{\citenamefont {Langowski}, \citenamefont {Noolandi},\ and\
  \citenamefont {Nickel}(1991)}]{lang:91}%
  \BibitemOpen
  \bibfield  {author} {\bibinfo {author} {\bibfnamefont {J.~B.}\ \bibnamefont
  {Langowski}}, \bibinfo {author} {\bibfnamefont {J.}~\bibnamefont {Noolandi}},
  \ and\ \bibinfo {author} {\bibfnamefont {B.}~\bibnamefont {Nickel}},\
  }\bibfield  {title} {\enquote {\bibinfo {title} {Stiff chain
  model---functional integral approach},}\ }\href {\doibase 10.1063/1.461106}
  {\bibfield  {journal} {\bibinfo  {journal} {J. Chem. Phys.}\ }\textbf
  {\bibinfo {volume} {95}},\ \bibinfo {pages} {1266} (\bibinfo {year}
  {1991})}\BibitemShut {NoStop}%
\bibitem [{\citenamefont {Winkler}, \citenamefont {Reineker},\ and\
  \citenamefont {Harnau}(1994)}]{wink:94}%
  \BibitemOpen
  \bibfield  {author} {\bibinfo {author} {\bibfnamefont {R.~G.}\ \bibnamefont
  {Winkler}}, \bibinfo {author} {\bibfnamefont {P.}~\bibnamefont {Reineker}}, \
  and\ \bibinfo {author} {\bibfnamefont {L.}~\bibnamefont {Harnau}},\
  }\bibfield  {title} {\enquote {\bibinfo {title} {Models and equilibrium
  properties of stiff molecular chains},}\ }\href {\doibase 10.1063/1.468239}
  {\bibfield  {journal} {\bibinfo  {journal} {J. Chem. Phys.}\ }\textbf
  {\bibinfo {volume} {101}},\ \bibinfo {pages} {8119} (\bibinfo {year}
  {1994})}\BibitemShut {NoStop}%
\bibitem [{\citenamefont {Ha}\ and\ \citenamefont {Thirumalai}(1995)}]{ha:95}%
  \BibitemOpen
  \bibfield  {author} {\bibinfo {author} {\bibfnamefont {B.~Y.}\ \bibnamefont
  {Ha}}\ and\ \bibinfo {author} {\bibfnamefont {D.}~\bibnamefont
  {Thirumalai}},\ }\bibfield  {title} {\enquote {\bibinfo {title} {A mean-field
  model for semiflexible chains},}\ }\href {\doibase 10.1063/1.470001}
  {\bibfield  {journal} {\bibinfo  {journal} {J. Chem. Phys.}\ }\textbf
  {\bibinfo {volume} {103}},\ \bibinfo {pages} {9408} (\bibinfo {year}
  {1995})}\BibitemShut {NoStop}%
\bibitem [{\citenamefont {Winkler}(2003)}]{wink:03}%
  \BibitemOpen
  \bibfield  {author} {\bibinfo {author} {\bibfnamefont {R.~G.}\ \bibnamefont
  {Winkler}},\ }\bibfield  {title} {\enquote {\bibinfo {title} {Deformation of
  semiflexible chains},}\ }\href {\doibase 10.1063/1.1537247} {\bibfield
  {journal} {\bibinfo  {journal} {J. Chem. Phys.}\ }\textbf {\bibinfo {volume}
  {118}},\ \bibinfo {pages} {2919} (\bibinfo {year} {2003})}\BibitemShut
  {NoStop}%
\bibitem [{\citenamefont {Harnau}, \citenamefont {Winkler},\ and\ \citenamefont
  {Reineker}(1995)}]{harn:95}%
  \BibitemOpen
  \bibfield  {author} {\bibinfo {author} {\bibfnamefont {L.}~\bibnamefont
  {Harnau}}, \bibinfo {author} {\bibfnamefont {R.~G.}\ \bibnamefont {Winkler}},
  \ and\ \bibinfo {author} {\bibfnamefont {P.}~\bibnamefont {Reineker}},\
  }\bibfield  {title} {\enquote {\bibinfo {title} {Dynamic properties of
  molecular chains with variable stiffness},}\ }\href {\doibase
  10.1063/1.469027} {\bibfield  {journal} {\bibinfo  {journal} {J. Chem.
  Phys.}\ }\textbf {\bibinfo {volume} {102}},\ \bibinfo {pages} {7750}
  (\bibinfo {year} {1995})}\BibitemShut {NoStop}%
\bibitem [{\citenamefont {Winkler}(2010)}]{wink:10}%
  \BibitemOpen
  \bibfield  {author} {\bibinfo {author} {\bibfnamefont {R.~G.}\ \bibnamefont
  {Winkler}},\ }\bibfield  {title} {\enquote {\bibinfo {title} {Conformational
  and rheological properties of semiflexible polymers in shear flow},}\ }\href
  {\doibase 10.1063/1.3497642} {\bibfield  {journal} {\bibinfo  {journal} {J.
  Chem. Phys.}\ }\textbf {\bibinfo {volume} {133}},\ \bibinfo {pages} {164905}
  (\bibinfo {year} {2010})}\BibitemShut {NoStop}%
\bibitem [{\citenamefont {Winkler}(2006)}]{wink:06.1}%
  \BibitemOpen
  \bibfield  {author} {\bibinfo {author} {\bibfnamefont {R.~G.}\ \bibnamefont
  {Winkler}},\ }\bibfield  {title} {\enquote {\bibinfo {title} {Semiflexible
  polymers in shear flow},}\ }\href {\doibase 10.1103/PhysRevLett.97.128301}
  {\bibfield  {journal} {\bibinfo  {journal} {Phys. Rev. Lett.}\ }\textbf
  {\bibinfo {volume} {97}},\ \bibinfo {pages} {128301} (\bibinfo {year}
  {2006})}\BibitemShut {NoStop}%
\bibitem [{\citenamefont {Eisenstecken}\ \emph {et~al.}(2017)\citenamefont
  {Eisenstecken}, \citenamefont {Ghavami}, \citenamefont {Mair}, \citenamefont
  {Gompper},\ and\ \citenamefont {Winkler}}]{eise:17.1}%
  \BibitemOpen
  \bibfield  {author} {\bibinfo {author} {\bibfnamefont {T.}~\bibnamefont
  {Eisenstecken}}, \bibinfo {author} {\bibfnamefont {A.}~\bibnamefont
  {Ghavami}}, \bibinfo {author} {\bibfnamefont {A.}~\bibnamefont {Mair}},
  \bibinfo {author} {\bibfnamefont {G.}~\bibnamefont {Gompper}}, \ and\
  \bibinfo {author} {\bibfnamefont {R.~G.}\ \bibnamefont {Winkler}},\
  }\bibfield  {title} {\enquote {\bibinfo {title} {Conformational and dynamical
  properties of semiflexible polymers in the presence of active noise},}\
  }\href {\doibase 10.1063/1.4996525} {\bibfield  {journal} {\bibinfo
  {journal} {AIP Conf. Proc.}\ }\textbf {\bibinfo {volume} {1871}},\ \bibinfo
  {pages} {050001} (\bibinfo {year} {2017})}\BibitemShut {NoStop}%
\bibitem [{\citenamefont {Kratky}\ and\ \citenamefont {Porod}(1949)}]{krat:49}%
  \BibitemOpen
  \bibfield  {author} {\bibinfo {author} {\bibfnamefont {O.}~\bibnamefont
  {Kratky}}\ and\ \bibinfo {author} {\bibfnamefont {G.}~\bibnamefont {Porod}},\
  }\bibfield  {title} {\enquote {\bibinfo {title} {R{\"o}ntgenuntersuchung
  gel{\"o}ster {F}adenmolek{\"u}le},}\ }\href {\doibase
  10.1002/recl.19490681203Ci} {\bibfield  {journal} {\bibinfo  {journal} {Recl.
  Trav. Chim. Pays-Bas}\ }\textbf {\bibinfo {volume} {68}},\ \bibinfo {pages}
  {1106} (\bibinfo {year} {1949})}\BibitemShut {NoStop}%
\bibitem [{\citenamefont {{Arag{\'{o}}n}}\ and\ \citenamefont
  {Pecora}(1985)}]{arag:85}%
  \BibitemOpen
  \bibfield  {author} {\bibinfo {author} {\bibfnamefont {S.~R.}\ \bibnamefont
  {{Arag{\'{o}}n}}}\ and\ \bibinfo {author} {\bibfnamefont {R.}~\bibnamefont
  {Pecora}},\ }\bibfield  {title} {\enquote {\bibinfo {title} {Dynamics of
  wormlike chains},}\ }\href {\doibase 10.1021/ma00152a014} {\bibfield
  {journal} {\bibinfo  {journal} {Macromolecules}\ }\textbf {\bibinfo {volume}
  {18}},\ \bibinfo {pages} {1868} (\bibinfo {year} {1985})}\BibitemShut
  {NoStop}%
\bibitem [{\citenamefont {Winkler}(2007)}]{wink:07.1}%
  \BibitemOpen
  \bibfield  {author} {\bibinfo {author} {\bibfnamefont {R.~G.}\ \bibnamefont
  {Winkler}},\ }\bibfield  {title} {\enquote {\bibinfo {title} {Diffusion and
  segmental dynamics of rodlike molecules by fluorescence correlation
  spectroscopy},}\ }\href {\doibase 10.1063/1.2753160} {\bibfield  {journal}
  {\bibinfo  {journal} {J. Chem. Phys.}\ }\textbf {\bibinfo {volume} {127}},\
  \bibinfo {pages} {054904} (\bibinfo {year} {2007})}\BibitemShut {NoStop}%
\bibitem [{\citenamefont {Winkler}\ and\ \citenamefont
  {Reineker}(1992)}]{wink:92}%
  \BibitemOpen
  \bibfield  {author} {\bibinfo {author} {\bibfnamefont {R.~G.}\ \bibnamefont
  {Winkler}}\ and\ \bibinfo {author} {\bibfnamefont {P.}~\bibnamefont
  {Reineker}},\ }\bibfield  {title} {\enquote {\bibinfo {title} {Finite size
  distribution and partition functions of {Gaussian} chains: Maximum entropy
  approach},}\ }\href {\doibase 10.1021/ma00051a026} {\bibfield  {journal}
  {\bibinfo  {journal} {Macromolecules}\ }\textbf {\bibinfo {volume} {25}},\
  \bibinfo {pages} {6891} (\bibinfo {year} {1992})}\BibitemShut {NoStop}%
\bibitem [{\citenamefont {Winkler}, \citenamefont {Keller},\ and\ \citenamefont
  {R{\"a}dler}(2006)}]{wink:06}%
  \BibitemOpen
  \bibfield  {author} {\bibinfo {author} {\bibfnamefont {R.~G.}\ \bibnamefont
  {Winkler}}, \bibinfo {author} {\bibfnamefont {S.}~\bibnamefont {Keller}}, \
  and\ \bibinfo {author} {\bibfnamefont {J.~O.}\ \bibnamefont {R{\"a}dler}},\
  }\bibfield  {title} {\enquote {\bibinfo {title} {Intramolecular dynamics of
  linear macromolecules by fluorescence correlation spectroscopy},}\ }\href
  {\doibase 10.1103/PhysRevE.73.041919} {\bibfield  {journal} {\bibinfo
  {journal} {Phys. Rev. E}\ }\textbf {\bibinfo {volume} {73}},\ \bibinfo
  {pages} {041919} (\bibinfo {year} {2006})}\BibitemShut {NoStop}%
\bibitem [{\citenamefont {{El Alaoui Faris}}\ \emph {et~al.}(2009)\citenamefont
  {{El Alaoui Faris}}, \citenamefont {Lacoste}, \citenamefont
  {P{\'e}cr{\'e}aux}, \citenamefont {Joanny}, \citenamefont {Prost},\ and\
  \citenamefont {Bassereau}}]{fari:09}%
  \BibitemOpen
  \bibfield  {author} {\bibinfo {author} {\bibfnamefont {M.~D.}\ \bibnamefont
  {{El Alaoui Faris}}}, \bibinfo {author} {\bibfnamefont {D.}~\bibnamefont
  {Lacoste}}, \bibinfo {author} {\bibfnamefont {J.}~\bibnamefont
  {P{\'e}cr{\'e}aux}}, \bibinfo {author} {\bibfnamefont {J.~F.}\ \bibnamefont
  {Joanny}}, \bibinfo {author} {\bibfnamefont {J.}~\bibnamefont {Prost}}, \
  and\ \bibinfo {author} {\bibfnamefont {P.}~\bibnamefont {Bassereau}},\
  }\bibfield  {title} {\enquote {\bibinfo {title} {Membrane tension lowering
  induced by protein activity},}\ }\href {\doibase
  10.1103/PhysRevLett.102.038102} {\bibfield  {journal} {\bibinfo  {journal}
  {Phys. Rev. Lett.}\ }\textbf {\bibinfo {volume} {102}},\ \bibinfo {pages}
  {038102} (\bibinfo {year} {2009})}\BibitemShut {NoStop}%
\bibitem [{\citenamefont {Turlier}\ \emph {et~al.}(2016)\citenamefont
  {Turlier}, \citenamefont {Fedosov}, \citenamefont {Audoly}, \citenamefont
  {Auth}, \citenamefont {Gov}, \citenamefont {Sykes}, \citenamefont {Joanny},
  \citenamefont {Gompper},\ and\ \citenamefont {Betz}}]{turl:16}%
  \BibitemOpen
  \bibfield  {author} {\bibinfo {author} {\bibfnamefont {H.}~\bibnamefont
  {Turlier}}, \bibinfo {author} {\bibfnamefont {D.~A.}\ \bibnamefont
  {Fedosov}}, \bibinfo {author} {\bibfnamefont {B.}~\bibnamefont {Audoly}},
  \bibinfo {author} {\bibfnamefont {T.}~\bibnamefont {Auth}}, \bibinfo {author}
  {\bibfnamefont {N.~S.}\ \bibnamefont {Gov}}, \bibinfo {author} {\bibfnamefont
  {C.}~\bibnamefont {Sykes}}, \bibinfo {author} {\bibfnamefont {J.~F.}\
  \bibnamefont {Joanny}}, \bibinfo {author} {\bibfnamefont {G.}~\bibnamefont
  {Gompper}}, \ and\ \bibinfo {author} {\bibfnamefont {T.}~\bibnamefont
  {Betz}},\ }\bibfield  {title} {\enquote {\bibinfo {title} {Equilibrium
  physics breakdown reveals the active nature of red blood cell flickering},}\
  }\href {\doibase 10.1038/nphys362} {\bibfield  {journal} {\bibinfo  {journal}
  {Nat. Phys.}\ }\textbf {\bibinfo {volume} {12}},\ \bibinfo {pages} {513}
  (\bibinfo {year} {2016})}\BibitemShut {NoStop}%
\bibitem [{\citenamefont {Vutukuri}\ \emph {et~al.}(2020)\citenamefont
  {Vutukuri}, \citenamefont {Hoore}, \citenamefont {Abaurrea-Velasco},
  \citenamefont {van Buren}, \citenamefont {Dutto}, \citenamefont {Auth},
  \citenamefont {Fedosov}, \citenamefont {Gompper},\ and\ \citenamefont
  {Vermant}}]{vutu:19}%
  \BibitemOpen
  \bibfield  {author} {\bibinfo {author} {\bibfnamefont {H.~R.}\ \bibnamefont
  {Vutukuri}}, \bibinfo {author} {\bibfnamefont {M.}~\bibnamefont {Hoore}},
  \bibinfo {author} {\bibfnamefont {C.}~\bibnamefont {Abaurrea-Velasco}},
  \bibinfo {author} {\bibfnamefont {L.}~\bibnamefont {van Buren}}, \bibinfo
  {author} {\bibfnamefont {A.}~\bibnamefont {Dutto}}, \bibinfo {author}
  {\bibfnamefont {T.}~\bibnamefont {Auth}}, \bibinfo {author} {\bibfnamefont
  {D.~A.}\ \bibnamefont {Fedosov}}, \bibinfo {author} {\bibfnamefont
  {G.}~\bibnamefont {Gompper}}, \ and\ \bibinfo {author} {\bibfnamefont
  {J.}~\bibnamefont {Vermant}},\ }\bibfield  {title} {\enquote {\bibinfo
  {title} {Sculpting vesicles with active particles: Less is more},}\
  }\href@noop {} {\bibfield  {journal} {\bibinfo  {journal} {Nature}\ ,\
  \bibinfo {pages} {in press}} (\bibinfo {year} {2020})},\ \Eprint
  {http://arxiv.org/abs/1911.02381} {arXiv:1911.02381 [cond-mat.soft]}
  \BibitemShut {NoStop}%
\bibitem [{\citenamefont {Takatori}\ and\ \citenamefont
  {Sahu}(2020)}]{taka:20}%
  \BibitemOpen
  \bibfield  {author} {\bibinfo {author} {\bibfnamefont {S.~C.}\ \bibnamefont
  {Takatori}}\ and\ \bibinfo {author} {\bibfnamefont {A.}~\bibnamefont
  {Sahu}},\ }\bibfield  {title} {\enquote {\bibinfo {title} {Active contact
  forces drive nonequilibrium fluctuations in membrane vesicles},}\ }\href
  {\doibase 10.1103/PhysRevLett.124.158102} {\bibfield  {journal} {\bibinfo
  {journal} {Phys. Rev. Lett.}\ }\textbf {\bibinfo {volume} {124}},\ \bibinfo
  {pages} {158102} (\bibinfo {year} {2020})}\BibitemShut {NoStop}%
\bibitem [{\citenamefont {Loubet}, \citenamefont {Seifert},\ and\ \citenamefont
  {Lomholt}(2012)}]{loub:12}%
  \BibitemOpen
  \bibfield  {author} {\bibinfo {author} {\bibfnamefont {B.}~\bibnamefont
  {Loubet}}, \bibinfo {author} {\bibfnamefont {U.}~\bibnamefont {Seifert}}, \
  and\ \bibinfo {author} {\bibfnamefont {M.~A.}\ \bibnamefont {Lomholt}},\
  }\bibfield  {title} {\enquote {\bibinfo {title} {Effective tension and
  fluctuations in active membranes},}\ }\href {\doibase
  10.1103/PhysRevE.85.031913} {\bibfield  {journal} {\bibinfo  {journal} {Phys.
  Rev. E}\ }\textbf {\bibinfo {volume} {85}},\ \bibinfo {pages} {031913}
  (\bibinfo {year} {2012})}\BibitemShut {NoStop}%
\bibitem [{\citenamefont {Harnau}, \citenamefont {Winkler},\ and\ \citenamefont
  {Reineker}(1996)}]{harn:96}%
  \BibitemOpen
  \bibfield  {author} {\bibinfo {author} {\bibfnamefont {L.}~\bibnamefont
  {Harnau}}, \bibinfo {author} {\bibfnamefont {R.~G.}\ \bibnamefont {Winkler}},
  \ and\ \bibinfo {author} {\bibfnamefont {P.}~\bibnamefont {Reineker}},\
  }\bibfield  {title} {\enquote {\bibinfo {title} {Dynamic structure factor of
  semiflexible macromolecules in dilute solution},}\ }\href {\doibase
  0.1063/1.471297} {\bibfield  {journal} {\bibinfo  {journal} {J. Chem. Phys.}\
  }\textbf {\bibinfo {volume} {104}},\ \bibinfo {pages} {6355} (\bibinfo {year}
  {1996})}\BibitemShut {NoStop}%
\bibitem [{\citenamefont {Rotne}\ and\ \citenamefont {Prager}(1969)}]{rotn:69}%
  \BibitemOpen
  \bibfield  {author} {\bibinfo {author} {\bibfnamefont {J.}~\bibnamefont
  {Rotne}}\ and\ \bibinfo {author} {\bibfnamefont {S.}~\bibnamefont {Prager}},\
  }\bibfield  {title} {\enquote {\bibinfo {title} {Variational treatment of
  hydrodynamic interaction in polymers},}\ }\href {\doibase 10.1063/1.1670977}
  {\bibfield  {journal} {\bibinfo  {journal} {J. Chem. Phys}\ }\textbf
  {\bibinfo {volume} {50}},\ \bibinfo {pages} {4831} (\bibinfo {year}
  {1969})}\BibitemShut {NoStop}%
\bibitem [{\citenamefont {Yamakawa}(1970)}]{yama:70}%
  \BibitemOpen
  \bibfield  {author} {\bibinfo {author} {\bibfnamefont {H.}~\bibnamefont
  {Yamakawa}},\ }\bibfield  {title} {\enquote {\bibinfo {title} {Transport
  properties of polymer chains in dilute solution: Hydrodynamic interaction},}\
  }\href {\doibase 10.1063/1.1673799} {\bibfield  {journal} {\bibinfo
  {journal} {J. Chem. Phys.}\ }\textbf {\bibinfo {volume} {53}},\ \bibinfo
  {pages} {436} (\bibinfo {year} {1970})}\BibitemShut {NoStop}%
\bibitem [{\citenamefont {Jain}\ \emph {et~al.}(2012)\citenamefont {Jain},
  \citenamefont {Sunthar}, \citenamefont {D{\"u}nweg},\ and\ \citenamefont
  {Prakash}}]{jain:12.1}%
  \BibitemOpen
  \bibfield  {author} {\bibinfo {author} {\bibfnamefont {A.}~\bibnamefont
  {Jain}}, \bibinfo {author} {\bibfnamefont {P.}~\bibnamefont {Sunthar}},
  \bibinfo {author} {\bibfnamefont {B.}~\bibnamefont {D{\"u}nweg}}, \ and\
  \bibinfo {author} {\bibfnamefont {J.~R.}\ \bibnamefont {Prakash}},\
  }\bibfield  {title} {\enquote {\bibinfo {title} {Optimization of a
  brownian-dynamics algorithm for semidilute polymer solutions},}\ }\href
  {\doibase 10.1103/PhysRevE.85.066703} {\bibfield  {journal} {\bibinfo
  {journal} {Phys. Rev. E}\ }\textbf {\bibinfo {volume} {85}},\ \bibinfo
  {pages} {066703} (\bibinfo {year} {2012})}\BibitemShut {NoStop}%
\bibitem [{\citenamefont {Dhont}(1996)}]{dhon:96}%
  \BibitemOpen
  \bibfield  {author} {\bibinfo {author} {\bibfnamefont {J.~K.~G.}\
  \bibnamefont {Dhont}},\ }\href@noop {} {\emph {\bibinfo {title} {An
  Introduction to Dynamics of Colloids}}}\ (\bibinfo  {publisher} {Elsevier},\
  \bibinfo {address} {Amsterdam},\ \bibinfo {year} {1996})\BibitemShut
  {NoStop}%
\bibitem [{\citenamefont {McNamara}\ and\ \citenamefont
  {Zanetti}(1988)}]{mcna:88}%
  \BibitemOpen
  \bibfield  {author} {\bibinfo {author} {\bibfnamefont {G.~R.}\ \bibnamefont
  {McNamara}}\ and\ \bibinfo {author} {\bibfnamefont {G.}~\bibnamefont
  {Zanetti}},\ }\bibfield  {title} {\enquote {\bibinfo {title} {Use of the
  {Boltzmann} equation to simulate lattice-gas automata},}\ }\href {\doibase
  10.1103/PhysRevLett.61.2332} {\bibfield  {journal} {\bibinfo  {journal}
  {Phys. Rev. Lett.}\ }\textbf {\bibinfo {volume} {61}},\ \bibinfo {pages}
  {2332} (\bibinfo {year} {1988})}\BibitemShut {NoStop}%
\bibitem [{\citenamefont {Shan}\ and\ \citenamefont {Chen}(1993)}]{shan:93}%
  \BibitemOpen
  \bibfield  {author} {\bibinfo {author} {\bibfnamefont {X.}~\bibnamefont
  {Shan}}\ and\ \bibinfo {author} {\bibfnamefont {H.}~\bibnamefont {Chen}},\
  }\bibfield  {title} {\enquote {\bibinfo {title} {Lattice {Boltzmann} model
  for simulating flows with multiple phases and components},}\ }\href {\doibase
  10.1103/PhysRevE.47.1815} {\bibfield  {journal} {\bibinfo  {journal} {Phys.
  Rev. E}\ }\textbf {\bibinfo {volume} {47}},\ \bibinfo {pages} {1815}
  (\bibinfo {year} {1993})}\BibitemShut {NoStop}%
\bibitem [{\citenamefont {Succi}(2001)}]{succ:01}%
  \BibitemOpen
  \bibfield  {author} {\bibinfo {author} {\bibfnamefont {S.}~\bibnamefont
  {Succi}},\ }\href@noop {} {\emph {\bibinfo {title} {The lattice Boltzmann
  equation: for fluid dynamics and beyond}}}\ (\bibinfo  {publisher} {Oxford
  University Press},\ \bibinfo {year} {2001})\BibitemShut {NoStop}%
\bibitem [{\citenamefont {D{\"u}nweg}\ and\ \citenamefont
  {Ladd}(2009)}]{duen:09}%
  \BibitemOpen
  \bibfield  {author} {\bibinfo {author} {\bibfnamefont {B.}~\bibnamefont
  {D{\"u}nweg}}\ and\ \bibinfo {author} {\bibfnamefont {A.~C.}\ \bibnamefont
  {Ladd}},\ }\bibfield  {title} {\enquote {\bibinfo {title} {Lattice
  {B}oltzmann simulations of soft matter systems},}\ }\href {\doibase
  10.1007/978-3-540-87706-6{\_}2} {\bibfield  {journal} {\bibinfo  {journal}
  {Adv. Polym. Sci.}\ }\textbf {\bibinfo {volume} {221}},\ \bibinfo {pages}
  {89} (\bibinfo {year} {2009})}\BibitemShut {NoStop}%
\bibitem [{\citenamefont {Hoogerbrugge}\ and\ \citenamefont
  {Koelman}(1992)}]{hoog:92}%
  \BibitemOpen
  \bibfield  {author} {\bibinfo {author} {\bibfnamefont {P.~J.}\ \bibnamefont
  {Hoogerbrugge}}\ and\ \bibinfo {author} {\bibfnamefont {J.~M. V.~A.}\
  \bibnamefont {Koelman}},\ }\bibfield  {title} {\enquote {\bibinfo {title}
  {Simulating microscopic hydrodynamics phenomena with dissipative particle
  dynamics},}\ }\href {\doibase 10.1209/0295-5075/19/3/001} {\bibfield
  {journal} {\bibinfo  {journal} {Europhys. Lett.}\ }\textbf {\bibinfo {volume}
  {19}},\ \bibinfo {pages} {155} (\bibinfo {year} {1992})}\BibitemShut
  {NoStop}%
\bibitem [{\citenamefont {Espa{{\~n}}ol}\ and\ \citenamefont
  {Warren}(1995)}]{espa:95}%
  \BibitemOpen
  \bibfield  {author} {\bibinfo {author} {\bibfnamefont {P.}~\bibnamefont
  {Espa{{\~n}}ol}}\ and\ \bibinfo {author} {\bibfnamefont {P.~B.}\ \bibnamefont
  {Warren}},\ }\bibfield  {title} {\enquote {\bibinfo {title} {Statistical
  mechanics of dissipative particle dynamics},}\ }\href {\doibase
  10.1209/0295-5075/30/4/001} {\bibfield  {journal} {\bibinfo  {journal}
  {Europhys. Lett.}\ }\textbf {\bibinfo {volume} {30}},\ \bibinfo {pages} {191}
  (\bibinfo {year} {1995})}\BibitemShut {NoStop}%
\bibitem [{\citenamefont {Malevanets}\ and\ \citenamefont
  {Kapral}(1999)}]{male:99}%
  \BibitemOpen
  \bibfield  {author} {\bibinfo {author} {\bibfnamefont {A.}~\bibnamefont
  {Malevanets}}\ and\ \bibinfo {author} {\bibfnamefont {R.}~\bibnamefont
  {Kapral}},\ }\bibfield  {title} {\enquote {\bibinfo {title} {Mesoscopic model
  for solvent dynamics},}\ }\href {\doibase 10.1063/1.478857} {\bibfield
  {journal} {\bibinfo  {journal} {J. Chem. Phys.}\ }\textbf {\bibinfo {volume}
  {110}},\ \bibinfo {pages} {8605} (\bibinfo {year} {1999})}\BibitemShut
  {NoStop}%
\bibitem [{\citenamefont {Kapral}(2008)}]{kapr:08}%
  \BibitemOpen
  \bibfield  {author} {\bibinfo {author} {\bibfnamefont {R.}~\bibnamefont
  {Kapral}},\ }\bibfield  {title} {\enquote {\bibinfo {title} {Multiparticle
  collision dynamics: Simulations of complex systems on mesoscale},}\ }\href
  {\doibase 10.1002/9780470371572.ch2} {\bibfield  {journal} {\bibinfo
  {journal} {Adv. Chem. Phys.}\ }\textbf {\bibinfo {volume} {140}},\ \bibinfo
  {pages} {89} (\bibinfo {year} {2008})}\BibitemShut {NoStop}%
\bibitem [{\citenamefont {Gompper}\ \emph {et~al.}(2009)\citenamefont
  {Gompper}, \citenamefont {Ihle}, \citenamefont {Kroll},\ and\ \citenamefont
  {Winkler}}]{gomp:09}%
  \BibitemOpen
  \bibfield  {author} {\bibinfo {author} {\bibfnamefont {G.}~\bibnamefont
  {Gompper}}, \bibinfo {author} {\bibfnamefont {T.}~\bibnamefont {Ihle}},
  \bibinfo {author} {\bibfnamefont {D.~M.}\ \bibnamefont {Kroll}}, \ and\
  \bibinfo {author} {\bibfnamefont {R.~G.}\ \bibnamefont {Winkler}},\
  }\bibfield  {title} {\enquote {\bibinfo {title} {Multi-particle collision
  dynamics: A particle-based mesoscale simulation approach to the hydrodynamics
  of complex fluids},}\ }\href {\doibase 10.1007/978-3-540-87706-6{\_1}}
  {\bibfield  {journal} {\bibinfo  {journal} {Adv. Polym. Sci.}\ }\textbf
  {\bibinfo {volume} {221}},\ \bibinfo {pages} {1} (\bibinfo {year}
  {2009})}\BibitemShut {NoStop}%
\bibitem [{\citenamefont {Theers}\ \emph
  {et~al.}(2016{\natexlab{b}})\citenamefont {Theers}, \citenamefont {Westphal},
  \citenamefont {Gompper},\ and\ \citenamefont {Winkler}}]{thee:16}%
  \BibitemOpen
  \bibfield  {author} {\bibinfo {author} {\bibfnamefont {M.}~\bibnamefont
  {Theers}}, \bibinfo {author} {\bibfnamefont {E.}~\bibnamefont {Westphal}},
  \bibinfo {author} {\bibfnamefont {G.}~\bibnamefont {Gompper}}, \ and\
  \bibinfo {author} {\bibfnamefont {R.~G.}\ \bibnamefont {Winkler}},\
  }\bibfield  {title} {\enquote {\bibinfo {title} {From local to hydrodynamic
  friction in {B}rownian motion: A multiparticle collision dynamics simulation
  study},}\ }\href {\doibase 10.1103/PhysRevE.93.032604} {\bibfield  {journal}
  {\bibinfo  {journal} {Phys. Rev. E}\ }\textbf {\bibinfo {volume} {93}},\
  \bibinfo {pages} {032604} (\bibinfo {year} {2016}{\natexlab{b}})}\BibitemShut
  {NoStop}%
\bibitem [{\citenamefont {Winkler}(2016{\natexlab{b}})}]{wink:16.1}%
  \BibitemOpen
  \bibfield  {author} {\bibinfo {author} {\bibfnamefont {R.~G.}\ \bibnamefont
  {Winkler}},\ }\bibfield  {title} {\enquote {\bibinfo {title} {Low {R}eynolds
  number hydrodynamics and mesoscale simulations},}\ }\href {\doibase
  10.1140/epjst/e2016-60087-9} {\bibfield  {journal} {\bibinfo  {journal} {Eur.
  Phys. J. Spec. Top.}\ }\textbf {\bibinfo {volume} {225}},\ \bibinfo {pages}
  {2079} (\bibinfo {year} {2016}{\natexlab{b}})}\BibitemShut {NoStop}%
\bibitem [{\citenamefont {Allen}\ and\ \citenamefont
  {Tildesley}(1987)}]{alle:87}%
  \BibitemOpen
  \bibfield  {author} {\bibinfo {author} {\bibfnamefont {M.~P.}\ \bibnamefont
  {Allen}}\ and\ \bibinfo {author} {\bibfnamefont {D.~J.}\ \bibnamefont
  {Tildesley}},\ }\href@noop {} {\emph {\bibinfo {title} {Computer Simulation
  of Liquids}}}\ (\bibinfo  {publisher} {Clarendon Press},\ \bibinfo {address}
  {Oxford},\ \bibinfo {year} {1987})\BibitemShut {NoStop}%
\bibitem [{\citenamefont {Ermak}\ and\ \citenamefont
  {McCammon}(1978)}]{erma:78}%
  \BibitemOpen
  \bibfield  {author} {\bibinfo {author} {\bibfnamefont {D.~L.}\ \bibnamefont
  {Ermak}}\ and\ \bibinfo {author} {\bibfnamefont {J.}~\bibnamefont
  {McCammon}},\ }\bibfield  {title} {\enquote {\bibinfo {title} {Brownian
  dynamics with hydrodynamic interactions},}\ }\href {\doibase
  10.1063/1.436761} {\bibfield  {journal} {\bibinfo  {journal} {J. Chem.
  Phys.}\ }\textbf {\bibinfo {volume} {69}},\ \bibinfo {pages} {1352} (\bibinfo
  {year} {1978})}\BibitemShut {NoStop}%
\bibitem [{\citenamefont {Zimm}(1956)}]{zimm:56}%
  \BibitemOpen
  \bibfield  {author} {\bibinfo {author} {\bibfnamefont {B.~H.}\ \bibnamefont
  {Zimm}},\ }\bibfield  {title} {\enquote {\bibinfo {title} {Dynamics of
  polymer molecules in dilute solution: Viscoelasticity, flow birefringence and
  dielectric loss},}\ }\href {\doibase 10.1063/1.1742462} {\bibfield  {journal}
  {\bibinfo  {journal} {J. Chem. Phys.}\ }\textbf {\bibinfo {volume} {24}},\
  \bibinfo {pages} {269} (\bibinfo {year} {1956})}\BibitemShut {NoStop}%
\bibitem [{\citenamefont {Kroy}\ and\ \citenamefont {Frey}(1997)}]{kroy:97}%
  \BibitemOpen
  \bibfield  {author} {\bibinfo {author} {\bibfnamefont {K.}~\bibnamefont
  {Kroy}}\ and\ \bibinfo {author} {\bibfnamefont {E.}~\bibnamefont {Frey}},\
  }\bibfield  {title} {\enquote {\bibinfo {title} {Dynamic scattering from
  solutions of semiflexible polymers},}\ }\href {\doibase
  10.1103/PhysRevE.55.3092} {\bibfield  {journal} {\bibinfo  {journal} {Phys.
  Rev. E}\ }\textbf {\bibinfo {volume} {55}},\ \bibinfo {pages} {3092}
  (\bibinfo {year} {1997})}\BibitemShut {NoStop}%
\bibitem [{\citenamefont {Winkler}(1999)}]{wink:99}%
  \BibitemOpen
  \bibfield  {author} {\bibinfo {author} {\bibfnamefont {R.~G.}\ \bibnamefont
  {Winkler}},\ }\bibfield  {title} {\enquote {\bibinfo {title} {Analytical
  calculation of the relaxation dynamics of partially stretched flexible chain
  molecules: Necessity of a wormlike chain description},}\ }\href {\doibase
  10.1103/PhysRevLett.82.1843} {\bibfield  {journal} {\bibinfo  {journal}
  {Phys. Rev. Lett.}\ }\textbf {\bibinfo {volume} {82}},\ \bibinfo {pages}
  {1843} (\bibinfo {year} {1999})}\BibitemShut {NoStop}%
\bibitem [{\citenamefont {Anand}\ and\ \citenamefont {Singh}(2020)}]{anan:20}%
  \BibitemOpen
  \bibfield  {author} {\bibinfo {author} {\bibfnamefont {S.~K.}\ \bibnamefont
  {Anand}}\ and\ \bibinfo {author} {\bibfnamefont {S.~P.}\ \bibnamefont
  {Singh}},\ }\bibfield  {title} {\enquote {\bibinfo {title} {Conformation and
  dynamics of a self-avoiding active flexible polymer},}\ }\href {\doibase
  10.1103/PhysRevE.101.030501} {\bibfield  {journal} {\bibinfo  {journal}
  {Phys. Rev. E}\ }\textbf {\bibinfo {volume} {101}},\ \bibinfo {pages}
  {030501} (\bibinfo {year} {2020})}\BibitemShut {NoStop}%
\bibitem [{\citenamefont {Petrov}\ \emph {et~al.}(2006)\citenamefont {Petrov},
  \citenamefont {Ohrt}, \citenamefont {Winkler},\ and\ \citenamefont
  {Schwille}}]{petr:06}%
  \BibitemOpen
  \bibfield  {author} {\bibinfo {author} {\bibfnamefont {E.~P.}\ \bibnamefont
  {Petrov}}, \bibinfo {author} {\bibfnamefont {T.}~\bibnamefont {Ohrt}},
  \bibinfo {author} {\bibfnamefont {R.~G.}\ \bibnamefont {Winkler}}, \ and\
  \bibinfo {author} {\bibfnamefont {P.}~\bibnamefont {Schwille}},\ }\bibfield
  {title} {\enquote {\bibinfo {title} {Diffusion and segmental dynamics of
  double-stranded {DNA}},}\ }\href {\doibase 10.1103/PhysRevLett.97.258101}
  {\bibfield  {journal} {\bibinfo  {journal} {Phys. Rev. Lett.}\ }\textbf
  {\bibinfo {volume} {97}},\ \bibinfo {pages} {258101} (\bibinfo {year}
  {2006})}\BibitemShut {NoStop}%
\bibitem [{\citenamefont {Grier}(2003)}]{grie:03}%
  \BibitemOpen
  \bibfield  {author} {\bibinfo {author} {\bibfnamefont {D.~G.}\ \bibnamefont
  {Grier}},\ }\bibfield  {title} {\enquote {\bibinfo {title} {A revolution in
  optical manipulation},}\ }\href {\doibase 10.1038/nature01935} {\bibfield
  {journal} {\bibinfo  {journal} {Nature}\ }\textbf {\bibinfo {volume} {424}},\
  \bibinfo {pages} {810} (\bibinfo {year} {2003})}\BibitemShut {NoStop}%
\bibitem [{\citenamefont {Ganguly}\ \emph {et~al.}(2012)\citenamefont
  {Ganguly}, \citenamefont {Williams}, \citenamefont {Palacios},\ and\
  \citenamefont {Goldstein}}]{gang:12}%
  \BibitemOpen
  \bibfield  {author} {\bibinfo {author} {\bibfnamefont {S.}~\bibnamefont
  {Ganguly}}, \bibinfo {author} {\bibfnamefont {L.~S.}\ \bibnamefont
  {Williams}}, \bibinfo {author} {\bibfnamefont {I.~M.}\ \bibnamefont
  {Palacios}}, \ and\ \bibinfo {author} {\bibfnamefont {R.~E.}\ \bibnamefont
  {Goldstein}},\ }\bibfield  {title} {\enquote {\bibinfo {title} {Cytoplasmic
  streaming in \emph{{D}rosophila} oocytes varies with kinesin activity and
  correlates with the microtubule cytoskeleton architecture},}\ }\href
  {\doibase 10.1073/pnas.1203575109} {\bibfield  {journal} {\bibinfo  {journal}
  {Proc. Natl. Acad. Sci. USA}\ }\textbf {\bibinfo {volume} {109}},\ \bibinfo
  {pages} {15109} (\bibinfo {year} {2012})}\BibitemShut {NoStop}%
\bibitem [{\citenamefont {Ennomani}\ \emph {et~al.}(2016)\citenamefont
  {Ennomani}, \citenamefont {Letort}, \citenamefont {Gu{\'e}rin}, \citenamefont
  {Martiel}, \citenamefont {Cao}, \citenamefont {N{\'e}d{\'e}lec},
  \citenamefont {De~La~Cruz}, \citenamefont {Th{\'e}ry},\ and\ \citenamefont
  {Blanchoin}}]{enno:16}%
  \BibitemOpen
  \bibfield  {author} {\bibinfo {author} {\bibfnamefont {H.}~\bibnamefont
  {Ennomani}}, \bibinfo {author} {\bibfnamefont {G.}~\bibnamefont {Letort}},
  \bibinfo {author} {\bibfnamefont {C.}~\bibnamefont {Gu{\'e}rin}}, \bibinfo
  {author} {\bibfnamefont {J.-L.}\ \bibnamefont {Martiel}}, \bibinfo {author}
  {\bibfnamefont {W.}~\bibnamefont {Cao}}, \bibinfo {author} {\bibfnamefont
  {F.}~\bibnamefont {N{\'e}d{\'e}lec}}, \bibinfo {author} {\bibfnamefont
  {E.~M.}\ \bibnamefont {De~La~Cruz}}, \bibinfo {author} {\bibfnamefont
  {M.}~\bibnamefont {Th{\'e}ry}}, \ and\ \bibinfo {author} {\bibfnamefont
  {L.}~\bibnamefont {Blanchoin}},\ }\bibfield  {title} {\enquote {\bibinfo
  {title} {Architecture and connectivity govern actin network contractility},}\
  }\href {\doibase https://doi.org/10.1016/j.cub.2015.12.069} {\bibfield
  {journal} {\bibinfo  {journal} {Curr. Biol.}\ }\textbf {\bibinfo {volume}
  {26}},\ \bibinfo {pages} {616} (\bibinfo {year} {2016})}\BibitemShut
  {NoStop}%
\bibitem [{\citenamefont {Wu}\ \emph {et~al.}(2017)\citenamefont {Wu},
  \citenamefont {Hishamunda}, \citenamefont {Chen}, \citenamefont {DeCamp},
  \citenamefont {Chang}, \citenamefont {Fern{\'a}ndez-Nieves}, \citenamefont
  {Fraden},\ and\ \citenamefont {Dogic}}]{wu:17}%
  \BibitemOpen
  \bibfield  {author} {\bibinfo {author} {\bibfnamefont {K.-T.}\ \bibnamefont
  {Wu}}, \bibinfo {author} {\bibfnamefont {J.~B.}\ \bibnamefont {Hishamunda}},
  \bibinfo {author} {\bibfnamefont {D.~T.~N.}\ \bibnamefont {Chen}}, \bibinfo
  {author} {\bibfnamefont {S.~J.}\ \bibnamefont {DeCamp}}, \bibinfo {author}
  {\bibfnamefont {Y.-W.}\ \bibnamefont {Chang}}, \bibinfo {author}
  {\bibfnamefont {A.}~\bibnamefont {Fern{\'a}ndez-Nieves}}, \bibinfo {author}
  {\bibfnamefont {S.}~\bibnamefont {Fraden}}, \ and\ \bibinfo {author}
  {\bibfnamefont {Z.}~\bibnamefont {Dogic}},\ }\bibfield  {title} {\enquote
  {\bibinfo {title} {Transition from turbulent to coherent flows in confined
  three-dimensional active fluids},}\ }\href {\doibase 10.1126/science.aal1979}
  {\bibfield  {journal} {\bibinfo  {journal} {Science}\ }\textbf {\bibinfo
  {volume} {355}},\ \bibinfo {pages} {eaal1979} (\bibinfo {year}
  {2017})}\BibitemShut {NoStop}%
\bibitem [{\citenamefont {Belmonte}, \citenamefont {Leptin},\ and\
  \citenamefont {N{\'e}d{\'e}lec}(2017)}]{belm:17}%
  \BibitemOpen
  \bibfield  {author} {\bibinfo {author} {\bibfnamefont {J.~M.}\ \bibnamefont
  {Belmonte}}, \bibinfo {author} {\bibfnamefont {M.}~\bibnamefont {Leptin}}, \
  and\ \bibinfo {author} {\bibfnamefont {F.}~\bibnamefont {N{\'e}d{\'e}lec}},\
  }\bibfield  {title} {\enquote {\bibinfo {title} {A theory that predicts
  behaviors of disordered cytoskeletal networks},}\ }\href {\doibase
  10.15252/msb.20177796} {\bibfield  {journal} {\bibinfo  {journal} {Mol. Syst.
  Biol.}\ }\textbf {\bibinfo {volume} {13}},\ \bibinfo {pages} {941} (\bibinfo
  {year} {2017})}\BibitemShut {NoStop}%
\bibitem [{\citenamefont {Huber}\ \emph {et~al.}(2018)\citenamefont {Huber},
  \citenamefont {Suzuki}, \citenamefont {Kr{\"u}ger}, \citenamefont {Frey},\
  and\ \citenamefont {Bausch}}]{hube:18}%
  \BibitemOpen
  \bibfield  {author} {\bibinfo {author} {\bibfnamefont {L.}~\bibnamefont
  {Huber}}, \bibinfo {author} {\bibfnamefont {R.}~\bibnamefont {Suzuki}},
  \bibinfo {author} {\bibfnamefont {T.}~\bibnamefont {Kr{\"u}ger}}, \bibinfo
  {author} {\bibfnamefont {E.}~\bibnamefont {Frey}}, \ and\ \bibinfo {author}
  {\bibfnamefont {A.~R.}\ \bibnamefont {Bausch}},\ }\bibfield  {title}
  {\enquote {\bibinfo {title} {Emergence of coexisting ordered states in active
  matter systems},}\ }\href {\doibase 10.1126/science.aao5434} {\bibfield
  {journal} {\bibinfo  {journal} {Science}\ }\textbf {\bibinfo {volume}
  {361}},\ \bibinfo {pages} {255} (\bibinfo {year} {2018})}\BibitemShut
  {NoStop}%
\bibitem [{\citenamefont {Guillamat}\ \emph {et~al.}(2018)\citenamefont
  {Guillamat}, \citenamefont {Kos}, \citenamefont {Hardo{\"u}in}, \citenamefont
  {Ign{\'e}s-Mullol}, \citenamefont {Ravnik},\ and\ \citenamefont
  {Sagu{\'e}s}}]{guil:18}%
  \BibitemOpen
  \bibfield  {author} {\bibinfo {author} {\bibfnamefont {P.}~\bibnamefont
  {Guillamat}}, \bibinfo {author} {\bibfnamefont {{\v Z}.}~\bibnamefont {Kos}},
  \bibinfo {author} {\bibfnamefont {J.}~\bibnamefont {Hardo{\"u}in}}, \bibinfo
  {author} {\bibfnamefont {J.}~\bibnamefont {Ign{\'e}s-Mullol}}, \bibinfo
  {author} {\bibfnamefont {M.}~\bibnamefont {Ravnik}}, \ and\ \bibinfo {author}
  {\bibfnamefont {F.}~\bibnamefont {Sagu{\'e}s}},\ }\bibfield  {title}
  {\enquote {\bibinfo {title} {Active nematic emulsions},}\ }\href {\doibase
  10.1126/sciadv.aao1470} {\bibfield  {journal} {\bibinfo  {journal} {Sci.
  Adv.}\ }\textbf {\bibinfo {volume} {4}},\ \bibinfo {pages} {eaao1470}
  (\bibinfo {year} {2018})}\BibitemShut {NoStop}%
\bibitem [{\citenamefont {Peruani}, \citenamefont {Deutsch},\ and\
  \citenamefont {B{\"a}r}(2006)}]{peru:06}%
  \BibitemOpen
  \bibfield  {author} {\bibinfo {author} {\bibfnamefont {F.}~\bibnamefont
  {Peruani}}, \bibinfo {author} {\bibfnamefont {A.}~\bibnamefont {Deutsch}}, \
  and\ \bibinfo {author} {\bibfnamefont {M.}~\bibnamefont {B{\"a}r}},\
  }\bibfield  {title} {\enquote {\bibinfo {title} {Nonequilibrium clustering of
  self-propelled rods},}\ }\href {\doibase 10.1103/PhysRevE.74.030904}
  {\bibfield  {journal} {\bibinfo  {journal} {Phys. Rev. E}\ }\textbf {\bibinfo
  {volume} {74}},\ \bibinfo {pages} {030904} (\bibinfo {year}
  {2006})}\BibitemShut {NoStop}%
\bibitem [{\citenamefont {Ginelli}\ \emph {et~al.}(2010)\citenamefont
  {Ginelli}, \citenamefont {Peruani}, \citenamefont {B{\"a}r},\ and\
  \citenamefont {Chat{\'e}}}]{gine:10}%
  \BibitemOpen
  \bibfield  {author} {\bibinfo {author} {\bibfnamefont {F.}~\bibnamefont
  {Ginelli}}, \bibinfo {author} {\bibfnamefont {F.}~\bibnamefont {Peruani}},
  \bibinfo {author} {\bibfnamefont {M.}~\bibnamefont {B{\"a}r}}, \ and\
  \bibinfo {author} {\bibfnamefont {H.}~\bibnamefont {Chat{\'e}}},\ }\bibfield
  {title} {\enquote {\bibinfo {title} {Large-scale collective properties of
  self-propelled rods},}\ }\href {\doibase 10.1103/PhysRevLett.104.184502}
  {\bibfield  {journal} {\bibinfo  {journal} {Phys. Rev. Lett}\ }\textbf
  {\bibinfo {volume} {104}},\ \bibinfo {pages} {184502} (\bibinfo {year}
  {2010})}\BibitemShut {NoStop}%
\bibitem [{\citenamefont {Wensink}\ \emph {et~al.}(2012)\citenamefont
  {Wensink}, \citenamefont {Dunkel}, \citenamefont {Heidenreich}, \citenamefont
  {Drescher}, \citenamefont {Goldstein}, \citenamefont {L{\"o}wen},\ and\
  \citenamefont {Yeomans}}]{wens:12}%
  \BibitemOpen
  \bibfield  {author} {\bibinfo {author} {\bibfnamefont {H.~H.}\ \bibnamefont
  {Wensink}}, \bibinfo {author} {\bibfnamefont {J.}~\bibnamefont {Dunkel}},
  \bibinfo {author} {\bibfnamefont {S.}~\bibnamefont {Heidenreich}}, \bibinfo
  {author} {\bibfnamefont {K.}~\bibnamefont {Drescher}}, \bibinfo {author}
  {\bibfnamefont {R.~E.}\ \bibnamefont {Goldstein}}, \bibinfo {author}
  {\bibfnamefont {H.}~\bibnamefont {L{\"o}wen}}, \ and\ \bibinfo {author}
  {\bibfnamefont {J.~M.}\ \bibnamefont {Yeomans}},\ }\bibfield  {title}
  {\enquote {\bibinfo {title} {Meso-scale turbulence in living fluids},}\
  }\href {\doibase 10.1073/pnas.1202032109} {\bibfield  {journal} {\bibinfo
  {journal} {Proc. Natl. Acad. Sci. USA}\ }\textbf {\bibinfo {volume} {109}},\
  \bibinfo {pages} {14308} (\bibinfo {year} {2012})}\BibitemShut {NoStop}%
\bibitem [{\citenamefont {Wensink}\ and\ \citenamefont
  {L{\"o}wen}(2012)}]{wens:12.1}%
  \BibitemOpen
  \bibfield  {author} {\bibinfo {author} {\bibfnamefont {H.~H.}\ \bibnamefont
  {Wensink}}\ and\ \bibinfo {author} {\bibfnamefont {H.}~\bibnamefont
  {L{\"o}wen}},\ }\bibfield  {title} {\enquote {\bibinfo {title} {Emergent
  states in dense systems of active rods: from swarming to turbulence},}\
  }\href {\doibase 10.1088/0953-8984/24/46/464130} {\bibfield  {journal}
  {\bibinfo  {journal} {J. Phys.: Condens. Matter}\ }\textbf {\bibinfo {volume}
  {24}},\ \bibinfo {pages} {460130} (\bibinfo {year} {2012})}\BibitemShut
  {NoStop}%
\bibitem [{\citenamefont {Abkenar}\ \emph {et~al.}(2013)\citenamefont
  {Abkenar}, \citenamefont {Marx}, \citenamefont {Auth},\ and\ \citenamefont
  {Gompper}}]{abke:13}%
  \BibitemOpen
  \bibfield  {author} {\bibinfo {author} {\bibfnamefont {M.}~\bibnamefont
  {Abkenar}}, \bibinfo {author} {\bibfnamefont {K.}~\bibnamefont {Marx}},
  \bibinfo {author} {\bibfnamefont {T.}~\bibnamefont {Auth}}, \ and\ \bibinfo
  {author} {\bibfnamefont {G.}~\bibnamefont {Gompper}},\ }\bibfield  {title}
  {\enquote {\bibinfo {title} {Collective behavior of penetrable self-propelled
  rods in two dimensions},}\ }\href {\doibase 10.1103/PhysRevE.88.062314}
  {\bibfield  {journal} {\bibinfo  {journal} {Phys. Rev. E}\ }\textbf {\bibinfo
  {volume} {88}},\ \bibinfo {pages} {062314} (\bibinfo {year}
  {2013})}\BibitemShut {NoStop}%
\bibitem [{\citenamefont {Abaurrea~Velasco}\ \emph {et~al.}(2018)\citenamefont
  {Abaurrea~Velasco}, \citenamefont {Abkenar}, \citenamefont {Gompper},\ and\
  \citenamefont {Auth}}]{abau:18}%
  \BibitemOpen
  \bibfield  {author} {\bibinfo {author} {\bibfnamefont {C.}~\bibnamefont
  {Abaurrea~Velasco}}, \bibinfo {author} {\bibfnamefont {M.}~\bibnamefont
  {Abkenar}}, \bibinfo {author} {\bibfnamefont {G.}~\bibnamefont {Gompper}}, \
  and\ \bibinfo {author} {\bibfnamefont {T.}~\bibnamefont {Auth}},\ }\bibfield
  {title} {\enquote {\bibinfo {title} {Collective behavior of self-propelled
  rods with quorum sensing},}\ }\href {\doibase 10.1103/PhysRevE.98.022605}
  {\bibfield  {journal} {\bibinfo  {journal} {Phys. Rev. E}\ }\textbf {\bibinfo
  {volume} {98}},\ \bibinfo {pages} {022605} (\bibinfo {year}
  {2018})}\BibitemShut {NoStop}%
\bibitem [{\citenamefont {B{\"a}r}\ \emph {et~al.}(2020)\citenamefont
  {B{\"a}r}, \citenamefont {Gro{\ss}mann}, \citenamefont {Heidenreich},\ and\
  \citenamefont {Peruani}}]{baer:20}%
  \BibitemOpen
  \bibfield  {author} {\bibinfo {author} {\bibfnamefont {M.}~\bibnamefont
  {B{\"a}r}}, \bibinfo {author} {\bibfnamefont {R.}~\bibnamefont
  {Gro{\ss}mann}}, \bibinfo {author} {\bibfnamefont {S.}~\bibnamefont
  {Heidenreich}}, \ and\ \bibinfo {author} {\bibfnamefont {F.}~\bibnamefont
  {Peruani}},\ }\bibfield  {title} {\enquote {\bibinfo {title} {Self-propelled
  rods: Insights and perspectives for active matter},}\ }\href {\doibase
  10.1146/annurev-conmatphys-031119-050611} {\bibfield  {journal} {\bibinfo
  {journal} {Annu. Rev. Condens. Matter Phys.}\ }\textbf {\bibinfo {volume}
  {11}},\ \bibinfo {pages} {441} (\bibinfo {year} {2020})}\BibitemShut
  {NoStop}%
\bibitem [{\citenamefont {Liverpool}, \citenamefont {Maggs},\ and\
  \citenamefont {Ajdari}(2001)}]{live:01}%
  \BibitemOpen
  \bibfield  {author} {\bibinfo {author} {\bibfnamefont {T.~B.}\ \bibnamefont
  {Liverpool}}, \bibinfo {author} {\bibfnamefont {A.~C.}\ \bibnamefont
  {Maggs}}, \ and\ \bibinfo {author} {\bibfnamefont {A.}~\bibnamefont
  {Ajdari}},\ }\bibfield  {title} {\enquote {\bibinfo {title} {Viscoelasticity
  of solutions of motile polymers},}\ }\href {\doibase
  10.1103/PhysRevLett.86.4171} {\bibfield  {journal} {\bibinfo  {journal}
  {Phys. Rev. Lett.}\ }\textbf {\bibinfo {volume} {86}},\ \bibinfo {pages}
  {4171} (\bibinfo {year} {2001})}\BibitemShut {NoStop}%
\bibitem [{\citenamefont {Jiang}\ and\ \citenamefont
  {Hou}(2014{\natexlab{a}})}]{jian:14}%
  \BibitemOpen
  \bibfield  {author} {\bibinfo {author} {\bibfnamefont {H.}~\bibnamefont
  {Jiang}}\ and\ \bibinfo {author} {\bibfnamefont {Z.}~\bibnamefont {Hou}},\
  }\bibfield  {title} {\enquote {\bibinfo {title} {Motion transition of active
  filaments: rotation without hydrodynamic interactions},}\ }\href {\doibase
  10.1039/C3SM52291A} {\bibfield  {journal} {\bibinfo  {journal} {Soft Matter}\
  }\textbf {\bibinfo {volume} {10}},\ \bibinfo {pages} {1012} (\bibinfo {year}
  {2014}{\natexlab{a}})}\BibitemShut {NoStop}%
\bibitem [{\citenamefont {Chelakkot}\ \emph {et~al.}(2014)\citenamefont
  {Chelakkot}, \citenamefont {Gopinath}, \citenamefont {Mahadevan},\ and\
  \citenamefont {Hagan}}]{chel:14}%
  \BibitemOpen
  \bibfield  {author} {\bibinfo {author} {\bibfnamefont {R.}~\bibnamefont
  {Chelakkot}}, \bibinfo {author} {\bibfnamefont {A.}~\bibnamefont {Gopinath}},
  \bibinfo {author} {\bibfnamefont {L.}~\bibnamefont {Mahadevan}}, \ and\
  \bibinfo {author} {\bibfnamefont {M.~F.}\ \bibnamefont {Hagan}},\ }\bibfield
  {title} {\enquote {\bibinfo {title} {Flagellar dynamics of a connected chain
  of active, polar, {B}rownian particles},}\ }\href {\doibase
  10.1098/rsif.2013.0884} {\bibfield  {journal} {\bibinfo  {journal} {J. R.
  Soc. Interf.}\ }\textbf {\bibinfo {volume} {11}},\ \bibinfo {pages}
  {20130884} (\bibinfo {year} {2014})}\BibitemShut {NoStop}%
\bibitem [{\citenamefont {Isele-Holder}, \citenamefont {Elgeti},\ and\
  \citenamefont {Gompper}(2015)}]{isel:15}%
  \BibitemOpen
  \bibfield  {author} {\bibinfo {author} {\bibfnamefont {R.~E.}\ \bibnamefont
  {Isele-Holder}}, \bibinfo {author} {\bibfnamefont {J.}~\bibnamefont
  {Elgeti}}, \ and\ \bibinfo {author} {\bibfnamefont {G.}~\bibnamefont
  {Gompper}},\ }\bibfield  {title} {\enquote {\bibinfo {title} {Self-propelled
  worm-like filaments: spontaneous spiral formation, structure, and
  dynamics},}\ }\href {\doibase 10.1039/C5SM01683E} {\bibfield  {journal}
  {\bibinfo  {journal} {Soft Matter}\ }\textbf {\bibinfo {volume} {11}},\
  \bibinfo {pages} {7181} (\bibinfo {year} {2015})}\BibitemShut {NoStop}%
\bibitem [{\citenamefont {Duman}\ \emph {et~al.}(2018)\citenamefont {Duman},
  \citenamefont {Isele-Holder}, \citenamefont {Elgeti},\ and\ \citenamefont
  {Gompper}}]{duma:18}%
  \BibitemOpen
  \bibfield  {author} {\bibinfo {author} {\bibfnamefont {O.}~\bibnamefont
  {Duman}}, \bibinfo {author} {\bibfnamefont {R.~E.}\ \bibnamefont
  {Isele-Holder}}, \bibinfo {author} {\bibfnamefont {J.}~\bibnamefont
  {Elgeti}}, \ and\ \bibinfo {author} {\bibfnamefont {G.}~\bibnamefont
  {Gompper}},\ }\bibfield  {title} {\enquote {\bibinfo {title} {Collective
  dynamics of self-propelled semiflexible filaments},}\ }\href {\doibase
  10.1039/C8SM00282G} {\bibfield  {journal} {\bibinfo  {journal} {Soft Matter}\
  }\textbf {\bibinfo {volume} {14}},\ \bibinfo {pages} {4483} (\bibinfo {year}
  {2018})}\BibitemShut {NoStop}%
\bibitem [{\citenamefont {Prathyusha}, \citenamefont {Henkes},\ and\
  \citenamefont {Sknepnek}(2018)}]{prat:18}%
  \BibitemOpen
  \bibfield  {author} {\bibinfo {author} {\bibfnamefont {K.~R.}\ \bibnamefont
  {Prathyusha}}, \bibinfo {author} {\bibfnamefont {S.}~\bibnamefont {Henkes}},
  \ and\ \bibinfo {author} {\bibfnamefont {R.}~\bibnamefont {Sknepnek}},\
  }\bibfield  {title} {\enquote {\bibinfo {title} {Dynamically generated
  patterns in dense suspensions of active filaments},}\ }\href {\doibase
  10.1103/PhysRevE.97.022606} {\bibfield  {journal} {\bibinfo  {journal} {Phys.
  Rev. E}\ }\textbf {\bibinfo {volume} {97}},\ \bibinfo {pages} {022606}
  (\bibinfo {year} {2018})}\BibitemShut {NoStop}%
\bibitem [{\citenamefont {Anand}\ and\ \citenamefont {Singh}(2018)}]{anan:18}%
  \BibitemOpen
  \bibfield  {author} {\bibinfo {author} {\bibfnamefont {S.~K.}\ \bibnamefont
  {Anand}}\ and\ \bibinfo {author} {\bibfnamefont {S.~P.}\ \bibnamefont
  {Singh}},\ }\bibfield  {title} {\enquote {\bibinfo {title} {Structure and
  dynamics of a self-propelled semiflexible filament},}\ }\href {\doibase
  10.1103/PhysRevE.98.042501} {\bibfield  {journal} {\bibinfo  {journal} {Phys.
  Rev. E}\ }\textbf {\bibinfo {volume} {98}},\ \bibinfo {pages} {042501}
  (\bibinfo {year} {2018})}\BibitemShut {NoStop}%
\bibitem [{\citenamefont {Peterson}, \citenamefont {Hagan},\ and\ \citenamefont
  {Baskaran}(2020)}]{pete:20}%
  \BibitemOpen
  \bibfield  {author} {\bibinfo {author} {\bibfnamefont {M.~S.~E.}\
  \bibnamefont {Peterson}}, \bibinfo {author} {\bibfnamefont {M.~F.}\
  \bibnamefont {Hagan}}, \ and\ \bibinfo {author} {\bibfnamefont
  {A.}~\bibnamefont {Baskaran}},\ }\bibfield  {title} {\enquote {\bibinfo
  {title} {Statistical properties of a tangentially driven active filament},}\
  }\href {\doibase 10.1088/1742-5468/ab6097} {\bibfield  {journal} {\bibinfo
  {journal} {J. Stat. Mech. Theor. Exp.}\ }\textbf {\bibinfo {volume} {2020}},\
  \bibinfo {pages} {013216} (\bibinfo {year} {2020})}\BibitemShut {NoStop}%
\bibitem [{\citenamefont {Risken}(1989)}]{risk:89}%
  \BibitemOpen
  \bibfield  {author} {\bibinfo {author} {\bibfnamefont {H.}~\bibnamefont
  {Risken}},\ }\href@noop {} {\emph {\bibinfo {title} {The Fokker-Planck
  Equation}}}\ (\bibinfo  {publisher} {Springer},\ \bibinfo {address}
  {Berlin},\ \bibinfo {year} {1989})\BibitemShut {NoStop}%
\bibitem [{\citenamefont {Philipps}, \citenamefont {Gompper},\ and\
  \citenamefont {Winkler}(2020)}]{phil:20}%
  \BibitemOpen
  \bibfield  {author} {\bibinfo {author} {\bibfnamefont {C.}~\bibnamefont
  {Philipps}}, \bibinfo {author} {\bibfnamefont {G.}~\bibnamefont {Gompper}}, \
  and\ \bibinfo {author} {\bibfnamefont {R.~G.}\ \bibnamefont {Winkler}},\
  }\bibfield  {title} {\enquote {\bibinfo {title} {Analytical studies of polar
  active filaments},}\ }\href@noop {} {\bibfield  {journal} {\bibinfo
  {journal} {in progress}\ } (\bibinfo {year} {2020})}\BibitemShut {NoStop}%
\bibitem [{\citenamefont {Isele-Holder}\ \emph {et~al.}(2016)\citenamefont
  {Isele-Holder}, \citenamefont {Jager}, \citenamefont {Saggiorato},
  \citenamefont {Elgeti},\ and\ \citenamefont {Gompper}}]{isel:16}%
  \BibitemOpen
  \bibfield  {author} {\bibinfo {author} {\bibfnamefont {R.~E.}\ \bibnamefont
  {Isele-Holder}}, \bibinfo {author} {\bibfnamefont {J.}~\bibnamefont {Jager}},
  \bibinfo {author} {\bibfnamefont {G.}~\bibnamefont {Saggiorato}}, \bibinfo
  {author} {\bibfnamefont {J.}~\bibnamefont {Elgeti}}, \ and\ \bibinfo {author}
  {\bibfnamefont {G.}~\bibnamefont {Gompper}},\ }\bibfield  {title} {\enquote
  {\bibinfo {title} {Dynamics of self-propelled filaments pushing a load},}\
  }\href {\doibase 10.1039/C6SM01094F} {\bibfield  {journal} {\bibinfo
  {journal} {Soft Matter}\ }\textbf {\bibinfo {volume} {12}},\ \bibinfo {pages}
  {8495} (\bibinfo {year} {2016})}\BibitemShut {NoStop}%
\bibitem [{\citenamefont {Anand}\ and\ \citenamefont {Singh}(2019)}]{anan:19}%
  \BibitemOpen
  \bibfield  {author} {\bibinfo {author} {\bibfnamefont {S.~K.}\ \bibnamefont
  {Anand}}\ and\ \bibinfo {author} {\bibfnamefont {S.~P.}\ \bibnamefont
  {Singh}},\ }\bibfield  {title} {\enquote {\bibinfo {title} {Behavior of
  active filaments near solid-boundary under linear shear flow},}\ }\href
  {\doibase 10.1039/C9SM00027E} {\bibfield  {journal} {\bibinfo  {journal}
  {Soft Matter}\ }\textbf {\bibinfo {volume} {15}},\ \bibinfo {pages} {4008}
  (\bibinfo {year} {2019})}\BibitemShut {NoStop}%
\bibitem [{\citenamefont {Jiang}\ and\ \citenamefont
  {Hou}(2014{\natexlab{b}})}]{jian:14.1}%
  \BibitemOpen
  \bibfield  {author} {\bibinfo {author} {\bibfnamefont {H.}~\bibnamefont
  {Jiang}}\ and\ \bibinfo {author} {\bibfnamefont {Z.}~\bibnamefont {Hou}},\
  }\bibfield  {title} {\enquote {\bibinfo {title} {Hydrodynamic interaction
  induced spontaneous rotation of coupled active filaments},}\ }\href {\doibase
  10.1039/C4SM01734J} {\bibfield  {journal} {\bibinfo  {journal} {Soft Matter}\
  }\textbf {\bibinfo {volume} {10}},\ \bibinfo {pages} {9248} (\bibinfo {year}
  {2014}{\natexlab{b}})}\BibitemShut {NoStop}%
\bibitem [{\citenamefont {Clopes~Llahi}, \citenamefont {Gompper},\ and\
  \citenamefont {Winkler}(2020)}]{clop:20}%
  \BibitemOpen
  \bibfield  {author} {\bibinfo {author} {\bibfnamefont {J.}~\bibnamefont
  {Clopes~Llahi}}, \bibinfo {author} {\bibfnamefont {G.}~\bibnamefont
  {Gompper}}, \ and\ \bibinfo {author} {\bibfnamefont {R.~G.}\ \bibnamefont
  {Winkler}},\ }\bibfield  {title} {\enquote {\bibinfo {title} {Swimming
  behavior of squirmer dumbbelles},}\ }\href@noop {} {\bibfield  {journal}
  {\bibinfo  {journal} {in preparation}\ } (\bibinfo {year}
  {2020})}\BibitemShut {NoStop}%
\bibitem [{\citenamefont {Ishikawa}(2019)}]{ishi:19}%
  \BibitemOpen
  \bibfield  {author} {\bibinfo {author} {\bibfnamefont {T.}~\bibnamefont
  {Ishikawa}},\ }\bibfield  {title} {\enquote {\bibinfo {title} {Stability of a
  dumbbell micro-swimmer},}\ }\href {\doibase 10.3390/mi10010033} {\bibfield
  {journal} {\bibinfo  {journal} {Micromachines}\ }\textbf {\bibinfo {volume}
  {10}},\ \bibinfo {pages} {33} (\bibinfo {year} {2019})}\BibitemShut {NoStop}%
\bibitem [{\citenamefont {Matas~Navarro}\ and\ \citenamefont
  {Pagonabarraga}(2010)}]{nava:10}%
  \BibitemOpen
  \bibfield  {author} {\bibinfo {author} {\bibfnamefont {R.}~\bibnamefont
  {Matas~Navarro}}\ and\ \bibinfo {author} {\bibfnamefont {I.}~\bibnamefont
  {Pagonabarraga}},\ }\bibfield  {title} {\enquote {\bibinfo {title}
  {Hydrodynamic interaction between two trapped swimming model
  micro-organisms},}\ }\href {\doibase 10.1140/epje/i2010-10654-7} {\bibfield
  {journal} {\bibinfo  {journal} {Eur. Phys. J. E}\ }\textbf {\bibinfo {volume}
  {33}},\ \bibinfo {pages} {27} (\bibinfo {year} {2010})}\BibitemShut {NoStop}%
\bibitem [{\citenamefont {Kirkwood}\ and\ \citenamefont
  {Riseman}(1948)}]{kirk:48}%
  \BibitemOpen
  \bibfield  {author} {\bibinfo {author} {\bibfnamefont {J.~G.}\ \bibnamefont
  {Kirkwood}}\ and\ \bibinfo {author} {\bibfnamefont {J.}~\bibnamefont
  {Riseman}},\ }\bibfield  {title} {\enquote {\bibinfo {title} {The intrinsic
  viscosities and diffusion constants of flexible macromolecules in
  solution},}\ }\href {\doibase 10.1063/1.1746947} {\bibfield  {journal}
  {\bibinfo  {journal} {J. Chem. Phys.}\ }\textbf {\bibinfo {volume} {16}},\
  \bibinfo {pages} {565} (\bibinfo {year} {1948})}\BibitemShut {NoStop}%
\bibitem [{\citenamefont {Saintillan}, \citenamefont {Shelley},\ and\
  \citenamefont {Zidovska}(2018)}]{saint:18}%
  \BibitemOpen
  \bibfield  {author} {\bibinfo {author} {\bibfnamefont {D.}~\bibnamefont
  {Saintillan}}, \bibinfo {author} {\bibfnamefont {M.~J.}\ \bibnamefont
  {Shelley}}, \ and\ \bibinfo {author} {\bibfnamefont {A.}~\bibnamefont
  {Zidovska}},\ }\bibfield  {title} {\enquote {\bibinfo {title} {Extensile
  motor activity drives coherent motions in a model of interphase chromatin},}\
  }\href {\doibase 10.1073/pnas.1807073115} {\bibfield  {journal} {\bibinfo
  {journal} {Proc. Natl. Acad. Sci. USA}\ }\textbf {\bibinfo {volume} {115}},\
  \bibinfo {pages} {11442} (\bibinfo {year} {2018})}\BibitemShut {NoStop}%
\bibitem [{\citenamefont {Wysocki}, \citenamefont {Winkler},\ and\
  \citenamefont {Gompper}(2014)}]{wyso:14}%
  \BibitemOpen
  \bibfield  {author} {\bibinfo {author} {\bibfnamefont {A.}~\bibnamefont
  {Wysocki}}, \bibinfo {author} {\bibfnamefont {R.~G.}\ \bibnamefont
  {Winkler}}, \ and\ \bibinfo {author} {\bibfnamefont {G.}~\bibnamefont
  {Gompper}},\ }\bibfield  {title} {\enquote {\bibinfo {title} {Cooperative
  motion of active {B}rownian spheres in three-dimensional dense
  suspensions},}\ }\href {\doibase 10.1209/0295-5075/105/48004} {\bibfield
  {journal} {\bibinfo  {journal} {EPL}\ }\textbf {\bibinfo {volume} {105}},\
  \bibinfo {pages} {48004} (\bibinfo {year} {2014})}\BibitemShut {NoStop}%
\bibitem [{\citenamefont {Stenhammar}\ \emph {et~al.}(2014)\citenamefont
  {Stenhammar}, \citenamefont {Marenduzzo}, \citenamefont {Allen},\ and\
  \citenamefont {Cates}}]{sten:14}%
  \BibitemOpen
  \bibfield  {author} {\bibinfo {author} {\bibfnamefont {J.}~\bibnamefont
  {Stenhammar}}, \bibinfo {author} {\bibfnamefont {D.}~\bibnamefont
  {Marenduzzo}}, \bibinfo {author} {\bibfnamefont {R.~J.}\ \bibnamefont
  {Allen}}, \ and\ \bibinfo {author} {\bibfnamefont {M.~E.}\ \bibnamefont
  {Cates}},\ }\bibfield  {title} {\enquote {\bibinfo {title} {Phase behaviour
  of active {Brownian} particles: the role of dimensionality},}\ }\href
  {\doibase 10.1039/C3SM52813H} {\bibfield  {journal} {\bibinfo  {journal}
  {Soft Matter}\ }\textbf {\bibinfo {volume} {10}},\ \bibinfo {pages} {1489}
  (\bibinfo {year} {2014})}\BibitemShut {NoStop}%
\bibitem [{\citenamefont {Wysocki}, \citenamefont {Winkler},\ and\
  \citenamefont {Gompper}(2016)}]{wyso:16}%
  \BibitemOpen
  \bibfield  {author} {\bibinfo {author} {\bibfnamefont {A.}~\bibnamefont
  {Wysocki}}, \bibinfo {author} {\bibfnamefont {R.~G.}\ \bibnamefont
  {Winkler}}, \ and\ \bibinfo {author} {\bibfnamefont {G.}~\bibnamefont
  {Gompper}},\ }\bibfield  {title} {\enquote {\bibinfo {title} {Propagating
  interfaces in mixtures of active and passive {B}rownian particles},}\ }\href
  {\doibase 10.1088/1367-2630/aa529d} {\bibfield  {journal} {\bibinfo
  {journal} {New J. Phys.}\ }\textbf {\bibinfo {volume} {18}},\ \bibinfo
  {pages} {123030} (\bibinfo {year} {2016})}\BibitemShut {NoStop}%
\bibitem [{\citenamefont {Cates}\ and\ \citenamefont
  {Tailleur}(2015)}]{cate:15}%
  \BibitemOpen
  \bibfield  {author} {\bibinfo {author} {\bibfnamefont {M.~E.}\ \bibnamefont
  {Cates}}\ and\ \bibinfo {author} {\bibfnamefont {J.}~\bibnamefont
  {Tailleur}},\ }\bibfield  {title} {\enquote {\bibinfo {title}
  {Motility-induced phase separation},}\ }\href {\doibase
  10.1146/annurev-conmatphys-031214-014710} {\bibfield  {journal} {\bibinfo
  {journal} {Annu. Rev. Condens. Matter Phys.}\ }\textbf {\bibinfo {volume}
  {6}},\ \bibinfo {pages} {219} (\bibinfo {year} {2015})}\BibitemShut {NoStop}%
\bibitem [{\citenamefont {Digregorio}\ \emph {et~al.}(2018)\citenamefont
  {Digregorio}, \citenamefont {Levis}, \citenamefont {Suma}, \citenamefont
  {Cugliandolo}, \citenamefont {Gonnella},\ and\ \citenamefont
  {Pagonabarraga}}]{digr:18}%
  \BibitemOpen
  \bibfield  {author} {\bibinfo {author} {\bibfnamefont {P.}~\bibnamefont
  {Digregorio}}, \bibinfo {author} {\bibfnamefont {D.}~\bibnamefont {Levis}},
  \bibinfo {author} {\bibfnamefont {A.}~\bibnamefont {Suma}}, \bibinfo {author}
  {\bibfnamefont {L.~F.}\ \bibnamefont {Cugliandolo}}, \bibinfo {author}
  {\bibfnamefont {G.}~\bibnamefont {Gonnella}}, \ and\ \bibinfo {author}
  {\bibfnamefont {I.}~\bibnamefont {Pagonabarraga}},\ }\bibfield  {title}
  {\enquote {\bibinfo {title} {Full phase diagram of active {B}rownian disks:
  From melting to motility-induced phase separation},}\ }\href {\doibase
  10.1103/PhysRevLett.121.098003} {\bibfield  {journal} {\bibinfo  {journal}
  {Phys. Rev. Lett.}\ }\textbf {\bibinfo {volume} {121}},\ \bibinfo {pages}
  {098003} (\bibinfo {year} {2018})}\BibitemShut {NoStop}%
\bibitem [{\citenamefont {Saintillan}\ and\ \citenamefont
  {Shelley}(2008)}]{sain:08}%
  \BibitemOpen
  \bibfield  {author} {\bibinfo {author} {\bibfnamefont {D.}~\bibnamefont
  {Saintillan}}\ and\ \bibinfo {author} {\bibfnamefont {M.~J.}\ \bibnamefont
  {Shelley}},\ }\bibfield  {title} {\enquote {\bibinfo {title} {Instabilities
  and pattern formation in active particle suspensions: {K}inetic theory and
  continuum simulations},}\ }\href {\doibase 10.1103/PhysRevLett.100.178103}
  {\bibfield  {journal} {\bibinfo  {journal} {Phys. Rev. Lett}\ }\textbf
  {\bibinfo {volume} {100}},\ \bibinfo {pages} {178103} (\bibinfo {year}
  {2008})}\BibitemShut {NoStop}%
\bibitem [{\citenamefont {Vliegenthart}\ \emph {et~al.}(2020)\citenamefont
  {Vliegenthart}, \citenamefont {Ravichandran}, \citenamefont {Ripoll},
  \citenamefont {Auth},\ and\ \citenamefont {Gompper}}]{vlie:19}%
  \BibitemOpen
  \bibfield  {author} {\bibinfo {author} {\bibfnamefont {G.}~\bibnamefont
  {Vliegenthart}}, \bibinfo {author} {\bibfnamefont {A.}~\bibnamefont
  {Ravichandran}}, \bibinfo {author} {\bibfnamefont {M.}~\bibnamefont
  {Ripoll}}, \bibinfo {author} {\bibfnamefont {T.}~\bibnamefont {Auth}}, \ and\
  \bibinfo {author} {\bibfnamefont {G.}~\bibnamefont {Gompper}},\ }\bibfield
  {title} {\enquote {\bibinfo {title} {Filamentous active matter: Band
  formation, bending, buckling, and defects},}\ }\href@noop {} {\bibfield
  {journal} {\bibinfo  {journal} {Sci. Adv.}\ ,\ \bibinfo {pages} {in press}}
  (\bibinfo {year} {2020})},\ \Eprint {http://arxiv.org/abs/1902.07904}
  {arXiv:1902.07904 [cond-mat.soft]} \BibitemShut {NoStop}%
\bibitem [{\citenamefont {Copeland}\ and\ \citenamefont
  {Weibel}(2009)}]{cope:09}%
  \BibitemOpen
  \bibfield  {author} {\bibinfo {author} {\bibfnamefont {M.~F.}\ \bibnamefont
  {Copeland}}\ and\ \bibinfo {author} {\bibfnamefont {D.~B.}\ \bibnamefont
  {Weibel}},\ }\bibfield  {title} {\enquote {\bibinfo {title} {Bacterial
  swarming: a model system for studying dynamic self-assembly},}\ }\href
  {\doibase 10.1039/B812146J} {\bibfield  {journal} {\bibinfo  {journal} {Soft
  Matter}\ }\textbf {\bibinfo {volume} {5}},\ \bibinfo {pages} {1174} (\bibinfo
  {year} {2009})}\BibitemShut {NoStop}%
\bibitem [{\citenamefont {Kearns}(2010)}]{kear:10}%
  \BibitemOpen
  \bibfield  {author} {\bibinfo {author} {\bibfnamefont {D.~B.}\ \bibnamefont
  {Kearns}},\ }\bibfield  {title} {\enquote {\bibinfo {title} {A field guide to
  bacterial swarming motility},}\ }\href {\doibase 10.1038/nrmicro2405}
  {\bibfield  {journal} {\bibinfo  {journal} {Nat. Rev. Microbiol.}\ }\textbf
  {\bibinfo {volume} {8}},\ \bibinfo {pages} {634} (\bibinfo {year}
  {2010})}\BibitemShut {NoStop}%
\bibitem [{\citenamefont {Cates}\ and\ \citenamefont
  {MacKintosh}(2011)}]{cate:11}%
  \BibitemOpen
  \bibfield  {author} {\bibinfo {author} {\bibfnamefont {M.~E.}\ \bibnamefont
  {Cates}}\ and\ \bibinfo {author} {\bibfnamefont {F.~C.}\ \bibnamefont
  {MacKintosh}},\ }\bibfield  {title} {\enquote {\bibinfo {title} {Active soft
  matter},}\ }\href {\doibase 10.1039/C1SM90014E} {\bibfield  {journal}
  {\bibinfo  {journal} {Soft Matter}\ }\textbf {\bibinfo {volume} {7}},\
  \bibinfo {pages} {3050} (\bibinfo {year} {2011})}\BibitemShut {NoStop}%
\bibitem [{\citenamefont {Duclos}\ \emph {et~al.}(2020)\citenamefont {Duclos},
  \citenamefont {Adkins}, \citenamefont {Banerjee}, \citenamefont {Peterson},
  \citenamefont {Varghese}, \citenamefont {Kolvin}, \citenamefont {Baskaran},
  \citenamefont {Pelcovits}, \citenamefont {Powers}, \citenamefont {Baskaran},
  \citenamefont {Toschi}, \citenamefont {Hagan}, \citenamefont {Streichan},
  \citenamefont {Vitelli}, \citenamefont {Beller},\ and\ \citenamefont
  {Dogic}}]{ducl:20}%
  \BibitemOpen
  \bibfield  {author} {\bibinfo {author} {\bibfnamefont {G.}~\bibnamefont
  {Duclos}}, \bibinfo {author} {\bibfnamefont {R.}~\bibnamefont {Adkins}},
  \bibinfo {author} {\bibfnamefont {D.}~\bibnamefont {Banerjee}}, \bibinfo
  {author} {\bibfnamefont {M.~S.~E.}\ \bibnamefont {Peterson}}, \bibinfo
  {author} {\bibfnamefont {M.}~\bibnamefont {Varghese}}, \bibinfo {author}
  {\bibfnamefont {I.}~\bibnamefont {Kolvin}}, \bibinfo {author} {\bibfnamefont
  {A.}~\bibnamefont {Baskaran}}, \bibinfo {author} {\bibfnamefont {R.~A.}\
  \bibnamefont {Pelcovits}}, \bibinfo {author} {\bibfnamefont {T.~R.}\
  \bibnamefont {Powers}}, \bibinfo {author} {\bibfnamefont {A.}~\bibnamefont
  {Baskaran}}, \bibinfo {author} {\bibfnamefont {F.}~\bibnamefont {Toschi}},
  \bibinfo {author} {\bibfnamefont {M.~F.}\ \bibnamefont {Hagan}}, \bibinfo
  {author} {\bibfnamefont {S.~J.}\ \bibnamefont {Streichan}}, \bibinfo {author}
  {\bibfnamefont {V.}~\bibnamefont {Vitelli}}, \bibinfo {author} {\bibfnamefont
  {D.~A.}\ \bibnamefont {Beller}}, \ and\ \bibinfo {author} {\bibfnamefont
  {Z.}~\bibnamefont {Dogic}},\ }\bibfield  {title} {\enquote {\bibinfo {title}
  {Topological structure and dynamics of three-dimensional active nematics},}\
  }\href {\doibase 10.1126/science.aaz4547} {\bibfield  {journal} {\bibinfo
  {journal} {Science}\ }\textbf {\bibinfo {volume} {367}},\ \bibinfo {pages}
  {1120} (\bibinfo {year} {2020})}\BibitemShut {NoStop}%
\bibitem [{\citenamefont {Das}\ and\ \citenamefont
  {Cacciuto}(2019)}]{das:19.2}%
  \BibitemOpen
  \bibfield  {author} {\bibinfo {author} {\bibfnamefont {S.}~\bibnamefont
  {Das}}\ and\ \bibinfo {author} {\bibfnamefont {A.}~\bibnamefont {Cacciuto}},\
  }\bibfield  {title} {\enquote {\bibinfo {title} {Deviations from blob scaling
  theory for active {B}rownian filaments confined within cavities},}\ }\href
  {\doibase 10.1103/PhysRevLett.123.087802} {\bibfield  {journal} {\bibinfo
  {journal} {Phys. Rev. Lett.}\ }\textbf {\bibinfo {volume} {123}},\ \bibinfo
  {pages} {087802} (\bibinfo {year} {2019})}\BibitemShut {NoStop}%
\bibitem [{\citenamefont {Stenhammar}\ \emph {et~al.}(2015)\citenamefont
  {Stenhammar}, \citenamefont {Wittkowski}, \citenamefont {Marenduzzo},\ and\
  \citenamefont {Cates}}]{sten:15}%
  \BibitemOpen
  \bibfield  {author} {\bibinfo {author} {\bibfnamefont {J.}~\bibnamefont
  {Stenhammar}}, \bibinfo {author} {\bibfnamefont {R.}~\bibnamefont
  {Wittkowski}}, \bibinfo {author} {\bibfnamefont {D.}~\bibnamefont
  {Marenduzzo}}, \ and\ \bibinfo {author} {\bibfnamefont {M.~E.}\ \bibnamefont
  {Cates}},\ }\bibfield  {title} {\enquote {\bibinfo {title} {Activity-induced
  phase separation and self-assembly in mixtures of active and passive
  particles},}\ }\href {\doibase 10.1103/PhysRevLett.114.018301} {\bibfield
  {journal} {\bibinfo  {journal} {Phys. Rev. Lett.}\ }\textbf {\bibinfo
  {volume} {114}},\ \bibinfo {pages} {018301} (\bibinfo {year}
  {2015})}\BibitemShut {NoStop}%
\bibitem [{\citenamefont {Rogel~Rodriguez}\ \emph {et~al.}(2020)\citenamefont
  {Rogel~Rodriguez}, \citenamefont {Alarc{\'o}n}, \citenamefont {Martinez},
  \citenamefont {Ram{\'\i}rez},\ and\ \citenamefont {Valeriani}}]{roge:20}%
  \BibitemOpen
  \bibfield  {author} {\bibinfo {author} {\bibfnamefont {D.}~\bibnamefont
  {Rogel~Rodriguez}}, \bibinfo {author} {\bibfnamefont {F.}~\bibnamefont
  {Alarc{\'o}n}}, \bibinfo {author} {\bibfnamefont {R.}~\bibnamefont
  {Martinez}}, \bibinfo {author} {\bibfnamefont {J.}~\bibnamefont
  {Ram{\'\i}rez}}, \ and\ \bibinfo {author} {\bibfnamefont {C.}~\bibnamefont
  {Valeriani}},\ }\bibfield  {title} {\enquote {\bibinfo {title} {Phase
  behaviour and dynamical features of a two-dimensional binary mixture of
  active/passive spherical particles},}\ }\href {\doibase 10.1039/C9SM01803D}
  {\bibfield  {journal} {\bibinfo  {journal} {Soft Matter}\ }\textbf {\bibinfo
  {volume} {16}},\ \bibinfo {pages} {1162} (\bibinfo {year}
  {2020})}\BibitemShut {NoStop}%
\bibitem [{\citenamefont {Nikola}\ \emph {et~al.}(2016)\citenamefont {Nikola},
  \citenamefont {Solon}, \citenamefont {Kafri}, \citenamefont {Kardar},
  \citenamefont {Tailleur},\ and\ \citenamefont {Voituriez}}]{niko:16}%
  \BibitemOpen
  \bibfield  {author} {\bibinfo {author} {\bibfnamefont {N.}~\bibnamefont
  {Nikola}}, \bibinfo {author} {\bibfnamefont {A.~P.}\ \bibnamefont {Solon}},
  \bibinfo {author} {\bibfnamefont {Y.}~\bibnamefont {Kafri}}, \bibinfo
  {author} {\bibfnamefont {M.}~\bibnamefont {Kardar}}, \bibinfo {author}
  {\bibfnamefont {J.}~\bibnamefont {Tailleur}}, \ and\ \bibinfo {author}
  {\bibfnamefont {R.}~\bibnamefont {Voituriez}},\ }\bibfield  {title} {\enquote
  {\bibinfo {title} {Active particles with soft and curved walls: Equation of
  state, ratchets, and instabilities},}\ }\href {\doibase
  10.1103/PhysRevLett.117.098001} {\bibfield  {journal} {\bibinfo  {journal}
  {Phys. Rev. Lett.}\ }\textbf {\bibinfo {volume} {117}},\ \bibinfo {pages}
  {098001} (\bibinfo {year} {2016})}\BibitemShut {NoStop}%
\bibitem [{\citenamefont {Xia}\ \emph {et~al.}(2019{\natexlab{a}})\citenamefont
  {Xia}, \citenamefont {Shen}, \citenamefont {Tian},\ and\ \citenamefont
  {Chen}}]{xia:19}%
  \BibitemOpen
  \bibfield  {author} {\bibinfo {author} {\bibfnamefont {Y.-q.}\ \bibnamefont
  {Xia}}, \bibinfo {author} {\bibfnamefont {Z.-l.}\ \bibnamefont {Shen}},
  \bibinfo {author} {\bibfnamefont {W.-d.}\ \bibnamefont {Tian}}, \ and\
  \bibinfo {author} {\bibfnamefont {K.}~\bibnamefont {Chen}},\ }\bibfield
  {title} {\enquote {\bibinfo {title} {Unfolding of a diblock chain and its
  anomalous diffusion induced by active particles},}\ }\href {\doibase
  10.1063/1.5095850} {\bibfield  {journal} {\bibinfo  {journal} {J. Chem.
  Phys.}\ }\textbf {\bibinfo {volume} {150}},\ \bibinfo {pages} {154903}
  (\bibinfo {year} {2019}{\natexlab{a}})}\BibitemShut {NoStop}%
\bibitem [{\citenamefont {Xia}\ \emph {et~al.}(2019{\natexlab{b}})\citenamefont
  {Xia}, \citenamefont {Tian}, \citenamefont {Chen},\ and\ \citenamefont
  {Ma}}]{xia:19.1}%
  \BibitemOpen
  \bibfield  {author} {\bibinfo {author} {\bibfnamefont {Y.-Q.}\ \bibnamefont
  {Xia}}, \bibinfo {author} {\bibfnamefont {W.-D.}\ \bibnamefont {Tian}},
  \bibinfo {author} {\bibfnamefont {K.}~\bibnamefont {Chen}}, \ and\ \bibinfo
  {author} {\bibfnamefont {Y.-Q.}\ \bibnamefont {Ma}},\ }\bibfield  {title}
  {\enquote {\bibinfo {title} {Globule--stretch transition of a self-attracting
  chain in the repulsive active particle bath},}\ }\href {\doibase
  10.1039/C8CP05976D} {\bibfield  {journal} {\bibinfo  {journal} {Phys. Chem.
  Chem. Phys.}\ }\textbf {\bibinfo {volume} {21}},\ \bibinfo {pages} {4487}
  (\bibinfo {year} {2019}{\natexlab{b}})}\BibitemShut {NoStop}%
\bibitem [{\citenamefont {Fily}, \citenamefont {Baskaran},\ and\ \citenamefont
  {Hagan}(2014)}]{fily:14}%
  \BibitemOpen
  \bibfield  {author} {\bibinfo {author} {\bibfnamefont {Y.}~\bibnamefont
  {Fily}}, \bibinfo {author} {\bibfnamefont {A.}~\bibnamefont {Baskaran}}, \
  and\ \bibinfo {author} {\bibfnamefont {M.~F.}\ \bibnamefont {Hagan}},\
  }\bibfield  {title} {\enquote {\bibinfo {title} {Dynamics of self-propelled
  particles under strong confinement},}\ }\href {\doibase 10.1039/C4SM00975D}
  {\bibfield  {journal} {\bibinfo  {journal} {Soft Matter}\ }\textbf {\bibinfo
  {volume} {10}},\ \bibinfo {pages} {5609} (\bibinfo {year}
  {2014})}\BibitemShut {NoStop}%
\bibitem [{\citenamefont {Grosberg}\ and\ \citenamefont
  {Joanny}(2015)}]{gros:15}%
  \BibitemOpen
  \bibfield  {author} {\bibinfo {author} {\bibfnamefont {A.~Y.}\ \bibnamefont
  {Grosberg}}\ and\ \bibinfo {author} {\bibfnamefont {J.~F.}\ \bibnamefont
  {Joanny}},\ }\bibfield  {title} {\enquote {\bibinfo {title} {Nonequilibrium
  statistical mechanics of mixtures of particles in contact with different
  thermostats},}\ }\href {\doibase 10.1103/PhysRevE.92.032118} {\bibfield
  {journal} {\bibinfo  {journal} {Phys. Rev. E}\ }\textbf {\bibinfo {volume}
  {92}},\ \bibinfo {pages} {032118} (\bibinfo {year} {2015})}\BibitemShut
  {NoStop}%
\bibitem [{\citenamefont {Brangwynne}, \citenamefont {Mitchison},\ and\
  \citenamefont {Hyman}(2011)}]{bran:11}%
  \BibitemOpen
  \bibfield  {author} {\bibinfo {author} {\bibfnamefont {C.~P.}\ \bibnamefont
  {Brangwynne}}, \bibinfo {author} {\bibfnamefont {T.~J.}\ \bibnamefont
  {Mitchison}}, \ and\ \bibinfo {author} {\bibfnamefont {A.~A.}\ \bibnamefont
  {Hyman}},\ }\bibfield  {title} {\enquote {\bibinfo {title} {Active
  liquid-like behavior of nucleoli determines their size and shape in {\em
  xenopus laevis} oocytes},}\ }\href {\doibase 10.1073/pnas.1017150108}
  {\bibfield  {journal} {\bibinfo  {journal} {Proc. Natl. Acad. Sci. USA}\
  }\textbf {\bibinfo {volume} {108}},\ \bibinfo {pages} {4334} (\bibinfo {year}
  {2011})}\BibitemShut {NoStop}%
\bibitem [{\citenamefont {Falahati}\ and\ \citenamefont
  {Wieschaus}(2017)}]{fala:17}%
  \BibitemOpen
  \bibfield  {author} {\bibinfo {author} {\bibfnamefont {H.}~\bibnamefont
  {Falahati}}\ and\ \bibinfo {author} {\bibfnamefont {E.}~\bibnamefont
  {Wieschaus}},\ }\bibfield  {title} {\enquote {\bibinfo {title} {Independent
  active and thermodynamic processes govern the nucleolus assembly in vivo},}\
  }\href {\doibase 10.1073/pnas.1615395114} {\bibfield  {journal} {\bibinfo
  {journal} {Proc. Natl. Acad. Sci. USA}\ }\textbf {\bibinfo {volume} {114}},\
  \bibinfo {pages} {1335} (\bibinfo {year} {2017})}\BibitemShut {NoStop}%
\bibitem [{\citenamefont {Babinchak}\ and\ \citenamefont
  {Surewicz}(2020)}]{babi:20}%
  \BibitemOpen
  \bibfield  {author} {\bibinfo {author} {\bibfnamefont {W.~M.}\ \bibnamefont
  {Babinchak}}\ and\ \bibinfo {author} {\bibfnamefont {W.~K.}\ \bibnamefont
  {Surewicz}},\ }\bibfield  {title} {\enquote {\bibinfo {title} {Liquid--liquid
  phase separation and its mechanistic role in pathological protein
  aggregation},}\ }\href {\doibase 10.1016/j.jmb.2020.03.004} {\bibfield
  {journal} {\bibinfo  {journal} {J. Mol. Biol.}\ } (\bibinfo {year} {2020}),\
  10.1016/j.jmb.2020.03.004}\BibitemShut {NoStop}%
\bibitem [{\citenamefont {Bates}\ and\ \citenamefont
  {Fredrickson}(1990)}]{bate:90}%
  \BibitemOpen
  \bibfield  {author} {\bibinfo {author} {\bibfnamefont {F.~S.}\ \bibnamefont
  {Bates}}\ and\ \bibinfo {author} {\bibfnamefont {G.~H.}\ \bibnamefont
  {Fredrickson}},\ }\bibfield  {title} {\enquote {\bibinfo {title} {Block
  copolymer thermodynamics: Theory and experiment},}\ }\href {\doibase
  10.1146/annurev.pc.41.100190.002521} {\bibfield  {journal} {\bibinfo
  {journal} {Annu. Rev. Phys. Chem.}\ }\textbf {\bibinfo {volume} {41}},\
  \bibinfo {pages} {525} (\bibinfo {year} {1990})}\BibitemShut {NoStop}%
\bibitem [{\citenamefont {Mai}\ and\ \citenamefont {Eisenberg}(2012)}]{mai:12}%
  \BibitemOpen
  \bibfield  {author} {\bibinfo {author} {\bibfnamefont {Y.}~\bibnamefont
  {Mai}}\ and\ \bibinfo {author} {\bibfnamefont {A.}~\bibnamefont
  {Eisenberg}},\ }\bibfield  {title} {\enquote {\bibinfo {title} {Self-assembly
  of block copolymers},}\ }\href {\doibase 10.1039/C2CS35115C} {\bibfield
  {journal} {\bibinfo  {journal} {Chem. Soc. Rev.}\ }\textbf {\bibinfo {volume}
  {41}},\ \bibinfo {pages} {5969} (\bibinfo {year} {2012})}\BibitemShut
  {NoStop}%
\bibitem [{\citenamefont {Cavagna}\ and\ \citenamefont
  {Giardina}(2014)}]{cava:14}%
  \BibitemOpen
  \bibfield  {author} {\bibinfo {author} {\bibfnamefont {A.}~\bibnamefont
  {Cavagna}}\ and\ \bibinfo {author} {\bibfnamefont {I.}~\bibnamefont
  {Giardina}},\ }\bibfield  {title} {\enquote {\bibinfo {title} {Bird flocks as
  condensed matter},}\ }\href {\doibase
  10.1146/annurev-conmatphys-031113-133834} {\bibfield  {journal} {\bibinfo
  {journal} {Annu. Rev. Condens. Matter Phys.}\ }\textbf {\bibinfo {volume}
  {5}},\ \bibinfo {pages} {183} (\bibinfo {year} {2014})}\BibitemShut {NoStop}%
\bibitem [{\citenamefont {Popkin}(2016)}]{popk:16}%
  \BibitemOpen
  \bibfield  {author} {\bibinfo {author} {\bibfnamefont {G.}~\bibnamefont
  {Popkin}},\ }\bibfield  {title} {\enquote {\bibinfo {title} {The physics of
  life},}\ }\href {\doibase 10.1038/529016a} {\bibfield  {journal} {\bibinfo
  {journal} {Nature}\ }\textbf {\bibinfo {volume} {529}},\ \bibinfo {pages}
  {16} (\bibinfo {year} {2016})}\BibitemShut {NoStop}%
\bibitem [{\citenamefont {Kolmogorov}\ \emph {et~al.}(1991)\citenamefont
  {Kolmogorov}, \citenamefont {Levin}, \citenamefont {Hunt}, \citenamefont
  {Phillips},\ and\ \citenamefont {Williams}}]{kolm:91}%
  \BibitemOpen
  \bibfield  {author} {\bibinfo {author} {\bibfnamefont {A.~N.}\ \bibnamefont
  {Kolmogorov}}, \bibinfo {author} {\bibfnamefont {V.}~\bibnamefont {Levin}},
  \bibinfo {author} {\bibfnamefont {J.~C.~R.}\ \bibnamefont {Hunt}}, \bibinfo
  {author} {\bibfnamefont {O.~M.}\ \bibnamefont {Phillips}}, \ and\ \bibinfo
  {author} {\bibfnamefont {D.}~\bibnamefont {Williams}},\ }\bibfield  {title}
  {\enquote {\bibinfo {title} {The local structure of turbulence in
  incompressible viscous fluid for very large {R}eynolds numbers},}\ }\href
  {\doibase 10.1098/rspa.1991.0075} {\bibfield  {journal} {\bibinfo  {journal}
  {Proc. R. Soc. A}\ }\textbf {\bibinfo {volume} {434}},\ \bibinfo {pages} {9}
  (\bibinfo {year} {1991})}\BibitemShut {NoStop}%
\bibitem [{\citenamefont {L{\'o}pez}\ \emph {et~al.}(2015)\citenamefont
  {L{\'o}pez}, \citenamefont {Gachelin}, \citenamefont {Douarche},
  \citenamefont {Auradou},\ and\ \citenamefont {Cl{\'e}ment}}]{lope:15}%
  \BibitemOpen
  \bibfield  {author} {\bibinfo {author} {\bibfnamefont {H.~M.}\ \bibnamefont
  {L{\'o}pez}}, \bibinfo {author} {\bibfnamefont {J.}~\bibnamefont {Gachelin}},
  \bibinfo {author} {\bibfnamefont {C.}~\bibnamefont {Douarche}}, \bibinfo
  {author} {\bibfnamefont {H.}~\bibnamefont {Auradou}}, \ and\ \bibinfo
  {author} {\bibfnamefont {E.}~\bibnamefont {Cl{\'e}ment}},\ }\bibfield
  {title} {\enquote {\bibinfo {title} {Turning bacteria suspensions into
  superfluids},}\ }\href {\doibase 10.1103/PhysRevLett.115.028301} {\bibfield
  {journal} {\bibinfo  {journal} {Phys. Rev. Lett}\ }\textbf {\bibinfo {volume}
  {115}},\ \bibinfo {pages} {028301} (\bibinfo {year} {2015})}\BibitemShut
  {NoStop}%
\bibitem [{\citenamefont {Martinez}\ \emph {et~al.}(2020)\citenamefont
  {Martinez}, \citenamefont {Cl{\'e}ment}, \citenamefont {Arlt}, \citenamefont
  {Douarche}, \citenamefont {Dawson}, \citenamefont {Schwarz-Linek},
  \citenamefont {Creppy}, \citenamefont {{\v S}kult{\'e}ty}, \citenamefont
  {Morozov}, \citenamefont {Auradou},\ and\ \citenamefont {Poon}}]{mart:20.1}%
  \BibitemOpen
  \bibfield  {author} {\bibinfo {author} {\bibfnamefont {V.~A.}\ \bibnamefont
  {Martinez}}, \bibinfo {author} {\bibfnamefont {E.}~\bibnamefont
  {Cl{\'e}ment}}, \bibinfo {author} {\bibfnamefont {J.}~\bibnamefont {Arlt}},
  \bibinfo {author} {\bibfnamefont {C.}~\bibnamefont {Douarche}}, \bibinfo
  {author} {\bibfnamefont {A.}~\bibnamefont {Dawson}}, \bibinfo {author}
  {\bibfnamefont {J.}~\bibnamefont {Schwarz-Linek}}, \bibinfo {author}
  {\bibfnamefont {A.~K.}\ \bibnamefont {Creppy}}, \bibinfo {author}
  {\bibfnamefont {V.}~\bibnamefont {{\v S}kult{\'e}ty}}, \bibinfo {author}
  {\bibfnamefont {A.~N.}\ \bibnamefont {Morozov}}, \bibinfo {author}
  {\bibfnamefont {H.}~\bibnamefont {Auradou}}, \ and\ \bibinfo {author}
  {\bibfnamefont {W.~C.~K.}\ \bibnamefont {Poon}},\ }\bibfield  {title}
  {\enquote {\bibinfo {title} {A combined rheometry and imaging study of
  viscosity reduction in bacterial suspensions},}\ }\href {\doibase
  10.1073/pnas.1912690117} {\bibfield  {journal} {\bibinfo  {journal} {Proc.
  Natl. Acad. Sci. USA}\ }\textbf {\bibinfo {volume} {117}},\ \bibinfo {pages}
  {2326} (\bibinfo {year} {2020})}\BibitemShut {NoStop}%
\bibitem [{\citenamefont {Mart{{\'\i}}n-G{{\'o}}mez}, \citenamefont {Gompper},\
  and\ \citenamefont {Winkler}(2018)}]{mart:18.1}%
  \BibitemOpen
  \bibfield  {author} {\bibinfo {author} {\bibfnamefont {A.}~\bibnamefont
  {Mart{{\'\i}}n-G{{\'o}}mez}}, \bibinfo {author} {\bibfnamefont
  {G.}~\bibnamefont {Gompper}}, \ and\ \bibinfo {author} {\bibfnamefont
  {R.~G.}\ \bibnamefont {Winkler}},\ }\bibfield  {title} {\enquote {\bibinfo
  {title} {Active {B}rownian filamentous polymers under shear flow},}\ }\href
  {\doibase 10.3390/polym10080837} {\bibfield  {journal} {\bibinfo  {journal}
  {Polymers}\ }\textbf {\bibinfo {volume} {10}},\ \bibinfo {pages} {837}
  (\bibinfo {year} {2018})}\BibitemShut {NoStop}%
\bibitem [{\citenamefont {Deblais}, \citenamefont {Woutersen},\ and\
  \citenamefont {Bonn}(2020)}]{bebl:20}%
  \BibitemOpen
  \bibfield  {author} {\bibinfo {author} {\bibfnamefont {A.}~\bibnamefont
  {Deblais}}, \bibinfo {author} {\bibfnamefont {S.}~\bibnamefont {Woutersen}},
  \ and\ \bibinfo {author} {\bibfnamefont {D.}~\bibnamefont {Bonn}},\
  }\bibfield  {title} {\enquote {\bibinfo {title} {Rheology of entangled active
  polymer-like t. tubifex worms},}\ }\href {\doibase
  10.1103/PhysRevLett.124.188002} {\bibfield  {journal} {\bibinfo  {journal}
  {Phys. Reev. Lett.}\ }\textbf {\bibinfo {volume} {124}},\ \bibinfo {pages}
  {188002} (\bibinfo {year} {2020})}\BibitemShut {NoStop}%
\end{thebibliography}
%

%

\end{document}